\documentclass[11pt]{book}

\usepackage[paperwidth=6.0in,paperheight=9.0in,left=0.5in,right=0.5in,top=.8in,bottom=0.75in]{geometry}

\usepackage{fancyhdr}
\usepackage{amsmath}
\usepackage{amsfonts}
\usepackage{amssymb}
\usepackage{amsthm}
\usepackage{latexsym}
\usepackage{listings}
\usepackage{makeidx}
\usepackage{graphicx}
\usepackage{pdfpages}
\usepackage{xcolor}

\makeindex

\theoremstyle{plain}
\theoremstyle{definition}
\newtheorem{definition}{Definition}[section]

\numberwithin{equation}{section}

\newcommand{\be}{\begin{equation*}}
\newcommand{\ee}{\end{equation*}}
\newcommand{\ben}{\begin{equation}}
\newcommand{\een}{\end{equation}}
\newcommand{\bal}{\begin{aligned}}
\newcommand{\eal}{\end{aligned}}
\newcommand{\ba}{\begin{eqnarray}}
\newcommand{\ea}{\end{eqnarray}}

\newcommand{\mb}{\mathbb}
\newcommand{\mbf}{\mathbf}

\newcommand{\lra}{\longrightarrow}
\newcommand{\lla}{\longleftarrow}

\begin{document}
	
\parindent0pt
\parskip10pt
\pagestyle{plain}

\definecolor{mycolor}{rgb}{0.8,0.8,1.0}
%
\pagestyle{empty}
\text{~}
\vspace{15mm}
\begin{center}
\rule{120mm}{.6mm} \\
\rule[3.7mm]{120mm}{.2mm} \\
\textbf{\Huge \textsf{Programs as the Language of Science}} \\
\rule{120mm}{.2mm} \\
\vspace{120mm}
\text{\textsl{\Large G. Pantelis}} \\
\end{center}

\newpage

\nopagecolor

\text{~}

\vspace{10mm}

Programs as the Language of Science. \\
G. Pantelis. \\

\vspace{5mm}

Copyright\copyright 2018~G.Pantelis \\

ISBN: 978-9925-566-34-1
 
\frontmatter

\parindent0pt
\parskip10pt
\pagestyle{plain}

\chapter*{Preface}

This book draws upon a number of converging ideas that have emerged over recent decades from pioneering researchers involved with the construction of computer models in a wide area of scientific applications.
These ideas challenge the dominant paradigm where a computer model is constructed as an attempt to provide a discrete approximation of some continuum theory.

As discussed in the first chapter, there is an argument that supports a departure from the current paradigm towards the construction of discrete models based on simple deterministic rules.
Although still limited in their use in the sciences, these models are producing results that show promise and cannot be easily dismissed.
But one can take this one step further and argue that such discrete models not only provide alternative tools for simulation but in themselves can be used as a new language that describe real world systems.

The question arises as to how can a solid foundation be built for validating such discrete models, both as a simulation tool as well as a language that describes the laws that govern the application at the most fundamental level.
It appears that these two aspects of the model are highly linked and rely heavily upon a single overriding property, namely that of computability.

Encouraged by current trends in theoretical physics where information is regarded to be fundamental we can define the intrinsic properties of physical objects and their dynamic state in terms of integer vectors.
The elements of integer vectors can only take on a finite number of assigned values and are often subject to a conservation law of information. 
Thus there is high degree of compatibility with the discrete model and the machine upon which it is to be executed.
It seems plausible then that the laws that govern the computability of any model based on integer state variables can be directly linked to the allowable computational operations that map configuration states of the machine itself.

Another important issue is that a computer model based on integer state variables may involve algorithms that are not readily expressible in the notation of conventional mathematics.
This suggests a new paradigm in which a real world system is best described by a language of algorithms and programs rather than a language based on any conventional mathematical representation.
If this is the case then one should attempt to construct a language that is both simple enough to be adopted by those whose background is not rooted in the computer sciences and yet be powerful enough to be employed as a tool of analysis at a sufficiently high level.  

This book is primarily aimed at students and researchers in the mathematical sciences who have little or no knowledge of subjects in the computer sciences, although some familiarity with programming will be helpful. 
Specific discrete models will not be discussed in great detail since the focus is directed towards the basic operations of finite state arithmetic on a real world classical computer.
A simple language based on programs will be constructed for the purposes of analysis. 
From such a study it is hoped will emerge the basic theoretical tools that will lay down the foundations for both the construction and rigorous validation of this class of computer models.

Chapter \ref{ci} outlines the motivation behind the material in this book.
Chapters \ref{cps}-\ref{appl} are largely dedicated to the construction of the formal language based on programs.
When constructing a new language the reader will be bombarded with many definitions before the language can actually be used in analysis.
This cannot be avoided and the reader will need to make some effort to acquire some understanding of these definitions and the motivations behind them.
Therefore the material contained in Chapters \ref{cps}-\ref{appl} require some patience and perseverance on behalf of the reader.

In Chapter \ref{int} a number of basic properties of arithmetic on a deterministic machine with finite memory will be derived.
The reader should be alerted to the subtleties in which the axioms are structured and how they differ from the axioms of commutative rings.
While the focus is on basic algebraic identities and inequalities, all proofs are provided for completion.  
The reader may wish to go through some of these proofs to get a feel of how the formal language works.
When the reader is satisfied that they have a sufficient understanding of the process the reader may then wish to skim through the remaining proofs of this chapter.
In Chapter \ref{vec} we begin with some basic properties of integer vector programs and then go on to examine the computability of fully discrete dynamical systems.

In Chapter \ref{cpcap} we examine some basic properties of the construction rules that were presented in Chapters \ref{cps}-\ref{appl}.
In Chapter \ref{cabstract} we take a more abstract approach by exploring some properties of our formal system that lead to completeness.
It includes a demonstration of how our formal system can be employed for applications requiring higher levels of abstraction.   
The book concludes with Chapter \ref{cfsis} that addresses some of the unresolved issues that surround formal systems and their role in the scientific method.

\parskip0pt

\tableofcontents

\mainmatter

\parindent0pt
\parskip10pt

\pagestyle{fancy}
\renewcommand{\chaptermark}[1]{\markboth{#1}{}}
\renewcommand{\sectionmark}[1]{\markright{\thesection\ #1}}
\fancyhf{} 
\fancyhead[LE,RO]{\bfseries\thepage}
\fancyhead[LO]{\bfseries\rightmark}
\fancyhead[RE]{\bfseries\leftmark}
\renewcommand{\headrulewidth}{0.5pt}
\renewcommand{\footrulewidth}{0pt}
\addtolength{\headheight}{0.5pt} 
\fancypagestyle{plain}{%
	\fancyhead{} 
	\renewcommand{\headrulewidth}{0pt} 
}

\chapter{Introduction.}\label{ci}

\section{Discrete versus Continuous Mathematics.} 

Debates surrounding continuous versus discrete mathematics arise in many subdisciplines of the mathematical sciences.
Where this debate is of paramount importance can be found in the general area of computer modeling of real world applications.
While we will not be discussing any specific computer model in great detail, it is appropriate that we at least start with some background to this topic since it largely represents the motivation behind much of the discourse presented throughout this book.

Computer modeling has become an important component of many scientific studies.
However, despite its widespread use in the sciences, computer modeling still has a reputation of taking on aspects of an art rather than an exact science.
The reasons for this reputation are somewhat historic.
While this field has provided many useful results and enhanced insights into a wide area of scientific research there still remain weaknesses in the validation of computer models at a theoretical level.

The difficulties largely originate from two, not entirely unrelated, sources.
The controversy surrounding the existence of the real numbers, $\mb{R}$, is a philosophical debate that has been around since antiquity.
More recently, the existence of infinite sets, that are the basis of many mathematical abstractions, is also one that is contested in the philosophical arena.
From the perspective of raw computations on a real world deterministic computer there is no philosophical issue.
A machine neither recognizes an infinitesimal nor infinite sets in general.

Pure mathematicians are motivated by structures that have an elegant representation in the platonic world of ideal forms and have very little interest in real world applications.
In this realm the construction of the reals along with other mathematical abstractions involving infinite sets cannot be readily dismissed.
While largely motivated by theoretical interest, the extensive products of the efforts of pure mathematicians over the years have nevertheless been found to be useful in the sciences.

It has been widely accepted that mathematics is the language of science but with this come compromises that are often overlooked.
Mathematical structures are abstract constructions.
In order to maintain some consistency, the physical system under investigation often needs to be idealized to suit the language that is employed to describe it.
As with the construction of the language upon which it is based, the constructions of models of the physical world under such idealizations can lead to theories that take on the appearance of elegance.
Unfortunately, the abstractions embedded in the descriptive language of contemporary mathematics can also lead to complications, as will now be discussed.

Following in the footsteps of many other scientists, computer modelers have often adopted the language of continuous mathematics without question.
The constraints of the finite resources of the computer and its inability to recognize the reals and infinite sets in general have led to the acceptance of the computer as a tool of approximation.
This in turn has led to the emergence of the discipline of numerical analysis that is largely dedicated to providing a rigorous foundation for approximation.

As an example, let us look at the wide application area of hydrodynamics.
In particular we want to focus on the traditional procedures for model validation.
We are concerned here with validation in the theoretical sense and this should not be confused with model validation that involves comparing simulation results with empirical data.

Hydrodynamics is largely based on conservation laws that are expressed in the form of second order partial differential equations (PDEs).
The most widely used hydrodynamic computer models involve some kind of discrete system of equations that is meant to represent an approximation of the continuum theory.
We can view this as a map that transforms a system of PDEs to a system of difference equations (DEs),
\ben\label{ctod}
\text{Continuum model (PDEs)} \longrightarrow \text{Discrete model (DEs)}
\een
Under this map the continuum model is regarded as the template that represents the exact description of the physical system being modeled.
The discrete model attempts to approximate the continuum model by employing some type of discretization scheme.
These include finite difference methods, spectral methods and other variants of discretization.
Associated with any discretization scheme is a characteristic spatial resolution, $\Delta x$, and in the case of time dependent problems a temporal resolution, $\Delta t$. 

The map (\ref{ctod}) associated with the continuum to discrete model is regarded as valid if all of the following conditions are met. 

\begin{itemize}
	
	\item The continuum model is well posed. The classical definition of a well posed system is based on the existence and uniqueness of a solution along with continuous dependence on the initial data.
	
	\item Stability of the solution of the discrete system of equations.
	
	\item The discrete system of equations converge to the continuum equations in the limit, $\Delta x \rightarrow 0$, $\Delta t \rightarrow 0$ (consistency).
	
	\item The solution of the discrete system converges to the solution of the continuum equations in the limit, $\Delta x \rightarrow 0$, $\Delta t \rightarrow 0$ (convergence).
	
\end{itemize}

There are many mathematical theories that have been developed to address these items.
To name a few are finite difference methods that are often based on Taylor expansions, weak or generalized formulations of PDEs and solution methods and stability analysis of DEs. 
Despite the extensive theoretical work on the subject, except for some special cases, a complete rigorous theoretical validation is rarely achievable.

The above conditions establish validity on theoretical grounds but do not entirely address the practical aspects of machine computation.
The exact solutions of the theoretical discrete system will not necessarily coincide with that of the discrete system implemented on the machine.
This is because we are forced to employ floating point arithmetic.
Thus round off errors introduce another complication that needs to be considered if a complete model validation is demanded.    

Matters become even worse when we find that often it is difficult to establish that the continuum model itself is well posed.
A case in point is the Navier-Stokes equation upon which fluid mechanics is based.
It is currently unknown whether there exist smooth unique solutions of the Navier-Stokes equation in three dimensional space that satisfy typical boundary and initial data. 
 
First, it is worthwhile examining the origins of the continuum model.
The most fundamental continuum theories are based on the Euler equations of an ideal fluid.
Here the laws of fluid flow are derived largely from the primary properties of a continuum.
In applications the Euler equations are replaced by equations that include terms associated with viscosity, heat transfer and other phenomena that may be deemed important for the particular problem being considered.
This class of semi-empirically based continuum equations of hydrodynamics are derived from a particle model through the application of Boltzmann equations that employ continuous distributions.
We can represent this as a map under the action of Boltzmann statistics 
\ben\label{dtoc}
\text{Particle model} \longrightarrow \text{Continuum model}
\een
Thus the continuum model itself is derived from a microscopic scale discrete system.
The combination of the maps (\ref{ctod}) and (\ref{dtoc}) is then a two step procedure of discrete to continuous back to discrete.
A striking feature of this procedure is that the properties of the two discrete systems are very different. 

In light of these observations one may ask whether it is not better to compare the computer model directly with the kinetic particle model.
Both are discrete systems and the existence of solutions of the formulation associated with the particle model is much more tractable.

Fortunately, this idea has, in some sense, been around since the advent of the first computers.
Cellular automata have been used to model many complex systems ranging from applications in the biological sciences to networks of information flow.
In more recent decades has emerged the use of cellular automata in the area of hydrodynamics.
The popularity of cellular automata in this application area has waxed and wained over the years but the results that they produce cannot be readily dismissed and are worthy of continued examination. 

Cellular automata in fluid dynamics are discrete rule based algorithms that attempt to mimic the particle model.
As such they can be directly translated into a computer program.
The fluid medium is discretized into a lattice and within each cell of the lattice there are only two possible states, $[0~1]$.
The dynamics of the system is governed by a collection of simple deterministic rules of cell pair interactions. 
The primary and defining feature of such systems is the conservation of information that reflects the physical law of conservation of mass.

For cellular automata to be effective the spatial domain needs to be discretized into a very large number of small cells.
But even with the smallest possible refinement on the most powerful computers, the characteristic size of the cells is still very much larger than the mean free path of the particles that it is meant to simulate.
This suggests that allowing only two states per cell may be inadequate. 

Issues of scale inconsistencies also arise in the formulation of continuum models.
In the early days of computational fluid dynamics it was found that the discrete models based on the Navier-Stokes equation did not perform well for large Reynolds numbers.
(Large Reynolds number flows are associated with the onset of hydrodynamic instabilities leading to turbulence.)
The earliest attempts to remedy this situation involved the introduction of a turbulence viscosity term that was identical in form to the molecular viscosity term.
While there were some improvements in simulation results, the introduction of the turbulence viscosity constant appeared to be inadequate to capture observed flows where a high degree of accuracy was required.
This led to the area of large eddy simulation models where the constant turbulence viscosity coefficient was replaced by a variable, usually based on a function of the deformation tensor (see for instance \cite{liu01}).
The so called Smagorinsky model \cite{smag} introduced in the 1960s remains the most popular model whose variants are still in use to this day.

The problem with fluid turbulence models based on large eddy simulation is their dependence on artificial parameters that need to be readjusted for each specific application.
The presence of these artificial parameters is an indication that the continuum model exhibits some type of inconsistency.
One can identify this inconsistency as arising from the fact that the continuum model is ill defined in the sense that it fails be scale invariant.

The important properties that would be demanded from a reformulation is that it be scale invariant and be independent of artificial parameters that require tuning to specific applications.
One way to do this is to accept that the dependent variables of the continuum equations are filtered variables that must not only be dependent on space and time but on a new independent variable associated with scale.
One introduces the following conditions that define consistency based on scale invariance \cite{pan99}-\cite{pan09}.\index{scale invariance}

\begin{itemize}
	
	\item The macroscopic formulation is described by a system of equations that represent conservation laws of the filtered variables and contain residual terms that capture all of the dissipative and dispersive effects that are associated with microscopic scale fluctuations.
	The macroscopic scale formulation must be form invariant with respect to scale.
	
	\item The dependent variables of the macroscopic formulation must also satisfy the filter equations that are expressed as second order partial differential equations rather than integrals.
	It is convenient to use the space-scale heat equation for this purpose because its solutions can be associated with a Gaussian type spatial filter.
	The filter equations provide a continuous relationship of the filtered variables with respect to scale.
	
	\item In the limit of increasing spatial resolution the residual terms vanish and the macroscopic equations collapse to the fully resolved continuum equations. 
	
\end{itemize}

The third item becomes problematic for applications such as in fluid dynamics where there remains the uncertainty of the existence of any meaningful solutions of the fully resolved continuum equations.
There have been attempts to address this by abandoning all together any notion of a fully resolved system (see for example \cite{pan09}-\cite{hoff}).

Close examination of the conditions for consistency of scale invariance suggests that a continuum formulation can be removed altogether. 
An important observation is that the scale parameter is proportional to the square of the desired characteristic spatial resolution, $\Delta x$.
Since this can be directly associated with the spatial resolution of the discretization that is employed in the computer model it appears reasonable to consider the possibility that one could discard the continuum model and adapt the above conditions for consistency based on scale invariance of an entirely discrete formulation.

To this end we must rely on the working hypothesis that the intrinsic properties and dynamic state of objects of the physical world can, at all scales, be defined in terms of information.
Under this regime the conservation laws of the continuum theories are replaced by laws that govern the conservation of information in some form.
Such a rule based algorithm involves finite state arithmetic as is reflective of computations on a real world deterministic machine.
Here there is a major philosophical shift in that the discrete model is both the computer model and the template that defines the laws that govern the physical system being simulated.
In this paradigm the traditional notions of validation by way of consistency, convergence and stability of a discrete computer model of a continuum theory is bypassed by the single property of computability of a program built upon sequentially ordered statements.
By pursuing this path it will eventually become apparent that, by necessity, the discourse is transferred from the language of mathematical equations to the language of rule based algorithms and computer programs.

It should be stressed that the issue being discussed here is one of a choice of the most efficient language that can be employed to model the physical world.
Whether we possess a language that is rich enough to allow us to completely describe the physical world will remain a controversial issue.
We can, however, be encouraged by current trends in theoretical physics where there is an increasing tendency towards formulating physical laws in a language based on information, where physical objects and their dynamic state are represented by integers.

This is related to the general area of digital physics of which early proponents include Zuse \cite{zuse} and Jaynes \cite{jay01}-\cite{jay02}.
More recent proponents of digital physics include Wolfram \cite{wolf} who explores the universe of the most elementary computer programs.
Allied to this subject are the deeper mathematical investigations of Chaitin (see for example \cite{chait01}-\cite{chait11}) who is credited as a major pioneer of algorithmic information theory.
Of particular note is Chaitin's interest in the work dating back to Leibniz \cite{leib} who, apart from being the earliest known discoverer of binary arithmetic, appears to have explored early notions of complexity and how complexity can be employed to ultimately construct a formal definition of what actually constitutes a scientific theory.
Using this as a starting point, Chaitin goes on to explain how this leads to the interesting idea that theories of science are best expressed in a language of programs.
Some aspects of these works, along with the more controversial views of Zeildberger on finitism and discrete versus continuous mathematics (see for example \cite{zeil01}-\cite{zeil02}), are highly influential in some of the ideas presented throughout this book. 

Continuum theories have so far served us very well, providing insights in many branches of scientific research.
But their limitations in providing closed form solutions for many complex systems, and hence the need to introduce discrete approximations along with their inherent problems, are increasingly becoming recognized.
In a future where greater rigor is demanded in modeling complex systems, alternatives need to be investigated.
A language based on programs and its association with discrete mathematics appear to provide a good candidate for such an alternative.     

It will be premature here to embark on a detailed review of methodologies associated with the construction of fully discrete models of specific real world applications.
Before we can do this we must first reassess the very foundations of basic arithmetic on a deterministic machine with finite memory.
Indeed, the axioms that dictate the basic rules of machine arithmetic will play an important role in defining the laws that govern the construction of the discrete model.
If we are to seriously take the discrete model as the defining language that describes a real world application then it is not unreasonable to expect that the conventional laws of physics will emerge as manifestations of the more fundamental laws of allowable finite state computations. 
In this book we will explore some of these ideas, starting with the most elementary laws that govern machine arithmetic.
From such a study it is hoped will emerge a platform from which a formal and rigorous approach to computer modeling can be constructed.

To this end we adopt an alternative to the traditional approach of proofs based on abstract mathematical structures.
We start by introducing an inference scheme based on the so called \emph{program extension rule}.
This formal system is a departure from the traditional formal schemes of proof theory in that it is largely constructed from the rules that govern the allowable computational operations on a real world computer that is constrained by finite memory storage.
The language that we will choose is one of representing formal statements as programs.
Having constructed the foundational tools of inference we will then explore the constraints imposed by a deterministic machine with finite memory and the most elementary operations of arithmetic that can be performed on it.
This will lead us to the exploration of computability of fully discrete dynamical systems as they are directly implemented on a real world deterministic computer. 

\section{Machine arithmetic.}

Our main objective is to construct a formal language from which we can validate computer models on the basis computability.
In the previous section we discussed discrete models based on simple deterministic rules but practical application often requires the necessity to reformulate these models on lattices that represent larger scales.
It follows that the transformation from the microscopic to the macroscopic scale is associated with a transition from a rule based algorithm to one that is largely based on numerical computations.  

It should be kept in mind that our goal for establishing computability is much higher than just avoiding underflows and overflows of the numerical computations.
We start with the hypothesis that the fully discrete model, and hence the operational parameters that characterize the machine upon which the model is to be executed, are reflective of the underlying structure of the real world system.
Under this hypothesis we are raising the status of the conditions of computability of a model by associating them with the laws that govern the dynamics of the real world system.

In any attempt to construct a tool for the validation of programs largely based on numerical computation one first looks to the basic foundations of arithmetic starting with the axioms of rings and fields (see for example \cite{burk}, \cite{mclane}).
Unfortunately, when encountering machine arithmetic one will eventually observe a departure from the elementary rules of arithmetic upon which one has been accustomed.
To explain some aspects of these departures one may delve deeper into analysis through topics such as modulo arithmetic and finite fields \cite{mclane}, but these too fall short of addressing many of the problems that are encountered when dealing with machine computations.

One promising approach that provides rigor through direct numerical computations can be found in interval arithmetic (see for example \cite{moor}, \cite{ale}).
This has found wide use in computations attempting to approximate continuum theories by way of floating point arithmetic.
For discrete based models where integer or fixed precision rational solutions are desired, we can define discrete interval arithmetic in a similar way but with some important differences.

To tackle this problem in its entirety one soon finds the need to investigate topics in a much wider area, many of which are found in the realm of the computer sciences.
In particular, the initial motivation of program verification evolves into an area involving inference methods in a more general sense.  

Traditional studies of computers and computation often start by constructing a theoretical model that reflects some properties of real world computers.
Such examples can be found in Turing machines along with abstractions of programming languages themselves such as lambda calculus \cite{baren} leading to the study of logic and the important link between proofs and programs.
The latter in turn leads one into the subject of proof theory.
This is a wide area of study of which an excellent coverage can be found in \cite{buss}.

The formal systems in the general area of proof theory were primarily developed to address important theoretical problems in logic and were not optimally designed for practical implementation in a machine environment. 
The approach taken here is to construct a formal language such that the rules of inference are not dictated by an external abstract theory of logic but rather on the allowable computational operations on a real world computer.
As a consequence there will be a need to abandon some of the expressiveness of formal systems found in current proof theory.  
Motivated by a more practical approach to program verification, the language is presented in a form that is less abstract than traditional studies of theoretical computers and programing based on the lambda calculus.

These methods will be described in the context of the software package VPC\index{VPC} (Verification of Program Computation) in its current phase of development.
While the source code of VPC will not be presented here, an effort will be made throughout this book to describe its internal algorithms in sufficient detail so that the reader will be equipped to construct their own version.

In the construction of our formal language the following properties are of primary importance.

\begin{itemize}
	
	\item \emph{Simplicity}. The language should be simple and accessible to those of various backgrounds outside of the computer sciences.
	
	\item \emph{Analysis.} The simplicity of the language should not compromise its power to be employed as a tool of analysis at a sufficiently high level.
	
	\item \emph{Proof assistance.} As a language based on programs it should be readily implemented on a machine platform.
	As such, automated procedures can be constructed that assist in the generation of proofs.
	This assistance comes in the form of (i) generating on screen real time constructions of formulations that remove the laborious and error prone task of writing down symbols on paper and (ii) a step by step guidance of valid options in a proof construction.
	
	\item \emph{Compatibility.} A language that is specially designed to address the issues of computability on a real world computer, with particular focus on models based on finite state arithmetic.
	It is also advantageous to converse in a language that closely resembles the actual code that will ultimately represent the computer model.
	
	\item \emph{Expressiveness.} The language will largely deal with objects as subtypes of strings that are immediately recognized by the machine.
	Consequently, it will be a low level language that will lack the expressiveness found in standard formal systems of proof theory.
	However, it should possess the properties that it can be used as a primitive upon which theories demanding higher levels of abstractions can be built.   
	
	\item \emph{Improvements.} The language should be flexible enough to be open for future developments that extends its scope for analysis.
	Towards the more ambitious goal found in the area of artificial intelligence there exists the potential to explore avenues that ultimately lead to complete automation.
	
\end{itemize}

\chapter{Program Structure.}\label{cps}

\section{Types.}

We shall deal with objects and types, where each object has a type\index{type}.
In a machine environment each object is identified by a single string.
Different string structures are associated by their type.

\underline{Properties of types.}

\begin{itemize}

\item Object $a$ has type $\mbf{t}$ is denoted by $a:\mbf{t}$. Types will always be written in bold symbols.

\item An object may also be dependent on other objects.
We write $a(b_1,~\ldots,~b_n)$ to mean that the object $a$ depends on the objects or parameters $b_1,~\ldots,~b_n$, where $a,~b_1,~\ldots,~b_n$ need not be of the same type.

\item Types may be subtypes of types\index{subtype}.
Type $\mbf{s}$ is a subtype of $\mbf{t}$ is denoted by
\be
\mbf{s} <: \mbf{t}
\ee
Subtypes have the property that if $a:\mbf{r}$ and $\mbf{r}<:\mbf{s}$ then $a:\mbf{s}$.
It follows that if $\mbf{r} <: \mbf{s}$ and $\mbf{s} <: \mbf{t}$ then $\mbf{r} <: \mbf{t}$.

\item Types may also be dependent on objects.
We write $\mbf{t}[a_1~\ldots~a_n]$ to mean that type $\mbf{t}$ depends on the parameters or objects $a_1,\ldots,a_n$.
Parameter dependent types are subtypes of their generic type, i.e.
\be
\mbf{t}[a_1~\ldots~a_n]<:\mbf{t}
\ee

\end{itemize}

Note that if $\mbf{s}[a_1~\ldots~a_n]$ and $\mbf{t}[b_1~\ldots~b_m]$ are parameter dependent types and $\mbf{s} <: \mbf{t}$ it does not necessarily follow that $\mbf{s}[a_1~\ldots~a_n]$ is a subtype of $\mbf{t}[b_1~\ldots~b_m]$.

\section{Alphabet and strings.}\label{sas}

Here we shall work in a machine environment based on a real world deterministic computer.
A real world deterministic computer is characterized by the properties of finite information storage along with a finite collection of well defined operations.
At any time the machine can exist in any one of a finite number of configuration states.
A program is a sequentially ordered list of instructions where each instruction attempts to map the current configuration state to a new configuration state.
The context in which we will choose to work can be defined by the following machine specific parameter constraints.
\ben\label{mparam}
\bal
nchar = & \text{~number of characters in the alphabet.} \\
nstr = & \text{~maximum number of characters in any string.} \\
nlst = & \text{~maximum number of elements of a list.} \\
nint = & \text{~maximum absolute machine integer.} \\
\eal
\een
We define the list of machine parameters by
\ben\label{pl}
mach = [nchar~nstr~nlst~nint]
\een
(All lists will be enclosed by square brackets and elements of a list are separated by a space).
The choice of the parameters of $mach$ are very much dependent on the physical storage capacity of the machine.
As outlined in the next section, the maximum list length, $nlst$, applies also to multidimensional arrays since they can be stored as partitioned lists. 

We will explore the computational processes that are entirely confined within a machine environment, $\mathfrak{M}$, under the machine specific constraints, $mach$.
Where it is necessary to stress this context we will write
\be
\mathfrak{M}(mach)
\ee

We start by defining the alphabet as a collection of symbols or characters
\be
s(1),\ldots,s(nchar)
\ee
The alphabet that we will work with consists of the following characters.

\begin{itemize}

\item Letters.
\be
\bal
a~b~\ldots~z \\
\eal
\ee

\item Digits.
\be
1~2~3~4~5~6~7~8~9~0
\ee

\item Special characters.
\be
-~[~]~|
\ee

\end{itemize}

\underline{Strings.}\index{string}

\begin{enumerate}

\item A string of the alphabet is a sequence of characters
\be
s(i_1) s(i_2) \ldots s(i_j), \qquad 1 \leq i_1,\ldots,i_j \leq nchar,~1 \leq j \leq nstr
\ee

\item A single string defines an object and is given a generic type denoted by $\mbf{strng}$.
Different string structures are identified as subtypes of $\mbf{strng}$.

\item There are two important subtypes of strings.

\begin{itemize}

\item $\mbf{char}<:\mbf{strng}$, alphanumeric strings comprised of any combination of letters and digits with the first character always being a letter.

\item $\mbf{int}<:\mbf{strng}$, signed integers comprised of digits preceded by a sign $\pm$.

\end{itemize}

\end{enumerate}

Other subtypes will be defined as they are encountered throughout this book.
Sometimes we will allow a space to be included in an individual string.
In such a case the space will be regarded as a special character.

\vspace{5mm}

\underline{Alphanumeric strings.}
Alphanumeric strings\index{alphanumeric string} are assigned the type $\mbf{char}$ and are often used to represent names of programs and variable names of elements of the input/output (I/O)\index{I/O lists} lists of programs.
Variable names of the elements of I/O lists of programs serve as place holders for assigned values that are defined as specific types within the program.
We write $a:\mbf{char}$ to stress that $a$ is a dummy variable that represents an alphanumeric string.
Upon entry to a program we may also write $a:\mbf{t}$ to mean that the alphanumeric string represented by the dummy variable $a$ has been assigned a value of type $\mbf{t}$.
The assigned value can be any object of a well defined type.

\underline{Equality\index{equality}.}
There is an important distinction that needs to be made with the notion of equality.

\begin{itemize}

\item If $a$ and $b$ are dummy variables representing two strings we write $a=b$ to mean that the two strings are identical.
The sense in which equality is being used here will always be assumed unless otherwise stated.

\item We may also write $a=b$ to mean that the assigned value of the alphanumeric string represented by the dummy variable $a$ is identical to the assigned value of the alphanumeric string represented by the dummy variable $b$.
The sense in which equality is used here will always be stated to avoid confusion.

\end{itemize}

\vspace{5mm}

\underline{Machine numbers.}
An object of type $\mbf{int}$ is a string that can be assigned any one of the integer values
\be
0, \pm 1,\ldots, \pm nint
\ee
We shall make extensive use of the following subtypes of $\mbf{int}$.
\be
\begin{array}{ll}
\mbf{int0} & \text{$a:\mbf{int0}$ denotes $a:\mbf{int}$ and $0 \leq a \leq nint$} \\
\mbf{int1} & \text{$a:\mbf{int1}$ denotes $a:\mbf{int}$ and $0 < a \leq nint$} \\  
\end{array}
\ee
We adopt the usual convention of dropping the prefix $+$ sign when dealing with positive integers.

One of our objectives is to describe the program VPC as a tool for analysis and verification of numerical computation. 
For the purpose of demonstration only we will restrict much of the outline to machine integer arithmetic.
It will be seen later that most of the results using machine integers can also be applied to fixed precision rational numbers.
It should be kept in mind that VPC has a much wider area of application that includes floating point arithmetic.
The reasons for excluding floating point arithmetic is based on the anticipated paradigm shift in computer modeling as discussed in the introduction of Chapter 1.

\section{Lists.}\label{sl}

Throughout we shall work with lists\index{list} rather than sets.
Many properties of lists, such as list intersections and sublists, will have strong similarities with those used in set theory.
For this reason much of the notation used in set theory will be adopted for lists.
Since we are working in an environment $\mathfrak{M}(mach)$ all lists will be of finite length.

\underline{Type.}
\be
\begin{array}{ll}
	\mbf{lst} & \text{generic type list of unspecified length.} \\
	\mbf{lst}[n] & \text{type list with $n:\mbf{int0}$ elements, $\mbf{lst}[n] <: \mbf{lst}$.} \\
\end{array}
\ee 

\underline{Properties of lists.}

\begin{itemize}
	
	\item
	A list $a:\mbf{lst}[n]$, has the representation
	\be
	~[a(1)~\ldots~a(n)]=[a(i)]_{i=1}^n
	\ee
	\emph{We use a space instead of a comma to separate elements of a list}.
	The notation
	\be
	a(i) \in a
	\ee
	means that $a(i)$ is an element of the list $a$.
	The object $n:\mbf{int0}$ is referred to as the length of the list $a$.
	We write
	\be
	n=length[a]
	\ee
	where $length$\index{length function} is the list length function.
	
	\item Elements of lists are objects that may be identified by a single string or may themselves be lists (see next section).
	
	\item An empty list $a:\mbf{lst}[0]$ is denoted by $a=[~]$.
	If, under the list representation $a=[a(i)]_{i=1}^n$, we have $n=0$ then it is understood that $a$ is the empty list. 
	
	\item For a list, $a$, of unit length we will sometimes write $a$ and $[a]$ to mean the same thing, i.e.
	\be
	a=[a], \quad a:\mbf{lst}[1]
	\ee
	
	\item Elements in a list need not all be assigned values of the same type.
	
	\item \emph{List equality.} If $a:\mbf{lst}[n]$ and $b:\mbf{lst}[n]$, $n:\mbf{int0}$, and $a(i)=b(i)$, $i=1,\ldots,n$, we write $a=b$.
	We use equality in both senses of identity of strings and the values assigned to the strings.
	Throughout, unless otherwise stated, equality will be assumed to be in the sense of the former, i.e. in the sense of the identity of strings.
	Whenever the equality is used in the sense of assigned values it will be stated as such.
	
\end{itemize}

\underline{List operations.}

\begin{itemize}
	
	\item \emph{List concatenation}. If $a=[a(1)~\ldots~a(m)]$ and $b=[b(1)~\ldots~b(n)]$ are two lists then the concatenation\index{concatenation} of $a$ and $b$ yields the list $c:\mbf{lst}[m+n]$ given by
	\be
	\bal
	c= & [a~b] = [[a(i)]_{i=1}^m~[b(i)]_{i=1}^n] \\
	= & [[a(1)~\ldots~a(m)]~[b(1)~\ldots~b(n)]] \\
	= & [a(1)~\ldots~a(m)~b(1)~\ldots~b(n)] \\
	\eal
	\ee 
	The internal square brackets that act as delimiters for the lists $a$ and $b$ may be removed.
	
	\item \emph{List intersection.} If $a=[a(1)~\ldots~a(m)]$ and $b=[b(1)~\ldots~b(n)]$ then the list intersection of $a$ and $b$ yields a new list $c:\mbf{lst}[k]$, $k \leq min[m~n]$, where $c=[c(1)~\ldots~c(k)]$ contains all of the elements that are common to both $a$ and $b$.
	We write
	\be
	c=a \cap b
	\ee
	to mean that $c$ is the list intersection of $a$ and $b$.
	Whenever a list intersection is constructed, the sequential order of the elements of $c$ are in the same hierarchy of the sequential order that they appear in $a$.
	
	\item \emph{Removal of repeated elements of a list.} If $a=[a(1)~\ldots~a(m)]$ has repeated elements we can construct a new list $b=[a(i_1)~\ldots~a(i_n)]$, $i_1 <~i_2 < \ldots <~i_n$, $n \leq m$, by removing repeated elements as follows.
	Reading the list $a$ from left to right, whenever an element is encountered that coincides with a preceding element of $a$ then that element is discarded.
	In other words, each element of $b$ contains all non-repeated elements of $a$ and the first occurrence of a repeated element of the list $a$, as read from left to right, maintaining the order in which they appear in $a$. 
	We write\index{unique function}
	\be
	b = unique[a]
	\ee
	to mean that $b$ is obtained by extracting repeated elements of $a$ by this procedure.
	
	\item \emph{List subtraction.} Suppose that $a:\mbf{lst}[m]$ and $b:\mbf{lst}[n]$.
	We can construct a new list $c$ obtained by extracting from $a$ those elements found in $b$.
	The new list maintains the sequential order found in $a$, i.e. $c=[a(i_1)~\ldots~a(i_k)]$, $i_1 <~i_2 \ldots <~i_k$, $k \leq m$, where $a(i_1)~\ldots~a(i_k)$ are all of the elements of $a$ not found in $b$.
	We write
	\be
	c=a \setminus b
	\ee
	to denote the new list constructed in this way.
	
	\item \emph{Element substitution.}\index{element substitution}
	For a list $a=[a(i)]_{i=1}^m$ we write $a(a(i) \to b)$ to denote substitution of the element $a(i) \in a$ with $b$, i.e.
	\be
	a(a(i) \to b) = [a(1)~\ldots~a(i-1)~b~a(i+1)~\ldots~a(m)]
	\ee
	
	\item \emph{Empty list extraction.} Suppose that $a=[a(i)]_{i=1}^n:\mbf{lst}[n]$, $n:\mbf{int1}$, contains an element $a(k) \in a$ that is an empty list, i.e. $a(k)=[~]$.
	We may extract the empty list element and write  
	\be
	a=[a(1)~\ldots~a(k-1)~a(k+1)~\ldots~a(n)]
	\ee
	After empty list extraction we can automatically redefine $a:\mbf{lst}[n-1]$. 
	
\end{itemize}

\underline{Sublists.}\index{sublist}
Because of its importance, the notion of a sublist affords a more formal definition.

\begin{definition}(Sublist.)\label{sublists}
	A list $b:\mbf{lst}$ is a sublist of list $a:\mbf{lst}$ if every element of $b$ is an element of $a$, i.e. if $x \in b$ then $x \in a$.
	We write $b \subseteqq a$ to mean that $b$ is a sublist of $a$.
	There are two cases that need to be distinguished.
	
	\begin{itemize}
		
		\item If $b \subseteqq a$ and $a \setminus b \ne [~]$ then we say that $b$ is a strict sublist of $a$.
		We write $b \varsubsetneqq a$ to stress that $b$ is a strict sublist of $a$.
		
		\item If $b \subseteqq a$ and $a \subseteqq b$ we say that the two lists are equivalent and write $a \equiv b$.
		
	\end{itemize}
	The empty list, $[~]$, is regarded as a sublist of all lists.
	
\end{definition}

\vspace{5mm}

\underline{Notes.}

\begin{itemize}
	
	\item If $q:\mbf{lst}[m]$ is a sublist of $p:\mbf{lst}[n]$ it does not necessarily follow that $m \leq n$. Consider the case $q=[a~b~b~a]$ and $p=[a~b~c]$.
	In this example $q$ is a strict sublist of $p$ yet $length[q]=4>length[p]=3$. 
	
	\item Similarly, two equivalent lists need not have the same length.
	For example $p=[b~b~c~a]$ and $q=[a~b~c]$.
	Here $q \subseteqq p$ and $p \subseteqq q$, hence $q \equiv p$.
	
\end{itemize}

\section{Arrays as lists.}\label{sarray}\index{array}

\underline{Integer vectors.}\index{vector}\index{integer vector}
An integer vector of dimension $m:\mbf{int1}$ is a list $\mbf{lst}[m]$ whose elements have all been assigned the values of type $\mbf{int}$.
We use the notation $\mbf{vec}[m]$ to denote the type integer vector with the understanding that the vector dimension, $m$, is type $\mbf{int1}$.
When all of the elements of the vector $\mbf{vec}[m]$ have been assigned the value of type $\mbf{int1}$ we write $\mbf{vec1}[m]$.

Sometimes we will refer to an object of type $\mbf{vec}[m]$ as an \emph{integer state vector}\index{state vector} or simply a \emph{state vector}.
When discussing fully discrete dynamical systems a state vector will be used to refer to an integer vector that is structured in such a way that it contains the minimum amount of information that is needed to completely describe the intrinsic properties and dynamic state of an object in the system at any given time.
(In this context the term state vector should not be confused with that defined in quantum mechanics.)

\underline{Arrays.}
The definition of a list can be extended by allowing any element, $a(i) \in a$, of a list $a=[a(1)~\ldots~a(n)]$ to be a list.
Elements of a list that are not lists are called \emph{atomic elements}\index{atomic elements}.
Atomic elements that have type assignments $\mbf{int}$ (integers) or $\mbf{rat}$ (rational numbers) are called scalars.\index{scalar} 

Lists of lists can be represented as arrays.
We now demonstrate how arrays can be stored as partitioned lists.

When discussing arrays of general dimensions in a conventional mathematical language indexing can become cumbersome to write down.
We will use some shorthand notation.

The array dimension list \index{dimension list} is defined by
\be
\begin{array}{ll}
	l=[l(1)~\ldots~l(m)], \qquad & l(i):\mbf{int1},~i=1,\ldots,m,~m:\mbf{int1} \\
\end{array}
\ee
Array dimension lists are fixed and will be used to define dimensions of arrays.
The rank \index{rank} of an array is equal to the length of its dimension list.
We write
\be
a:\mbf{lst}[l], \quad l:\mbf{vec1}[m],~m:\mbf{int1}
\ee
to mean that $a$ is an array of rank $m$ with dimension list $l$.
In expanded form an element of an array, $a:\mbf{lst}[l],~l:\mbf{vec1}[m]$, can be written as $a(i)$, where $i:\mbf{vec}[m]$ is the index list\label{index list} given by
\be
\begin{array}{ll}
	i=[i(1)~\ldots~i(m)], \qquad & 1 \leq i(j) \leq l(j),~j=1,\ldots,m \\
\end{array}
\ee
Sometimes it will be convenient to partition the dimension lists.
We can define an array, $a:\mbf{lst}[l,l^\prime]$, where we have combined two dimension lists, $l:\mbf{vec1}[m],~m:\mbf{int1}$ and $l^\prime:\mbf{vec1}[m^\prime],~m^\prime:\mbf{int1}$.
The rank of the array $a$ is $m+m^\prime$ and has the element representation $a(i,i^\prime)=a(i(1),\ldots,i(m),i^\prime(1),\ldots,i^\prime(m^\prime))$.

\underline{List partitions.}
Arrays are stored as lists with a specific partition.
The position of an element, $a(i),~i:\mbf{vec}[m]$, of an array, $a:\mbf{lst}[l],~l:\mbf{vec1}[m]$, in a single list is given by the list index
\be
i(1) + (i(2)-1)*l(1) + \cdots + (i(m)-1)*l(m-1)*l(m-2)* \cdots *l(1)
\ee
In this way we can always redefine an array, $a:\mbf{lst}[l],~l:\mbf{vec1}[m],~m:\mbf{int1}$, as the list
\be
a:\mbf{lst}[s], \quad s=l(1)*l(2)* \cdots *l(m)
\ee
where $s:\mbf{int1}$ is a scalar.

\underline{Matrices.}
We can also express an array as a matrix.
A matrix can be thought of as an array of rank two and is given the type $\mbf{lst}[s~t]$, where $s,t:\mbf{int1}$.
If $a:\mbf{lst}[l~k],~l:\mbf{vec1}[m],~k:\mbf{vec1}[n]$, whose elements are given by $a(i,j),~i:\mbf{vec}[m],~j:\mbf{vec}[n]$, then we can construct the matrix $c:\mbf{lst}[s~t]$, $s,t:\mbf{int1}$, where
\be
\bal
s = & l(1)*l(2)* \cdots *l(m) \\
t = & k(1)*k(2)* \cdots *k(n) \\
\eal
\ee
Each element $c(p,q),~p,q:\mbf{int1},$ of the matrix $c:\mbf{lst}[s~t],~s,t:\mbf{int1}$, can be obtained from the array $a$ through the order index functions 
\be
\bal
p(i) = & i(1) + (i(2)-1)*l(1) + \cdots + \\
& \hspace{35mm} (i(m)-1)*l(m-1)*l(m-2)* \cdots *l(1) \\
q(j) = & j(1) + (j(2)-1)*k(1) + \cdots + \\
& \hspace{35mm} (j(n)-1)*k(n-1)*k(n-2)* \cdots *k(1) \\
\eal
\ee

\underline{Elements of arrays.} Consider the list
\be
x=[x(i)]_{i=1}^m
\ee
Suppose that each element $x(i)$ is itself a list
\be
x(i)=[x(i,j)]_{j=1}^{n(i)}
\ee
The list, $x$, has the expanded form
\ben\label{aas01}
\bal
x = & [x(i)]_{i=1}^m \\
= & [[x(i,j)]_{j=1}^{n(i)}]_{i=1}^m \\
= & [ [x(1,1)~ \ldots ~x(1,n(1))]~ \ldots ~[x(m,1)~ \ldots ,x(m,n(m))] ] \\
= & [ x(1,1)~ \ldots ~x(1,n(1))~ \ldots ~x(m,1)~ \ldots ~x(m,n(m)) ] \\
\eal
\een
where the internal square brackets can be removed in the final expansion.
We may write
\ben\label{aas02}
x(i,j) \in x, \quad x(i) \subseteqq x
\een
to mean $x(i):\mbf{lst}$ is a sublist of $x:\mbf{lst}$ and $x(i,j)$ is an individual element of the list $x=[[x(i,j)]_{j=1}^{n(i)}]_{i=1}^m$ as represented by the expanded form of the last identity of the above concatenation (\ref{aas01}).

Sometimes we may write
\ben\label{aas03}
x(i) \in x
\een
to mean that $x(i):\mbf{lst}$ is an element of the list of lists $x=[x(i)]_{i=1}^m:\mbf{lst}$.
The meaning here will be evident that we are treating $x$ as a list of elements that are type $\mbf{lst}$.
Otherwise we would write $x(i) \subseteqq x$ with the meaning that the list of elements of $x(i)$ is a sublist of the expanded list of elements $x$.
Where there might be confusion the sense in which an element of a list of lists is defined will be stated.

We may continue the expansion of (\ref{aas01}) in the case that each $x(i,j)$ is itself a list.
In a similar way, defining elements and sublists of the list $x$ will depend on the context that we choose.  

\underline{Notes.}

\begin{itemize}

	\item Here we have borrowed the definitions of the rank and dimensions of an array from the programming language Fortran.
	This differs from the definition of the rank of a matrix as employed in mathematics.

	\item We have defined an integer vector to be just a finite list of integers.
	In Chapter \ref{vec} we will examine such objects under the operations of addition, subtraction and scalar multiplication.
	However, because we a working in a machine environment, $\mathfrak{M}(mach)$, it will become evident that such objects fail to obey many of the rules associated with contemporary vector spaces.
	
	\item An integer vector will be referred to as a state vector  when it is used to define the minimum amount of information necessary to completely describe the intrinsic properties of an object and its dynamic state at any given time in a real world system.
	Here we use time in a discrete sense that can be associated with the machine operation of mapping the current configuration state of the system to a new configuration state.
	The time defined in this sense need not be that associated with the physical time.
	Moreover, the objects and the lattice of a dynamical system need not be assumed to be directly related to physical objects in physical space.
	In this way we may consider models of dynamical systems where the notions of physical objects along with physical space-time are emergent.         
	
\end{itemize} 

\section{Programs.}\label{programs}

Programs are made up of strings or lists of strings with a well defined structure and are assigned the type denoted by $\mbf{prgm}$.
Program names are assigned the type $\mbf{pname}$ and are subtypes of alphanumeric strings, i.e. $\mbf{pname} <: \mbf{char}$.

We start by defining an atomic program.

\begin{definition}\label{prog}(Atomic program.)\index{atomic program}
An atomic program has the representation
\be
p~x~y
\ee
with the allocation of types of its component parts given by
\be
\begin{array}{ll}
p~x~y:\mbf{prgm}  & \text{program} \\
p:\mbf{pname} & \text{program name} \\
x:\mbf{lst} & \text{input list} \\
y:\mbf{lst} & \text{output list} \\
\end{array}
\ee
The program name $p$ and the lists $x$ and $y$ are separated by a space and the expression $p~x~y$ is treated as a single string of type $\mbf{prgm}$.
An atomic program $p~x~y$ satisfies all of the following conditions.

\begin{itemize}

\item Elements of the I/O lists
\be
x=[x(i)]_{i=1}^{nx},~y=[y(i)]_{i=1}^{ny}, \quad nx:\mbf{int1},~ny:\mbf{int0}
\ee
are alphanumeric variable names (type $\mbf{char}$) that serve as placeholders for assigned values.
The output list, $y$, may be the empty list.

\item The type of the assigned value of every element of the I/O lists is checked within the program.
If there is a type violation of any value assigned element then the program halts with an error message

\item The variable names of the elements of the output list, $y$, are distinct, i.e.
\be
y(i) \neq y(j), \qquad i \neq j,~1 \leq i,j \leq ny
\ee

\item No element of the input list, $x$, can have a variable name that coincides with a variable name of an element of the output list, $y$, i.e.
\be
x \cap y = [~]
\ee
\end{itemize}
\end{definition}

Programs are constructed from lists of atomic programs.
Both atomic programs and program lists have the generic type $\mbf{prgm}$.
An atomic program can be distinguished from a program list by the subtype $\mbf{atm} <: \mbf{prgm}$.

\vspace{3mm}   

\begin{definition}\label{prog1}(Program list.)\index{program list}
A program can be represented by a list 
\ben\label{plist}
\bal
~[p(k)~x(k)~y(k)]_{k=1}^n \\
\eal
\een
for some $n:\mbf{int1}$, where each triplet $p(k)~x(k)~y(k):\mbf{prgm}$, $k=1,\ldots,n$, is an atomic program so that $p(k):\mbf{pname}$ and $x(k),y(k):\mbf{lst}$.
The I/O lists, $x(k)$ and $y(k)$, of the atomic programs, $p(k)~x(k)~y(k)$, $k=1,\ldots,n$, have the expanded form
\ben\label{iolists1}
\bal
& x(k)=[x(k,i)]_{i=1}^{nx(k)}, \quad nx(k):\mbf{int1} \\
& y(k)=[y(k,i)]_{i=1}^{ny(k)}, \quad ny(k):\mbf{int0} \\
\eal
\een
The atomic programs, $p(k)~x(k)~y(k)$, $k=1,\ldots,n$, are referred to as \emph{subprograms} of the program list.\index{subprogram}
The program list, (\ref{plist}), satisfies all of the following conditions.

\begin{itemize}

\item The variable names of the elements of the subprogram output lists are distinct, i.e. 
\be
y(k) \cap y(l) = [~] , \quad k,l=1,\ldots,n ,~k \neq l
\ee

\item For each $k=1,\ldots,n$, the variable names of the elements of the list $x(k)$ must not coincide with a variable name of the elements of the lists $y(l)$ for $l=k,\ldots,n$, i.e.
\be
x(k) \cap [y(l)]_{l=k}^n = [~], \quad k=1,\ldots,n 
\ee

\end{itemize}

We write
\be
[p(k)~x(k)~y(k)]_{k=1}^n:\mbf{prgm}[n]
\ee
to indicate that $[p(k)~x(k)~y(k)]_{k=1}^n$ is a program with list length $n:\mbf{int1}$.
The parameter dependent type $\mbf{prgm}[n]$ is a subtype of the generic type $\mbf{prgm}$, i.e. $\mbf{prgm}[n] <: \mbf{prgm}$.

\end{definition}

\underline{Matrix representations of programs.} Both atomic programs and program lists have the generic type $\mbf{prgm}$.
The same notation, $p~x~y$, for program lists, (\ref{plist}), is used and can be thought of as having the matrix representation
\ben\label{pmatrix}
p~x~y =
\left [
\begin{array}{l}
	p(1) \\
	\vdots \\
	p(n) \\
\end{array}
\right ]
\left [
\begin{array}{lll}
	x(1,1)& \ldots & x(1,nx) \\
	\vdots &  & \vdots \\
	x(n,1)& \ldots & x(n,nx) \\
\end{array}
\right ]
\left [
\begin{array}{lll}
	y(1,1)& \ldots & y(1,ny) \\
	\vdots &  & \vdots \\
	y(n,1)& \ldots & y(n,ny) \\
\end{array}
\right ]
\een
where we have defined
\be
\bal
nx = & \text{maximum length of the input lists of all atomic programs} \\
ny = & \text{maximum length of the output lists of all atomic programs} \\
\eal
\ee 
Blanks are inserted for elements of the arrays $x$ and $y$ that do not contain variables.
Here $p$ can be thought of in terms of the traditional transpose of a vector that is a vertical list of the atomic program names of its subprograms.
From the perspective of machine memory a vertical list is indistinguishable from a horizontal list so that we can simply write $p=[p(k)]_{k=1}^n:\mbf{lst}[n]$. 
The I/O lists are also arrays with
\be
x:\mbf{lst}[n~nx],~y:\mbf{lst}[n~ny]
\ee
The notation $p~x~y:\mbf{prgm}[n]$ defines the type program list of length $n$ that can be expressed as both (\ref{plist}) and (\ref{pmatrix}).

Often we shall simply represent a program as a vertical list
\ben\label{vplist}
\begin{array}{l}
	p(1)~x(1)~y(1) \\
	\hspace{10mm} \vdots \\
	p(n)~x(n)~y(n) \\ 
\end{array}
\een
\emph{It is important to note throughout that when dealing with program lists each $x(k)$ and $y(k)$, $k=1,\ldots,n$, are themselves lists.}

The execution of the program (\ref{vplist}), for a given value assigned input, is completed when all of the subprograms, $p(k)~x(k)~y(k)$, $k=1,\ldots,n$, have been executed in the sequential order from top to bottom in the vertical program list.

For $n=1$ we simply drop the list representation so that $p~x~y$ represents an atomic program.
We may also write $\mbf{atm}=\mbf{prgm}[1]$.
The case $n=0$ means that the program is the empty program list, denoted by $[~]$ or the alphanumeric name $ep$.

\underline{I/O value assignments.}\index{value assignment} The elements of the I/O lists of a program, $p~x~y$, are alphanumeric strings that serve as placeholders for assigned values.
The action of assigning a value to an alphanumeric string that represents an element of an I/O list involves the allocation of an address where the assigned value of the alphanumeric string and its type are stored.
These assigned values can be integers, fixed precision rational numbers or other objects of a well defined type.
 
An element of an I/O list may also be assigned a value that comes in the form of a list or an array.
In such a case it is understood that the assigned value is a subtype of $\mbf{lst}$ of a specified length.

\underline{Constants.}\index{constants}
We need to make a distinction between common variables and constants.
For each type there may exist special objects of that type that appear as fixed assigned values in an input list of a program.

Throughout we use the notation
\ben\label{var}
var=[var(i)]_{i=1}^{nvar}, \quad nvar:\mbf{int1},~var(i):\mbf{char}
\een
to denote the list of variable names used as elements of I/O lists that are distinguished from the names of constants that are elements of the list
\ben\label{cst}
cst=[cst(i)]_{i=1}^{ncst}, \quad ncst:\mbf{int0},~cst(i):\mbf{char}
\een
Each element, $cst(i)$, of the list of constants, $cst$, is an alphanumeric string that is assigned a fixed value.
The fixed value assignments of the elements of $cst$ need not be of the same type. 

\underline{I/O dependency condition.}\index{I/O dependency condition}
Input variable names of a program may be elements of $var$ or $cst$ while output variable names can only be elements of $var$.  
When an atomic program is placed in a program list it can have an input variable name that coincides with an I/O variable name of a subprogram that precedes it in the program list.
Such an input variable is said to be a \emph{bound~variable}\index{bound variable}.
An input variable may also be bound to another input variable in the same program or bound to a constant.
An input variable that is not bound is said to be a \emph{free~variable}\index{free variable}.

We distinguish between different kinds of variable bindings for program lists.
For a program $p~x~y:\mbf{prgm}[n]$, $n \geq 1$, the I/O lists have the expanded form
\ben\label{iolists}
\bal
& x=[x(k)]_{k=1}^n=[[x(k,i)]_{i=1}^{nx(i)}]_{k=1}^n \\
& y=[y(k)]_{k=1}^n=[[y(k,i)]_{i=1}^{ny(i)}]_{k=1}^n \\
\eal
\een
The possible variable bindings are as follows.
\begin{itemize}
	
\item Input variable bound to a constant.
\ben\label{iibd0}
x(k,i) \in cst
\een
	
\item Input variable bound to another input variable within a subprogram.
\ben\label{iibd1}
x(k,i) = x(k,j), \quad x(k,j) \notin cst,~j < i
\een

\item Input variable bound to an input variable of a preceding subprogram.
\ben\label{iibd2}
x(k,i) = x(l,j), \quad x(l,j) \notin cst,~l < k
\een

\item Input variable bound to an output variable of a preceding subprogram.
\ben\label{iobnd}
x(k,i) = y(l,j), \quad l < k
\een

\end{itemize}

The sequential order of subprograms that are not subject to the bound variables of the kind (\ref{iobnd}) can be interchanged in a program list.
The sequential order of subprograms that are subject to the bound variables of the kind (\ref{iobnd}) cannot be interchanged.
We shall often refer to this property as the \emph{I/O dependency condition}.
The I/O dependency condition is a consequence of our functional program list representation that disallows reassigning values to a variable name.
This differs from imperative programming languages where it is common to reassign values to a variable name within a program.
The I/O dependency condition plays a crucial role on how program lists can be manipulated.

The list of primary input variables for a program list $p~x~y:\mbf{prgm}[n]$, $n \geq 1$, is defined by
\ben\label{ub}
piv[x]=unique[x \setminus y]
\een
that lists all of the input variable names that are not bound to output variables.\index{primary input variables}
The unique function is used to remove repetitions.
We will always discuss the computability of programs with respect to the value assignments of their primary input variables.
  
We define
\ben\label{free}
free[x]=piv[x] \setminus cst
\een
The list $free[x]$ is not meant to represent a list of free variables since its elements may be bounded in the sense of (\ref{iibd1}) or (\ref{iibd2}).
The elements of the list $free[x]$ are primary input variables that are free to be assigned any value of their respective type.

The variable name of each element of the output list of a program is unique.
The elements of the list
\ben\label{intcalc}
y \cap x
\een
can be regarded as output variables that are employed as intermediate calculations in a program list.
Output variables that are elements of the list
\ben\label{primout}
y \setminus (y \cap x)
\een
can be thought of as the primary output variables of the program.
Programs written in some imperative language often discard variables that are employed as intermediate calculations returning only the primary output variables.
In our functional program format such variables are not discarded.

\underline{The read/print programs.}
A program list (\ref{vplist}) should be thought of as being a core program embedded in a larger program that can be represented by the vertical list
\ben \label{core}
\begin{array}{l}
	read~[file1]~[free[x]] \\
	p(1)~x(1)~y(1) \\
	\hspace{10mm} \vdots \\
	p(n)~x(n)~y(n) \\
	print~[y]~[file2] \\ 
\end{array}
\een
The variable name $file1$ is associated with a file where the value and type assignments of each element of $free[x]$ is stored.
It can be assumed that all elements of $cst$ are assigned their values and types in an initializing file that can be accessed by a program whenever a constant variable is encountered in a subprogram input list.

The variable name $file2$ is associated with a file where the output, $y$, is written.
By accessing an input data file, $file1$, the program
\be
read~[file1]~[free[x]]
\ee
assigns to each element of the variables of the free primary input list a value and a type consistent with the entry type checking of the subprograms of the program (\ref{vplist}).
The assigned values and types of the elements of the output list, $y$, are generated through the dual actions of value and type assignments contained within the subprograms of the program.

The program
\be
print~[y]~[file2]
\ee
prints the value assigned output list, $y$, to a file, $file2$.
If an execution error is encountered in the program the execution is halted and an error message is printed to a file and/or screen.
For the purposes of analysis the core program, $[p(k)~x(k)~y(k)]_{k=1}^n$, will always be considered in isolation with the understanding that the value and type assignments of the elements of the input list, $free[x]$, have been prescribed by the initializing program $read~[file1]~[free[x]]$.

\underline{Elements of program lists.}\index{program list elements}
It is convenient to rewrite (\ref{pmatrix}) as an augmented matrix
\be
p~x~y = \left [
\begin{array}{l|lll|lll}
	p(1) & x(1,1)& \ldots & x(1,nx) & y(1,1)& \ldots & y(1,ny) \\
	\vdots & \vdots &  & \vdots & \vdots &  & \vdots \\
	p(n) & x(n,1)& \ldots & x(n,nx) & y(n,1)& \ldots & y(n,ny) \\
\end{array}
\right ]
\ee
In this form each row, $[p(k)~x(k)~y(k)]$, can be regarded as an element of the program $p~x~y$ represented as a vertical list.

A program list represented by (\ref{plist}) is a list of ordered triplets
\be
(program~name)~[input~list]~[output~list]
\ee
We will always define individual elements of a program list to be the strings of triplets $p(k)~x(k)~y(k)$, $k=1,\ldots,n$, that represent the subprograms of $p~x~y$.
We write
\be
p(k)~x(k)~y(k) \in p~x~y,~1 \leq k \leq n
\ee
to mean that the triplet $p(k)~x(k)~y(k)$ is an individual element of the program list $p~x~y$.
The notion of a sublist of a program is defined in terms of the elements of a program list in this sense.

When reading a program represented as a vertical list the machine will recognize each subprogram, defined by the triplet $p(k)~x(k)~y(k)$, as an individual string so that the internal spaces that separate the program name and the I/O lists will be regarded as special characters of that string.
Each triplet has a well defined structure so that the machine will have no trouble in recognizing the internal spaces that separate the components of the triplet as special characters of the string that represents an atomic program.

\underline{Notes.}

\begin{itemize}

\item If the empty program is encountered in the execution of a program list then the program does not halt and execution proceeds to the next subprogram of the list.
Subprograms of a program list that are empty programs can be immediately removed by the process of an empty list extraction.

\item  Haskell is a common functional programming language that is employed in the computer sciences.
It exploits the lambda calculus formulation and can be employed as a proof checker using the inference rules of propositional and first order logic. 
An introductory coverage of Haskell can be found in \cite {hall}. 

Our programs based on Definitions \ref{prog} and \ref{prog1} are structurally quite different from those of lambda calculus.
Other objects also defined in this book will differ from those presented to students in the computer sciences. 

\item At this point, a reader who has a background in the computer sciences might regard the I/O dependency condition as an unnecessary complication that is treated more efficiently in the lambda calculus.
It can only be recommended here that the reader persevere.
An attempt will be made to demonstrate that the I/O dependency condition is quite manageable and that our language will contain some useful features for the purposes that it is has been designed.

\end{itemize}

\section{Computability.}\label{sc}\index{computability}

\underline{Execution error.}
Within all programs type checking is performed on the assigned values of all elements of the I/O lists.
Execution errors\index{execution error} are based on type violations\index{type violation}.
A program will halt with an execution error if during its execution there is a type violation of any assigned value of the elements of its I/O lists.
Otherwise, the execution of a program is completed when all subprograms of the program list have been successfully executed in the sequential order that they appear in the list.

In the next section we will introduce atomic programs that also check for the satisfaction of a relation between a pair of elements of its input list.
In such cases an execution error will also include the case where the relation is not satisfied. 

In a later chapter we will also introduce program disjunctions.
Disjunctions essentially split the execution of a single program list into a number of parallel program lists.
These parallel program lists can be associated with operands of the disjunction.
If at least one of the operand programs of the disjunction does not contain a type violation then all type violations that exist in the other operand programs are overridden and the main program will not halt with an execution error.

A more formal definition of an execution error will be postponed until we have introduced disjunctions.
For the moment it will suffice to regard an execution error to be solely associated with the encounter of a type violation in a single program list.   

\underline{Computable programs.}
Our main objective is to construct computer models that can be validated by establishing computability.
By this it is meant that a program will eventually halt without encountering an execution error and return a value assigned output.

In general there may exist programs for which we will be unable to rigorously establish computability or noncomputability for that matter.
In the computer sciences undecidability is highlighted by the halting problem, although this is almost exclusively discussed in the context of abstract computers such as Turing machines with infinite tapes.
Undecidability also arises in mathematics where it is often regarded as troublesome and a reminder of a limitation of mathematics.
In a more general context of the scientific method, undecidability is an accepted concession where the best that can be hoped for is a process of continual revision from which will emerge theories with expanded scope of applicability.
These issues will be addressed in more detail in the final chapter of this book.
With these issues in mind the following definition of computability will be sufficient for the most part.

\begin{definition}\label{comp}(Computability.)
A program $p~x~y$ is said to be computable, with respect to the value assignments of its primary input variables, $piv[x]$, if upon execution it eventually halts without encountering an execution error based on type violations.
A computable program returns the value assigned output, $y$, where $y$ may be the empty list.
\end{definition}

\underline{Terms and assignment maps.}\index{term}\index{assignment map}
Programs can have an empty output list.
Programs with an empty output list are often associated with the sole task of checking the types of the value assignments of the elements of their input list.

An atomic program, $p~x~y:\mbf{atm}$, with a nonempty output list, $y$, will be referred to as a value assignment program to be distinguished from a type assignment program.
Where there is no confusion in the context we refer to value assignment programs as simply assignment programs.

Assignment programs are often associated with arithmetic calculations but may also involve algorithms that cannot be concisely expressed in the conventional mathematical notation.
The internal algorithm of an atomic program can be thought of as a sequence of instructions, including arithmetic computations, written in some imperative language.

We introduce the type $\mbf{term} <: \mbf{strng}$.
Objects of type $\mbf{term}$ are alphanumeric character strings that also include the special characters $[~]$.
Terms take the general form
\be
\text{(term name)}[\text{list of variables}]
\ee
Terms will often be used to represent functions.
We have already used terms to express the functions $length$, $unique$, $piv$ and $free$.
We have also used terms to express parameter dependent types where the term name is given in bold characters.

Given a list $x:\mbf{lst}[n]$ the expression
\be
y:=f[x]=f[x(1)~\ldots~x(n)]
\ee
means that $y$ is assigned the values assigned to the elements of $x$ through  a function or map $f:x \mapsto y$.
The object
\be
f[x(1)~\ldots~x(n)] : \mbf{term}
\ee
can be associated with the classical function notation $f(x(1),\ldots,x(n))$.
Note that spaces are used instead of commas to separate the arguments of the term.

For a value assignment program, $p~x~y$, the term notation
\be
y:=p[x]=p[x(1)~\ldots~x(n)]
\ee
is used to indicate that the program $p~x~y$ can be associated with the map $p:x \mapsto y$, where for convenience the name of the term is the same as the associated program name.
We will often refer to the expression $y:=p[x]$ as the assignment map of the value assignment program $p~x~y$.
If $length[y]=n > 1$ it will be understood that $p[x]$ represents a list of terms of length $n$.
Sometimes we will just represent the assignment map of a program, $p~x~y$, by $p: \mbf{s} \to \mbf{t}$ to indicate a map from objects of type $\mbf{s}$ to objects of type $\mbf{t}$.
While the association of programs with maps and functions will be useful it should not be taken too formally since our approach will be mainly syntactic based on the manipulation of strings.

In Chapter \ref{cabstract} we consider programs where the assigned values of the output list are character strings of type $\mbf{term}$.
This will be useful in applications of VPC where higher levels of abstractions are employed.
Abstract objects are assigned abstract types that can represent maps from sets to sets or predicates of sets. 

\underline{False programs.}
In the context of value assignment programs, the associated function can be thought of as a partial function.
There are cases where an object has all of the structural properties of a program but will not be computable for any type compatible valued input list.  

\begin{definition}(False program.)
A program is said to be a false program if there does not exist a value assigned input list such that the program is computable.
A false program is assigned the type $\mbf{false}$, where $\mbf{false}<:\mbf{prgm}$.
\end{definition}

In the definition of a false program it is stated that $\mbf{false}<:\mbf{prgm}$.
This means that for an object to have the type assignment $\mbf{false}$ it must first have the structure of a program under Definitions \ref{prog} and \ref{prog1}.
The statement that a program will always halt as a result of an error in syntax is not considered meaningful in this context since such an object cannot be assigned the type $\mbf{prgm}$.

\section{Atomic programs.}\label{saps}\index{atomic program}

Programs are built up from lists of atomic programs.
The internal algorithms of atomic programs will be understood to be constructed from some imperative language and is not seen by the machine during constructions of program lists.
Because the internal algorithm of an atomic program is not accessible to the machine in explicit form it will be necessary to supply a collection of rules or axioms that describe it.
It is through these axioms that the machine will be able to recognize the main computational operations associated with the internal algorithm of an atomic program.

\begin{definition}(Atomic program.)\label{atomic}
An atomic program is a subtype of program type, $\mbf{atm}<:\mbf{prgm}$.
An atomic program $p~x~y:\mbf{atm}$ must include type checking for the assigned values of every element of its I/O lists.
If for any value assigned element of the I/O lists, $x$ and $y$, there is a type violation the program halts prematurely as a type violation error.
Otherwise the atomic program returns the assigned valued output, $y$, where $y$ may be the empty list.
Atomic programs may call other atomic programs but each atomic program introduces a new computational operation.
\end{definition}

Atomic programs can be partitioned into the three subtypes of type and relation checking, value assignment and type assignment.

\begin{definition}(Type checking programs.)\index{type checking program}
A type checking program, $p~x~y$, is an atomic program with the following properties.
\begin{itemize}

\item The output list $y$ is the empty list so that type checking programs have the representation $p~x~[~]$.

\item The type of the assigned values of every element of the input list is checked upon entry.

\item If a type violation is encountered the program halts prematurely with a type violation error.

\end{itemize}
A type checking program is assigned the type $p~x~y:\mbf{chck}$, where $\mbf{chck} <: \mbf{atm}$.

\end{definition}

Type checking within a program is an action that checks the type of the assigned value of a given variable.
Type checking may also include the checking of some relation between its input variables.
For example, type checking for valued assigned variables that are integers, say $a:\mbf{int}$ and $b:\mbf{int}$, may include a check for value assigned equality, $a=b$, or value assigned inequality, $a<b$.
In other words a type violation error will include failure of at least one of the actions of type checking, $a:\mbf{int}$, $b:\mbf{int}$, and the value assigned equality or inequality.

\begin{definition}(Value assignment programs.)\index{value assignment program}
	Value assignment programs are atomic programs that combine all of the actions of entry type checking, value assignment and type assignment. 
	A value assignment program, $p~x~y$, has the following properties.
	
	\begin{itemize}
		
		\item The type of the assigned values of every element of the input list, $x$, is checked upon entry.
		
		\item If there is a type violation of at least one element of the input list the program halts prematurely with a type violation error.
		
		\item If, upon entry, there are no type violations, a value assignment program then attempts to assign a value to each element of the output list through the action of an assignment map.
		
		\item If there is a type violation of an assigned value of an element of the output list the program halts with a type violation error.
		
		\item If there are no type violations each element of the output list, $y$, is assigned a value and a type consistent with the value assignment.
		
	\end{itemize}
	
	A value assignment program has the type $p~x~y:\mbf{asgn}$, where $\mbf{asgn} <: \mbf{atm}$.
	
\end{definition}

Object $b$ has type $\mbf{t}$ is denoted by $b:\mbf{t}$.
Throughout we will use the notation
\be
b::\mbf{t}
\ee
to denote the action of assigning object $b$ the type $\mbf{t}$.\index{type assignment}

\underline{Abstract types.}
There will be situations where objects of some specific type will be assigned a new type.
These newly assigned types will often be referred to as abstract types\index{abstract type}. 

\begin{definition}(Type assignment programs.)\index{type assignment program}
A type assignment program, $p~x~y$, is an atomic program with the following properties.

\begin{itemize}

\item Type assignment programs do not introduce new variables nor do they modify the assigned values of the input variables.
They admit input objects that have already been assigned a value with a well defined type.
Hence type assignment programs have an empty output list, $y$, with the representation $p~x~[~]$.

\item The input list will contain an object that is the target of the new type assignment.
The input list may also include additional objects that serve as parameters upon which the new subtype is dependent.

\item The type of the assigned values of every element of the input list is checked upon entry.
The check is performed on the type already assigned to each input variable upon entry and not the type that is to be assigned.

\item If there is a type violation the program halts prematurely with a type violation error.

\item If, upon entry, there are no type violations, a type assignment program then assigns to the target element of the input list a new type.   

\item Once the target variable is assigned the new type it is internally stored in memory as such so that if the target variable is encountered as input in a subprogram of a larger program list it is identified by that assigned type. 
 
\end{itemize}

A type assignment program has the type $p~x~y:\mbf{tasgn}$, where $\mbf{tasgn} <: \mbf{atm}$.

\end{definition}

For each atomic abstract type assignment program, type $\mbf{tasgn}$, there must be an associated type checking program, type $\mbf{chck}$.
The important distinction between programs of type $\mbf{tasgn}$ and programs of type $\mbf{asgn}$ is that the latter create new variables and assign a value to these variables.
Type $\mbf{tasgn}$ programs do not create new variables and do not modify the values assigned to the existing variables that appear in their input list.
The properties of objects associated with abstract types can only be recognized by the machine through a supplied collection of axioms.

\underline{Pseudo-atomic programs.}\index{pseudo-atomic program}
Sometimes it might be convenient to define a program as atomic even though it could otherwise be represented as a program list.
Such programs are said to be \emph{pseudo-atomic programs.}

Consider the program list
\ben\label{pl001}
[p(k)~x(k)~y(k)]_{k=1}^n
\een
where, as usual, $x(k)$ and $y(k)$ are lists and we write
\be
x=[x(k)]_{k=1}^n,~y=[y(k)]_{k=1}^n
\ee
Let
\be
p^\prime~x^\prime~y^\prime
\ee
be defined as the pseudo-atomic program of the program list (\ref{pl001}).
The input list, $x^\prime$, of the pseudo-atomic program, $p^\prime~x^\prime~y^\prime$, is given by
\be
x^\prime= unique[x \setminus [y~cst]]=piv[x] \setminus cst=free[x]
\ee
where $cst$ is the list of constants, and the output list, $y^\prime$, is given by
\be
y^\prime= y \setminus [x \cap y]
\ee

We see that the input list, $x^\prime$, of the pseudo-atomic program, $p^\prime~x^\prime~y^\prime$, removes all constants and input variable names associated with intermediate calculations in the program list (\ref{pl001}).
Repetitions of variable names are removed by the $unique$ function.
The output list, $y^\prime$, of the pseudo-atomic program, $p^\prime~x^\prime~y^\prime$, removes all output variable names that are employed as intermediate calculations of the program list, (\ref{pl001}), retaining only the primary output variable names.

While the internal algorithm of the pseudo-atomic program, $p^\prime~x^\prime~y^\prime$, is based on a program list of atomic programs, the actual program list (\ref{pl001}) is not seen by the machine.
As with standard atomic programs, because the internal algorithm of a pseudo-atomic program is not accessible to the machine in explicit form it will be necessary to supply a collection of rules or axioms that describe it.

For pseudo-atomic programs associated with long program lists the collection of rules or axioms that need to be supplied might be very large.
The convenience of defining a pseudo-atomic program has to be weighed against the number of rules that have to accompany it.    

\underline{Notes.}

\begin{itemize} 

\item All programs will be constructed from atomic programs through the construction rules to be presented in the following chapters.
Hence all programs will contain the action of type checking for the assigned values of all elements of their I/O lists.

\item Due to the I/O dependency condition there is no general rule that allows the repetition of subprograms of a program list.
However, repetition of subprograms with an empty list output is allowed and will not effect the computability of the program list.
For computational efficiency such repeated subprograms are redundant and should be avoided. 

\end{itemize}

\chapter{Construction Rules.}\label{ccr}

\section{Introduction.}

Our objective is to construct a formal language from which we can determine the computability of programs, with particular interest in programs that are designed to model fully discrete dynamical systems.
Adopting a language based on programs under the constraints of a machine environment, $\mathfrak{M}(mach)$, will require a departure from conventional languages employed in formal systems of proof theory.
Here we will lay down a collection of program construction rules that better reflect the operational constraints of our language based on programs on a working platform $\mathfrak{M}(mach)$.

To this end we will largely deal with objects that are recognized by the machine from their string structure.
These include subtypes of strings or lists of strings such as machine integer scalars and vectors and programs.
As a result we will be dealing with a low level language that will lack the expressiveness found in many formal systems of proof theory.
However, we will demand that the language possess the power of analysis at a sufficiently high level.
We will also demand that the language be soundly rooted as a primitive on top of which theories requiring higher levels of abstractions can be built.

The construction rules\index{construction rules} form the basis of the program VPC and can be regarded as the general inference rules that are applied to specific applications.
An application, $S$, sometimes referred to as a theory, comes with its own collection of atomic programs and axioms that serve as initializing input data to VPC.
Programs of $S$ are constructed inductively from these atomic programs as program lists.
Later we will include program constructions that are based on disjunctions.
Computability of the programs of $S$ is defined in terms of the value assignments of program I/O lists and governed by the axioms associated with the application, $S$, in conjunction with the construction rules.

\underline{Shorthand notation.} From this point on throughout this book we will often use the shorthand notation of representing programs with lower case letters so that, for example, by
\be
a:\mbf{prgm}
\ee
it is understood that the dummy variable $a$ is assigned the value of a string, or list of strings, of subtype $\mbf{prgm}$.
More generally we regard $a$ as being assigned the value of type $\mbf{prgm}[n]$, for some $n:\mbf{int0}$.
We refer to the I/O lists of $a$ by the notation
\be
\bal
x_a = & [x(k)]_{k=1}^n \\
y_a = & [y(k)]_{k=1}^n \\
\eal
\ee
where $x(k)$ and $y(k)$, $k=1,\ldots,n$, are I/O lists of the atomic programs, $p(k)~x(k)~y(k)$, $k=1,\ldots,n$, that are subprograms of the program list.
From the perspective of machine hardware the value assignment $a::\mbf{prgm}[n]$ involves the allocation of an address that links the dummy variable name $a:\mbf{char}$ to an object type $\mbf{prgm}[n]$ that is stored in memory as a list of atomic programs.

Of particular interest will be sublists of programs, i.e. a program $b:\mbf{prgm}[m]$ will be a sublist of the program $a:\mbf{prgm}[n]$, written $b \subseteqq a$, if every subprogram of $b$ is a subprogram of $a$.

The concatenation of two programs $a:\mbf{prgm}[n]$ and $b:\mbf{prgm}[m]$,
\be
~[a~b]
\ee
is often performed to construct a program of type $\mbf{prgm}[n+m]$.
The validity of a program concatenation will be subject to the compatibility of I/O variable names as outlined in Definitions \ref{prog} and \ref{prog1}.
This means that the type assignment $[a~b]::\mbf{prgm}$ is not automatic and will be invalid if there is a violation of the I/O dependency condition.

\section{Program extensions.}

The main idea behind our formal system is to construct computable programs as extensions of programs that are known to be computable.
The following definition formalizes this idea. 

\begin{definition}\label{ce}(Program extension.)\index{program extension}
	A program $c:\mbf{prgm}$ is called an \textbf{extension} of the program $p:\mbf{prgm}$ and assigned the subtype
	\be
	c:\mbf{ext}[p]
	\ee
	if all of the following conditions are satisfied.
	
	\begin{enumerate}
		
        \item $[p~c]:\mbf{prgm}$. 
		
		\item The input list, $x_c$, of the program $c$ cannot introduce new variable names other than constants, i.e.
		\be
		x_c \subseteqq [x_p~y_p~cst]
		\ee
		where $x_p$ and $y_p$, respectively, are the input and output lists, respectively, of the program $p$ and $cst$ is the list of constants.  
		
		\item If $p$ is computable with respect to the assigned values of its primary input variables, $piv[x_p]$, then the program $s=[p~c]$ is also computable for the same value assigned input.

	\end{enumerate}
	We write $c:\mbf{ext}[p]$ to stress that $c$ is an extension associated with $p$, where $\mbf{ext}[p]<:\mbf{prgm}$.
	The program $s=[p~c]$, such that $c:\mbf{ext}[p]$, is simply referred to as an \textbf{extended program}.\index{extended program}
\end{definition}

Condition 1 of Definition \ref{ce} requires that $[p~c]$ be a program.
This condition is necessary to ensure that there is no conflict of variable names of the elements of the output list of the program $c$ with the variable names of the elements of the I/O lists of the program $p$, i.e.
\be
y_c \cap [x_p~y_p] = [~]
\ee
The primary input variables, $piv[x_s]$, of the extended program $s=[p~c]$ can only differ from the primary input variables $piv[x_p]$ by constants not contained in $x_p$, i.e. $piv[x_s] \subseteqq [piv[x_p]~cst]$.
It is in this sense that computability of $s$ is dependent on the valued assigned list $piv[x_p]$.

\vspace{5mm}

\begin{definition}\label{ice}(Irreducible program extension.)\index{irreducible program extension}
	A program $c:\mbf{prgm}$ is called an \textbf{irreducible extension} of the program $p:\mbf{prgm}$ and assigned the subtype
	\be
	c:\mbf{iext}[p]
	\ee
	if the following conditions are satisfied.
	
	\begin{enumerate}
		
		\item $c:\mbf{ext}[p]$.
		
		\item The program $p$ is irreducible in the following sense. There does not exist a program $r=[q~c]$ such that $q \varsubsetneqq p$ ($q$ is a strict sublist of $p$) and $c:\mbf{ext}[q]$.
		
	\end{enumerate}
	The program $s=[p~c]$, such that $c:\mbf{iext}[p]$, is said to be an \textbf{irreducible extended program}.\index{irreducible extended program}
	The programs $p$ and $c$, respectively, are said to be the premise and conclusion, respectively, of the irreducible extended program $s$.
	The hierarchy of subtypes is $\mbf{iext}[p]<:\mbf{ext}[p]$.
\end{definition}

\section{Higher order programs.}\label{hop}

Programs are strings, or lists of strings, with a well defined structure and may serve as assigned values of elements of an I/O list of a program.
A program will be said to be a higher order program if the assigned values of the elements of its I/O lists are of type $\mbf{prgm}$.
Higher order programs\index{higher order program} essentially recognize programs as strings, or lists of strings, with a particular structure, namely that outlined in Definitions \ref{prog} and \ref{prog1}.

We can partition type $\mbf{prgm}$ objects into the subtypes $\mbf{prgm}^{(k)}$, $k:\mbf{int0}$.
A type $\mbf{prgm}^{(k)}$, $k \geq 1$, object is a $k$-order program whose I/O lists contain elements that can be assigned values that are type $\mbf{prgm}^{(k-1)}$ objects.

The elements of the I/O lists of zeroth-order programs, type $\mbf{prgm}^{(0)}$ objects, are assigned values such as machine integer scalars and vectors, but exclude type $\mbf{prgm}$ objects.
The elements of the I/O lists of first-order programs, type $\mbf{prgm}^{(1)}$ objects, are assigned values of strings of a specific subtype, namely zeroth-order programs.
\emph{First-order programs do not recognize the value assignments of the I/O lists of the zeroth-order programs.}

We will often omit the superscript that indicates the order of the program type and just write $\mbf{prgm}$.
The order of the program type of an object should be evident in the context that it appears.

\underline{Higher order atomic programs.} Higher order programs will be constructed from the atomic programs whose names are given in the tables below.
Some of the atomic programs are based on definitions that will be presented later in the text.  
All atomic programs check the entry type of the input variables.
If there is a type violation of a value assigned input or output variable the atomic program halts with an error message.

\emph{Program type} $\mbf{chck}$.

\begin{tabular}{|l|l|}
	\hline
	\textbf{Syntax} & \textbf{Type checks} \\
	\hline \hline
	$typep~[p]~[~]$ & $p:\mbf{prgm}$ \\
	\hline
	$eqp~[p~q]~[~]$ & $p,q:\mbf{prgm}$, $p \equiv q$ \\
	\hline
	$eqio~[p~q]~[~]$ & $p,q:\mbf{prgm}$, $p \equiv_{io} q$ \\
	\hline
	$sub~[p~q]~[~]$ & $p,q:\mbf{prgm}$, $p \subseteqq q$ \\
	\hline
\end{tabular}

\emph{Program type} $\mbf{asgn}$.

\begin{tabular}{|l|l|l|l|}
	\hline
	\textbf{Syntax} & \textbf{Type checks} & \textbf{Assignment} & \textbf{Type} \\
	                &                 & \textbf{map}        & \textbf{assignment} \\
	\hline \hline
	$conc~[a~b]~[c]$ & $a,b:\mbf{prgm}$, & $c:=[a~b]$ & $c::\mbf{prgm}$ \\
	 & $y_b \cap [x_a~y_a] = [~]$ & & \\
	\hline
	$disj~[a~b]~[d]$ & $a,b:\mbf{prgm}$, & $d:=a~|~b$ & $d::\mbf{atm}$ \\
	 & $free[x_a] \equiv free[x_b]$, & & \\
     & $y_a \setminus x_a=y_b \setminus x_b$ & & \\
	\hline
\end{tabular}

\underline{Abstract type checking and assignment.}

\emph{Program type} $\mbf{chck}$.

\begin{tabular}{|l|l|}
	\hline
	\textbf{Syntax} & \textbf{Type checks} \\
	\hline \hline
	$ext~[p~c]~[~]$ & $p:\mbf{prgm}$, $c:\mbf{ext}[p]$ \\
	\hline
	$false~[p]~[~]$ & $p:\mbf{false}$ \\
	\hline
\end{tabular}

\newpage 

\emph{Program type} $\mbf{tasgn}$.

\begin{tabular}{|l|l|l|}
	\hline
	\textbf{Syntax} & \textbf{Type checks} & \textbf{Type assignment} \\
	\hline \hline
	$aext~[p~c]~[~]$ & $p,c:\mbf{prgm}$,~$[p~c]:\mbf{prgm}$, & $c::\mbf{ext}[p]$ \\
      & $x_c \subseteqq [x_p~y_p~cst]$ & \\
	\hline
	$afalse~[p]~[~]$ & $p:\mbf{prgm}$ & $p::\mbf{false}$ \\
	\hline
\end{tabular}

\parskip10pt

\underline{Notes.}

\begin{itemize}
	
	 \item In this book we will largely make use of zeroth and first-order programs.
	 However, our formal system can be extended to include statements of higher order constructs that make use of objects of type $\mbf{prgm}^{(k)}$, $k \geq 2$.
	 We will discuss this in a little more detail in a later chapter.
	 For the most part we will remain focused on the most basic foundations of our formal system and leave the details of these higher order constructs for future development.
	
	\item An application or theory comes with its own atomic programs.
	In most of our applications we will regard the above atomic programs to be first-order programs.
	The dummy input variables of atomic first-order programs are assigned the values of zeroth-order programs that are associated with a specific application.
	In general the input zeroth-order programs may be lists of the application specific atomic programs, i.e. they are type $\mbf{prgm}[n]$ for some $n:\mbf{int0}$.
	The dimensions, $n$, of the program list is assumed to be recognized upon entry to the first-order program and does not appear explicitly in the input list.
	This kind of dimension free formulation for arrays in general will be adopted throughout this book and is discussed in more detail Chapter \ref{vec}. 
	The adoption of a dimension free formulation is done out of convenience rather than necessity and requires a slight change in a model for array memory storage where an array variable name is linked to an address that includes its dimensions along with the value and type assignments of its elements. 
		
	\item An application specific axiom or theorem of the form $[p~c]$, $c:\mbf{iext}[p]$, is stored in the file \emph{axiom.dat}.
	By default the type assignment $c:\mbf{ext}[p]$ is made.
	Otherwise a program can only acquire the type $\mbf{ext}[p]$ through the type assignment program named $aext$.
	
	\item Objects of type $\mbf{false}$ that are defined as constants of type $\mbf{prgm}$ are stored in an initializing file.
	Otherwise a program can only acquire the type $\mbf{false}$ through the type assignment program named $afalse$ that appears in the axioms and theorems of falsity stored in the file $\emph{axiom.dat}$.

\end{itemize}

\section{Extended program derivation.}\index{extended program derivation}

\begin{definition}(Extended program derivation.)
	An \textbf{extended program derivation}, $s$, with respect to the program $[q~c]:\mbf{prgm}$ is an assignment $s:=[p~c]$ subject to the conditions $q \subseteqq p$, $c:\mbf{ext}[q]$ and $s:\mbf{prgm}$.
	It is constructed from the higher order program defined by
	\ben\label{epd}
	[sub~[q~p]~[~]~ext~[q~c]~[~]~conc~[p~c]~[s]]
	\een
\end{definition}

\underline{Derivations and proofs.}\index{derivation}\index{proof}
A program $[p_i]_{i=1}^m$, $p_i:\mbf{prgm}$, $i=1, \ldots,m$, is called a derivation if it is constructed from a sequence of extended program derivations.
Let $s_k=[p_i]_{i=1}^k$ and consider the following iteration.

\begin{itemize}
	
	\item The program $s_n=[p_i]_{i=1}^n$, for some $n<m$, serves as a list of premises of the derivation.
	
	\item For each iteration $i=n+1,\ldots,m$, the statement $p_i$, in an extended program derivation
	\be
	[sub~[q_i~s_{i-1}]~[~]~ext~[q_i~p_i]~[~]~conc~[s_{i-1}~p_i]~[s_i]]
	\ee
	is introduced from some known extension $p_i:\mbf{ext}[q_i]$ such that $q_i \subseteqq s_{i-1}$.
	
\end{itemize}

The program $s_m = [p_i]_{i=1}^m$ is called a \emph{derivation program}\index{derivation program} or simply a derivation.
A derivation program may be called a proof if its final statement is of particular interest in relation to its premise program.
An irreducible extended program that is extracted from a proof is called a \emph{theorem}\index{theorem}.
An irreducible extended program for which no derivation is known will be called an \emph{axiom}\index{axiom} if it belongs to a collection of generators from which theorems can be derived. 

An axiom and theorem can be represented as a program list $[q~c]$, where $c:\mbf{iext}[q]$.
If $q$ represents the program list $[p(k)~x(k)~y(k)]_{k=1}^n$ and $c$ represents the program $p(n+1)~x(n+1)~y(n+1)$, the axiom/theorem can be written as a vertical list  
\ben\label{atrep}
\begin{array}{l}
	p(1)~x(1)~y(2) \\
	~~~~~~~~~\vdots \\
	p(n)~x(n)~y(n) \\
	\hline
	p(n+1)~x(n+1)~y(n+1) \\
\end{array}
\quad
\een
where the conclusion program is placed below the horizontal line to distinguish it from the premise program.
(Keep in mind that each $x(k)$ and $y(k)$ is a list.) 
By Condition 2 of Definition \ref{ce} we must have
\be
x(n+1) \subseteqq [[x(k)]_{k=1}^n~[y(k)]_{k=1}^n~cst]
\ee
Once constructed, the axiom/theorem, $[q~c]$, can be embedded as a subprogram in a larger program list. 

There is a need to adopt some convention that reflects the constraints imposed by working in an environment $\mathfrak{M}(mach)$, where the list of machine parameters is given by $mach = [nchar~nstr~nlst~nint]$.
Proofs will always be expressed as program lists.
An irreducible extended program can only be considered to be a theorem in $\mathfrak{M}(mach)$ if the derivation program of its proof has a list length at most $nlst$.
An irreducible extended program that has no proof in $\mathfrak{M}(mach)$ but has a proof in a larger machine $\mathfrak{M}(mach^\prime)$ with maximum list length $nlst^\prime > nlst$ can only be regarded as a potential axiom in $\mathfrak{M}(mach)$.

\vspace{5mm}

\underline{Notes.}

\begin{itemize}
	
	\item The last statement, $conc~[p~c]~[s]$, in the definition of an extended program derivation (\ref{epd}) is not necessarily computable if the first two statements are computable.
	A necessary condition for the extended program derivation (\ref{epd}) to be computable is that $[p~c]:\mbf{prgm}$.
	The program concatenation $[p~c]$ may fail if for instance $y_c \cap [x_p~y_p] \neq [~]$.
	When attempting to construct $[p~c]$ from an extended program derivation we will always be able to choose the variable names of the elements of the output list, $y_c$, of the program $c$ such that they do not conflict with the variable names of the elements of the I/O lists, $x_p$ and $y_p$, of the program $p$.		
	
	\item The representation of axioms/theorems in the form (\ref{atrep}) has some similarity with the format of logic programming.
	However, our formal system will differ from the general area of logic programming and its typed extensions in certain crucial respects.
	
	\item It should be evident by now that our formal system is aimed at constructing proofs that are based on the computability of programs of real world computation.
	This should be distinguished from the traditional formal languages of proof theory where proofs are based on the semantics of truth value assignments.
	The general category of computability logic has emerged in recent years as a redevelopment of logic \cite{jap}.
    Our formal system can be put into the same category of computability logic but taking a very different approach to that described in \cite{jap}.
		
\end{itemize}

\section{The program extension rule.}\label{stper}

Derivations and proofs are constructed from rules that are expressed as higher order irreducible extended programs using the atomic programs of Section \ref{hop}.
We start with the principle inference rule called the program extension rule\index{program extension rule}.

\emph{Program extension rule.}
\be\label{per}
\begin{array}{l}
	sub~[q~p]~[~] \\
	ext~[q~c]~[~] \\
	conc~[p~c]~[s] \\
	\hline
	aext~[p~c]~[~] \\
\end{array}
\quad \textsf{per} 
\ee
The three statements that make up the premise program represents an extended program derivation, (\ref{epd}). 
The program extension rule, \textsf{per}, states that if $s:=[p~c]$ is an extended program derivation with respect to the extension $c:\mbf{ext}[q]$ then it follows that $c$ is also an extension of $p$.
The conclusion program of the program extension rule is a type assignment $c::\mbf{ext}[p]$.

The following construction rule states that once assigned, the property of an extension is retained.
In other words, once a program, $c$, has been assigned the type $\mbf{ext}[p]$, for some program $p$, it is stored in memory as such so that it is recognized as that type whenever it is accessed by any following subprogram of a higher-order program list. 

\emph{Retention of subtype assignment.}
\be
\begin{array}{l}
	aext~[p~c]~[~] \\
	\hline
	ext~[p~c]~[~] \\
\end{array}
\quad \textsf{cr1}
\ee

By definition, for each program, $c$, that is an extension of a program, $p$, there exists an extended program $s=[p~c]$.
The following rule constructs the program $s$ given $c:\mbf{ext}[p]$.

\emph{Extended program construction.}
\be
\begin{array}{l}
	ext~[p~c]~[~] \\
	\hline
	conc~[p~c]~[s] \\
\end{array}
\quad \textsf{cr2}
\ee

The formal system based on the program extension rule, \textsf{per}, along with \textsf{cr1}-\textsf{cr2} and the additional construction rules that will follow, will be referred to as PECR\index{PECR} (Program Extension Construction Rules).
The formal system PECR can be regarded as the rules of inference that are designed to be applied on a working platform $\mathfrak{M}(mach)$ and forms the basis of the program VPC.

\underline{Notes.}

\begin{itemize}
	
	\item We will often deal with objects that are subtypes of strings or lists of strings such as machine integer scalars, vectors and programs. 
	All of these objects have a well defined string structure as specified by their definitions and are recognized by the machine.
	
	The program $aext~[p~c]~[~]$ makes the type assignment $c::\mbf{ext}[p]$.
	A machine can readily verify Conditions 1 and 2 of Definition \ref{ce} from its string structure.
	However, from the perspective of a feasible computation, the machine has no general way of recognizing that a concatenation of programs has the property associated with computability as outlined in Condition 3 of Definition \ref{ce}.
	Consequently, a machine can only interpret objects of type $\mbf{ext}$ through the properties embedded in the construction rules.
	In this sense the type $\mbf{ext}$ can be referred to as an abstract type\index{abstract type}.  
	 	
\end{itemize}

\section{Program and I/O equivalence.}
Program equivalence\index{program equivalence} refers to programs that may appear to have a different structure but are functionally identical.
Program equivalence will be defined in terms of sublists and are associated with program lists whose subprograms appear in a different sequential order.
This definition will be extended later to include disjunctions.

\begin{definition}\label{equiv0}(Program equivalence.)
Two programs $p:\mbf{prgm}$ and $q:\mbf{prgm}$ are said to be program equivalent provided that $p \subseteqq q$ and $q \subseteqq p$.
Program equivalence is denoted by $p \equiv q$ and is reflexive, symmetric and transitive.

\end{definition}

The second important kind of equivalence refers to programs where the variable names of the elements of their I/O lists differ but can be associated by some degree of functionality.

\begin{definition}\label{equivio}(I/O equivalence.)\index{I/O equivalence}
Consider two programs with the list representations $[p(i)~x(i)~y(i)]_{i=1}^n$ and $[p(i)~\bar{x}(i)~\bar{y}(i)]_{i=1}^n$.
Let
\be
x(i)=[x(i,j)]_{j=1}^{nx(i)},~~~\bar{x}(i)=[\bar{x} (i,j) ]_{j=1}^{nx(i)}, \qquad nx(i):\mbf{int1},~i=1,\ldots,n
\ee
\be
y(i)=[y(i,j)]_{j=1}^{ny(i)},~~~\bar{y}(i)=[\bar{y}(i,j) ]_{j=1}^{ny(i)}, \qquad ny(i):\mbf{int0},~i=1,\ldots,n
\ee
The program $[p(i)~\bar{x}(i)~\bar{y}(i)]_{i=1}^n$ is I/O equivalent to the program \\
$[p(i)~x(i)~y(i)]_{i=1}^n$ provided that all of the following conditions are satisfied.

\begin{itemize}

\item If $x(i,k) = x(j,l)$ then $\bar{x}(i,k) = \bar{x}(j,l)$, \\
$1 \leq k \leq nx(i)$, $1 \leq l \leq nx(j)$, $1 \leq j \leq i$, $1 \leq i \leq n$.

\item If $x(i,k) = y(j,l)$ then $\bar{x}(i,k) = \bar{y}(j,l)$, \\
$1 \leq k \leq nx(i)$, $1 \leq l \leq ny(j)$, $1 \leq j \leq i-1$, $2 \leq i \leq n$.
\item If $x(i,k)$ is a constant then $\bar{x}(i,k)$ is the same constant, \\
$1 \leq k \leq nx(i)$, $1 \leq i \leq n$.

\end{itemize}
We write
\be
[p(i)~\bar{x}(i)~\bar{y}(i)]_{i=1}^n \equiv_{io} [p(i)~x(i)~y(i)]_{i=1}^n
\ee
to mean that $[p(i)~\bar{x}(i)~\bar{y}(i)]_{i=1}^n$ is I/O equivalent to the program \\
$[p(i)~x(i)~y(i)]_{i=1}^n$.

\end{definition}

I/O equivalence does not satisfy the property of symmetry.  

\underline{Notes.}

\begin{itemize}

\item In the definition of program equivalence the program lists of $p$ and $q$ will usually be permutations of each other.
This includes the case where $p$ and $q$ are identical programs.  
However, there is the possibility that the lengths of the program lists of $p$ and $q$ are not the same (see the notes of Section 2.3).
This is because repetitions of subprograms of a program list are allowed for subprograms with an empty list output.
Such repetitions introduce redundancies and should be removed before a program equivalence check is performed.
   
\end{itemize}

\section{Additional construction rules.}

In applications, proofs are constructed by the recursive application of the program extension rule through an extended program derivation.
In VPC there are internal procedures that employ some additional rules that are listed below.
These rules largely follow from the definitions.
In the next chapter we will also include rules associated with false programs and disjunctions. 

The empty list program\index{empty program} is an assigned value of the alphanumeric variable name $ep$ and can be regarded as a constant for type $\mbf{prgm}$ objects.

\emph{I/O equivalence.}
\be
\begin{array}{l}
	typep~[p]~[~] \\
	\hline
	eqio~[p~p]~[~] \\
\end{array}
\quad \textsf{cr3a} \hspace{10mm}
\begin{array}{l}
	eqio~[p~q]~[~] \\
	eqio~[q~r]~[~] \\
	\hline
	eqio~[p~r]~[~] \\
\end{array}
\quad \textsf{cr3b}
\ee
\be
\begin{array}{l}
	ext~[q~c]~[~] \\
	conc~[q~c]~[r] \\
	conc~[p~d]~[s] \\
	eqio~[p~q]~[~] \\
	eqio~[s~r]~[~] \\
	\hline
	aext~[p~d]~[~] \\
\end{array}
\quad \textsf{cr3c}
\ee

\emph{Program equivalence.}
\be
\begin{array}{l}
	typep~[p]~[~]\\
	\hline
	eqp~[p~p]~[~] \\
\end{array}
\quad \textsf{cr4a} \hspace{10mm}
\begin{array}{l}
	eqp~[p~q]~[~] \\
	\hline
	eqp~[q~p]~[~] \\
\end{array}
\quad \textsf{cr4b}
\ee

\emph{Sublists.}
\be
\begin{array}{l}
	eqp~[p~q]~[~] \\
	\hline
	sub~[p~q]~[~] \\
\end{array}
\quad \textsf{cr5a} \hspace{10mm}
\begin{array}{l}
	sub~[p~q]~[~] \\
	sub~[q~p]~[~] \\
	\hline
	eqp~[p~q]~[~] \\
\end{array}
\quad \textsf{cr5b}
\ee
\be
\begin{array}{l}
	sub~[q~p]~[~] \\
	sub~[p~r]~[~] \\
	\hline
	sub~[q~r]~[~] \\
\end{array}
\quad \textsf{cr5c}
\ee

\emph{Sublists of concatenations.}

\be
\begin{array}{l}
	conc~[p~q]~[s] \\
	\hline
	sub~[p~s]~[~] \\
\end{array}
\quad \textsf{cr6a} \hspace{10mm}
\begin{array}{l}
	conc~[p~q]~[s] \\
	\hline
	sub~[q~s]~[~] \\
\end{array}
\quad \textsf{cr6b}
\ee
\be	
\begin{array}{l}
	conc~[p~q]~[r] \\
	sub~[p~s]~[~] \\
    sub~[q~s]~[~] \\
	\hline
	sub~[r~s]~[~] \\
\end{array}
\quad \textsf{cr6c}
\ee

\emph{Concatenation with the empty program (right)}.
\be
\begin{array}{l}
	typep~[p]~[~] \\
	\hline
	conc~[p~ep]~[s] \\
\end{array}
\quad \textsf{cr7a} \hspace{10mm}
\begin{array}{l}
	conc~[p~ep]~[s] \\
	\hline
	eqp~[s~p]~[~] \\
\end{array}
\quad \textsf{cr7b}
\ee

\emph{Concatenation with the empty program (left)}.
\be
\begin{array}{l}
	typep~[p]~[~] \\
	\hline
	conc~[ep~p]~[s] \\
\end{array}
\quad \textsf{cr8a} \hspace{10mm}
\begin{array}{l}
	conc~[ep~p]~[s] \\
	\hline
	eqp~[s~p]~[~] \\
\end{array}
\quad \textsf{cr8b}
\ee

\emph{Non-atomic extensions.}
\be 
\begin{array}{l}
	ext~[p~c]~[~] \\
	eqp~[c~d]~[~] \\
	\hline
	aext~[p~d]~[~] \\
\end{array}
\quad \textsf{cr9a} \hspace{10mm}
\begin{array}{l}
	ext~[p~a]~[~] \\
	ext~[p~b]~[~] \\
	conc~[a~b]~[s] \\
	\hline
	aext~[p~s]~[~] \\
\end{array}
\quad \textsf{cr9b}
\ee

To these axioms we include common application rules such as I/O type axioms and the substitution rule.
These are presented in Chapter \ref{appl}.

\underline{Notes.}

\begin{itemize}
	
\item The substitution rule can be applied as an axiom to the programs named $eqp$ and $conc$.

\item Program equivalence is also transitive, i.e.
\be
\begin{array}{l}
	eqp~[p~q]~[~] \\
	eqp~[q~r]~[~] \\
	\hline
	eqp~[p~r]~[~] \\
\end{array}
\ee
This is not included as an axiom since it follows from the substitution rule (see Section \ref{ssr}).

\item As stated in \textsf{cr3a}-\textsf{cr3b}, I/O equivalence is reflexive and transitive but is not symmetric.

\item The I/O equivalence program $eqio~[p~q]~[~]$ does not satisfy the substitution rule as a general rule.

\item Rule \textsf{cr5c} states that sublists are transitive.
Sublists also satisfy the property of reflexivity
\be
\begin{array}{l}
	typep~[p]~[~] \\
	\hline
	sub~[p~p]~[~] \\
\end{array}
\ee
This is not stated as an axiom because it can be derived from \textsf{cr4a} and \textsf{cr5a}.

\item Note that \textsf{cr9a} is valid for extensions that are atomic since identical programs are program equivalent.

\item We will always generate proofs such that the conclusion program of an extended program derivation is an atomic program.
The construction rules \textsf{cr9a}-\textsf{cr9b} state that this need not always be the case and an extension can be a program that has a list representation.
Extensions can be built up as program lists by repeated applications of \textsf{cr9b}.
The construction rule \textsf{cr9b} can be made
redundant if extensions that can be represented by program lists are always replaced by pseudo-atomic programs.

\item It should also be noted that if we insist that program extensions must always be atomic then the fourth statement in axiom \textsf{cr3c} can be removed.

\item Similar to the rule \textsf{cr9a} is the rule for equivalent premise programs
\be 
\begin{array}{l}
	ext~[p~c]~[~] \\
	eqp~[p~q]~[~] \\
	\hline
	aext~[q~c]~[~] \\
\end{array}
\ee
This is not stated as an axiom because it can be derived (see Chapter \ref{cpcap}).

\item It is important to note that the substitution rule should not be directly applied to the program $ext~[p~c]~[~]$ as an axiom.
In Chapter \ref{cpcap} it will be shown that it does satisfy the substitution rule as a derivation.

\end{itemize}

\section{Options file.}\label{sof}

In VPC, the program, $c$, of an extended program derivation
\ben\label{epd1}
[sub~[q~p]~[~]~ext~[q~c]~[~]~conc~[p~c]~[s]]
\een
is regarded as an extension if the program $[q~c]$ is program and I/O equivalent to an axiom or theorem that is stored in the file \emph{axiom.dat}.
In other words, derivations in VPC are constructed only with respect to irreducible extended programs $[q~c]$, $c:\mbf{iext}[q]$.
The program extension rule, \textsf{per}, is less restrictive and requires that $c:\mbf{ext}[p]$.
Generality of the application of the program extension rule under this process is not lost.
To see this more clearly consider the following.

Suppose that we are given an extended program derivation (\ref{epd1}) such that $c:\mbf{ext}[q]$ is not an irreducible extension.
Then there must exist a program $\bar{q} \varsubsetneqq q$ ($\bar{q}$ is a strict sublist of $q$) such that $c:\mbf{ext}[\bar{q}]$ and $t=[q~c]$ was obtained from the extended program derivation
\ben\label{epd2}
[sub~[\bar{q}~q]~[~]~ext~[\bar{q}~c]~[~]~conc~[q~c]~[t]]
\een
Since $\bar{q} \varsubsetneqq q \subseteqq p$ we can also construct an extended program derivation
\ben\label{epd3}
[sub~[\bar{q}~p]~[~]~ext~[\bar{q}~c]~[~]~conc~[p~c]~[r]]
\een
and $r$ is identical to the program $s$.
We can now apply the same argument to $\bar{q}$ and so on until we are left with a derivation of $[p~c]$ with respect to a program that is an irreducible extension.

During proof construction, VPC accesses a file \emph{axiom.dat} that initially stores all of the axioms of the application associated with the specific theory under investigation.
Any program, $c$, associated with an extended program $[q~c]$ that is stored as an axiom/theorem in the file \emph{axiom.dat} acquires the type $\mbf{ext}[q]$ by default.
Otherwise a program $c$ can only acquire the type $\mbf{ext}[q]$ through the type assignment program $aext~[q~c]~[~]$.

As proofs are completed the theorems that are extracted from them are also stored in \emph{axiom.dat}.
The program $[q~c]$ of an extended program derivation (\ref{epd1}) is identified as an extended program if it can be matched to an axiom/theorem stored in the file \emph{axiom.dat}.
The matching procedure relies on the program equivalence rules \textsf{cr9a} and \textsf{thm7} (derived in Chapter \ref{cpcap}) along with the I/O equivalence rule \textsf{cr3c}.
In this way each axiom/theorem stored in the file \emph{axiom.dat} acts as a template from which programs of an application can be identified as extensions.

The procedure of generating a new statement of the current derivation program, $p$, can be outlined in the following procedure.
Suppose that $q$ is a sublist of $p$.
An extended program $[q~c]$, $c:\mbf{ext}[q]$, is constructed as follows.

\emph{Step 1.} Find $q^\prime$ such that $q^\prime \equiv q$ and $q^\prime \equiv_{io} \bar{q}$, for some $[\bar{q}~\bar{c}]$, $\bar{c}:\mbf{iext}[\bar{q}]$, stored as an axiom/theorem in the file \emph{axiom.dat}.

\emph{Step 2.} Construct $c^\prime$ such that $[q^\prime~c^\prime]:\mbf{prgm}$ and $[q^\prime~c^\prime] \equiv_{io} [\bar{q}~\bar{c}]$.

\emph{Step 3.} Make the assignment $c:=c^\prime$.

(In a more general context, if $c^\prime$ is non-atomic then any $c$ such that $c \equiv c^\prime$ can be used.) 

In Section \ref{sspotcr} it will be rigorously shown that the program $[q~c]$ constructed in this way is indeed an extended program such that $c=\mbf{ext}[q]$.
Since $q \subseteqq p$ we have by the program extension rule, \textsf{per}, that $[p~c]$ is also an extended program such that $c:\mbf{ext}[p]$ so that the program $c$ may be appended to the current derivation program, $p$.

\underline{Interactive proofs.} At each step of a proof construction, VPC determines the conclusion programs of all possible extended program derivations that can be obtained from the main derivation program with respect to the axioms and theorems that are currently stored in the file \emph{axiom.dat}.
The conclusions of these extended program derivations are listed in an options file, \emph{options.dat}, from which the user may select to append to the main derivation program.
The process is repeated until the proof is completed.

If at any point of a proof construction the user inserts a statement that is not currently stored as an option in the file, \emph{options.dat}, then VPC will halt with an execution error message.

Each option in the options file includes the axiom/theorem label and the associated labels of the subprograms that make up the sublists of the current derivation program that can be matched to a premise program of an axiom/theorem stored in \emph{axiom.dat}.
The procedure is one of extracting all sublists of the current derivation program that are program and I/O equivalent to a premise program of the axioms/theorems stored in \emph{axiom.dat}.    

Extractions of sublists from the current derivation program based on a raw search of all possible permutations followed by an I/O equivalence matching algorithm can be computationally expensive.
VPC employs special techniques that speed up this process by detecting and eliminating unsuccessful matches before a complete sublist extraction and I/O equivalence check is performed.
This significantly reduces the computations making the enumeration of all possible extended program derivations quite manageable.

\section{Connection List.}\label{connection}\index{connection list}

During a proof construction each derived statement comes with an axiom/theorem label and a connection list that records the origin of the statement.

\begin{definition}(Connection list.)\label{clist}
	For each subprogram $p(i)~x(i)~y(i)$ of the derivation program list $[p(i)~x(i)~y(i)]_{i=1}^m$ that is obtained from an extended program derivation is generated a list that contains the line labels of the premises used to obtain that subprogram.
	If $ncl(i)$ is the length of the premise program list of the axiom/theorem labeled $a(i)$ that is employed to infer the statement $p(i)~x(i)~y(i)$ then the associated connection list, $cl(i):\mbf{lst}[ncl(i)]$, is of the form
	\be
	cl(i)=[cl(i,1)~\ldots~cl(i,ncl(i))]
	\ee
	where $1 \leq cl(i,1),~\ldots~,cl(i,ncl(i)) \leq i-1$ are the line labels of the subprograms that make up the sublist
	\be
	~[p(cl(i,j))~x(cl(i,j))~y(cl(i,j))]_{j=1}^{ncl(i)}
	\ee
	that is program and I/O equivalent to the premise program of the axiom/theorem, labeled $a(i)$.
	Each derived statement of a proof is accompanied by an axiom label followed by a connection list.
	The connection list is empty for any subprogram that belongs to the premise of the program derivation.
\end{definition}

Consider the derivation program $p=[p_i]_{i=1}^m$, where $p=[p_i]_{i=1}^n$ is the list of premises and $[p_i]_{i=n+1}^m$ are the statements obtained by a sequence of $m-n$ extended program derivations.
In VPC, derivation programs are output as a vertical list with four columns.
The first column contains the statement label (line number), the second column contains the statement itself and the third and fourth columns contain the axiom/theorem label followed by the connection list.
Statements that are premises of the main derivation program have an empty connection list, i.e. $cl(i)=[~],~i=1,\ldots,n$.
The general output layout can be illustrated as follows. 
\ben \label{proofprgm}
\begin{array}{llll}
	Line  & Statement & Axiom/theorem & Connection~list \\
	number  &  & label & \\
	1      & p_1       & & \\
	\vdots & \vdots    & & \\
	n      & p_n       & & \\
	n+1    & p_{n+1}   & a(n+1) & [cl(n+1,j)]_{j=1}^{ncl(n+1)} \\
	\vdots & \vdots    & \vdots & \vdots \\
	m      & p_{m}     & a(m) & [cl(m,j)]_{j=1}^{ncl(m)} \\
\end{array}
\een
Here $ncl(i)$ is the length of the premise program of the axiom/theorem labeled $a(i)$, $i=n+1,\ldots,m$.
The elements of the connection list, $1 \leq cl(i,1),\ldots,cl(i,ncl(i)) \leq i-1$, are statement labels of the sublist of the program $[p_l]_{l=1}^{i-1}$ that is program and I/O equivalent to the premise program of the axiom/theorem labeled $a(i)$, $i=n+1,\ldots,m$.

\emph{In order that the derivation program $p$ be designated as a proof each variable name of the input list of $p_m$ must coincide with a non-constant variable name that appears in the I/O lists of the premise program of, $p$, or be a constant.
	Otherwise the derivation program $p$ is not a proof.} 

\underline{Extraction of theorems from proofs.}
A derivation program has the form $p=[p]_{i=1}^m$, where $[p_i]_{i=1}^n$ is the list of premises of the proof and $[p_i]_{i=n+1}^m$ are the statements obtained by a sequence of $m-n$ extended program derivations. 
We now describe an algorithm that extracts the theorem, $[[p_i]_{i=1}^n~p_m]$, from the derivation program $p$.  

The connection list of the derived statement, $p_i \in p$, is given by \\
\\
$~~~~~~~~~~~~~~~~cl(i) = [cl(i,j)]_{j=1}^{ncl(i)}, \quad i=n+1,\ldots,m$ \\
\\
where each list $cl(i):\mbf{lst}[ncl(i)]$, $i=n+1,\ldots,m$, contains the line labels, $cl(i,j):\mbf{int1}$, $j=1,\ldots,ncl(i)$, of the statements that were used to infer the program $p_i$.
The pseudo-code (\ref{algclr}), below, traces the conclusion program, $p_m$, back to the elements of the premise program of $p$ through the connection lists.
We refer to this process as the \emph{connection list reduction}\index{connection list reduction}.
The pseudo-code (\ref{algclr}) represents an algorithm that can be written in some imperative language that allows variable names to be reassigned values.
The lengths of the integer lists $b$ and $r$ are dynamically updated during each iteration.  
The function $unique$ is used to remove repetitions of elements of the updated lists.  
After completion of the $do$-loop the list $b$ will contain only labels associated with the premises.

Note that the list $b$ is not updated if it does not contain any common elements with the connection list $cl(i)$.
This indicates that the statement $p_i$ is redundant in the proof of the theorem.
The program appends the line label $i$ to the list $r$ and then continues the cycle of the $do$-loop.
The final list $r$ contains labels of all redundant premises and statements in a proof.

If there are redundant premise statements then $[[p]_{i=1}^n~p_m]$ will not be an irreducible extended program and hence will not be a theorem.
The proof can be reconstructed by discarding the redundant premises and derived statements that correspond to the labels that appear in the list $r$.

\ben\label{algclr}
\begin{array}{l}
	\hline
	\text{algorithm for connection list reduction} \\
	\hline
	b:=cl(m) \\
	r:=[1~\ldots~n] \\
	\text{do $i=m-1,\ldots,n+1$} \\	
	\hspace{5mm} \text{if $i \in b$ then} \\
	\hspace{10mm} b:=b \setminus [i] \\
	\hspace{10mm} b:=unique[[b~cl(i)]] \\
	\hspace{5mm} \text{else} \\
	\hspace{10mm} r:=[r~[i]] \\
	\hspace{5mm} \text{end if} \\
	\text{end do} \\
	r:=r \setminus b	\\
	\hline
\end{array}
\een

\section{Theorem connection lists.}\label{thmconn}

In any proof, the connection list of each statement gives knowledge of the axiom or theorem used to infer that statement along with its dependence on the preceding statements of the derivation program.
Provided that there are no redundant statements, the procedure of connection list reduction traces the conclusion of the proof back to the premises of the derivation program.

In a similar way we can define connection lists for theorems that can be employed to trace back a theorem's dependence on the axioms of the theory\index{theory}.
We shall refer to this process as a \emph{theorem connection list reduction}\index{theorem connection list reduction}.

A theory, $S$, is defined by a collection of atomic programs, a list of constants and a list of axioms
\be
ax=[ax(i)]_{i=1}^{nax}, \quad nax:\mbf{int0},~ax(i):\mbf{prgm}
\ee
Suppose that at any given time there are $nth:\mbf{int1}$ derived theorems.
Let
\be
th=[th(i)]_{i=1}^{nth}, \quad nth:\mbf{int1},~th(i):\mbf{prgm}
\ee
be the current list of theorems.
Each element, $ax(i)$, of the axiom list, $ax$, and each element, $th(i)$, of the theorem list, $th$, is an irreducible extended program of the form $[p~c]$, where $c:\mbf{iext}[p]$ is an irreducible extension of $p$.
While $ax$ and $th$ are lists of programs they are not in themselves meant to be programs.

For each axiom $ax(i) \in ax$ there corresponds an axiom label $axl(i):\mbf{char}$ that is stored in the axiom label list
\be
axl=[axl(i)]_{i=1}^{nax}
\ee
Similarly, for each theorem $th(i) \in th$ there corresponds a theorem label $thl(i):\mbf{char}$ that is stored in the theorem label list
\be
thl=[thl(i)]_{i=1}^{nth}
\ee

In addition, each theorem $th(i) \in th$ comes with a theorem connection list, $tcl(i):\mbf{lst}[ntcl(i)]$, for some $ntcl(i):\mbf{int1}$, that takes the form
\be
tcl(i)=[tcl(i,j)]_{j=1}^{ntcl(i)}
\ee
where each element, $tcl(i,j):\mbf{char}$, of the list, $tcl(i)$, is either an axiom or theorem label.
The theorem connection list $tcl(i)$ is the list of all axiom/theorem labels employed in the proof of the theorem $th(i)$.
In terms of the format of a derivation program expressed by (\ref{proofprgm}) the theorem connection list includes all of the axiom/theorem labels of column 3 with repetitions removed.

The following psuedo-code is the theorem connection list reduction that traces the dependence of the proof of the theorem, $th(i) \in th$, to the axioms of the theory.
The lengths of the character lists $b$, $c$ and $d$ are dynamically updated at each iteration.

The term, $locelt[b(i)~thl]$, locates the label $b(i)$ in the list $thl$.
At each iteration (outer loop), each element, $b(i) \in b$, that is identical to some theorem label, $thl(j)$, is replaced by the theorem connection list, $tcl(j)$, of that theorem.
Repetitions of elements in the list are removed by the $unique$ function. 
In this way the length of the list $b$ is dynamically updated at each iteration.
When all elements of the list $b$ are axiom labels the outer loop is exited.

\ben\label{algtclr}
\begin{array}{l}
	\hline
	\text{algorithm for the theorem connection list} \\
	\text{reduction of theorem $th(i)$} \\
	\hline
	b:=tcl(i) \\
	\text{do} \\
	\hspace{5mm} \text{if~$b \setminus axl = [~]~exit$} \\
	\hspace{5mm} c:=b \\
	\hspace{5mm} d:=[~] \\	
	\hspace{5mm} \text{do $i=1,\ldots,length[b]$} \\  
	\hspace{10mm} \text{if $b(i) \in thl$ then} \\
	\hspace{15mm} j:=locelt[b(i)~thl] \\
	\hspace{15mm} c:=c \setminus [b(i)] \\
	\hspace{15mm} d:=unique[[d~tcl(j)]] \\
	\hspace{10mm} \text{end if} \\
	\hspace{5mm} \text{end do} \\
	\hspace{5mm} b:=unique[[c~d]] \\
	\text{end do} \\
	\hline
\end{array}
\een

\chapter{Disjunctions and False Programs}\label{cdafp}

\section{Axioms/theorems of falsity}\label{satof}

While a program $p:\mbf{false}$ will halt with an execution error for any value assigned input it does not necessarily follow that the extended program derivation
\ben\label{epd10}
[sub~[q~p]~[~]~ext~[q~c]~[~]~conc~[p~c]~[s]]
\een
will also halt with an execution error.
The reason for this is that (\ref{epd10}) is a list of first-order programs so that type checking is based on program structure and does not recognize value assignments of the I/O lists of the zeroth-order programs $q,p,c$ and $s$.
Hence there is nothing stopping us from allowing the program $p$ of an extended program derivation (\ref{epd10}) to be of type $\mbf{false}$.
If (\ref{epd10}) does not halt with an execution error then the derived object $s$ will be a program but will also be of subtype $\mbf{false}$.
In this section we will demonstrate how extended program derivations can be used to identify false programs.

In general, if a program $p:\mbf{false}$ is irreducible in the sense that there does not exist a program $q \varsubsetneqq p$ ($q$ is a strict sublist of $p$) such that $q:\mbf{false}$ then the statement $p:\mbf{false}$ represents an axiom or theorem of falsity.
Axioms and theorems of falsity are higher order constructs of irreducible extended programs with an empty list premise and can also be written as
\ben\label{false}
\begin{array}{l}
	~~ \\
	\hline
	false~[p]~[~] \\
\end{array}
\een
(See Section \ref{shop} for more details.)

The zeroth-order programs that appear as input to axioms of falsity, are application specific.
As such they are user supplied and can be regarded as constants for type $\mbf{prgm}$ objects (more specifically type $\mbf{prgm}^{(0)}$ objects) associated with the application.  

To the construction rules we introduce the additional rules of falsity.

\emph{Sublist falsity rule.}

\be
\begin{array}{l}
	sub~[q~p]~[~] \\
	false~[q]~[~] \\
	\hline
	afalse~[p]~[~] \\
\end{array}
\quad \textsf{flse1}	
\ee

\emph{Retention of subtype assignment.}
\be
\begin{array}{l}
	afalse~[p]~[~] \\
	\hline
	false~[p]~[~] \\
\end{array}
\quad \textsf{flse2}	
\ee

Consider a premise program $[p_i]_{i=1}^n$.
Since we have allowed the premise program to be of type $\mbf{false}$ we may iteratively generate the program $p=[p_i]_{i=1}^m$, where $[p_i]_{i=n+1}^m$ are the statements obtained by $m-n$ extended program derivations.
If $[p_i]_{i=1}^n:\mbf{false}$ the iteration should continue until a sublist of $[p_i]_{i=1}^m$ can be matched to a false program defined by an axiom or theorem of falsity stored in \emph{axiom.dat}.
When this occurs we abandon the usual extended program derivation format and infer that the premise program $[p_i]_{i=1}^n$ is type $\mbf{false}$.

In VPC the output of the derivation program will appear in the form of the following vertical list.
\be
\begin{array}{llll}
	Line  & Statement & Axiom/theorem & Connection~list \\
	number  &  & label & \\
	1      & p_1       & & \\
	\vdots & \vdots    & & \\
	n      & p_n       & & \\
	n+1    & p_{n+1}   & a(n+1) & [cl(n+1,j)]_{j=1}^{ncl(n+1)} \\
	\vdots & \vdots    & \vdots & \vdots \\
	m      & p_{m}     & a(m) & [cl(m,j)]_{j=1}^{ncl(m)} \\
	m+1     & false  & a(m+1) & [cl(m+1,j)]_{j=1}^{ncl(m+1)} \\
\end{array}
\ee
where the connection lists of the subprograms of the premise are empty lists, i.e. $cl(i)=[~],~i=1,\ldots,n$.
Here $ncl(i)$, $i=n+1,\ldots,m+1$, is the length of the premise program of the axiom/theorem labeled $a(i)$, and
\be
1 \leq cl(i,1),\ldots,cl(i,ncl(i)) \leq i-1
\ee
are statement labels of the sublist of the program $[p_l]_{l=1}^{i-1}$ that is program and I/O equivalent to the premise program of the axiom/theorem labeled $a(i)$.

The first $m$ lines are in the standard derived proof format.
The addition of the final statement, $false$, means that the standard proof format is to be abandoned from which we infer that $[p_i]_{i=1}^n:\mbf{false}$.
We may extract from this statement a theorem of falsity of the form (\ref{false}), provided that the premise program $[p_i]_{i=1}^n$ is irreducible in the sense that there are no strict sublists of $[p_i]_{i=1}^n$ that are of type $\mbf{false}$.

\underline{Notes.}

\begin{itemize}
	
	\item It should be noted that for any $p:\mbf{false}$ the first-order program $false~[p]~[~]$ is computable while the zeroth-order program $p$ is not computable for any value assigned input of $p$.
	
	\item There is an important consequence of allowing the program $p$ of an extended program derivation (\ref{epd10}) to be of type $\mbf{false}$.
	If we accept the program extension rule, \textsf{per}, without exception we must conclude that there exist extensions $c:\mbf{ext}[p]$ such that $p:\mbf{false}$.
	Careful reading of the definition for a program extension, Definition \ref{ce}, does not disallow such a possibility.
	The definition only states that if the premise $p$ is computable for a given value assigned input then it is guaranteed that $[p~c]$ is computable for the same value assigned input.
	
	\item Given that the premise program of an extended program derivation could be of type $\mbf{false}$ one should avoid terminating a derivation before a conclusion leads to a statement of falsity.
	More will be said on this in a later chapter.
	
	\item As discussed earlier, objects that are subtypes of strings or lists of strings such as machine integer scalars, vectors and programs are recognized by the machine from their string structure.
	This is not the case with objects of type $\mbf{ext}$, where the machine can only acquire an interpretation of such objects through the properties embedded in the constructions rules. 
	
	Type $\mbf{false}$ can also be put into the same class of abstract types as $\mbf{ext}$.
	The input zeroth-order programs that are the target of axioms of falsity are assigned the type $\mbf{false}$ by default.
	Otherwise an object can only acquire the type $\mbf{false}$ by inference using the type assignment program $afalse$.
	
	\item It is important to note that the substitution rule should not be directly applied as an axiom to the program $false$.
	The statement
	\be
	\begin{array}{l}
		false~[p]~[~] \\
		eqp~[p~q]~[~] \\
		\hline
		false~[q]~[~] \\
	\end{array}
	\ee
	is a theorem that can be derived from \textsf{cr5a} and \textsf{flse1}-\textsf{flse2}.
	
\end{itemize}

\section{Disjunctions.}\label{sdisj}\index{disjunction}

In conventional theories of logic, disjunctions have an important role to play in the expressiveness and manipulation of formal statements.
Program disjunctions have a more basic role in that they effectively split a program into several parallel programs, where each parallel program is associated with an operand of the disjunction contained within the main program.
Once a disjunction has been split, extended program derivations can be performed independently on each operand program.

\begin{definition}\label{disj}(Disjunction.)
	A disjunction, $d$, of two programs, $a,b:\mbf{prgm}$, is written as 
	\be
	d=a~|~b
	\ee
	where $a$ and $b$ are called the operands of the disjunction.
	The programs $a$ and $b$ need not be atomic.
	We must have
	\begin{enumerate}
		\item $x_d \equiv free[x_a] \equiv free[x_b]$
						
		\item $y_d = y_a \setminus x_a=y_b \setminus x_b$
	\end{enumerate}
	where $y_d$ may be the empty list.
	The disjunction, $d$, is computable for the value assigned input list, $x_d$, if at least one of the operand programs, $a$ and $b$, is computable.
	Otherwise the disjunction will halt with a disjunction violation error.
	Disjunctions that are computable in this sense are said to override type violation errors.
\end{definition}

In PECR the program $disj~[a~b]~[d]$ constructs the program, $d=a~|~b$, as an atomic disjunction program of the operand programs $a$ and $b$.
Notice that the input list, $x_d$, of the disjunction program, $d$, removes all variables that are bound to output variables and constants as they appear in the input lists of the operand programs $a$ and $b$.
Also, the output list, $y_d$, of the disjunction program, $d$, removes all output variables that are used as intermediate calculations in the operand programs. 
It is for these reasons that we define the disjunction $d=a~|b$ that is constructed by the program $disj~[a~b]~[d]$ as an atomic program even though $a$ and $b$ need not be atomic.   

\underline{Program equivalence.}
So far we have defined program equivalence in terms of the sequential order of subprograms in a program list. 
We now extend the definition of program equivalence to include disjunctions.

\begin{definition}\label{equiv}(Program equivalence.)\index{program equivalence}
Two programs $u:\mbf{prgm}$ and $v:\mbf{prgm}$ are said to be program equivalent if any of the following conditions are satisfied.
\begin{itemize}

\item $u \subseteqq v$ and $v \subseteqq u$.

\item $u =a~|~b$, $v =b~|~a$.

\item $d =a~|~b$, $u=[p~d~q]$, $v =[p~a~q]~|~[p~b~q]$.

\item $u=v~|~b$, $b:\mbf{false}$             

\end{itemize}
where $p$ and/or $q$ may be the empty list programs.
Program equivalence is denoted by $u \equiv v$ and is reflexive, symmetric and transitive.

\end{definition}

\underline{Operand programs.}
A program, $s$, containing a disjunction can be expressed in the general form $s=[p~d~q]$, where $d=a~|~b$, and $p$ and/or $q$ may be the empty list program.
The program $s$ can be split into the programs $[p~a~q]$ and $[p~b~q]$, by the two step procedure
\be
[p~d~q] \to [ [p~a]~|~[p~b]~q] \to [p~a~q]~|~[p~b~q]
\ee
or
\be
[p~d~q] \to  [p~[a~q]~|~[b~q]] \to [p~a~q]~|~[p~b~q]
\ee
In Section \ref{sspotcr} it will be shown that these constructions can be derived from the left and right disjunction distribution rules to be presented in the next section.
 
The programs $[p~a~q]$ and $[p~b~q]$ will be referred to as the operand programs of $s$ based on the disjunction program $d=a~|~b$.
Extended program derivations can be performed independently on each operand program.
When independent derivations of the operand programs yield a common conclusion, say $c$, then the common conclusion can be contracted back onto the main program, $s$, to produce an extended program $[s~c],~c:\mbf{ext}[s]$.
The disjunction contraction rule demonstrates how this is done.
There are two additional disjunction contraction rules that involve type $\mbf{false}$ operand programs.
 
\underline{Disjunction contraction rules.}
To the existing construction rules we introduce the additional rule for programs containing disjunctions.

\emph{Disjunction contraction rule.}
\be
\begin{array}{l}
	ext~[a~c]~[~] \\
	ext~[b~c]~[~] \\
	disj~[a~b]~[d] \\
	\hline
	aext~[d~c]~[~] \\
\end{array}
\quad \textsf{dsj1}
\ee

The following contraction rules involve false programs.
They are not stated as axioms because they can be derived as theorems (see Section \ref{sspotcr}).

\emph{Disjunction contraction rule 2.}
\be
\begin{array}{l}
	false~[a]~[~] \\
	ext~[b~c]~[~] \\
	disj~[a~b]~[d] \\
	\hline
	aext~[d~c]~[~] \\
\end{array}
\ee

\emph{Disjunction contraction rule 3.}
\be
\begin{array}{l}
	false~[a]~[~] \\
	false~[b]~[~] \\
	disj~[a~b]~[d] \\
	\hline
	afalse~[d]~[~] \\
\end{array}
\ee

\underline{Execution errors.}
So far we have associated execution errors with type violations of a single program list containing no disjunctions.
We now give an extended definition of an execution error that includes programs containing disjunctions.

\begin{definition}(Execution error.)
	A program $s:\mbf{prgm}$ will halt with an execution error if any of the following occur.
	
	\begin{itemize}
		
		\item $s$ does not contain a disjunction and there is a type violation of at least one assigned value of the elements of its I/O lists.
		(Type violations may also include the failure of the satisfaction of a relation between a pair of input elements.)
		
		\item $s =a~|~b$, where the programs $a$ and $b$ do not contain any disjunction.
		The disjunction $s$ will halt with an execution error if both operands, $a$ and $b$, halt with an execution error.
		The disjunction $s$ will not halt with an execution error if at least one of the operand programs, $a$ and $b$, does not halt with an execution error.
		
	\end{itemize}
	
\end{definition}

An operand of a disjunction may itself be a disjunction.
In such a general form, an execution error of a disjunction can be detected by the recursive application of the second item of the above definition.

\underline{Disjunction splitting.}
Consider the program $s=[p~d~q]$, where $p=[p_i]_{i=1}^m$, $q=[q_i]_{i=1}^n$ and $d =a~|~b$. 
Based on the disjunction, $d$, the main program $s=[p~d~q]$ can be split into the two operand programs $u=[p~a~q]$ and $v=[p~b~q]$.
Suppose that we have independently applied extended program derivations to each of the operand programs $u$ and $v$ to obtain a common conclusion, $c$.
We may then contract the common conclusion, $c$, back onto the main program by applying the disjunction contraction rule. 
The procedure is depicted in the following table.
The connection lists associated with each statement have been omitted due to space restrictions. 

\be
\begin{array}{llllllll}
label  & [s~c]   &      & label   & [u~c]   & & label   & [v~c]   \\
       &         &      &         &         & &         &         \\
1      & p_1     &      & 1       & p_1     & & 1       & p_1     \\
\vdots & \vdots  &      & \vdots  & \vdots  & & \vdots  & \vdots  \\
m      & p_m     &      & m       & p_m     & & m       & p_m     \\
m+1    & d~*     & \lra & m+1     & a       & & m+1     & b       \\
m+2    & q_1     &      & m+2     & q_1     & & m+2     & q_1     \\      
\vdots & \vdots  &      & \vdots  & \vdots  & & \vdots  & \vdots  \\      
m+n+1  & q_n     &      & m+n+1   & q_n     & & m+n+1   & q_n     \\
m+n+2  & c       & \lla & m+n+2   & c       & & m+n+2   & c       \\
\end{array}
\ee

The asterisk next to the statement $d$ indicates that the disjunction splitting is based on the operands of that statement.
The right arrow, $\lra$, indicates that the user has requested that the main program be split into two operand programs at line $m+1$ of the main program based on the operands of the disjunction $d =a~|~b$.
The left arrow, $\lla$, indicates that the common conclusion, $c$, of the two operand programs, $u$ and $v$, is to be contracted back onto the main program at the line labeled $n+m+2$ of the main program by applying the disjunction contraction rule.

A contraction of the conclusion, $c$, to the main program can also occur if one of the operand programs leads to a conclusion $c$ and the other a conclusion $false$.
This case follows from the disjunction contraction rule 2.
If both operand programs are type $\mbf{false}$ then, by the disjunction contraction rule 3, the main program will be assigned the type $\mbf{false}$.

In mainstream mathematics derivations associated with each operand are often conducted as separate cases within a single proof.
The reason for this is that many of these derivations are not of sufficient interest in themselves to be considered as separate theorems.
When using VPC, derivations of proofs associated with each operand program must be conducted outside of the main derivation program containing the disjunction.
The theorems extracted from the separate operand program derivations are stored in the file \emph{axiom.dat}. 
The derivation of the proof associated with the main program containing the disjunction can then access the theorems associated with each operand program through the disjunction contraction rules.

Extracting and storing theorems associated with each operand program may lead to an accumulation of theorems in \emph{axiom.dat} that are often trivial and not of particular interest in themselves.
However, this should not be a problem for storage and retrieval purposes.
In VPC one may choose to label these theorems as lemmas to weaken their status.
There is sometimes an advantage in storing these individual operand cases as separate lemmas outside of the main proof because it is not uncommon that they can be employed again in other proofs.

\underline{Disjunction connection lists.} Suppose that under an extended program derivation, the conclusion, $c$, of the operands $u$ and $v$, respectively, was derived from axioms/theorems labeled $a$ and $a^\prime$, respectively.
Upon output the conclusion statement, $c$, of the derivation program will have an attached composite connection list of the form
\be
 disj~[a~a^\prime]
\ee
This differs from the standard connection lists in that $a$ and $a^\prime$ are axiom/theorem labels and not statement labels.
The label $disj$ indicates that the conclusion statement, $c$, was obtained under the disjunction contraction rules.

\underline{Redundancy in disjunction splitting.}
Suppose that under disjunction splitting, $v$ is a false program and $u$ is a computable program.
It is possible that we may find an extended program derivation leading to the conclusion of $v$ that coincides with the conclusion of $u$ before detecting the falsity of $v$.
We may then proceed to contract the common conclusion, say $c$, to the main derivation program containing the disjunction to obtain $[[p~d~q]~c]$.
We may suspect that this will lead to an error in our derivation of the main proof.
This will not be the case since, under the disjunction contraction rule 2, this would be the identical conclusion that would have been made if we had detected that $v:\mbf{false}$.

\underline{Notes.}

\begin{itemize}

\item There remains the possibility that all operand programs are of type $\mbf{false}$ and that derivations associated with both operand programs have been terminated prematurely with a common derived conclusion, say $c$.
We may then proceed to contract this common conclusion back onto the main program to obtain $[[p~d~q]~c]$.
By the disjunction contraction rule 3 the derivations associated with each operand program should have been continued until one arrives at a common conclusion $false$ so that the program $[p~d~q]$ is identified as type $\mbf{false}$.
This type of occurrence is related to the situation described in a note of the previous section and will be discussed further in a later chapter.

\end{itemize}

\section{Additional disjunction rules.}

The following are additional construction rules based on disjunctions.
The disjunction distributivity rules are split into left and right, each involving two independent existence axioms followed by an equivalence axiom.
As before, the empty program is denoted by $ep$ and can be regarded as a constant for type $\mbf{prgm}$ objects\index{empty program} associated with the application.

\emph{Disjunction Commutativity.}
\be
\begin{array}{l}
	disj~[a~b]~[s] \\
	\hline
	disj~[b~a]~[r] \\
\end{array}
\quad \textsf{dsj2a} \hspace{10mm}
\begin{array}{l}
	disj~[a~b]~[s] \\
	disj~[b~a]~[r] \\
	\hline
	eqp~[r~s]~[~] \\
\end{array}
\quad \textsf{dsj2b}
\ee

\emph{Disjunction distributivity (right).}
\be
\begin{array}{l}
	conc~[p~a]~[r] \\
	conc~[p~b]~[s] \\
	disj~[a~b]~[d] \\
	\hline
	disj~[r~s]~[v] \\
\end{array}
\quad \textsf{dsj3a} \hspace{10mm}
\begin{array}{l}
	conc~[p~a]~[r] \\
	conc~[p~b]~[s] \\
	disj~[a~b]~[d] \\
	\hline
	conc~[p~d]~[u] \\
\end{array}
\quad \textsf{dsj3b}
\ee
\be
\begin{array}{l}
conc~[p~a]~[r] \\
conc~[p~b]~[s] \\
disj~[a~b]~[d] \\
conc~[p~d]~[u] \\
disj~[r~s]~[v] \\
\hline
eqp~[u~v]~[~] \\
\end{array}
\quad \textsf{dsj3c}
\ee

\emph{Disjunction distributivity (left).}
\be
\begin{array}{l}
conc~[a~p]~[r] \\
conc~[b~p]~[s] \\
disj~[a~b]~[d] \\
\hline
disj~[r~s]~[v] \\
\end{array}
\quad \textsf{dsj4a} \hspace{10mm}
\begin{array}{l}
conc~[a~p]~[r] \\
conc~[b~p]~[s] \\
disj~[a~b]~[d] \\
\hline
conc~[d~p]~[u] \\
\end{array}
\quad \textsf{dsj4b}
\ee
\be
\begin{array}{l}
conc~[a~p]~[r] \\
conc~[b~p]~[s] \\
disj~[a~b]~[d] \\
conc~[d~p]~[u] \\
disj~[r~s]~[v] \\
\hline
eqp~[u~v]~[~] \\
\end{array}
\quad \textsf{dsj4c}
\ee

\emph{False operand program.}
\be
\begin{array}{l}
disj~[a~b]~[p] \\
false~[b]~[~] \\
\hline
eqp~[p~a]~[~] \\
\end{array}
\quad \textsf{dsj5}
\ee

\emph{Extension disjunction introduction.}
\be
\begin{array}{l}
	ext~[p~a]~[~] \\
	disj~[a~b]~[c] \\
	conc~[p~c]~[s] \\
	\hline
	aext~[p~c]~[~] \\
\end{array}
\quad \textsf{dsj6}
\ee

\underline{Notes.}

\begin{itemize}

\item The rules \textsf{per}, \textsf{cr1}-\textsf{cr9}, \textsf{flse1}-\textsf{flse2} and \textsf{dsj1}-\textsf{dsj6} along with the common application axioms such as the I/O type axioms and the substitution rule, are presented as irreducible extended higher-order programs.
They can be regarded as the axioms of a theory for the construction of programs as proofs in the context of the formal system PECR upon which VPC is based.
Later we will employ VPC as a self referencing tool to investigate certain properties of the construction  rules themselves.
To this end the rules \textsf{per}, \textsf{cr1}-\textsf{cr9}, \textsf{flse1}-\textsf{flse2} and \textsf{dsj1}-\textsf{dsj6} will be supplied as axioms in the file \emph{axiom.dat}.

\item A collection of constants that serve as input to the first-order programs associated with the construction rules are type $\mbf{prgm}$ objects that must be defined with respect to the application theory, $S$, to which the construction rules are being applied.
As a consequence the collection of constants called by first-order programs of the construction rules may differ among applications.
The empty list zeroth-order program, $ep$, is defined as a constant of type $\mbf{prgm}$ and will be common to all applications.
Other type $\mbf{prgm}$ objects that are constants include type $\mbf{false}$ objects that are associated with axioms of falsity specific to the application, $S$.

\end{itemize}

\chapter{Common applications axioms.}\label{appl}

\section{Introduction.}

The construction rules of PECR define the structural foundations of VPC.
They are general rules that should be distinguished from axioms that are supplied for specific applications.
An application is also referred to as a theory\index{theory}.

A theory, $S$, is defined by a list of atomic programs, $atom$, a list of axioms, $ax$, and a list of constants, $cst$.
\be
\bal
atom = & [atom(i)]_{i=1}^{nat}, \quad nat:\mbf{int1},~atom(i):\mbf{atm}<:\mbf{prgm} \\
ax = & [ax(i)]_{i=1}^{nax}, \quad nax:\mbf{int0},~ax(i):\mbf{prgm} \\
cst = & [cst(i)]_{i=1}^{ncst}, \quad ncst:\mbf{int0},~cst(i):\mbf{char} \\
\eal
\ee
Each element, $ax(i)$, of the list of axioms, $ax$, is an irreducible extended program of the form $[p~c]$, where $c:\mbf{iext}[p]$.
While each element of $atom$ and $ax$ is a program, the lists $atom$ and $ax$ are not in themselves meant to represent a program.

A theory, $S$, must also be dependent on the machine environment, $\mathfrak{M}(mach)$, in which it is being constructed, where
\be
mach = [nchar~nstr~nlst~nint]
\ee
are the machine parameter constraints defined by (\ref{mparam}).
We sometimes write
\be
S(atom,ax,cst,mach)
\ee
to emphasize the constraints imposed on the theory, $S$.
In this context the length of any program list is bound by $nlst$.
This includes the length of any derivation program.

The list of I/O variable names is denoted by
\be
\bal
var = & [var(i)]_{i=1}^{nvar}, \quad nvar:\mbf{int1},~var(i):\mbf{char} \\
\eal
\ee
Input variable names may be elements of $var$ or $cst$ while output variable names can only be elements of $var$.

The axioms of a theory, $S$, are stored in the file \emph{axiom.dat} that is accessed by VPC during program constructions.
In a later chapter we will use VPC as a self referencing tool to investigate certain properties of the construction rules themselves.
In this case the construction rules are inserted in the file \emph{axiom.dat} as axioms.

The list of axioms, $ax$, will differ for each application.
However, there are certain axioms that will have a common structure in all applications.
Of these are three classes of axioms, (1) I/O type axioms, (2) the substitution rule and (3) application specific axioms of falsity.
These common axioms are not treated in the same way as the other construction rules and are automated in VPC.
There are minor adaptations that may be necessary and the user is required to supply additional initial data to instruct VPC how to fine tune these axioms for the specific application under consideration.

In this book we will consider the following three main applications.

\begin{itemize}

\item Arithmetic over $\mbf{int}$ examines the properties of machine arithmetic under the elementary operations of addition, subtraction, multiplication and division.
The values assigned to elements of the I/O lists of integer programs are of type $\mbf{int}$.  

\item Integer vectors and discrete boxes.
The values assigned to elements of the I/O lists of integer vector and box programs are of a mixed type and include scalars of type $\mbf{int}$, integer vectors of type $\mbf{vec}$ and intervals/boxes of type $\mbf{box}$.  

\item Theory of programs as proofs in the context of our formal system PECR.
The atomic programs of this application are higher-order programs.
The values assigned to the elements of I/O lists of these higher order programs are also of type $\mbf{prgm}$ but of lower order.

\end{itemize}

In Chapter \ref{cabstract} we will demonstrate how PECR can be employed as a primitive on top of which theories based on higher levels of abstraction can be constructed.
The common application axioms (1) and (3) will still hold but auxiliary rules of abstraction will need to be supplied before these become active.
Since we are primarily interested in applications of computability of real world applications we will not pursue the employment of PECR for abstract theories in more detail beyond Chapter \ref{cabstract}.     

\section{I/O type axioms.}

An important property of all programs is that the type of the assigned values of all elements of the I/O lists are checked within the program.
Atomic programs of type $\mbf{chck}$ that have the sole task of checking the type of the assigned value of a single input variable will always be assigned the name $typeX$, where $X$ refers to some distinguishing lower case letters and/or numbers.

There are four kinds of type checking programs that are associated with the main applications discussed in the previous section.
Let
\ben \label{3.100}
\mathfrak{t}~[a]~[~] = \left \{ \begin{array}{ll}
	typei~[a]~[~], & \text{if}~a:\mbf{int} \\
	typev~[a]~[~], & \text{if}~a:\mbf{vec}[m],~m:\mbf{int1} \\
	typebx~[a]~[~], & \text{if}~a:\mbf{box}[m],~m:\mbf{int1} \\
	typep~[a]~[~], & \text{if}~a:\mbf{prgm}[m],~m:\mbf{int0} \\
\end{array}
\right .
\een
The input variables of the programs for vector type, $typev$, box type, $typebx$, and program type, $typep$, are parameter dependent types, for some list dimension, $m$.
Under our dimension free formulation, list dimensions do not appear in the I/O lists of programs.
The list dimension, $m$, for the input variables are set prior to entry to the programs and are identified within the program upon entry.

I/O type axioms give a conclusion of type for an assigned value of an element of an I/O list of a program.
For any program, $p~x~y$, there is no restriction that all of the value assignments of elements of its I/O lists are of the same type.
As such the type checking program in the conclusion must be type related to the element of the I/O list of the program $p~x~y$ that is being singled out.  

When using VPC, all program names and the type of each element of their associated I/O lists are specified by the user in an initializing setup file.
I/O type axioms are labeled by the letters \textsf{aio}.
They take the general form
\be
\begin{array}{l}
	[p~x~y]_{a \in [x~y]} \\
	\hline
	\mathfrak{t}~[a]~[~] \\
\end{array}
\quad \textsf{aio}
\ee

\section{Substitution rule.}\label{ssr}

We have already encountered the atomic equivalence program $eqp~[p~q]~[~]$, where $p$ and $q$ are variable names that have been assigned the values of programs.
Upon entry $eqp~[p~q]~[~]$ checks that $p$ and $q$ have been assigned values of type $\mbf{prgm}$ and then checks that they are equivalent, i.e. $p \equiv q$.
$eqp~[p~q]~[~]$ is symmetric, reflexive and transitive.
We note that by definition, program equivalence includes the case where $p$ and $q$ are identical programs.
For other applications there are equality checking programs that are similar in function.

Atomic programs of type $\mbf{chck}$ that have the additional task of checking the equivalence or equality of assigned values of pairs of input variables will always be assigned the name $eqX$, where $X$ represents some distinguishing lower case letters and/or numbers.
Let
\ben \label{3.200}
\mathfrak{e}~[a~b]~[~] = \left \{ \begin{array}{ll}
	eqi~[a~b]~[~], & \text{if}~a,b:\mbf{int} \\
    eqv~[a~b]~[~], & \text{if}~a,b:\mbf{vec}[m],~m:\mbf{int1} \\
	eqbx~[a~b]~[~], & \text{if}~a,b:\mbf{box}[m],~m:\mbf{int1} \\
	eqp~[a~b]~[~], & \text{if}~a,b:\mbf{prgm}[m],~m:\mbf{int0} \\
\end{array}
\right .
\een
The input variables of the programs for vector equality, $eqv$, box equality, $eqbx$, and program equivalence, $eqp$, are parameter dependent types, for some list dimension $m$.
The list dimension, $m$, for the input variables are set prior to entry to the programs and are identified within the program upon entry.

Using the notation
\be
\bal
& x=[x(i)]_{i=1}^{nx} \\
& y=[y(j)]_{j=1}^{ny} \\
\eal
\ee
for the I/O lists of an atomic program $p~x~y$ define
\be
\bal
& \bar{y}=[\bar{y}(j)]_{j=1}^{ny} \\
& \bar{x} = x(x(k) \to a),~1 \leq k \leq nx \\
\eal
\ee
The substitution rule is presented in two parts.
The first part is an existence axiom.
The second part is an equality axiom generated for each $j=1,\ldots,ny$, and is applicable only when $y$ is a nonempty list.
\be
\begin{array}{l}
p~x~y \\
\mathfrak{e}~[x(k)~a]~[~] \\
\hline
p~\bar{x}~\bar{y} \\
\end{array}
\quad \textsf{sr1} \hspace{10mm}
\begin{array}{l}
p~x~y \\
\mathfrak{e}~[x(k)~a]~[~] \\
p~\bar{x}~\bar{y} \\
\hline
\mathfrak{e}~[\bar{y}(j)~y(j)]~[~]
\end{array}
\quad \textsf{sr2} 
\ee
It is important to note that some atomic programs can be shown to satisfy the substitution rule from the other axioms.
The substitution rule should not be regarded as an axiom for such programs.
When setting up an application for VPC the user is required to supply the names of the programs for which the substitution rule is to be applied as an axiom.  

\underline{Repetition of subprograms.} We have already noted that we could have repetitions of subprograms in a program list for subprograms with an empty output list.
Sometimes we may wish to apply a rule that allows a repetition of an assignment program, but this can only be done by introducing new names for the output list variables.
The substitution rules may be applied to the case where $\mathfrak{e}~[x(k)~a]~[~]$ is replaced by $\mathfrak{e}~[x(k)~x(k)]~[~]$, for any $x(k) \in x$, to obtain the rules
\ben \label{identity}
\begin{array}{l}
	p~x~y \\
	\hline
	p~x~\bar{y} \\
\end{array}
\quad \hspace{10mm}
\begin{array}{l}
	p~x~y \\
	p~x~\bar{y} \\
	\hline
	\mathfrak{e}~[\bar{y}(j)~y(j)]~[~]
\end{array}
\een
where the second rule is generated for all $j=1,\ldots,ny$.
The combination of these two rules can be interpreted as an analogy to the identity axiom of classical logic.
However, in PECR they are not axioms since they are derivable from the substitution rule.
An example where these rules are employed in proofs is presented in Section \ref{saodb}.

\underline{Notes.}

\begin{itemize}
	
	\item The substitution rule should not be directly applied as an axiom to atomic programs of type $\mbf{chck}$ that involve abstract types.
	These include the atomic programs $ext$ and $false$.
	However, it can be shown that these programs satisfy the substitution rule as a derivation. 	
	
	\item There are programs that do not in general satisfy the substitution rule.
	The atomic program $eqio$ is one such program.
	
\end{itemize}

\section{Axioms of falsity.}

For each application, $S$, there will be known false programs that are supplied in the file \emph{axiom.dat} when initializing axioms of falsity.
These programs acquire the type $\mbf{false}$ by default and form the seeds from which theorems of falsity are generated for a theory, $S$.
All other false programs acquire the type $\mbf{false}$ through the type assignment program $afalse$ by way of inference.

Application specific axioms of falsity are higher order constructs.
They can be represented by
\ben \label{4.6.100}
\begin{array}{l}
~~~ \\
\hline
false~[q]~[~] \\
\end{array}
\een
where $q$ is assigned the fixed value of an object of type $\mbf{prgm}$ that is expressed in the form of an atomic program associated with the application, $S$.
One can think of (\ref{4.6.100}) as being equivalent to a second-order irreducible extended program where the premise is assigned the value of the empty first-order program and the conclusion program is assigned the value of $false~[q]~[~]$ (see Section \ref{shop}).
The prescribed program, $q$, must be defined as a constant for type $\mbf{prgm}$ objects associated with the theory, $S$.

\chapter{Arithmetic over $\mbf{int}$.}\label{int}

\section{Introduction.}\label{ring}

The objective here is to construct an axiomatic system for the elementary operations of integer arithmetic that reflect feasible computations of maps on configuration states in a machine environment, $\mathfrak{M}(mach)$.
To this end we work with objects of type $\mbf{int}$ that can be assigned any one of the integer values 
\be
0, \pm 1,\ldots, \pm nint,
\ee
where $nint$ is the maximum positive integer and is a machine dependent parameter.

An important feature of our formal system is that we replace the notion of sets with lists.
Nevertheless, our axiomatic system will be guided by the traditional axioms of commutative rings but with important differences.
It is useful to remind ourselves of these axioms.

\underline{Commutative rings.}\index{commutative rings}
A commutative ring $\{ \mathcal{R},+,* \}$ is a set $\mathcal{R}$ with two binary operations $+$ and $*$ subject to the following axioms.

\begin{itemize}

\item $\{ \mathcal{R},+,* \}$ is closed under the operation $+$, i.e. if $a$ and $b$ are elements of $\mathcal{R}$ then $a+b$ is also an element of $\mathcal{R}$.

\item The operation $+$ is commutative, i.e. if $a$ and $b$ are elements of $\mathcal{R}$ then $a+b=b+a$.

\item The operation $+$ is associative, i.e. if $a,b$ and $c$ are elements of $\mathcal{R}$ then $a+(b+c)=(a+b)+c$.

\item For any element $a$ of $\mathcal{R}$ there is a unique element of $\mathcal{R}$, denoted by $0$, called the zero element such that $a+0=a$.

\item For any element $a$ of $\mathcal{R}$ there is a unique element of $\mathcal{R}$, denoted by $-a$, called the additive inverse of $a$ such that $a+(-a)=0$.

\item $\{ \mathcal{R},+,* \}$ is closed under the operation $*$, i.e. if $a$ and $b$ are elements of $\mathcal{R}$ then $a*b$ is also an element of $\mathcal{R}$.

\item The operation $*$ is commutative, i.e. if $a$ and $b$ are elements of $\mathcal{R}$ then $a*b=b*a$.

\item The operation $*$ is associative, i.e. if $a,b$ and $c$ are elements of $\mathcal{R}$ then $a*(b*c)=(a*b)*c$.

\item The operation $*$ is distributive over the operation $+$, i.e. if $a,b$ and $c$ are elements of $\mathcal{R}$ then $a*(b+c)=a*b+a*c$.

\item For any element $a$ of $\mathcal{R}$ there is a unique element of $\mathcal{R}$, denoted by $1$, called the multiplicative identity such that $1*a=a$.

\item $0 \neq 1$.

\end{itemize}

The set of integers, $\mb{Z}$, with the usual operations of addition and multiplication is an example of a commutative ring.
 
Let $\{ \mathcal{R},+,* \}$ be a commutative ring.
An element $a$ of $\mathcal{R}$ has a multiplicative inverse $b$ contained in $\mathcal{R}$ if and only if $b*a=1$.
In such a case we write $b=a^{-1}$.

An ordered set is a set $\mathcal{R}$, together with a relation $<$ such that

\begin{itemize}

\item For any elements $x,y$ of $\mathcal{R}$, exactly one of $x<y$, $x=y$, $x>y$ holds. 

\item For any elements $x,y,z$ of $\mathcal{R}$, if $x<y$ and $y<z$ then $x<z$

\end{itemize}

A ring $\{ \mathcal{R},+,* \}$ is said to be an ordered ring if $\mathcal{R}$ is an ordered set such that 

\begin{itemize}

\item For any elements $x,y,z$ of $\mathcal{R}$, if $x<y$ then $x+z<y+z$.

\item For any elements $x,y$ of $\mathcal{R}$, if $x>0$ and $y>0$ then $x*y>0$.

\end{itemize}

\section{Atomic programs for arithmetic over $\mbf{int}$.}\label{apfaoi}

Derivations of the basic identities of arithmetic over fields and commutative rings are often presented to students in an introductory course to analysis.
A major difficulty when working with $\mbf{int}$ is the absence of closure for the operations of addition and multiplication.
While the derivations of the basic identities of arithmetic are elementary, it will be necessary to restate the axioms of arithmetic in the context of a machine environment $\mathfrak{M}(mach)$.
Here we shall take a constructive approach by introducing rules that address the operations of machine arithmetic that lend themselves to a more practical approach towards establishing computability.

For arithmetic on $\mbf{int}$ we make use of four constants.
\be
\bal
& cst=[-1~0~1~pr] \\
\eal
\ee
where the first three elements are assigned the fixed values of type $\mbf{int}$ and the parameter $pr$ is assigned the fixed value of type $\mbf{prgm}$.
The program $pr$ will be presented later.
 
Here we have departed slightly from the convention of representing all elements of program I/O lists by alphanumeric variable names by allowing some elements to be represented by the numeric constants $-1,0,1$.
To strictly adhere to the convention we could introduce special alphanumeric names for these constants.
For convenience we allow, as exceptions, these constants to appear in the input lists of programs in numeric form.

The atomic programs for arithmetic on $\mbf{int}$ are defined in the following tables.

\newpage 

\emph{Program type} $\mbf{chck}$.

\begin{tabular}{|l|l|}
	\hline
	\textbf{Syntax} & \textbf{Type checks} \\
	\hline \hline
	$typei~[a]~[~]$ & $a:\mbf{int}$ \\
	\hline
	$lt~[a~b]~[~]$ & $a,b:\mbf{int}$, $a < b$ \\
	\hline
	$eqi~[a~b]~[~]$ & $a,b:\mbf{int}$, $a = b$ \\
	\hline
\end{tabular}

The expressions $a<b$ and $a=b$ in the above tables are meant to represent inequality and equality with respect to the assigned values of $a$ and $b$.

\emph{Program type} $\mbf{asgn}$.

\begin{tabular}{|l|l|l|l|}
	\hline
	\textbf{Syntax} & \textbf{Type} & \textbf{Assignment} & \textbf{Type} \\
	                & \textbf{checks} & \textbf{map} & \textbf{assignment} \\
	\hline \hline
	$add~[a~b]~[c]$ & $a,b:\mbf{int}$ & $c:=a+b$ & $c::\mbf{int}$ \\
	\hline
	$mult~[a~b]~[c]$ & $a,b:\mbf{int}$ & $c:=a*b$ & $c::\mbf{int}$ \\
	\hline
	$div~[a~b]~[c]$ & $a,b:\mbf{int}$,~$b \neq 0$ & $c:=a/b$ & $c::\mbf{int}$ \\
    \hline
\end{tabular}

All atomic programs will be computable if the type checks are not violated.
Otherwise they halt with an execution error.

Type checking also includes the validity of the type assignments of the output variables.
This means that the type checks of the entry variables of the second column do not guarantee computability. 
For example, the type assignment $c::\mbf{int}$ in the program $div~[a~b]~[c]$ will be invalid if $b$ is not a nonzero integer multiple of $a$.
Numerical overflows are also a common cause of execution errors.

We will also make use of the following special programs associated with disjunctions.
Disjunctions are defined as atomic programs. 
\vspace{5mm}

\begin{tabular}{|c|}
	\hline
	Disjunction program names \\
	\hline
	$neq,~le,~abs,~trich$ \\
	\hline
\end{tabular}

\underline{Notes.}

\begin{itemize}
	
	\item The substitution rule can be applied as an axiom to all atomic programs of type $\mbf{chk}$ and $\mbf{asgn}$ except for the program $typei~[a]~[~]$ that can be shown to satisfy the substitution rule from a trivial derivation.
	The disjunction programs shown in the table will also satisfy the substitution rule by derivation.  
	
\end{itemize}

\section{Axioms of arithmetic over $\mbf{int}$.}\label{saoaoi}

Axioms for arithmetic over $\mbf{int}$ are labeled by the letters \textsf{axi} followed by a number and/or letter.
To these are appended the order axioms that are labeled by \textsf{ord} followed by a number and/or letter.
These axioms are stored in a file, \emph{axiom.dat}, that is accessed by VPC during proof construction.
Axioms based on I/O type axioms and the substitution rule are automated within the program VPC.

\underline{Equality axioms.}

\be
\begin{array}{l}
	typei~[a]~[~] \\
	\hline
	eqi~[a~a]~[~] \\
\end{array}
\quad \textsf{axi1a} \hspace{10mm}
\begin{array}{l}
	eqi~[a~b]~[~] \\
	\hline
	eqi~[b~a]~[~] \\
\end{array}
\quad \textsf{axi1b}
\ee
The equality program satisfies the property of transitivity
\be
\begin{array}{l}
	eqi~[a~b]~[~] \\
	eqi~[b~c]~[~] \\
	\hline
	eqi~[a~c]~[~] \\
\end{array}
\ee
This is not included as an axiom because it follows from the substitution rule.

\underline{Axioms of addition and multiplication.}

\emph{Commutativity of addition.}
\be
\begin{array}{l}
	add~[a~b]~[c] \\
	\hline
	add~[b~a]~[d] \\
\end{array}
\quad \textsf{axi2a} \hspace{10mm}
\begin{array}{l}
	add~[a~b]~[c] \\
	add~[b~a]~[d] \\
	\hline
	eqi~[d~c]~[~] \\
\end{array}
\quad \textsf{axi2b}
\ee

\emph{Associativity of addition.}
\be
\begin{array}{l}
	add~[a~b]~[d] \\
	add~[d~c]~[x] \\
	add~[b~c]~[e] \\
	\hline
	add~[a~e]~[y] \\
\end{array}
\quad \textsf{axi3a} \hspace{10mm}
\begin{array}{l}
	add~[a~b]~[d] \\
	add~[d~c]~[x] \\
	add~[b~c]~[e] \\
	add~[a~e]~[y] \\
	\hline
	eqi~[y~x]~[~] \\
\end{array}
\quad \textsf{axi3b}
\ee

\emph{Addition by zero.}
\be
\begin{array}{l}
	typei~[a]~[~] \\
	\hline
	add~[a~0]~[b] \\
\end{array}
\quad \textsf{axi4a} \hspace{10mm}
\begin{array}{l}
	add~[a~0]~[b] \\
	\hline
	eqi~[b~a]~[~] \\
\end{array}
\quad \textsf{axi4b}
\ee

\emph{Additive inverse.}
\be
\begin{array}{l}
	typei~[a]~[~] \\
	\hline
	mult~[-1~a]~[b] \\
\end{array}
\quad \textsf{axi5a} \hspace{10mm}
\begin{array}{l}
	mult~[-1~a]~[b] \\
	\hline
	add~[a~b]~[d] \\
\end{array}
\quad \textsf{axi5b}
\ee
\be
\begin{array}{l}
	mult~[-1~a]~[b] \\
	add~[a~b]~[d] \\
	\hline
	eqi~[d~0]~[~] \\
\end{array}
\quad \textsf{axi5c}
\ee

\emph{Commutativity of multiplication.}
\be
\begin{array}{l}
	mult~[a~b]~[c] \\
	\hline
	mult~[b~a]~[d] \\
\end{array}
\quad \textsf{axi6a} \hspace{10mm}
\begin{array}{l}
	mult~[a~b]~[c] \\
	mult~[b~a]~[d] \\
	\hline
	eqi~[d~c]~[~] \\
\end{array}
\quad \textsf{axi6b}
\ee

\emph{Associativity of multiplication.}
\be
\begin{array}{l}
	mult~[a~b]~[d] \\
	mult~[d~c]~[x] \\
	mult~[b~c]~[e] \\
	\hline
	mult~[a~e]~[y] \\
\end{array}
\quad \textsf{axi7a} \hspace{10mm}
\begin{array}{l}
	mult~[a~b]~[d] \\
	mult~[d~c]~[x] \\
	mult~[b~c]~[e] \\
	mult~[a~e]~[y] \\
	\hline
	eqi~[y~x]~[~] \\
\end{array}
\quad \textsf{axi7b}
\ee

\emph{Multiplication by unity.}
\be
\begin{array}{l}
	typei~[a]~[~] \\
	\hline
	mult~[1~a]~[b] \\
\end{array}
\quad \textsf{axi8a} \hspace{10mm}
\begin{array}{l}
	mult~[1~a]~[b] \\
	\hline
	eqi~[b~a]~[~] \\
\end{array}
\quad \textsf{axi8b}
\ee

\emph{Distribution law.}
\be
\begin{array}{l}
	add~[b~c]~[d] \\
	mult~[a~d]~[x] \\
	mult~[a~b]~[u] \\
	mult~[a~c]~[v] \\
	\hline
	add~[u~v]~[y] \\
\end{array}
\quad \textsf{axi9a} \hspace{10mm}
\begin{array}{l}
	mult~[a~b]~[u] \\
	mult~[a~c]~[v] \\
	add~[u~v]~[y] \\
	add~[b~c]~[d] \\
	\hline
	mult~[a~d]~[x] \\
\end{array}
\quad \textsf{axi9b}
\ee
\be 
\begin{array}{l}
	add~[b~c]~[d] \\
	mult~[a~d]~[x] \\
	mult~[a~b]~[u] \\
	mult~[a~c]~[v] \\
	add~[u~v]~[y] \\
	\hline
	eqi~[y~x]~[~] \\
\end{array}
\quad \textsf{axi9c}
\ee

\underline{Divisor.}
\be
\begin{array}{l}
	neq~[a~0]~[~] \\
	mult~[a~b]~[c] \\
	\hline
	div~[c~a]~[d] \\
\end{array}
\quad \textsf{axi10a} \hspace{5mm}
\begin{array}{l}
	mult~[a~b]~[c] \\
	div~[c~a]~[d] \\
	\hline
	eqi~[d~b]~[~] \\
\end{array}
\quad \textsf{axi10b}
\ee

\underline{Order axioms.}
\be
\begin{array}{l}
	lt~[a~b]~[~] \\
	add~[a~c]~[x] \\
	add~[b~c]~[y] \\
	\hline
	lt~[x~y]~[~] \\
\end{array}
\quad \textsf{ord1a} \hspace{5mm}
\begin{array}{l}
	lt~[a~b]~[~] \\
	lt~[c~d]~[~] \\
	add~[a~c]~[x] \\
	add~[b~d]~[y] \\
	\hline
    lt~[x~y]~[~] \\
\end{array}
\quad \textsf{ord1b}
\ee

\be
\begin{array}{l}
	lt~[a~b]~[~] \\
	lt~[0~c]~[~] \\
	mult~[a~c]~[x] \\
	mult~[b~c]~[y] \\
	\hline
	lt~[x~y]~[~] \\
\end{array}
\quad \textsf{ord2a} \hspace{5mm}
\begin{array}{l}
	lt~[a~b]~[~] \\
	lt~[c~0]~[~] \\
	mult~[a~c]~[x] \\
	mult~[b~c]~[y] \\
	\hline
	lt~[y~x]~[~] \\
\end{array}
\quad \textsf{ord2b}
\ee

\emph{Transitivity of inequality.}
\be
\begin{array}{l}
	lt~[a~b]~[~] \\
	lt~[b~c]~[~] \\
	\hline
	lt~[a~c]~[~] \\
\end{array}
\quad \textsf{ord3}
\ee

To the order axioms we include the following axiom that has an empty list premise 
\be
\begin{array}{l}
	~ \\
	\hline
	lt~[0~1]~[~] \\
\end{array}
\quad \textsf{ord4}
\ee

\underline{Axiom of falsity.} (higher order type checking axiom)
Axioms of falsity are higher order constructs.
We depart slightly from the convention of expressing the axiom in the form of a concatenation, $[p~c]$, $c:\mbf{iexp}[p]$, by simply assigning the type $p:\mbf{false}$ to the premise.
For arithmetic over $\mbf{int}$ we include the following axiom of falsity.

\be
\begin{array}{l}
	lt~[a~a]~[~] \\
	\hline
	false \\
\end{array}
\quad \textsf{ord5}
\ee
This can also be expressed as
\be
lt~[a~a]~[~] : \mbf{false}
\ee 

One can also think of axiom \textsf{ord5} as being equivalent to the higher order axiom with an empty premise
\be
\begin{array}{l}
	~ \\
	\hline
	false~[pr]~[~] \\
\end{array}
\ee
where the assigned value of $pr$ is an object of type $\mbf{prgm}$ and is given explicitly by
\be
pr:=lt~[a~a]~[~]
\ee
The object $pr$ is regarded as a constant of type $\mbf{prgm}$ associated with the application of integer arithmetic over $\mbf{int}$.

\underline{Notes.}

\begin{itemize}

    \item For any integer $a:\mbf{int}$ there is no attempt made to abstract its additive identity $a+0$, its multiplicative identity $1*a$ and its additive inverse $-1*a$.
    Here, the assigned values of $-1,0$ and $1$ are immediately recognized as type $\mbf{int}$ objects.
    
    \item Our definition of $\mbf{int}$ does not necassarily coincide with the actual integers of a real world computer.
    Most modern computers define the type integers in the range of $-2^N$ to $2^N-1$, for some $N$.
    For such machines we must define the type $\mbf{int}$ objects to be all machine integers in this range excluding the integer $-2^N$.
    Thus our type checking programs for $\mbf{int}$ will have to be constructed in such a way that they override the actual machine integers.  
    We do this because we want to retain axiom \textsf{axi5a}.
    Without this axiom the establishment of computability of arithmetic on $\mbf{int}$ will become too restrictive. 

\end{itemize}

\section{Extra conditional statements.}

The axioms of the previous section are presented in a style that has some similarity with logic programming.   
Having adjusted to this style of inference we are also forced to include extra conditional constraints in the premises to address the bounded computational domain that we are working in.

The most important departure from the axioms for addition and multiplication of fields and commutative rings is the absence of closure, i.e. $add~[a~b]~[c]$ and $mult~[a~b]~[c]$ do not necessarily follow from $typei~[a]~[~]$ and $typei~[b]~[~]$.
Consequently, the axioms for addition and multiplication are split into one or more existence parts followed by an identity axiom.
Any occurrence of statements involving $add$ and $mult$ in a program list must either have been inferred from the axioms or have simply been inserted as conditional statements in the premise program of an axiom or theorem.

Associativity of addition requires an existence axiom, \textsf{axi3a}, followed by an identity axiom, \textsf{axi3b}.
The existence part is necessary because $y=a+(b+c)$ does not necessarily follow from $x=(a+b)+c$.
As an example set $a=-nint,~b=nint,~c=1$.
We have $d=a+b=0:\mbf{int}$ and hence $x=(-nint+nint)+1=1:\mbf{int}$ but $e=b+c$, and hence $y=a+(b+c)$, is not of type $\mbf{int}$. 
In order that $y=a+(b+c):\mbf{int}$ we must include in the premise the extra conditional statement that $e=b+c:\mbf{int}$.

Similarly, associativity of multiplication requires an existence axiom, \textsf{axi7a}, followed by an identity axiom, \textsf{axi7b}.
The existence part is necessary because $y=a*(b*c)$ does not necessarily follow from $x=(a*b)*c$.
As an example set $a=0,~b=nint,~c=2$.
We have $d=a*b=0:\mbf{int}$ and hence $x=(0)*2=0:\mbf{int}$ but $e=b*c$ is not of type $\mbf{int}$. 
In order that $y=a*(b*c):\mbf{int}$ we must include in the premise the extra conditional statement that $e=b*c:\mbf{int}$.

The axiom of distributivity has two independent existence parts, \textsf{axi9a} and \textsf{axi9b}, followed by an identity axiom, \textsf{axi9c}.
The existence axiom, \textsf{axi9a}, is necessary because $y=a*b+a*c$ does not necessarily follow from $x=a*(b+c)$.
As an example set $a=nint,~b=nint,~c=-nint$.
We have $d=b+c=0:\mbf{int}$ and hence $x=nint*(-nint+nint)=0:\mbf{int}$ but neither $u=a*b$ and $v=a*c$, and hence $y=u+v$, are of type $\mbf{int}$. 
In order that $y=a*b+a*c:\mbf{int}$ we must include in the premise the extra conditional statement that $u$ and $v$ are of type $\mbf{int}$.

Similarly, the existence axiom, \textsf{axi9b}, is necessary because $x=a*(b+c)$ does not necessarily follow from $y=a*b+a*c$.
As an example set $a=0,~b=nint,~c=nint$.
We have $u=a*b=0:\mbf{int}$ and $v=a*c=0:\mbf{int}$ and hence their sum $y=0:\mbf{int}$.
But $d=b+c$ is not of type $\mbf{int}$.
In order that $x=a*(b+c):\mbf{int}$ we must include in the premise the extra conditional statement that $d=b+c$ is of type $\mbf{int}$.

Axiom \textsf{ord1a} is similar to the first ordered ring axiom except for the inclusion of conditional statements.
Axiom \textsf{ord1b} is included due to complications of working with arithmetic on \textsf{int}.
For similar reasons the order axioms \textsf{ord2a} and \textsf{ord2b} differ significantly from the second of the ordered ring axiom.

\section{Special disjunction programs.}\label{sdp}

For arithmetic over $\mbf{int}$ we will make use of the special disjunction programs $neq,~le$, $abs$ and $trich$. 

\underline{Not equal.} The program $neq~[a~b]~[~]$ is an atomic program associated with the disjunction
\be
neq~[a~b]~[~] = lt~[a~b]~[~]~|~lt~[b~a]~[~]
\ee

The following axioms are included. 
\be
\begin{array}{l}
	lt~[a~b]~[~] \\
	\hline
	neq~[a~b]~[~] \\
\end{array}
\quad \textsf{neq1} \hspace{10mm}
\begin{array}{l}
	lt~[b~a]~[~] \\
	\hline
	neq~[a~b]~[~] \\
\end{array}
\quad \textsf{neq2}
\ee

\underline{Less than or equal.} The program $le~[a~b]~[~]$ is an atomic program associated with the disjunction
\be
le~[a~b]~[~] = lt~[a~b]~[~]~|~eqi~[a~b]~[~]
\ee

The following axioms are included. 
\be
\begin{array}{l}
	lt~[a~b]~[~] \\
	\hline
	le~[a~b]~[~] \\
\end{array}
\quad \textsf{le1} \hspace{10mm}
\begin{array}{l}
	eqi~[a~b]~[~] \\
	\hline
	le~[a~b]~[~] \\
\end{array}
\quad \textsf{le2}
\ee

\underline{Absolute value.} The program $abs~[a]~[b]$ is an atomic program associated with the disjunction
\be
abs~[a]~[b] = [lt~[a~0]~[~]~mult~[-1~a]~[b]]~|~[le~[0~a]~[~]~mult~[1~a]~[b]]
\ee
The program $abs~[a]~[b]$ makes the assignment $b:=|a|$.

\underline{Trichotomy.} Having constructed the disjunction $le~[a~b]~[~]$ we can now state the trichotomy axiom as
\be
\begin{array}{l}
	typei~[a]~[~] \\
	typei~[b]~[~] \\
	\hline
	trich~[a~b]~[~] \\
\end{array}
\quad \textsf{ord6}
\ee
where we have introduced the disjunction
\be
trich~[a~b]~[~] = le~[a~b]~[~]~|~lt~[b~a]~[~]
\ee

\underline{Notes.}

\begin{itemize}	  
	
	\item The disjunction symbol, $|$, differs in some way from the classical connective, $\lor$, and some care needs to taken when making comparisons between disjunctions in PECR and those of classical logic.
	  
	By the construction rules \textsf{ord3} and \textsf{ord5} it follows that the program
	\be
	~[lt~[a~b]~[~]~lt~[b~a]]
	\ee
	is type $\mbf{false}$.
	In light of the sublist falsity rule, \textsf{flse1}, a concatenation of the program $neq~[a~b]~[~]$ with this false program will also be of type $\mbf{false}$.
	This means that the operands of the program $neq~[a~b]~[~]$ can be considered to be exclusive.
	
	Similarly, by \textsf{ord5} and the substitution rule it follows that the program
	\be
	~[lt~[a~b]~[~]~eqi~[b~a]]
	\ee
	is also type $\mbf{false}$.
	A concatenation of the program $le~[a~b]~[~]$ with this false program will also be of type $\mbf{false}$ and hence the operands of the program $le~[a~b]~[~]$ are exclusive.
	
	It follows that a concatenation of the program $trich~[a~b]~[~]$ with either of the above false programs will be of type $\mbf{false}$ so the operands of the program $trich~[a~b]~[~]$ are also exclusive.
	
	\item The axioms \textsf{le1}, \textsf{le2}, \textsf{neq1} and \textsf{neq2} are specific examples that are similar to the general classical rules of disjunction introduction.
	All disjunctions that are to be employed in a theory must be defined in an initializing input file to VPC.
	Axioms of disjunction introduction, like \textsf{le1}, \textsf{le2}, \textsf{neq1} and \textsf{neq2}, have to be manually inserted in the file \emph{axiom.dat} on a case by case basis.
	
	\item Symmetry of the disjunction named $neq$ follows from the axioms \textsf{neq1} and \textsf{neq2}.  
	
\end{itemize}

\section{Algebraic identities over $\mbf{int}$.}\label{saioi}

Derivations of proofs for arithmetic over $\mbf{int}$ can sometimes be much lengthier than their counterparts in commutative ring theory.
The main difficulty arises from the absence of closure of addition and multiplication.
As a consequence many proofs are actually dedicated to the establishment of existence.

The derivations presented in this and the following three sections are quite elementary and are analogous to identities and inequalities that are presented to students in an introductory course to analysis.
Despite the elementary nature of these results they will be presented here in some detail.
We do this for two reasons.
First, they provide a good starting point for the reader to acquire familiarity with the expressiveness of our formal system.
Second, because we are working in the environment $\mathfrak{M}(mach)$ there will be important departures of the derived theorems from their counterparts found in the theory of commutative rings.
The reader should closely examine the extra conditional statements that appear in some of the premises of the theorems to fully appreciate the constraints under which they hold.

At each step of a proof construction, VPC accesses the data file \emph{axiom.dat} that initially stores all of the axioms of the theory under consideration.
In the present context they are axioms \textsf{axi1}-\textsf{axi10}, \textsf{ord1}-\textsf{ord6} and the specific disjunction axioms \textsf{le1}, \textsf{le2}, \textsf{neq1} and \textsf{neq2}. 
As proofs are completed the theorems extracted from them are automatically appended to the file \emph{axiom.dat}.
All axioms and theorems that are stored in \emph{axiom.dat} are provided with a label.
Theorems are labeled by \textsf{thm} followed by a number.
Theorems that are of less interest in themselves but are derived for the purposes of use in other proofs are referred to as lemmas and labeled \textsf{lem} followed by a number.
Lemmas are often used when considering the separate cases of theorems containing disjunctions.

I/O type axioms are labeled \textsf{aio} and the substitution rule is labeled \textsf{sr} followed by a number.
These have a common structure for all applications and are automated in VPC.

The proofs presented below were generated interactively.
At each step of a proof, VPC determines all possible program extensions that can be derived from the current derivation program.
These are listed in an options file that the user can consult to select a desired conclusion program.
Each option includes the axiom/theorem label and the associated connection list.
The user then selects the desired option (conclusion program) to generate a new statement in the derivation program list.
The process is repeated until the proof is completed.
Crucial to the matching procedure of sublists of the derivation program with premise programs of the axioms/theorems stored in \emph{axiom.dat} are program and I/O equivalence.

Theorems and derivation programs are presented as vertical lists. 
The first entry of each line of a proof is the program label (equivalent to the program list element number) followed by the statement.
Following the statement is the connection list.
The connection list is preceded by the axiom/theorem label and contains the labels associated with the premises used to generate the current statement from an extended program derivation.
The absence of a connection list means that the statement is a premise of the derivation program.
When a proof is completed, VPC will extract and store the theorem after it checks for redundant premise statements and redundant steps in the proof.
If redundancies are detected VPC halts with an output that lists the redundant statements.   

Many theorems that are presented below come in pairs, the first part establishing existence and the second part establishing an identity.
The proofs are presented for demonstration purposes only and are not meant to represent the most efficient proof of the given theorem.
We start with some algebraic identities.
The listings are imported directly from the output file \emph{theorem.dat} generated by VPC.

Theorems \textsf{thm1} and \textsf{thm2} highlight the difficulties associated with arithmetic over $\mbf{int}$.
In the theory of fields and rings the identity $a=c-b$ follows trivially from the identity $c=a+b$.   
For arithmetic over $\mbf{int}$ more work is required.

In theorem \textsf{thm1} the existence of $c-b$ over $\mbf{int}$ is established from the premise that $c=a+b$ exists over $\mbf{int}$.
Theorem \textsf{thm2} establishes the identity $a=c-b$.
\begin{lstlisting}
Theorem thm1

add [a b] [c]
mult [-1 b] [d]
---------------
add [c d] [m]

Proof.
  1 add [a b] [c]
  2 mult [-1 b] [d]
  3 add [b d] [e]            axi5b [2]
  4 add [d b] [f]            axi2a [3]
  5 eqi [e 0] [ ]            axi5c [2 3]
  6 eqi [e f] [ ]            axi2b [4 3]
  7 eqi [0 f] [ ]            sr1 [6 5]
  8 add [b a] [g]            axi2a [1]
  9 eqi [g c] [ ]            axi2b [1 8]
 10 typei [a] [ ]            aio [1]
 11 add [a 0] [h]            axi4a [10]
 12 add [0 a] [i]            axi2a [11]
 13 add [f a] [j]            sr1 [12 7]
 14 add [d g] [k]            axi3a [4 13 8]
 15 add [d c] [l]            sr1 [14 9]
 16 add [c d] [m]            axi2a [15]

Theorem thm2

add [a b] [c]
mult [-1 b] [d]
add [c d] [m]
---------------
eqi [m a] [ ]

Proof.
  1 add [a b] [c]
  2 mult [-1 b] [d]
  3 add [c d] [m]
  4 add [b d] [e]            axi5b [2]
  5 add [d b] [f]            axi2a [4]
  6 eqi [e 0] [ ]            axi5c [2 4]
  7 eqi [e f] [ ]            axi2b [5 4]
  8 eqi [0 f] [ ]            sr1 [7 6]
  9 add [b a] [g]            axi2a [1]
 10 eqi [g c] [ ]            axi2b [1 9]
 11 typei [a] [ ]            aio [1]
 12 add [a 0] [h]            axi4a [11]
 13 add [0 a] [i]            axi2a [12]
 14 add [f a] [j]            sr1 [13 8]
 15 add [d g] [k]            axi3a [5 14 9]
 16 add [d c] [l]            axi2a [3]
 17 eqi [m l] [ ]            axi2b [16 3]
 18 eqi [l k] [ ]            sr2 [15 10 16]
 19 eqi [k j] [ ]            axi3b [5 14 9 15]
 20 eqi [l j] [ ]            sr1 [18 19]
 21 eqi [j i] [ ]            sr2 [13 8 14]
 22 eqi [l i] [ ]            sr1 [20 21]
 23 eqi [i h] [ ]            axi2b [12 13]
 24 eqi [l h] [ ]            sr1 [22 23]
 25 eqi [h a] [ ]            axi4b [12]
 26 eqi [l a] [ ]            sr1 [24 25]
 27 eqi [m a] [ ]            sr1 [17 26]
\end{lstlisting}
Theorem \textsf{thm3} shows that if the sums $a+b$ and $a+d$ exist over $\mbf{int}$ and are equal then $b=d$.
\begin{lstlisting}
Theorem thm3

add [a b] [c]
add [a d] [e]
eqi [c e] [ ]
-------------
eqi [b d] [ ]

Proof.
  1 add [a b] [c]
  2 add [a d] [e]
  3 eqi [c e] [ ]
  4 add [b a] [f]            axi2a [1]
  5 add [d a] [g]            axi2a [2]
  6 eqi [f c] [ ]            axi2b [1 4]
  7 eqi [g e] [ ]            axi2b [2 5]
  8 typei [a] [ ]            aio [1]
  9 mult [-1 a] [h]          axi5a [8]
 10 add [f h] [i]            thm1 [4 9]
 11 add [g h] [j]            thm1 [5 9]
 12 add [c h] [k]            sr1 [10 6]
 13 add [e h] [l]            sr1 [11 7]
 14 eqi [i b] [ ]            thm2 [4 9 10]
 15 eqi [j d] [ ]            thm2 [5 9 11]
 16 eqi [k i] [ ]            sr2 [10 6 12]
 17 eqi [l j] [ ]            sr2 [11 7 13]
 18 eqi [l k] [ ]            sr2 [12 3 13]
 19 eqi [k b] [ ]            sr1 [16 14]
 20 eqi [l b] [ ]            sr1 [18 19]
 21 eqi [b l] [ ]            axi1b [20]
 22 eqi [l d] [ ]            sr1 [17 15]
 23 eqi [b d] [ ]            sr1 [21 22]
\end{lstlisting}
Theorem thm4 is the multiplication version of thm3.
It shows that if $a*b$ and $a*d$ exist over $\mbf{int}$ and are equal and $a \neq 0$ then $b=d$.
\begin{lstlisting}
Theorem thm4

mult [a b] [c]
mult [a d] [e]
eqi [c e] [ ]
neq [a 0] [ ]
--------------
eqi [b d] [ ]

Proof.
  1 mult [a b] [c]
  2 mult [a d] [e]
  3 eqi [c e] [ ]
  4 neq [a 0] [ ]
  5 div [c a] [f]            axi10a [4 1]
  6 div [e a] [g]            axi10a [4 2]
  7 eqi [f b] [ ]            axi10b [1 5]
  8 eqi [g d] [ ]            axi10b [2 6]
  9 eqi [g f] [ ]            sr2 [5 3 6]
 10 eqi [g b] [ ]            sr1 [9 7]
 11 eqi [b d] [ ]            sr1 [8 10]
\end{lstlisting}
Theorems \textsf{thm5} and \textsf{thm6} provide another example that highlights the difficulties associated with arithmetic over $\mbf{int}$ where existence is not immediate.
In the theory of fields and commutative rings the existence of $0*a$ follows immediately from the closure of multiplication.
The proof of theorem \textsf{thm5} is a rather lengthy derivation dedicated just to the establishment that $0*a$ exists over $\mbf{int}$. 
This is followed by \textsf{thm6} that establishes the equality $0*a=0$.
\begin{lstlisting}
Theorem thm5

typei [a] [ ]
--------------
mult [0 a] [o]

Proof.
  1 typei [a] [ ]
  2 mult [1 a] [b]           axi8a [1]
  3 mult [a 1] [c]           axi6a [2]
  4 mult [-1 a] [d]          axi5a [1]
  5 mult [a -1] [e]          axi6a [4]
  6 add [a d] [f]            axi5b [4]
  7 typei [1] [ ]            aio [2]
  8 mult [-1 1] [g]          axi5a [7]
  9 mult [1 -1] [h]          axi6a [8]
 10 add [1 g] [i]            axi5b [8]
 11 eqi [h -1] [ ]           axi8b [9]
 12 eqi [g h] [ ]            axi6b [9 8]
 13 eqi [g -1] [ ]           sr1 [12 11]
 14 add [1 -1] [j]           sr1 [10 13]
 15 eqi [d e] [ ]            axi6b [5 4]
 16 add [a e] [k]            sr1 [6 15]
 17 eqi [c b] [ ]            axi6b [2 3]
 18 eqi [b a] [ ]            axi8b [2]
 19 eqi [c a] [ ]            sr1 [17 18]
 20 eqi [a c] [ ]            axi1b [19]
 21 add [c e] [l]            sr1 [16 20]
 22 mult [a j] [m]           axi9b [3 5 21 14]
 23 mult [j a] [n]           axi6a [22]
 24 eqi [j i] [ ]            sr2 [10 13 14]
 25 eqi [i 0] [ ]            axi5c [8 10]
 26 eqi [j 0] [ ]            sr1 [24 25]
 27 mult [0 a] [o]           sr1 [23 26]

Theorem thm6

mult [0 a] [o]
--------------
eqi [o 0] [ ]

Proof.
  1 mult [0 a] [o]
  2 typei [a] [ ]            aio [1]
  3 mult [1 a] [b]           axi8a [2]
  4 mult [a 1] [c]           axi6a [3]
  5 mult [-1 a] [d]          axi5a [2]
  6 mult [a -1] [e]          axi6a [5]
  7 add [a d] [f]            axi5b [5]
  8 typei [1] [ ]            aio [3]
  9 mult [-1 1] [g]          axi5a [8]
 10 mult [1 -1] [h]          axi6a [9]
 11 add [1 g] [i]            axi5b [9]
 12 eqi [h -1] [ ]           axi8b [10]
 13 eqi [g h] [ ]            axi6b [10 9]
 14 eqi [g -1] [ ]           sr1 [13 12]
 15 add [1 -1] [j]           sr1 [11 14]
 16 eqi [d e] [ ]            axi6b [6 5]
 17 add [a e] [k]            sr1 [7 16]
 18 eqi [c b] [ ]            axi6b [3 4]
 19 eqi [b a] [ ]            axi8b [3]
 20 eqi [c a] [ ]            sr1 [18 19]
 21 eqi [a c] [ ]            axi1b [20]
 22 add [c e] [l]            sr1 [17 21]
 23 mult [a j] [m]           axi9b [4 6 22 15]
 24 mult [j a] [n]           axi6a [23]
 25 eqi [j i] [ ]            sr2 [11 14 15]
 26 eqi [i 0] [ ]            axi5c [9 11]
 27 eqi [j 0] [ ]            sr1 [25 26]
 28 eqi [o n] [ ]            sr2 [24 27 1]
 29 eqi [n m] [ ]            axi6b [23 24]
 30 eqi [l m] [ ]            axi9c [15 23 4 6 22]
 31 eqi [m l] [ ]            axi1b [30]
 32 eqi [n l] [ ]            sr1 [29 31]
 33 eqi [l k] [ ]            sr2 [17 21 22]
 34 eqi [n k] [ ]            sr1 [32 33]
 35 eqi [k f] [ ]            sr2 [7 16 17]
 36 eqi [n f] [ ]            sr1 [34 35]
 37 eqi [f 0] [ ]            axi5c [5 7]
 38 eqi [n 0] [ ]            sr1 [36 37]
 39 eqi [o 0] [ ]            sr1 [28 38]
\end{lstlisting}
Theorem \textsf{thm7} shows that $-(-a)=a$.
Note that it follows from the axioms that the additive inverse of an object of type $\mbf{int}$ always exists.
Hence a necessary and sufficient condition for the computability of the premise of \textsf{thm7} is that $a:\mbf{int}$.
Given that upon entry $mult$ checks the type of the value assignments of its input lists, the computability of the premise program is guaranteed if $a:\mbf{int}$.
\begin{lstlisting}
Theorem thm7

mult [-1 a] [b]
mult [-1 b] [c]
---------------
eqi [c a] [ ]

Proof.
  1 mult [-1 a] [b]
  2 mult [-1 b] [c]
  3 add [a b] [d]            axi5b [1]
  4 add [b c] [e]            axi5b [2]
  5 eqi [d 0] [ ]            axi5c [1 3]
  6 eqi [e 0] [ ]            axi5c [2 4]
  7 eqi [0 e] [ ]            axi1b [6]
  8 eqi [d e] [ ]            sr1 [5 7]
  9 add [b a] [f]            axi2a [3]
 10 eqi [f d] [ ]            axi2b [3 9]
 11 eqi [f e] [ ]            sr1 [10 8]
 12 eqi [a c] [ ]            thm3 [9 4 11]
 13 eqi [c a] [ ]            axi1b [12]
\end{lstlisting}
Theorems \textsf{thm8} and \textsf{thm9} show that if $a*b$ exists over $\mbf{int}$ then $a*(-b)$ also exists over $\mbf{int}$ and is equal to the additive inverse of $a*b$, i.e. $a*(-b)=-(a*b)$.
\begin{lstlisting}
Theorem thm8

mult [a b] [c]
mult [-1 b] [d]
---------------
mult [a d] [i]

Proof.
  1 mult [a b] [c]
  2 mult [-1 b] [d]
  3 typei [c] [ ]            aio [1]
  4 mult [-1 c] [e]          axi5a [3]
  5 mult [b -1] [f]          axi6a [2]
  6 mult [c -1] [g]          axi6a [4]
  7 eqi [f d] [ ]            axi6b [2 5]
  8 mult [a f] [h]           axi7a [1 6 5]
  9 mult [a d] [i]           sr1 [8 7]

Theorem thm9

mult [a b] [c]
mult [-1 b] [d]
mult [a d] [i]
mult [-1 c] [e]
---------------
eqi [i e] [ ]

Proof.
  1 mult [a b] [c]
  2 mult [-1 b] [d]
  3 mult [a d] [i]
  4 mult [-1 c] [e]
  5 mult [b -1] [f]          axi6a [2]
  6 mult [c -1] [g]          axi6a [4]
  7 eqi [f d] [ ]            axi6b [2 5]
  8 eqi [g e] [ ]            axi6b [4 6]
  9 mult [a f] [h]           axi7a [1 6 5]
 10 eqi [h g] [ ]            axi7b [1 6 5 9]
 11 eqi [i h] [ ]            sr2 [9 7 3]
 12 eqi [h e] [ ]            sr1 [10 8]
 13 eqi [i e] [ ]            sr1 [11 12]
\end{lstlisting}
Theorems \textsf{thm10} and \textsf{thm11} show that if $a*b$ exists over $\mbf{int}$ then $(-a)*b$ also exists over $\mbf{int}$ and is equal to the additive inverse of $a*b$, i.e. $(-a)*b=-(a*b)$.
\begin{lstlisting}
Theorem thm10

mult [a b] [c]
mult [-1 a] [d]
---------------
mult [d b] [g]

Proof.
  1 mult [a b] [c]
  2 mult [-1 a] [d]
  3 mult [b a] [e]           axi6a [1]
  4 mult [b d] [f]           thm8 [3 2]
  5 mult [d b] [g]           axi6a [4]

Theorem thm11

mult [a b] [c]
mult [-1 a] [d]
mult [d b] [g]
mult [-1 c] [h]
---------------
eqi [g h] [ ]

Proof.
  1 mult [a b] [c]
  2 mult [-1 a] [d]
  3 mult [d b] [g]
  4 mult [-1 c] [h]
  5 eqi [h g] [ ]            axi7b [2 3 1 4]
  6 eqi [g h] [ ]            axi1b [5]
\end{lstlisting}
Theorems \textsf{thm12} and \textsf{thm13} show that if $a*b$ exists over $\mbf{int}$ then $(-a)*(-b)$ also exists over $\mbf{int}$ and is equal to $a*b$, i.e. $(-a)*(-b)=a*b$.
\begin{lstlisting}
Theorem thm12

mult [a b] [c]
mult [-1 a] [d]
mult [-1 b] [e]
---------------
mult [d e] [g]

Proof.
  1 mult [a b] [c]
  2 mult [-1 a] [d]
  3 mult [-1 b] [e]
  4 mult [a e] [f]           thm8 [1 3]
  5 mult [d e] [g]           thm10 [4 2]

Theorem thm13

mult [a b] [c]
mult [-1 a] [d]
mult [-1 b] [e]
mult [d e] [f]
---------------
eqi [f c] [ ]

Proof.
  1 mult [a b] [c]
  2 mult [-1 a] [d]
  3 mult [-1 b] [e]
  4 mult [d e] [f]
  5 typei [c] [ ]            aio [1]
  6 mult [-1 c] [g]          axi5a [5]
  7 typei [g] [ ]            aio [6]
  8 mult [-1 g] [h]          axi5a [7]
  9 eqi [h c] [ ]            thm7 [6 8]
 10 mult [a e] [i]           thm8 [1 3]
 11 eqi [i g] [ ]            thm9 [1 3 10 6]
 12 mult [-1 i] [j]          axi7a [2 4 10]
 13 eqi [j f] [ ]            axi7b [2 4 10 12]
 14 eqi [h j] [ ]            sr2 [12 11 8]
 15 eqi [h f] [ ]            sr1 [14 13]
 16 eqi [f c] [ ]            sr1 [9 15]
\end{lstlisting}

\section{Inequalities.}

The inequalities derived here are fairly straight forward.
The final derivation involves an application of the disjunction contraction rule.
With the use of disjunction splitting the proofs associated with the separate operand programs precede the proof of the main program containing the disjunction.
They correspond to the separate cases that are accessed by the proof of the main program containing the disjunction and are stored as lemmas.
Lemmas are labeled by \textsf{lem} followed by a number.
A statement followed by an asterisk $*$ indicates that disjunction splitting has been applied to the operands of that statement.
The connection list of the conclusion statement in theorems derived by the disjunction contraction rules will differ from the standard connection list format as outline in Section \ref{sdisj}.

Theorem \textsf{thm14} shows that if $a>0$ then $-a<0$ and theorem \textsf{thm15} shows that if $a<0$ then $-a>0$.
\begin{lstlisting}
Theorem thm14

lt [0 a] [ ]
mult [-1 a] [b]
---------------
lt [b 0] [ ]

Proof.
  1 lt [0 a] [ ]
  2 mult [-1 a] [b]
  3 add [a b] [c]            axi5b [2]
  4 eqi [c 0] [ ]            axi5c [2 3]
  5 typei [b] [ ]            aio [3]
  6 add [b 0] [d]            axi4a [5]
  7 eqi [d b] [ ]            axi4b [6]
  8 add [0 b] [e]            axi2a [6]
  9 eqi [e d] [ ]            axi2b [6 8]
 10 eqi [e b] [ ]            sr1 [9 7]
 11 lt [e c] [ ]             ord1a [1 8 3]
 12 lt [b c] [ ]             sr1 [11 10]
 13 lt [b 0] [ ]             sr1 [12 4]

Theorem thm15

lt [a 0] [ ]
mult [-1 a] [b]
---------------
lt [0 b] [ ]

Proof.
  1 lt [a 0] [ ]
  2 mult [-1 a] [b]
  3 add [a b] [c]            axi5b [2]
  4 eqi [c 0] [ ]            axi5c [2 3]
  5 typei [b] [ ]            aio [3]
  6 add [b 0] [d]            axi4a [5]
  7 eqi [d b] [ ]            axi4b [6]
  8 add [0 b] [e]            axi2a [6]
  9 eqi [e d] [ ]            axi2b [6 8]
 10 eqi [e b] [ ]            sr1 [9 7]
 11 lt [c e] [ ]             ord1a [1 3 8]
 12 lt [c b] [ ]             sr1 [11 10]
 13 lt [0 b] [ ]             sr1 [12 4]
\end{lstlisting}
We include the following result that will also be needed in later derivations.
Like axiom \textsf{ord4}, theorem \textsf{thm16} has an empty list premise.
\begin{lstlisting}
Theorem thm16

-------------
lt [-1 0] [ ]

Proof.
  1 lt [0 1] [ ]             ord4
  2 typei [1] [ ]            aio [1]
  3 mult [-1 1] [a]          axi5a [2]
  4 mult [1 -1] [b]          axi6a [3]
  5 eqi [b -1] [ ]           axi8b [4]
  6 eqi [a b] [ ]            axi6b [4 3]
  7 eqi [a -1] [ ]           sr1 [6 5]
  8 lt [a 0] [ ]             thm14 [1 3]
  9 lt [-1 0] [ ]            sr1 [8 7]
\end{lstlisting}
As a first application of the disjunction contraction rule we establish that if $a \neq 0$ and $a^2:\mbf{int}$ then $a^2>0$.
Theorem \textsf{thm17} is preceded by lemmas \textsf{lem1} and \textsf{lem2} that are associated with derivations based upon the operand programs that result from the disjunction splitting in theorem \textsf{thm17}.
\begin{lstlisting}
Lemma lem1

lt [0 a] [ ]
mult [a a] [b]
--------------
lt [0 b] [ ]

Proof.
  1 lt [0 a] [ ]
  2 mult [a a] [b]
  3 typei [a] [ ]            aio [1]
  4 mult [0 a] [c]           thm5 [3]
  5 eqi [c 0] [ ]            thm6 [4]
  6 lt [c b] [ ]             ord2a [1 1 4 2]
  7 lt [0 b] [ ]             sr1 [6 5]

Lemma lem2

lt [a 0] [ ]
mult [a a] [b]
--------------
lt [0 b] [ ]

Proof.
  1 lt [a 0] [ ]
  2 mult [a a] [b]
  3 typei [a] [ ]            aio [1]
  4 mult [0 a] [c]           thm5 [3]
  5 eqi [c 0] [ ]            thm6 [4]
  6 lt [c b] [ ]             ord2b [1 1 2 4]
  7 lt [0 b] [ ]             sr1 [6 5]
\end{lstlisting}
We now apply the disjunction contraction rule.
\begin{lstlisting}
Theorem thm17

neq [a 0] [ ]
mult [a a] [b]
--------------
lt [0 b] [ ]

Proof.
  1 neq [a 0] [ ] *
  2 mult [a a] [b]
  3 lt [0 b] [ ]             disj [lem2 lem1]
\end{lstlisting}
VPC splits the premise of theorem \textsf{thm17} into the two operand programs
\be
~[lt~[0~a]~[~]~mult~[a~a]~[b]]
\ee
and
\be
~[lt~[a~0]~[~]~mult~[a~a]~[b]]
\ee
A search is conducted for premises of the axioms/theorems stored in the file \emph{axiom.dat} that can be matched to sublists of each operand program and their conclusions stored in memory.
It then searches through the two collections of conclusions associated with each operand program and extracts those conclusions that are common to both.

\section{Non strict inequalities over $\mbf{int}$.}

Before moving onto absolute values we need to generalize some of the inequalities just derived by replacing the strict inequality $<$ with the non-strict inequality $\leq$.

The following two theorems involve mixed inequalities.
\begin{lstlisting}
Theorem thm18

lt [a b] [ ]
le [b c] [ ]
------------
lt [a c] [ ]

Proof.
  1 lt [a b] [ ]
  2 le [b c] [ ] *
  3 lt [a c] [ ]             disj [ord3 sr1]

Lemma lem3

eqi [a b] [ ]
lt [b c] [ ]
-------------
lt [a c] [ ]

Proof.
  1 eqi [a b] [ ]
  2 lt [b c] [ ]
  3 eqi [b a] [ ]            axi1b [1]
  4 lt [a c] [ ]             sr1 [2 3]

Theorem thm19

le [a b] [ ]
lt [b c] [ ]
------------
lt [a c] [ ]

Proof.
  1 le [a b] [ ] *
  2 lt [b c] [ ]
  3 lt [a c] [ ]             disj [ord3 lem3]
\end{lstlisting}

When combined, theorems \textsf{thm20} and \textsf{thm21} show that the non-strict inequality satisfies the substitution rule.
This is an example that demonstrates why the substitution rule should not be applied as an axiom to all programs. 
\begin{lstlisting}
Lemma lem4

eqi [a b] [ ]
eqi [b c] [ ]
-------------
le [a c] [ ]

Proof.
  1 eqi [a b] [ ]
  2 eqi [b c] [ ]
  3 eqi [a c] [ ]            sr1 [1 2]
  4 le [a c] [ ]             le2 [3]

Lemma lem5

lt [a b] [ ]
eqi [b c] [ ]
-------------
le [a c] [ ]

Proof.
  1 lt [a b] [ ]
  2 eqi [b c] [ ]
  3 lt [a c] [ ]             sr1 [1 2]
  4 le [a c] [ ]             le1 [3]

Theorem thm20

le [a b] [ ]
eqi [b c] [ ]
-------------
le [a c] [ ]

Proof.
  1 le [a b] [ ] *
  2 eqi [b c] [ ]
  3 le [a c] [ ]             disj [lem5 lem4]

Lemma lem6

eqi [a b] [ ]
eqi [a c] [ ]
-------------
le [c b] [ ]

Proof.
  1 eqi [a b] [ ]
  2 eqi [a c] [ ]
  3 eqi [c b] [ ]            sr1 [1 2]
  4 le [c b] [ ]             le2 [3]

Lemma lem7

lt [a b] [ ]
eqi [a c] [ ]
-------------
le [c b] [ ]

Proof.
  1 lt [a b] [ ]
  2 eqi [a c] [ ]
  3 lt [c b] [ ]             sr1 [1 2]
  4 le [c b] [ ]             le1 [3]

Theorem thm21

le [a b] [ ]
eqi [a c] [ ]
-------------
le [c b] [ ]

Proof.
  1 le [a b] [ ] *
  2 eqi [a c] [ ]
  3 le [c b] [ ]             disj [lem7 lem6]
\end{lstlisting}

Theorem \textsf{thm22} generalizes the order axiom of transitivity, \textsf{ord3}, and can be translated to the statement that if $a \leq b$ and $b \leq c$ then $a \leq c$.
\begin{lstlisting}
Lemma lem8

le [a b] [ ]
lt [b c] [ ]
------------
le [a c] [ ]

Proof.
  1 le [a b] [ ]
  2 lt [b c] [ ]
  3 lt [a c] [ ]             thm19 [1 2]
  4 le [a c] [ ]             le1 [3]

Theorem thm22

le [a b] [ ]
le [b c] [ ]
------------
le [a c] [ ]

Proof.
  1 le [a b] [ ]
  2 le [b c] [ ] *
  3 le [a c] [ ]             disj [lem8 thm20]
\end{lstlisting}

Theorem \textsf{thm23} generalizes theorem \textsf{thm14} and can be translated to the statement that if $c \geq 0$ then $-c \leq 0$.
\begin{lstlisting}
Lemma lem9

lt [0 c] [ ]
mult [-1 c] [d]
---------------
le [d 0] [ ]

Proof.
  1 lt [0 c] [ ]
  2 mult [-1 c] [d]
  3 lt [d 0] [ ]             thm14 [1 2]
  4 le [d 0] [ ]             le1 [3]

Lemma lem10

eqi [0 c] [ ]
mult [-1 c] [d]
---------------
le [d 0] [ ]

Proof.
  1 eqi [0 c] [ ]
  2 mult [-1 c] [d]
  3 eqi [c 0] [ ]            axi1b [1]
  4 mult [-1 0] [a]          sr1 [2 3]
  5 mult [0 -1] [b]          axi6a [4]
  6 eqi [a b] [ ]            axi6b [5 4]
  7 eqi [b 0] [ ]            thm6 [5]
  8 eqi [a 0] [ ]            sr1 [6 7]
  9 eqi [d a] [ ]            sr2 [4 1 2]
 10 le [d 0] [ ]             lem4 [9 8]

Theorem thm23

le [0 c] [ ]
mult [-1 c] [d]
---------------
le [d 0] [ ]

Proof.
  1 le [0 c] [ ] *
  2 mult [-1 c] [d]
  3 le [d 0] [ ]             disj [lem9 lem10]
\end{lstlisting}
Theorem \textsf{thm24} generalizes axiom \textsf{ord1a} and can be translated to the statement that if $a \leq b$ and the sums $a+c$ and $b+c$ exist over $\mbf{int}$ then $a+c \leq b+c$.
\begin{lstlisting}
Lemma lem11

lt [a b] [ ]
add [a c] [x]
add [b c] [y]
-------------
le [x y] [ ]

Proof.
  1 lt [a b] [ ]
  2 add [a c] [x]
  3 add [b c] [y]
  4 lt [x y] [ ]             ord1a [1 2 3]
  5 le [x y] [ ]             le1 [4]

Lemma lem12

eqi [a b] [ ]
add [a c] [x]
add [b c] [y]
-------------
le [x y] [ ]

Proof.
  1 eqi [a b] [ ]
  2 add [a c] [x]
  3 add [b c] [y]
  4 eqi [y x] [ ]            sr2 [2 1 3]
  5 eqi [x y] [ ]            axi1b [4]
  6 le [x y] [ ]             le2 [5]

Theorem thm24

le [a b] [ ]
add [a c] [x]
add [b c] [y]
-------------
le [x y] [ ]

Proof.
  1 le [a b] [ ] *
  2 add [a c] [x]
  3 add [b c] [y]
  4 le [x y] [ ]             disj [lem11 lem12]
\end{lstlisting}
Theorem \textsf{thm25} generalizes axiom \textsf{ord1b} and can be translated to the statement that if $a \leq b$ and $c \leq d$ and the sums $a+c$ and $b+d$ exist over $\mbf{int}$ then $a+c \leq b+d$.
\begin{lstlisting}
Lemma lem13

lt [a b] [ ]
eqi [c d] [ ]
add [a c] [x]
add [b d] [y]
-------------
le [x y] [ ]

Proof.
  1 lt [a b] [ ]
  2 eqi [c d] [ ]
  3 add [a c] [x]
  4 add [b d] [y]
  5 eqi [d c] [ ]            axi1b [2]
  6 add [b c] [e]            sr1 [4 5]
  7 eqi [e y] [ ]            sr2 [4 5 6]
  8 le [x e] [ ]             lem11 [1 3 6]
  9 le [x y] [ ]             thm20 [8 7]

Lemma lem14

lt [a b] [ ]
lt [c d] [ ]
add [a c] [x]
add [b d] [y]
-------------
le [x y] [ ]

Proof.
  1 lt [a b] [ ]
  2 lt [c d] [ ]
  3 add [a c] [x]
  4 add [b d] [y]
  5 lt [x y] [ ]             ord1b [1 2 3 4]
  6 le [x y] [ ]             le1 [5]

Lemma lem15

lt [a b] [ ]
le [c d] [ ]
add [a c] [x]
add [b d] [y]
-------------
le [x y] [ ]

Proof.
  1 lt [a b] [ ]
  2 le [c d] [ ] *
  3 add [a c] [x]
  4 add [b d] [y]
  5 le [x y] [ ]             disj [lem14 lem13]

Lemma lem16

eqi [a b] [ ]
le [c d] [ ]
add [a c] [x]
add [b d] [y]
-------------
le [x y] [ ]

Proof.
  1 eqi [a b] [ ]
  2 le [c d] [ ]
  3 add [a c] [x]
  4 add [b d] [y]
  5 add [c a] [e]            axi2a [3]
  6 add [c b] [f]            sr1 [5 1]
  7 add [d b] [g]            axi2a [4]
  8 le [f g] [ ]             thm24 [2 6 7]
  9 eqi [g y] [ ]            axi2b [4 7]
 10 eqi [f e] [ ]            sr2 [5 1 6]
 11 eqi [e x] [ ]            axi2b [3 5]
 12 eqi [f x] [ ]            sr1 [10 11]
 13 le [f y] [ ]             thm20 [8 9]
 14 le [x y] [ ]             thm21 [13 12]

Theorem thm25

le [a b] [ ]
le [c d] [ ]
add [a c] [x]
add [b d] [y]
-------------
le [x y] [ ]

Proof.
  1 le [a b] [ ] *
  2 le [c d] [ ]
  3 add [a c] [x]
  4 add [b d] [y]
  5 le [x y] [ ]             disj [lem15 lem16]
\end{lstlisting}

\section{Absolute values over $\mbf{int}$.}\label{savoi}

We start by showing that $|a| \geq 0$.
\begin{lstlisting}
Lemma lem17

lt [a 0] [ ]
mult [-1 a] [b]
---------------
le [0 b] [ ]

Proof.
  1 lt [a 0] [ ]
  2 mult [-1 a] [b]
  3 lt [0 b] [ ]             thm15 [1 2]
  4 le [0 b] [ ]             le1 [3]

Lemma lem18

le [0 a] [ ]
mult [1 a] [b]
--------------
le [0 b] [ ]

Proof.
  1 le [0 a] [ ]
  2 mult [1 a] [b]
  3 eqi [b a] [ ]            axi8b [2]
  4 eqi [a b] [ ]            axi1b [3]
  5 le [0 b] [ ]             thm20 [1 4]

Theorem thm26

abs [a] [b]
------------
le [0 b] [ ]

Proof.
  1 abs [a] [b] *
  2 le [0 b] [ ]             disj [lem17 lem18]
\end{lstlisting}
The next two theorems are examples where a disjunction splitting and contraction involves detecting a false program in one of the operand derivation programs.
Theorem \textsf{thm27} is equivalent to the statement that if $|a|=0$ then $a=0$.
It is preceded by two lemmas, \textsf{lem19} and \textsf{lem20}, that are associated with the two operand programs that result from disjunction splitting in \textsf{thm27}.
The premise of the first lemma, \textsf{lem19}, is type $\mbf{false}$.
Appealing to the disjunction contraction rule 2, the conclusion of the second lemma, \textsf{lem20}, is contracted back onto the main proof of \textsf{thm27}.
\begin{lstlisting}
Lemma lem19

lt [a 0] [ ]
mult [-1 a] [b]
eqi [b 0] [ ]
---------------
false

Proof.
  1 lt [a 0] [ ]
  2 mult [-1 a] [b]
  3 eqi [b 0] [ ]
  4 lt [0 b] [ ]             thm15 [1 2]
  5 lt [0 0] [ ]             sr1 [4 3]
  6 false                    ord5 [5]

Lemma lem20

mult [1 a] [b]
eqi [b 0] [ ]
--------------
eqi [a 0] [ ]

Proof.
  1 mult [1 a] [b]
  2 eqi [b 0] [ ]
  3 eqi [b a] [ ]            axi8b [1]
  4 eqi [a 0] [ ]            sr1 [2 3]

Theorem thm27

abs [a] [b]
eqi [b 0] [ ]
-------------
eqi [a 0] [ ]

Proof.
  1 abs [a] [b] *
  2 eqi [b 0] [ ]
  3 eqi [a 0] [ ]            disj [lem19 lem20]
\end{lstlisting}

Theorem \textsf{thm28} is the converse of \textsf{thm27} and is equivalent to the statement that if $a=0$ then $|a|=0$.
\begin{lstlisting}
Lemma lem21

lt [a 0] [ ]
eqi [a 0] [ ]
-------------
false

Proof.
  1 lt [a 0] [ ]
  2 eqi [a 0] [ ]
  3 lt [0 0] [ ]             sr1 [1 2]
  4 false                    ord5 [3]

Lemma lem22

mult [1 a] [b]
eqi [a 0] [ ]
--------------
eqi [b 0] [ ]

Proof.
  1 mult [1 a] [b]
  2 eqi [a 0] [ ]
  3 eqi [b a] [ ]            axi8b [1]
  4 eqi [b 0] [ ]            sr1 [3 2]

Theorem thm28

abs [a] [b]
eqi [a 0] [ ]
-------------
eqi [b 0] [ ]

Proof.
  1 abs [a] [b] *
  2 eqi [a 0] [ ]
  3 eqi [b 0] [ ]            disj [lem21 lem22]
\end{lstlisting}

We now prove that if $|a| \leq c$ then $-c \leq a \leq c$.
Theorems \textsf{thm29} and \textsf{thm30}, respectively, split this into the two parts leading to the conclusions $a \leq c$ and $-c \leq a$, respectively.
\begin{lstlisting}
Lemma lem23

lt [a 0] [ ]
mult [-1 a] [b]
le [b c] [ ]
---------------
le [a c] [ ]

Proof.
  1 lt [a 0] [ ]
  2 mult [-1 a] [b]
  3 le [b c] [ ]
  4 lt [0 b] [ ]             thm15 [1 2]
  5 lt [a b] [ ]             ord3 [1 4]
  6 lt [a c] [ ]             thm18 [5 3]
  7 le [a c] [ ]             le1 [6]

Lemma lem24

mult [1 a] [b]
le [b c] [ ]
--------------
le [a c] [ ]

Proof.
  1 mult [1 a] [b]
  2 le [b c] [ ]
  3 eqi [b a] [ ]            axi8b [1]
  4 le [a c] [ ]             thm21 [2 3]

Theorem thm29

abs [a] [b]
le [b c] [ ]
------------
le [a c] [ ]

Proof.
  1 abs [a] [b] *
  2 le [b c] [ ]
  3 le [a c] [ ]             disj [lem23 lem24]

Lemma lem25

mult [-1 a] [b]
lt [b c] [ ]
mult [-1 c] [d]
---------------
le [d a] [ ]

Proof.
  1 mult [-1 a] [b]
  2 lt [b c] [ ]
  3 mult [-1 c] [d]
  4 typei [b] [ ]            aio [2]
  5 mult [-1 b] [e]          axi5a [4]
  6 eqi [e a] [ ]            thm7 [1 5]
  7 mult [b -1] [f]          axi6a [5]
  8 eqi [f e] [ ]            axi6b [5 7]
  9 eqi [f a] [ ]            sr1 [8 6]
 10 mult [c -1] [g]          axi6a [3]
 11 lt [-1 0] [ ]            thm16
 12 lt [g f] [ ]             ord2b [2 11 7 10]
 13 eqi [d g] [ ]            axi6b [10 3]
 14 lt [d f] [ ]             lem3 [13 12]
 15 le [d a] [ ]             lem5 [14 9]

Lemma lem26

mult [-1 a] [b]
eqi [b c] [ ]
mult [-1 c] [d]
---------------
le [d a] [ ]

Proof.
  1 mult [-1 a] [b]
  2 eqi [b c] [ ]
  3 mult [-1 c] [d]
  4 eqi [c b] [ ]            axi1b [2]
  5 mult [-1 b] [e]          sr1 [3 4]
  6 eqi [e a] [ ]            thm7 [1 5]
  7 eqi [d e] [ ]            sr2 [5 2 3]
  8 le [d a] [ ]             lem4 [7 6]

Lemma lem27

mult [-1 a] [b]
le [b c] [ ]
mult [-1 c] [d]
---------------
le [d a] [ ]

Proof.
  1 mult [-1 a] [b]
  2 le [b c] [ ] *
  3 mult [-1 c] [d]
  4 le [d a] [ ]             disj [lem25 lem26]

Lemma lem28

le [0 a] [ ]
mult [1 a] [b]
le [b c] [ ]
mult [-1 c] [d]
---------------
le [d a] [ ]

Proof.
  1 le [0 a] [ ]
  2 mult [1 a] [b]
  3 le [b c] [ ]
  4 mult [-1 c] [d]
  5 le [a c] [ ]             lem24 [2 3]
  6 le [0 c] [ ]             thm22 [1 5]
  7 le [d 0] [ ]             thm23 [6 4]
  8 le [d a] [ ]             thm22 [7 1]

Theorem thm30

abs [a] [b]
le [b c] [ ]
mult [-1 c] [d]
---------------
le [d a] [ ]

Proof.
  1 abs [a] [b] *
  2 le [b c] [ ]
  3 mult [-1 c] [d]
  4 le [d a] [ ]             disj [lem27 lem28]
\end{lstlisting}
Theorem \textsf{thm31} proves the converse statement that if $-c \leq a \leq c$ then $|a| \leq c$.
\begin{lstlisting}
Lemma lem29

le [a c] [ ]
mult [1 a] [b]
--------------
le [b c] [ ]

Proof.
  1 le [a c] [ ]
  2 mult [1 a] [b]
  3 eqi [b a] [ ]            axi8b [2]
  4 eqi [a b] [ ]            axi1b [3]
  5 le [b c] [ ]             thm21 [1 4]

Theorem thm31

mult [-1 c] [d]
le [a c] [ ]
le [d a] [ ]
abs [a] [b]
---------------
le [b c] [ ]

Proof.
  1 mult [-1 c] [d]
  2 le [a c] [ ]
  3 le [d a] [ ]
  4 abs [a] [b] *
  5 le [b c] [ ]             disj [lem27 lem29]
\end{lstlisting}
When combined, theorems \textsf{thm32} and \textsf{thm33}, state that $-|a| \leq a \leq |a|$.
No disjunction splitting is required in the proofs. 
\begin{lstlisting}
Theorem thm32

abs [a] [b]
------------
le [a b] [ ]

Proof.
  1 abs [a] [b]
  2 typei [b] [ ]            aio [1]
  3 eqi [b b] [ ]            axi1a [2]
  4 le [b b] [ ]             le2 [3]
  5 le [a b] [ ]             thm29 [1 4]

Theorem thm33

abs [a] [b]
mult [-1 b] [c]
---------------
le [c a] [ ]

Proof.
  1 abs [a] [b]
  2 mult [-1 b] [c]
  3 typei [b] [ ]            aio [2]
  4 eqi [b b] [ ]            axi1a [3]
  5 le [b b] [ ]             le2 [4]
  6 le [c a] [ ]             thm30 [1 5 2
\end{lstlisting}
Theorem \textsf{thm34} states that if $|x|+|y|$ and $|x+y|$ exist over $\mbf{int}$ then $|x+y| \leq |x|+|y|$.
No disjunction splitting is required.
\begin{lstlisting}
Theorem thm34

abs [x] [u]
abs [y] [v]
add [u v] [w]
add [x y] [z]
abs [z] [p]
-------------
le [p w] [ ]

Proof.
  1 abs [x] [u]
  2 abs [y] [v]
  3 add [u v] [w]
  4 add [x y] [z]
  5 abs [z] [p]
  6 le [x u] [ ]             thm32 [1]
  7 le [y v] [ ]             thm32 [2]
  8 le [z w] [ ]             thm25 [6 7 4 3]
  9 typei [u] [ ]            aio [3]
 10 mult [-1 u] [a]          axi5a [9]
 11 typei [v] [ ]            aio [3]
 12 mult [-1 v] [b]          axi5a [11]
 13 le [a x] [ ]             thm33 [1 10]
 14 le [b y] [ ]             thm33 [2 12]
 15 typei [w] [ ]            aio [8]
 16 mult [-1 w] [c]          axi5a [15]
 17 add [a b] [d]            axi9a [3 16 10 12]
 18 eqi [d c] [ ]            axi9c [3 16 10 12 17]
 19 le [d z] [ ]             thm25 [13 14 17 4]
 20 le [c z] [ ]             thm21 [19 18]
 21 le [p w] [ ]             thm31 [16 8 20 5]
\end{lstlisting}

\underline{Notes.}

\begin{itemize}
	
	\item As with many of the derivations presented in this chapter, theorem \textsf{thm34} is weaker than its counterpart in the theory of commutative rings.
	The reason for this is that for arithmetic over $\mbf{int}$ the existence of $|x|+|y|$ is not guaranteed given the existence of $x+y$.
	In the premise of \textsf{thm34} we must also include the extra conditional statement that $|x|+|y|$ exist over $\mbf{int}$.   
	
	\item There are a few derivations of standard identities for absolute values that have been omitted.
	We leave as an exercise for the reader to establish the following.
	(i) $|-a|=|a|$, (ii) If $|a*b|:\mbf{int}$ then $|a*b|=|a|*|b|$, (iii) If $a^2:\mbf{int}$ then $|a|^2=a^2$.
	These have been omitted because their proofs can be rather lengthy due to the need to apply a few more applications of disjunction splitting.
	Otherwise they are fairly straight forward.  

\end{itemize}

\section{Arithmetic over $\mbf{rat}$.}

We could also work with the rationals, i.e. objects of type $\mbf{rat}$.
An object of type $\mbf{rat}$ can take on any one of the assigned values
\be
0, \pm \epsilon, \pm 2 \epsilon, \ldots , \pm nint * \epsilon
\ee
where $0<\epsilon << 1$ and $\epsilon$ is also a machine specific parameter.
The finite collection of rationals, $\mbf{rat}$, has a fixed resolution size so that arithmetic over $\mbf{rat}$ differs from floating point arithmetic.
Because of this all of the results of arithmetic over $\mbf{int}$ of the previous sections can be directly applied to $\mbf{rat}$.

We use the same atomic programs with the important modification that all type checks within the atomic programs that are associated with $\mbf{int}$ are replaced by $\mbf{rat}$.
We accept the same axioms \textsf{axi1}-\textsf{axi10} and \textsf{ord1}-\textsf{ord6} by replacing all references to type $\mbf{int}$ objects by type $\mbf{rat}$ objects.

From the ordered ring axioms it can be shown that if $0<x<y$ then $0 < \frac{1}{y} < \frac{1}{x}$.
The standard proof follows by first establishing that if $0<x$ then $0 < \frac{1}{x}$.
For an ordered ring we can derive the result that for any nonzero element $x$, $x^2>0$.
Hence we have $(\frac{1}{x})^2 >0$ and using the second ordered ring axiom we obtain $x (\frac{1}{x})^2 >0$ and the desired result follows.

Adapting theorem \textsf{thm17} to $\mbf{rat}$ we have that if $a:\mbf{rat}$, $a \neq 0$ and $a^2:\mbf{rat}$ then $a^2>0$.
But $a^2:\mbf{rat}$ does not necessarily follow from $a:\mbf{rat}$.
For this reason the standard proof that starts with the result $(\frac{1}{x})^2 >0$ cannot be used.

In the absence of a known proof, for arithmetic over $\mbf{rat}$ we include the additional order axioms
\be
\begin{array}{l}
	lt~[0~a]~[~] \\
	div~[1~a]~[x] \\
	\hline
	lt~[0~x]~[~] \\
\end{array}
\quad \textsf{ord7a} \hspace{10mm}
\begin{array}{l}
	lt~[0~a]~[~] \\
	lt~[a~b]~[~] \\
	div~[1~a]~[x] \\
	div~[1~b]~[y] \\
	\hline
	lt~[y~x]~[~] \\
\end{array}
\quad \textsf{ord7b}
\ee
Note that $div~[1~a]~[x]$ does not necessarily follow from $a:\mbf{rat}$ and $a \neq 0$.
Because of this the premises in the above axioms are conditional on the computability of the statements $div~[1~a]~[x]$ and $div~[1~b]~[y]$.

The previous sections largely addressed the prevention of arithmetic operations that lead to overflows.
Special care needs to be exercised when dealing with objects of type $\mbf{rat}$ because we are now faced with possible underflows as well as overflows.
This is because the absolute values of objects of $\mbf{rat}$ are not only bounded above by $\epsilon*nint$ but also have a finite resolution, $\epsilon$, that provides a lower bound on operations of multiplication of nonzero elements of $\mbf{rat}$.

Working over $\mbf{rat}$ is often desired because dynamical systems that model objects that are defined by integer state vectors can sometimes generate huge integers.
However, it is often the case that when constructing a model over $\mbf{rat}$ we have actually applied some scaling law to a dynamical system that has originally been posed over $\mbf{int}$.
When attempting to establish the computability of a model based over $\mbf{rat}$ it is usually safer to return to the original formulation and carry out the analysis on the original model based over $\mbf{int}$.

\newpage 

\underline{Notes.}

\begin{itemize}
	
	\item For fully discrete dynamical systems we have defined state vectors as a list of integers that represent the minimum amount of information needed to completely describe the intrinsic properties and dynamic state of an object in that system.
	There are applications where we might have to generalize the definition of a state vector to include elements that are rational numbers.
	When discussing dynamical systems we will, for brevity, always refer to state vectors as a list of integers.
	However, it should be kept in mind that, in light of the outline given in this section, there is no loss of generality if we are forced to employ state vectors whose elements contain rational numbers.
	    
\end{itemize}

\chapter{Vectors.}\label{vec}

\section{Vector programs.}

Our ultimate goal is to construct a theory from which we can rigorously validate computer models of dynamical systems based on rule based algorithms and finite state arithmetic.
In such models physical objects and their dynamic state are expressed in terms of state vectors that we have defined as being essentially represented by a list of integers. 
Before proceeding to dynamical systems we will present some basic properties of programs whose I/O lists contain elements that are integer vectors.
As discussed in Section \ref{sarray}, an integer vector of dimensions $m:\mbf{int1}$ has the type $\mbf{vec}[m]$.
For programs whose I/O lists contain elements that are vectors we adopt a dimension free format 

A dimension free format means that the dimensions of a vector will not be explicitly specified in the I/O lists of atomic programs.
Every atomic program will internally identify the dimensions of the vectors upon entry.
We can regard the dimensions of the initial vectors, along with value assignments of their elements, to be prescribed by the $read$  program under the general program structure (\ref{core}).
New value assigned vectors are generated from atomic assignment programs that internally set the dimensions of the new vectors.
Once the dimensions of a vector have been assigned they are stored in memory and accessed whenever that vector is employed as input to another program.

For example the assignment program
\be
p~[a]~[b] , \quad a:\mbf{vec}[m],~b:\mbf{vec}[n]
\ee
accepts as input the value assigned vector $a:\mbf{vec}[m]$ and under the associated assignment map returns the vector $b:\mbf{vec}[n]$.
It is understood that the assigned value and dimensions of the vector $a$ have been assigned prior to entry to the program $p~[a]~[b]$.
The program $p~[a]~[b]$ recognizes the dimensions of $a$ upon entry and checks that the assigned value of each element of $a$ is type $\mbf{int}$.
It then attempts to construct the array $b$ from the associated assignment map.
If successful the program returns as output the value assigned vector $b$ while simultaneously assigning the parameter dependent type $b::\mbf{vec}[n]$.
Once the dimensions of the vector $b$ have been assigned it may serve as an element of an input list of a subprogram that follows it in a program list.

Programs may also have I/O lists of the mixed type.
For example
\be
p~[s~a]~[b] , \quad s:\mbf{int},~a:\mbf{vec}[m],~b:\mbf{vec}[n]
\ee
accepts as input the value assigned scalar $s:\mbf{int}$ and the vector $a:\mbf{vec}[m]$ and under the associated assignment map returns the vector $b:\mbf{vec}[n]$.
The scalars themselves must obey the usual axioms of arithmetic over $\mbf{int}$.
This means that when setting up an application involving integer vectors we must also include the atomic programs and axioms associated with arithmetic over $\mbf{int}$.

\underline{Notes.}

\begin{itemize}
	\item Most conventional programming languages require that both the array variable name and its dimensions appear in the I/O lists of the subprogram.
	The adoption here of a dimension free formulation is made for convenience rather than necessity.
	If desired, the reader may modify all representations of the programs employed here to include the array dimensions in the I/O lists.
	A dimension free formulation requires a modification of the conventional model of array storage where allocating a memory address to an array variable name includes its parameter dependent type that incorporates its dimensions along with the value assignments of its vector elements.   
\end{itemize}

\section{Atomic programs.}

A vector $a:\mbf{vec}[m]$, $m:\mbf{int}1$, is an unpartitioned list $a=[a(i)]_{i=1}^m$, where $a(i):\mbf{int},~i=1,\ldots,m$, are the elements of the list.
The values of the elements and dimensions $m:\mbf{int1}$ of a vector of type $\mbf{vec}[m]$ that appear as an element of an input list of an atomic program are assigned prior to entry to the program.
Once assigned they are stored in memory at an address that is linked to the variable name of that vector.
Upon entry the atomic program searches the memory address linked to the input vector variable name and identifies its dimensions $m:\mbf{int1}$.
It then proceeds to check that the assigned values of the elements of the vector are type $\mbf{int}$.
If there is a type violation the atomic program halts with an execution error message.

\emph{Program type} $\mbf{chck}$.

\begin{tabular}{|l|l|}
	\hline
	\textbf{Syntax} & \textbf{Type checks} \\
	\hline \hline
	$typev~[a]~[~]$ & $a:\mbf{vec}[m]$,~$m:\mbf{int1}$ \\
	& $a(i):\mbf{int},~i=1,\ldots m$ \\
	\hline
	$eqv~[a~b]~[~]$ & $a,b:\mbf{vec}[m]$,~$m:\mbf{int1}$ \\
	& $a(i)=b(i),~i=1,\ldots m$ \\
	\hline
	$dim~[a~b]~[~]$ & $a,b:\mbf{vec}[m]$,~$m:\mbf{int1}$ \\
	\hline
\end{tabular} 

\emph{Program type} $\mbf{asgn}$.

\begin{tabular}{|l|l|l|l|}
	\hline
	\textbf{Syntax} & \textbf{Type checks} & \textbf{Assignment} & \textbf{Type} \\
	&  & \textbf{map} & \textbf{assignment} \\
	\hline \hline
	$addv~[a~b]~[c]$ & $a,b:\mbf{vec}[m]$, & $c:=a+b$ & $c::\mbf{vec}[m]$ \\
	& $m:\mbf{int1}$ & & \\
	\hline
	$smult~[s~a]~[b]$ & $s:\mbf{int}$, $a:\mbf{vec}[m]$,& $b:=s*a$ & $b::\mbf{vec}[m]$ \\
		& $m:\mbf{int1}$ & & \\
	\hline
	$zvec~[a]~[b]$ & $a:\mbf{vec}[m]$ & $b:=[b(i)]_{i=1}^m$, & $b::\mbf{vec}[m]$ \\
    & $m:\mbf{int1}$ & $b(i)=0$, & \\
	& & $i=1,\ldots,m$ & \\
	\hline
\end{tabular}

\underline{Notes.}

\begin{itemize}

\item The program $smult~[s~a]~[b]$ returns the scalar multiplication
\be
b:= s*[a(i)]_{i=1}^m = [s*a(i)]_{i=1}^m
\ee
where $s:\mbf{int}$ is a scalar and $a:\mbf{vec}[m]$.

\item Given any vector $a:\mbf{vec}[m]$ the program $zvec~[a]~[b]$ returns the zero vector, $b$, with the same dimensions as $a$.

\end{itemize}

\section{Axioms.}

As with scalar arithmetic on $\mbf{int}$, some of the premises of our axioms for integer vectors will contain extra conditional statements that reflect the constraints imposed by working in the environment $\mathfrak{M}(mach)$.

\emph{Equality.}

\be
\begin{array}{l}
	typev~[a]~[~] \\
	\hline
	eqv~[a~a]~[~] \\
\end{array}
\quad \textsf{axv1a} \hspace{10mm}
\begin{array}{l}
	eqv~[a~b]~[~] \\
	\hline
	eqv~[b~a]~[~] \\
\end{array}
\quad \textsf{axv1b}
\ee

The equality program satisfies the property of transitivity
\be
\begin{array}{l}
	eqv~[a~b]~[~] \\
	eqv~[b~c]~[~] \\
	\hline
	eqv~[a~c]~[~] \\
\end{array}
\ee
This is not included as an axiom because it follows from the substitution rule.

\emph{Commutativity of addition.}

\be
\begin{array}{l}
	addv~[a~b]~[c] \\
	\hline
	addv~[b~a]~[d] \\
\end{array}
\quad \textsf{axv2a} \hspace{10mm}
\begin{array}{l}
	addv~[a~b]~[c] \\
	addv~[b~a]~[d] \\
	\hline
	eqv~[d~c]~[~] \\
\end{array}
\quad \textsf{axv2b}
\ee

\emph{Associativity of addition.}
\be
\begin{array}{l}
	addv~[a~b]~[d] \\
	addv~[d~c]~[x] \\
	addv~[b~c]~[e] \\
	\hline
	addv~[a~e]~[y] \\
\end{array}
\quad \textsf{axv3a} \hspace{10mm}
\begin{array}{l}
	addv~[a~b]~[d] \\
	addv~[b~c]~[e] \\
	addv~[a~e]~[y] \\
	\hline
	addv~[d~c]~[x] \\
\end{array}
\quad \textsf{axv3b}
\ee
\be
\begin{array}{l}
	addv~[a~b]~[d] \\
	addv~[d~c]~[x] \\
	addv~[b~c]~[e] \\
	addv~[a~e]~[y] \\
	\hline
	eqv~[y~x]~[~] \\
\end{array}
\quad \textsf{axv3c}
\ee

\emph{Addition with the zero vector.}

\be
\begin{array}{l}
	typev~[a]~[~] \\
	\hline
	zvec~[a]~[b] \\
\end{array}
\quad \textsf{axv4a} \hspace{10mm}
\begin{array}{l}
	zvec~[a]~[b] \\
	\hline
	addv~[a~b]~[c] \\
\end{array}
\quad \textsf{axv4b}
\ee
\be
\begin{array}{l}
	zvec~[a]~[b] \\
	addv~[a~b]~[c] \\
	\hline
	eqv~[c~a]~[~] \\
\end{array}
\quad \textsf{axv4c}
\ee

\emph{Additive inverse.}
\be
\begin{array}{l}
	typev~[a]~[~] \\
	\hline
	smult~[-1~a]~[b] \\
\end{array}
\quad \textsf{axv5a} \hspace{10mm}
\begin{array}{l}
	smult~[-1~a]~[b] \\
	\hline
	addv~[a~b]~[c] \\
\end{array}
\quad \textsf{axv5b}
\ee
\be
\begin{array}{l}
	smult~[-1~a]~[b] \\
	addv~[a~b]~[c] \\
	zvec~[a]~[d] \\
	\hline
	eqv~[c~d]~[~] \\
\end{array}
\quad \textsf{axv5c}
\ee

\underline{Scalar multiplication axioms.}

\emph{Scalar multiplication of vector additions.}
\be
\begin{array}{l}
	addv~[a~b]~[c] \\
	smult~[s~c]~[d] \\
	smult~[s~a]~[e] \\
	smult~[s~b]~[f] \\
	\hline
	addv~[e~f]~[g] \\
\end{array}
\quad \textsf{axv6a} \hspace{10mm}
\begin{array}{l}
	addv~[a~b]~[c] \\
	smult~[s~a]~[e] \\
	smult~[s~b]~[f] \\
	addv~[e~f]~[g] \\
	\hline
	smult~[s~c]~[d] \\
\end{array}
\quad \textsf{axv6b}
\ee
\be
\begin{array}{l}
	addv~[a~b]~[c] \\
	smult~[s~c]~[d] \\
	smult~[s~a]~[e] \\
	smult~[s~b]~[f] \\
	addv~[e~f]~[g] \\
	\hline
	eqv~[d~g]~[~] \\
\end{array}
\quad \textsf{axv6c}
\ee

\emph{Addition of scalars.}
\be
\begin{array}{l}
	add~[r~s]~[t] \\
	smult~[t~a]~[b] \\
	smult~[r~a]~[c] \\
	smult~[s~a]~[d] \\
	\hline
	addv~[c~d]~[e] \\
\end{array}
\quad \textsf{axv7a} \hspace{10mm}
\begin{array}{l}
	add~[r~s]~[t] \\
	smult~[r~a]~[c] \\
	smult~[s~a]~[d] \\
	addv~[c~d]~[e] \\
	\hline
	smult~[t~a]~[b] \\
\end{array}
\quad \textsf{axv7b}
\ee
\be
\begin{array}{l}
	add~[r~s]~[t] \\
	smult~[t~a]~[b] \\
	smult~[r~a]~[c] \\
	smult~[s~a]~[d] \\
	addv~[c~d]~[e] \\
	\hline
	eqv~[e~b]~[~] \\
\end{array}
\quad \textsf{axv7c}
\ee

\emph{Multiplication of scalars.}
\be
\begin{array}{l}
	mult~[r~s]~[t] \\
	smult~[t~a]~[b] \\
	smult~[s~a]~[c] \\
	\hline
	smult~[r~c]~[d] \\
\end{array}
\quad \textsf{axv8a} \hspace{10mm}
\begin{array}{l}
	mult~[r~s]~[t] \\
	smult~[s~a]~[c] \\
	smult~[r~c]~[d] \\
	\hline
	smult~[t~a]~[b] \\
\end{array}
\quad \textsf{axv8b}
\ee
\be
\begin{array}{l}
	mult~[r~s]~[t] \\
	smult~[t~a]~[b] \\
	smult~[s~a]~[c] \\
	smult~[r~c]~[d] \\
	\hline
	eqv~[d~b]~[~] \\
\end{array}
\quad \textsf{axv8c}
\ee

\underline{Notes.}

\begin{itemize}
	
	\item For brevity we have omitted axioms associated with the usual vector dot product.
	It would be more efficient to incorporate these axioms in an extended theory that includes products of vectors with arrays of any rank $n \geq 1$.
 
\end{itemize}

\section{Compatibility.}

We have adopted a dimension free formulation which means that the dimensions of the vectors do not appear in the I/O lists of the vector atomic programs.
Within all vector atomic programs, type checking requires that the elements of all vectors be assigned a value of type $\mbf{int}$ along with compatibility of vector dimensions.
The following compatibility rules are included as axioms.  

\emph{Vector dimensions.}
\be
\begin{array}{l}
	dim~[a~b]~[~] \\
	\hline
	dim~[b~a]~[~] \\
\end{array}
\quad \textsf{dm1a} \hspace{10mm}
\begin{array}{l}
	dim~[a~b]~[~] \\
	dim~[b~c]~[~] \\
	\hline
	dim~[a~c]~[~] \\
\end{array}
\quad \textsf{dm1b}
\ee

\emph{Equality.}
\be
\begin{array}{l}
	eqv~[a~b]~[~] \\
	\hline
	dim~[b~a]~[~] \\
\end{array}
\quad \textsf{dm2}
\ee

\emph{Zero vector.}
\be
\begin{array}{l}
	zvec~[a]~[b] \\
	\hline
	dim~[b~a]~[~] \\
\end{array}
\quad \textsf{dm3a} \hspace{10mm}
\begin{array}{l}
	zvec~[a]~[b] \\
	zvec~[c]~[d] \\
	dim~[a~c]~[~] \\
	\hline
	eqv~[d~b]~[~] \\
\end{array}
\quad \textsf{dm3b}
\ee

\emph{Vector addition.}
\be
\begin{array}{l}
	addv~[a~b]~[c] \\
	\hline
	dim~[c~a]~[~] \\
\end{array}
\quad \textsf{dm4}
\ee

\emph{Scalar Multiplication.}
\be
\begin{array}{l}
	smult~[s~a]~[b] \\
	\hline
	dim~[b~a]~[~] \\
\end{array}
\quad \textsf{dm5}
\ee

It must be stressed again that we are adopting a dimension free formulation for convenience only.
There will be no change to our results if we choose to include the dimensions of the vector variables in the I/O lists of our atomic programs.
In such a case we can discard the axioms \textsf{dm1}-\textsf{dm5} and replace them with axioms of falsity that identify incompatibilities in vector dimensions. 

\section{Basic identities.}\label{sbia}

Here we will derive some basic results for integer vector addition and scalar multiplication subject to the constraints imposed by the environment $\mathfrak{M}(mach)$.
To the axioms presented in the previous sections of this chapter we will need to include the axioms of scalar integer arithmetic on $\mbf{int}$ presented in Chapter \ref{int}.

We start with a few preliminary results that will shorten proofs that follow.
The theorem labels are a continuation of those presented in Chapter \ref{int}.
Theorem \textsf{thm35} extends the compatibility axiom \textsf{dm4} for the second input element.
\begin{lstlisting}
Theorem thm35

addv [a b] [c]
--------------
dim [b a] [ ]

Proof.
  1 addv [a b] [c]
  2 dim [c a] [ ]            dm4 [1]
  3 addv [b a] [d]           axv2a [1]
  4 eqv [d c] [ ]            axv2b [1 3]
  5 dim [d b] [ ]            dm4 [3]
  6 dim [c b] [ ]            sr1 [5 4]
  7 dim [b c] [ ]            dm1a [6]
  8 dim [b a] [ ]            dm1b [7 2]
\end{lstlisting}
Theorems \textsf{thm36}-\textsf{thm37} extend the axiom of associativity of vector addition by making use of the commutativity of vector addition.
Since we are working with vectors in an environment $\mathfrak{M}(mach)$ we must first establish the existence of $(c+a)+b$ given the existence of $a+b$, $c+(a+b)$ and $c+a$.
Having established existence (\textsf{thm36}), theorem \textsf{thm37} shows that $c+(a+b)=(c+a)+b$.
\begin{lstlisting}
Theorem thm36

addv [a b] [d]
addv [c d] [e]
addv [c a] [f]
--------------
addv [f b] [m]

Proof.
  1 addv [a b] [d]
  2 addv [c d] [e]
  3 addv [c a] [f]
  4 addv [b a] [g]           axv2a [1]
  5 eqv [d g] [ ]            axv2b [4 1]
  6 addv [d c] [h]           axv2a [2]
  7 addv [g c] [i]           sr1 [6 5]
  8 addv [a c] [j]           axv2a [3]
  9 addv [b j] [k]           axv3a [4 7 8]
 10 addv [j b] [l]           axv2a [9]
 11 eqv [j f] [ ]            axv2b [3 8]
 12 addv [f b] [m]           sr1 [10 11]

Theorem thm37

addv [a b] [d]
addv [c d] [e]
addv [c a] [f]
addv [f b] [m]
--------------
eqv [m e] [ ]

Proof.
  1 addv [a b] [d]
  2 addv [c d] [e]
  3 addv [c a] [f]
  4 addv [f b] [m]
  5 eqv [e m] [ ]            axv3b [3 4 1 2]
  6 eqv [m e] [ ]            axv1b [5]
\end{lstlisting}
Theorem \textsf{thm38} shows that the vector sum $c+(-b)$ exists if the vector sum $c=a+b$ exists.
Theorem \textsf{thm39} establishes the identity $a=c+(-b)$.  
\begin{lstlisting}
Theorem thm38

addv [a b] [c]
smult [-1 b] [d]
----------------
addv [c d] [j]

Proof.
  1 addv [a b] [c]
  2 smult [-1 b] [d]
  3 addv [b d] [e]           axv5b [2]
  4 typev [b] [ ]            aio [1]
  5 zvec [b] [f]             axv4a [4]
  6 eqv [e f] [ ]            axv5c [2 3 5]
  7 typev [a] [ ]            aio [1]
  8 zvec [a] [g]             axv4a [7]
  9 addv [a g] [h]           axv4b [8]
 10 dim [b a] [ ]            thm35 [1]
 11 eqv [g f] [ ]            dm3b [5 8 10]
 12 eqv [f e] [ ]            axv1b [6]
 13 eqv [g e] [ ]            sr1 [11 12]
 14 addv [a e] [i]           sr1 [9 13]
 15 addv [c d] [j]           thm36 [3 14 1]

Theorem thm39

addv [a b] [c]
smult [-1 b] [d]
addv [c d] [j]
----------------
eqv [j a] [ ]

Proof.
  1 addv [a b] [c]
  2 smult [-1 b] [d]
  3 addv [c d] [j]
  4 addv [b d] [e]           axv5b [2]
  5 typev [b] [ ]            aio [1]
  6 zvec [b] [f]             axv4a [5]
  7 eqv [e f] [ ]            axv5c [2 4 6]
  8 typev [a] [ ]            aio [1]
  9 zvec [a] [g]             axv4a [8]
 10 addv [a g] [h]           axv4b [9]
 11 eqv [h a] [ ]            axv4c [9 10]
 12 dim [b a] [ ]            thm35 [1]
 13 eqv [g f] [ ]            dm3b [6 9 12]
 14 eqv [f e] [ ]            axv1b [7]
 15 eqv [g e] [ ]            sr1 [13 14]
 16 addv [a e] [i]           axv3a [1 3 4]
 17 eqv [i h] [ ]            sr2 [10 15 16]
 18 eqv [j i] [ ]            thm37 [4 16 1 3]
 19 eqv [i a] [ ]            sr1 [17 11]
 20 eqv [j a] [ ]            sr1 [18 19]
\end{lstlisting}
Theorem \textsf{thm40} establishes that if the vector sums $a+b$ and $a+d$ exist and are equal then $b=d$.
\begin{lstlisting}
Theorem thm40

addv [a b] [c]
addv [a d] [e]
eqv [c e] [ ]
--------------
eqv [b d] [ ]

Proof.
  1 addv [a b] [c]
  2 addv [a d] [e]
  3 eqv [c e] [ ]
  4 addv [b a] [f]           axv2a [1]
  5 addv [d a] [g]           axv2a [2]
  6 eqv [f c] [ ]            axv2b [1 4]
  7 eqv [g e] [ ]            axv2b [2 5]
  8 typev [a] [ ]            aio [1]
  9 smult [-1 a] [h]         axv5a [8]
 10 addv [f h] [i]           thm38 [4 9]
 11 addv [g h] [j]           thm38 [5 9]
 12 eqv [i b] [ ]            thm39 [4 9 10]
 13 eqv [j d] [ ]            thm39 [5 9 11]
 14 addv [c h] [k]           sr1 [10 6]
 15 addv [e h] [l]           sr1 [11 7]
 16 eqv [k i] [ ]            sr2 [10 6 14]
 17 eqv [l j] [ ]            sr2 [11 7 15]
 18 eqv [l k] [ ]            sr2 [14 3 15]
 19 eqv [k b] [ ]            sr1 [16 12]
 20 eqv [l d] [ ]            sr1 [17 13]
 21 eqv [l b] [ ]            sr1 [18 19]
 22 eqv [b d] [ ]            sr1 [20 21]
\end{lstlisting}
For any vector, $a$, the existence of the additive inverses $-a$ and $-(-a)$ are guaranteed.
Theorem \textsf{thm41} establishes that $-(-a)=a$.
\begin{lstlisting}
Theorem thm41

smult [-1 a] [b]
smult [-1 b] [c]
----------------
eqv [c a] [ ]

Proof.
  1 smult [-1 a] [b]
  2 smult [-1 b] [c]
  3 addv [a b] [d]           axv5b [1]
  4 typev [a] [ ]            aio [1]
  5 zvec [a] [e]             axv4a [4]
  6 eqv [d e] [ ]            axv5c [1 3 5]
  7 addv [b c] [f]           axv5b [2]
  8 typev [b] [ ]            aio [2]
  9 zvec [b] [g]             axv4a [8]
 10 eqv [f g] [ ]            axv5c [2 7 9]
 11 dim [b a] [ ]            dm5 [1]
 12 eqv [e g] [ ]            dm3b [9 5 11]
 13 eqv [g e] [ ]            axv1b [12]
 14 addv [b a] [h]           axv2a [3]
 15 eqv [h d] [ ]            axv2b [3 14]
 16 eqv [h e] [ ]            sr1 [15 6]
 17 eqv [f e] [ ]            sr1 [10 13]
 18 eqv [e f] [ ]            axv1b [17]
 19 eqv [h f] [ ]            sr1 [16 18]
 20 eqv [a c] [ ]            thm40 [14 7 19]
 21 eqv [c a] [ ]            axv1b [20]
\end{lstlisting}
It does not immediately follow from the axioms that the multiplication of the scalar $1$ with any vector exists. 
Before proving that it does exist we need the following result for scalar multiplication.
\begin{lstlisting}
Theorem thm42

mult [-1 -1] [a]
----------------
eqi [a 1] [ ]

Proof.
  1 mult [-1 -1] [a]
  2 typei [-1] [ ]           aio [1]
  3 mult [1 -1] [b]          axi8a [2]
  4 mult [-1 1] [c]          axi6a [3]
  5 mult [c -1] [d]          thm10 [3 4]
  6 mult [-1 c] [e]          axi6a [5]
  7 eqi [e 1] [ ]            thm7 [4 6]
  8 eqi [b -1] [ ]           axi8b [3]
  9 eqi [c b] [ ]            axi6b [3 4]
 10 eqi [c -1] [ ]           sr1 [9 8]
 11 eqi [a e] [ ]            sr2 [6 10 1]
 12 eqi [a 1] [ ]            sr1 [11 7]
\end{lstlisting}
Theorem \textsf{thm43} proves that the multiplication of the scalar $1$ with any vector does exist.
Theorem \textsf{thm44} establishes the identity $1*a=a$ for any vector $a$.
\begin{lstlisting}
Theorem thm43

typev [a] [ ]
---------------
smult [1 a] [f]

Proof.
  1 typev [a] [ ]
  2 smult [-1 a] [b]         axv5a [1]
  3 typev [b] [ ]            aio [2]
  4 smult [-1 b] [c]         axv5a [3]
  5 typei [-1] [ ]           aio [2]
  6 mult [-1 -1] [d]         axi5a [5]
  7 eqi [d 1] [ ]            thm42 [6]
  8 smult [d a] [e]          axv8b [6 2 4]
  9 smult [1 a] [f]          sr1 [8 7]

Theorem thm44

smult [1 a] [f]
---------------
eqv [f a] [ ]

Proof.
  1 smult [1 a] [f]
  2 typev [a] [ ]            aio [1]
  3 smult [-1 a] [b]         axv5a [2]
  4 typev [b] [ ]            aio [3]
  5 smult [-1 b] [c]         axv5a [4]
  6 typei [-1] [ ]           aio [3]
  7 mult [-1 -1] [d]         axi5a [6]
  8 eqi [d 1] [ ]            thm42 [7]
  9 smult [d a] [e]          axv8b [7 3 5]
 10 eqv [f e] [ ]            sr2 [9 8 1]
 11 eqv [c e] [ ]            axv8c [7 9 3 5]
 12 eqv [e c] [ ]            axv1b [11]
 13 eqv [f c] [ ]            sr1 [10 12]
 14 eqv [c a] [ ]            thm41 [3 5]
 15 eqv [f a] [ ]            sr1 [13 14]
\end{lstlisting}
Similarly, it does not immediately follow from the axioms that the multiplication of the scalar $0$ with any vector exists. 
Theorem \textsf{thm45} proves that it does exist and theorem \textsf{thm46} establishes that it is equal to the zero vector.
\begin{lstlisting}
Theorem thm45

typev [a] [ ]
---------------
smult [0 a] [k]

Proof.
  1 typev [a] [ ]
  2 smult [1 a] [b]          thm43 [1]
  3 eqv [b a] [ ]            thm44 [2]
  4 smult [-1 a] [c]         axv5a [1]
  5 addv [a c] [d]           axv5b [4]
  6 eqv [a b] [ ]            axv1b [3]
  7 addv [b c] [e]           sr1 [5 6]
  8 typei [1] [ ]            aio [2]
  9 mult [-1 1] [f]          axi5a [8]
 10 add [1 f] [g]            axi5b [9]
 11 mult [1 -1] [h]          axi6a [9]
 12 eqi [h f] [ ]            axi6b [9 11]
 13 eqi [h -1] [ ]           axi8b [11]
 14 eqi [f -1] [ ]           sr1 [13 12]
 15 eqi [g 0] [ ]            axi5c [9 10]
 16 add [1 -1] [i]           sr1 [10 14]
 17 eqi [i g] [ ]            sr2 [10 14 16]
 18 eqi [i 0] [ ]            sr1 [17 15]
 19 smult [i a] [j]          axv7b [16 2 4 7]
 20 smult [0 a] [k]          sr1 [19 18]

Theorem thm46

smult [0 a] [l]
zvec [a] [m]
---------------
eqv [l m] [ ]

Proof.
  1 smult [0 a] [l]
  2 zvec [a] [m]
  3 typev [a] [ ]            aio [1]
  4 smult [1 a] [b]          thm43 [3]
  5 eqv [b a] [ ]            thm44 [4]
  6 smult [-1 a] [c]         axv5a [3]
  7 addv [a c] [d]           axv5b [6]
  8 eqv [a b] [ ]            axv1b [5]
  9 addv [b c] [e]           sr1 [7 8]
 10 typei [1] [ ]            aio [4]
 11 mult [-1 1] [f]          axi5a [10]
 12 add [1 f] [g]            axi5b [11]
 13 mult [1 -1] [h]          axi6a [11]
 14 eqi [h f] [ ]            axi6b [11 13]
 15 eqi [h -1] [ ]           axi8b [13]
 16 eqi [f -1] [ ]           sr1 [15 14]
 17 eqi [g 0] [ ]            axi5c [11 12]
 18 add [1 -1] [i]           sr1 [12 16]
 19 eqi [i g] [ ]            sr2 [12 16 18]
 20 eqi [i 0] [ ]            sr1 [19 17]
 21 smult [i a] [j]          axv7b [18 4 6 9]
 22 eqv [d m] [ ]            axv5c [6 7 2]
 23 eqv [l j] [ ]            sr2 [21 20 1]
 24 eqv [e j] [ ]            axv7c [18 21 4 6 9]
 25 eqv [j e] [ ]            axv1b [24]
 26 eqv [e d] [ ]            sr2 [7 8 9]
 27 eqv [j d] [ ]            sr1 [25 26]
 28 eqv [j m] [ ]            sr1 [27 22]
 29 eqv [l m] [ ]            sr1 [23 28]
\end{lstlisting}

\underline{Notes.}

\begin{itemize}
	\item The elementary vector identities derived here are meant to serve as a starting point for a theory of finite state arithmetic in a machine environment, $\mathfrak{M}(mach)$.
	They demonstrate that, subject to extra conditional constraints, our vectors obey the usual rules of vector sums and scalar multiplication of conventional vectors of mathematics. 
	However, as with scalar arithmetic on $\mbf{int}$, we are disadvantaged by the absence of closure.
	Because of this much of the well developed theories of vector spaces will not hold in general.
	Conventional mathematical theorems derived from the notions of vector bases and subspaces are quite satisfying in an abstract sense but do not always translate to our constrained environment based on real world computations.
	Therefore, future research should explore alternative concepts and methods that are aimed at identifying useful properties of vectors in a machine environment.		
\end{itemize}  

\section{Vector inequalities.}\label{svi}

Inequalities for vectors and matrices usually involve positive scalars that are associated with some norm.
For instance one may define the norm, $\| a \|$, of an array, $a$, to be the maximum absolute value of the elements of the array.
Conventional $L_p$ norms for $p>1$ involve taking $p$-th roots of scalar quantities and are not appropriate when working with type $\mbf{int}$ and $\mbf{vec}[m]$ objects. 

Rather than deal with norms we will find it useful to define vector inequalities that involve a scalar inequality applied to each corresponding elements of two vectors.
For vectors we write
\be
u < v, \quad u,v:\mbf{vec}[m]
\ee
to mean that the inequality is satisfied by each corresponding element of the vectors, i.e.
\be
u(i) < v(i),~i=1,\dots,m
\ee
We use the non-strict inequality $\leq$ for vectors in a similar way.
This kind of inequality will be employed later when constructing discrete multidimensional intervals or boxes.

The table below presents the atomic programs that are used to check inequalities of state vectors.
In the description of these atomic programs the following notation for the arbitrary dimensions will be assumed.

A vector $a:\mbf{vec}[m]$, $m:\mbf{int1}$, is an unpartitioned list $a=[a(i)]_{i=1}^m$, where $a(i):\mbf{int},~i=1,\ldots,m$, are the elements of the list.
As before we adopt a dimension free formulation so that the dimensions, $m:\mbf{int1}$, of a vector of type $\mbf{vec}[m]$ that appears as an element of an input list of an atomic program are assigned prior to entry to the program.
Once assigned they are stored in memory at an address that is linked to the variable name of that vector.
Upon entry the atomic program searches the memory address linked to the input vector variable name and identifies its dimensions $m:\mbf{int1}$.
It then proceeds to check that the assigned values of the elements of the vector are type $\mbf{int}$.

\emph{Program type} $\mbf{chck}$.

\begin{tabular}{|l|l|}
	\hline
	\textbf{Syntax} & \textbf{Type checks} \\
	\hline \hline
	$ltv~[a~b]~[~]$ & $a,b:\mbf{vec}[m]$,~$m:\mbf{int1}$,~$a<b$ \\
	\hline
	$lev~[a~b]~[~]$ & $a,b:\mbf{vec}[m]$,~$m:\mbf{int1}$,~$a \leq b$ \\
	\hline
\end{tabular}

The strict inequality vector program, $ltv~[a~b]~[~]$, checks that strict inequality is satisfied by each element of $a$ and $b$.
It will often be more convenient to make use of the non-strict atomic vector inequality program, $lev~[a~b]~[~]$.
The atomic program $lev~[a~b]~[~]$ should not be confused with the more restrictive disjunction
\be
ltv~[a~b]~[~]~|~eqv~[a~b]~[~]
\ee

\underline{Inequality axioms.}

\emph{Transitivity of inequality.}
\be
\begin{array}{l}
	ltv~[a~b]~[~] \\
	ltv~[b~c]~[~] \\
	\hline
	ltv~[a~c]~[~] \\
\end{array}
\quad \textsf{ordv1}
\ee

\emph{Non-strict inequalities.}
The axioms for non-strict vector inequalities follow from the axioms and theorems of scalar arithmetic on $\mbf{int}$ applied to each vector element. 
\be
\begin{array}{l}
	ltv~[a~b]~[~] \\
	\hline
	lev~[a~b]~[~] \\
\end{array}
\quad \textsf{lev1a} \hspace{10mm}
\begin{array}{l}
	eqv~[a~b]~[~] \\
	\hline
	lev~[a~b]~[~] \\
\end{array}
\quad \textsf{lev1b}
\ee

\be
\begin{array}{l}
	lev~[a~b]~[~] \\
	lev~[b~c]~[~] \\
	\hline
	lev~[a~c]~[~] \\
\end{array}
\quad \textsf{lev2}
\ee

We will also include the axiom
\be
\begin{array}{l}
	lev~[a~b]~[~] \\
	lev~[b~a]~[~] \\
	\hline
	eqv~[b~a]~[~] \\
\end{array}
\quad \textsf{lev3}
\ee

\underline{Notes.}

\begin{itemize}
	\item We could derive the above order axioms for vectors directly from the order axioms for scalar arithmetic if we extend our theory of vectors to include atomic programs that identify elements of vectors.
\end{itemize}

\section{Discrete intervals and boxes.}\label{si}\index{discrete interval}\index{discrete box}

\underline{Discrete interval.} Consider the vector assignment program
\be
f~[v]~[w], \quad v,w:\mbf{vec}[m],~m:\mbf{int1}
\ee
We are interested in obtaining bounds on the primary output variable, $w$, of the program $f~[v]~[w]$ over a discrete domain.
Discrete domains can be constructed using discrete intervals and boxes. 
We start by introducing the notion of a discrete interval over $\mbf{int}$.

A discrete interval over $\mbf{int}$ can be represented by the two element list
\be
[a~b], \qquad a,b:\mbf{int},~a \leq b
\ee
where $a$ and $b$, respectively, are the lower and upper bounds, respectively, of the interval.
Note that $[a~b]$ is a two element list that represents the larger list $[a~a+1~\ldots~b]$ that contains all elements of $\mbf{int}$ between and including $a$ and $b$.
So that the machine can recognize its distinction from a standard list, a two element list $p=[a~b]$ that represents an interval over $\mbf{int}$ will be assigned the parameter dependent type $p:\mbf{box}[1]$.
The dependence of the type $\mbf{box}[1]$ on the parameter $1$ indicates that an interval is a $1$-dimensional box.
This notation is employed because shortly we will extend the notion of intervals to $m$-dimensional boxes that have type $\mbf{box}[m]$. 

We say that $v:\mbf{int}$ is an element contained in the interval $[a~b]:\mbf{box}[1]$ to mean that $v \in [a~a+1~\ldots~b]$.
An interval represented by $[a~b]:\mbf{box}[1]$ is an \emph{interval enclosure}\index{interval enclosure} of the interval represented by $[c~d]:\mbf{box}[1]$ when $[c~c+1~\ldots~d] \subseteqq [a~a+1~\dots~b]$.
For this to hold we simply require that $a \leq c$ and $d \leq b$.
An interval enclosure $[a~b]:\mbf{box}[1]$ of $[c~d]:\mbf{box}[1]$ is denoted by $[c~d] \subseteq [a~b]$, where the symbol $\subseteq$ is to be distinguished from the symbol $\subseteqq$ that is used for sublists of standard lists.
A single point interval $[a~a]:\mbf{box}[1]$ contains only the single element $a:\mbf{int}$.

Let $p=[v_1~v_n]:\mbf{box}[1]$ be the interval that represents the standard list $[v_1 \dots v_n]$, for some $n:\mbf{int1}$.
A map $f: \mbf{int} \to \mbf{int}$ under the restriction to $p$ generates the standard list $q=[w_1 \ldots w_n]$, where each $w_i$ is the evaluation $w_i:=f[v_i]$, $i=1,\ldots,n$.
We can define the interval that represents the tightest bound of $f: \mbf{int} \to \mbf{int}$ under the restriction to $p$ by
\be
R(f,p) = [min[q]~max[q]]
\ee   
where $min[q]$ and $max[q]$, respectively, are the minimum and maximum values, respectively, of the list $q$. 
We will refer to $R(f,p)$ as the discrete range, or simply the range, of $f$ over the interval $p$.

Let $c:\mbf{int}$ and $d:\mbf{int}$, respectively, be any lower and any upper bound, respectively, of $f:\mbf{int} \to \mbf{int}$ over the interval $p=[a~b]$.
We write
\ben \label{int100}
B(f,p)=[c~d], \quad c,d:\mbf{int}
\een 
to represent an interval that bounds $f:\mbf{int} \to \mbf{int}$ over the interval represented by $p$.
Here, $c$ and $d$, respectively, need not be the greatest lower bound and lowest upper bound, respectively, of $f$ over $p$.
In general we have
\ben \label{int101}
R(f,p) \subseteq B(f,p)
\een 
Given $p:\mbf{box}[1]$, the aim is to find a suitable interval $q=B(f,p)$ such that $q$ is a sufficiently tight enclosure of $R(f,p)$.

If the assignment map $f:\mbf{int} \to \mbf{int}$ can be expressed as a function of simple operations of arithmetic then the following rules for interval addition, subtraction and multiplication might sometimes be useful in computing $B(f,p)$.\index{interval arithmetic}
\ben\label{iarith01}
\bal
& [a~b]+[c~d] = [a+c~b+d] \\
& [a~b]-[c~d] = [a-d~b-c] \\
& [a~b]*[c~d] = [e~f] \\
& e= min[a*c~a*d~b*c~b*d] \\
& f= max[a*c~a*d~b*c~b*d] \\
\eal
\een
Since we are working over $\mbf{int}$, the above rules are conditional on the existence of the sums and multiplications of the interval bounds.
For discrete intervals the operation of interval division is not well defined. 

Interval arithmetic is not distributive and satisfies the weaker rule
\be
r*(p+q) \subseteq r*p + r*q, \quad p,q,r:\mbf{box}[1]
\ee
One of the major drawbacks of constructing $B(f,p)$ using interval arithmetic is that it is often the case that for a given interval $p:\mbf{box}[1]$ the construction of $B(f,p)$ will be a large over estimate of $R(f,p)$.
One can construct a tighter enclosure by splitting the interval, $p$, into smaller intervals.

\underline{Discrete boxes.}
A discrete box is a generalization of a one-dimensional interval and is also assigned the generic type $\mbf{box}$.
In particular, an $m$-dimensional box is assigned the subtype $\mbf{box}[m]$ and can be represented by the two element list
\be
p=[a~b], \quad a,b:\mbf{vec}[m],~m:\mbf{int1}
\ee 
where now the bounds of the $m$-dimensional interval are represented by vectors and
\be
a(i) \leq b(i), \quad i=1,\ldots,m
\ee
Once a two element list, $[a~b]$, $a,b:\mbf{vec}[m]$, has been assigned the type $\mbf{box}[m]$ it is recognized by the machine to represent a box with the structure
\ben\label{box01}
[a(1)~b(1)] \times \ldots \times [a(m)~b(m)]
\een
where each $[a(i)~b(i)]$, $i=1,\ldots,m$, is an interval of type $\mbf{box}[1]$ (an interval is a one-dimensional box).

We introduce the functions $lbx[p]$ and $ubx[p]$, respectively, that give the lower and upper bound, respectively, of the box $p$, so that if $p=[a~b]$, $a,b:\mbf{vec}[m]$, then $lbx[p]=a$ and $ubx[p]=b$.
For a map $f:\mbf{vec}[m] \to \mbf{vec}[m]$ we use the same notation for the discrete range of $f$ over the box $p:\mbf{box}[m]$ as $R(f,p)$ and its box enclosure by $B(f,p)$.

\section{Atomic box programs.} \label{sintap}

The following atomic programs can be included in applications employing boxes.
The assigned values of the bounds and the scalar $m:\mbf{int1}$ representing the dimensions of the parameter dependent types of the elements of the input lists are assigned prior to entry to the program.
Upon entry the atomic program identifies the parameter dependent type of these input elements.

\emph{Program type} $\mbf{chck}$.

\begin{tabular}{|l|l|}
	\hline
	\textbf{Syntax} & \textbf{Type checks} \\
	\hline \hline
	$typebx~[p]~[~]$ & $p:\mbf{box}[m]$ \\
	\hline
	$eqbx~[p~q]~[~]$ & $p,q:\mbf{box}[m]$, $p=q$ \\
	\hline
	$eltbx~[v~p]~[~]$ & $v:\mbf{vec}[m]$, $p:\mbf{box}[m]$ \\
	& $lbx[p] \leq v \leq ubx[p]$ \\
	\hline
	$subbx~[q~p]~[~]$ & $p,q:\mbf{box}[m]$, $q \subseteq p$ \\
	& $lbx[p] \leq lbx[q]$, $ubx[q] \leq ubx[p]$ \\
	\hline
\end{tabular}

The following atomic assignment programs are associated with box bounds.

\newpage 

\emph{Program type} $\mbf{asgn}$.

\begin{tabular}{|l|l|l|l|}
	\hline
	\textbf{Syntax} & \textbf{Type checks} & \textbf{Assignment} & \textbf{Type} \\
	&  & \textbf{map} & \textbf{assignment} \\
	\hline \hline
	$lbx~[p]~[a]$ & $p:\mbf{box}[m]$ & $a:=lbx[p]$ & $a::\mbf{vec}[m]$ \\
	\hline
	$ubx~[p]~[b]$ & $p:\mbf{box}[m]$ & $b:=ubx[p]$ & $b::\mbf{vec}[m]$ \\
	\hline
	$box~[a~b]~[p]$ & $a,b:\mbf{vec}[m]$, $a \leq b$ & $p:=[a~b]$ & $p::\mbf{box}[m]$ \\
	\hline
\end{tabular}

Note that the non-strict inequalities, $\leq$, that appear in the above tables are vector inequalities as defined in Section \ref{svi}.

\section{Axioms of discrete boxes.}\label{saodb}

The following axioms should be sufficient for much of the work required to construct discrete boxes.
They are not meant to be exhaustive.
When working with discrete boxes we need to include the axioms of integer scalars of Chapter \ref{int} and integer vectors presented earlier in this chapter.

\emph{Reflexivity.}
\be
\begin{array}{l}
	typebx~[p]~[~] \\
	\hline
	eqbx~[p~p]~[~] \\
\end{array}
\quad \textsf{bx1}
\ee

\emph{Box bounds.}
\be
\begin{array}{l}
	typebx~[p]~[~] \\
	\hline
	lbx~[p]~[a] \\
\end{array}
\quad \textsf{bx2a} \hspace{10mm}
\begin{array}{l}
	typebx~[p]~[~] \\
	\hline
	ubx~[p]~[b] \\
\end{array}
\quad \textsf{bx2b}
\ee
\be
\begin{array}{l}
	lbx~[p]~[a] \\
	ubx~[p]~[b] \\
	\hline
	lev~[a~b]~[~] \\
\end{array}
\quad \textsf{bx2c} \hspace{10mm}
\begin{array}{l}
	lbx~[p]~[a] \\
	ubx~[p]~[b] \\
	lbx~[q]~[c] \\
	ubx~[q]~[d] \\
	eqv~[c~a]~[~] \\
	eqv~[d~b]~[~] \\
	\hline
	eqbx~[q~p]~[~] \\
\end{array}
\quad \textsf{bx2d}
\ee

\emph{Box elements.}
\be
\begin{array}{l}
	lbx~[p]~[a] \\
	eltbx~[v~p]~[~] \\
	\hline
	lev~[a~v]~[~] \\
\end{array}
\quad \textsf{bx3a} \hspace{10mm}
\begin{array}{l}
	ubx~[p]~[b] \\
	eltbx~[v~p]~[~] \\
	\hline
	lev~[v~b]~[~] \\
\end{array}
\quad \textsf{bx3b}
\ee
\be
\begin{array}{l}
	lbx~[p]~[a] \\
	ubx~[p]~[b] \\
	lev~[a~v]~[~] \\
	lev~[v~b]~[~] \\
	\hline
	eltbx~[v~p]~[~] \\
\end{array}
\quad \textsf{bx3c}
\ee

\emph{Box enclosures.}
\be
\begin{array}{l}
	lbx~[p]~[a] \\
	lbx~[q]~[c] \\
	subbx~[q~p]~[~] \\
	\hline
	lev~[a~c]~[~] \\
\end{array}
\quad \textsf{bx4a} \hspace{10mm}
\begin{array}{l}
	ubx~[p]~[b] \\
	ubx~[q]~[d] \\
	subbx~[q~p]~[~] \\
	\hline
	lev~[d~b]~[~] \\
\end{array}
\quad \textsf{bx4b}
\ee
\be
\begin{array}{l}
	lbx~[p]~[a] \\
	lbx~[q]~[c] \\
	lev~[a~c]~[~] \\
	ubx~[p]~[b] \\
	ubx~[q]~[d] \\
	lev~[d~b]~[~] \\
	\hline
	subbx~[q~p]~[~] \\
\end{array}
\quad \textsf{bx4c}
\ee

\emph{Box construction.}
\be
\begin{array}{l}
	lev~[a~b]~[~] \\
	\hline
	box~[a~b]~[p] \\
\end{array}
\quad \textsf{bx5a}
\ee
\be
\begin{array}{l}
	box~[a~b]~[p] \\
	lbx~[p]~[c] \\
	\hline
	eqv~[c~a]~[~] \\
\end{array}
\quad \textsf{bx5b} \hspace{10mm}
\begin{array}{l}
	box~[a~b]~[p] \\
	ubx~[p]~[c] \\
	\hline
	eqv~[c~b]~[~] \\
\end{array}
\quad \textsf{bx5c}
\ee

\section{Basic properties of boxes.}
Before proceeding, it is worthwhile to check that the above axioms are strong enough to allow us to derive some basic properties of boxes.
We start by showing that box enclosures satisfy the properties of reflexivity, theorem \textsf{thm47}, and transitivity, theorem \textsf{thm48}.

\begin{lstlisting}
Theorem thm47

typebx [p] [ ]
---------------
subbx [p p] [ ]

Proof.
  1 typebx [p] [ ]
  2 eqbx [p p] [ ]           bx1 [1]
  3 lbx [p] [a]              bx2a [1]
  4 lbx [p] [b]              sr1 [3 2]
  5 eqv [b a] [ ]            sr2 [3 2 4]
  6 lev [b a] [ ]            lev1b [5]
  7 ubx [p] [c]              bx2b [1]
  8 ubx [p] [d]              sr1 [7 2]
  9 eqv [d c] [ ]            sr2 [7 2 8]
 10 lev [d c] [ ]            lev1b [9]
 11 subbx [p p] [ ]          bx4c [4 3 6 7 8 10]

Theorem thm48

subbx [p q] [ ]
subbx [q r] [ ]
---------------
subbx [p r] [ ]

Proof.
  1 subbx [p q] [ ]
  2 subbx [q r] [ ]
  3 typebx [p] [ ]           aio [1]
  4 lbx [p] [a]              bx2a [3]
  5 ubx [p] [b]              bx2b [3]
  6 typebx [q] [ ]           aio [1]
  7 lbx [q] [c]              bx2a [6]
  8 ubx [q] [d]              bx2b [6]
  9 lev [c a] [ ]            bx4a [7 4 1]
 10 lev [b d] [ ]            bx4b [8 5 1]
 11 typebx [r] [ ]           aio [2]
 12 lbx [r] [e]              bx2a [11]
 13 ubx [r] [f]              bx2b [11]
 14 lev [e c] [ ]            bx4a [12 7 2]
 15 lev [d f] [ ]            bx4b [13 8 2]
 16 lev [e a] [ ]            lev2 [14 9]
 17 lev [b f] [ ]            lev2 [10 15]
 18 subbx [p r] [ ]          bx4c [12 4 16 13 5 17]
\end{lstlisting}
Note that the proof of theorem \textsf{thm47} makes use of repetitions of assignment programs where, by necessity, the output variable names are changed (lines 3-4 and lines 7-8).
These can be linked to the first part of the identity rule (\ref{identity}) that follows from the application of the substitution rule \textsf{sr1} using the equality program $eqbx~[p~p]~[~]$ (see Section \ref{ssr}).
The output variables for each pair of assignment programs are subsequently equated by the substitution rule \textsf{sr2}.    

We also have
\begin{lstlisting}
Theorem thm49

subbx [p q] [ ]
subbx [q p] [ ]
---------------
eqbx [q p] [ ]

Proof.
  1 subbx [p q] [ ]
  2 subbx [q p] [ ]
  3 typebx [p] [ ]           aio [1]
  4 lbx [p] [a]              bx2a [3]
  5 ubx [p] [b]              bx2b [3]
  6 typebx [q] [ ]           aio [1]
  7 lbx [q] [c]              bx2a [6]
  8 ubx [q] [d]              bx2b [6]
  9 lev [c a] [ ]            bx4a [7 4 1]
 10 lev [a c] [ ]            bx4a [4 7 2]
 11 eqv [c a] [ ]            lev3 [10 9]
 12 lev [d b] [ ]            bx4b [5 8 2]
 13 lev [b d] [ ]            bx4b [8 5 1]
 14 eqv [d b] [ ]            lev3 [13 12]
 15 eqbx [q p] [ ]           bx2d [4 5 7 8 11 14]
\end{lstlisting}

The following shows that an element of a box will also be an element of its enclosure. 
\begin{lstlisting}
Theorem thm50

eltbx [v p] [ ]
subbx [p q] [ ]
---------------
eltbx [v q] [ ]

Proof.
  1 eltbx [v p] [ ]
  2 subbx [p q] [ ]
  3 typebx [p] [ ]           aio [1]
  4 lbx [p] [a]              bx2a [3]
  5 ubx [p] [b]              bx2b [3]
  6 typebx [q] [ ]           aio [2]
  7 lbx [q] [c]              bx2a [6]
  8 ubx [q] [d]              bx2b [6]
  9 lev [c a] [ ]            bx4a [7 4 2]
 10 lev [b d] [ ]            bx4b [8 5 2]
 11 lev [a v] [ ]            bx3a [4 1]
 12 lev [v b] [ ]            bx3b [5 1]
 13 lev [c v] [ ]            lev2 [9 11]
 14 lev [v d] [ ]            lev2 [12 10]
 15 eltbx [v q] [ ]          bx3c [7 8 13 14]
\end{lstlisting}

\section{Dynamical Systems.}\label{sfds}

We are interested in computer models that can be posed as dynamical systems where the intrinsic properties of real world objects and their dynamic state are defined by integer state vectors.
In particular we shall focus exclusively on dynamical systems whose properties are constrained by the machine, $\mathfrak{M}(mach)$, on which the dynamical system is executed.
In this regime we will always deal with dynamical systems that can be represented by the map
\ben\label{fddsmap}
\Phi : \mb{T} \times \mb{M} \to \mb{M}
\een
where we set $\mb{T} = \mbf{int}$ or $\mb{T} = \mbf{int0}$ and objects of the state space, $\mb{M}=\mbf{vec}[m]$, are integer vectors of dimensions $m:\mbf{int1}$. (There is no loss of generality if the variables of the state space are originally constructed as integer arrays of higher rank since they are essentially partitioned lists that can be redefined as state vectors as outlined in Section \ref{sarray}.) 

It will often be convenient to set the ordered list of objects of type $\mb{T}=\mbf{int0}$.
If $t:\mbf{int0}$ is the time parameter and $v:\mbf{vec}[m]$, $m:\mbf{int1}$, is the initial state then the map $\Phi$ satisfies the properties
\be
\bal
\Phi(0,v) = & v \\
\Phi(t_2,\Phi(t_1,v)) = & \Phi(t_1+t_2,v), \qquad t_1,~t_2,~t_1+t_2:\mbf{int0}
\eal
\ee
where, for the moment, we are using conventional mathematical notation. 

We will often refer to such systems as \emph{fully discrete systems}\index{fully discrete system} to stress that $\Phi$ is a map over the lists of finite objects that are associated with the types $\mb{T} \times \mb{M} = \mbf{int} \times \mbf{vec}[m]$.
Dynamical systems are usually defined in a more general sense that include the following.

\emph{Real dynamical system:}\index{real dynamical system} Here $\mb{T}$ is an open interval of the set of real numbers, $\mb{R}$, and $\mb{M}$ is a manifold locally diffeomorphic to a Banach space.
For the case $\mb{T}=\mb{R}$ the system is called global.
If $\Phi$ is continuously differentiable then the system is said to be a differentiable dynamical system.

\emph{Discrete dynamical system:}\index{discrete dynamical system} As with a real dynamical system, $\mb{M}$ is a manifold locally diffeomorphic to a Banach space but $\mb{T}$ is the set of integers, $\mb{Z}$.

\emph{Cellular automata:}\index{cellular automata} Our definition of a dynamical system closely resembles cellular automata.
Cellular automata, in the most common form, are characterized by $\mb{T}=\mb{Z}$ and $\mb{M}$ is a finite lattice of integer state vectors.
The most fundamental cellular automata are ones where each element of the state vector can only take on the values zero or one in each cell of the lattice. 

A survey of the current literature indicates that most of the analysis on nonlinear systems is carried out in the context of real and discrete dynamical systems.
There is no single global method of analysis and the choice of the theoretical tools that are used depend on the specific properties of the system under investigation.

The behavior of the solutions in the vicinity of fixed points\index{fixed points} are of particular interest.
Fixed points can act as local attractors where a trajectory of $v$ can enter a basin of attraction about the fixed point and remain within that region.
Once captured within a basin of attraction the trajectory of $v$ need not converge to the fixed point.
If the trajectory does converge to the fixed point the attractor is said to be locally asymptotically stable.
If convergence to the fixed point is independent of the initial condition then the fixed point is said to be globally asymptotically stable.

While these continuous based methods provide useful insights into the properties of complex solutions generated by nonlinear systems they can be incompatible with tests of computability in our formal system.
This means that we need to find other methods of analysis that target the specific issues that arise when working on $\mathfrak{M}(mach)$.

A fully discrete dynamical system can be represented by the difference equation 
\ben\label{int010}
\bal
v^{(t)} = f[v^{(t-1)}], \quad t=1,\ldots,n
\eal
\een
where $v^{(t)}:\mbf{vec}[m]$, $t=0,1,\dots,n$, for some $m:\mbf{int1}$, and $f:\mbf{vec}[m] \to \mbf{vec}[m]$.
By prescribing an initial state vector $v^{(0)}:\mbf{vec}[m]$, the state vectors $v^{(t)}:\mbf{vec}[m]$, $t=1,\dots,n$, that are generated by (\ref{int010}) form a sequence that defines the evolution of the state vector, $v$, from the initial state, $v^{(0)}$, in discrete time, $t$.

For brevity we will only consider the system of first-order difference equations, (\ref{int010}). 
There is no loss of generality here because higher order difference systems can always be written in the form of (\ref{int010}) by introducing new variables. 

The system of difference equations (\ref{int010}) is commonly thought of as being associated with an assignment map $f:\mbf{vec}[m] \to \mbf{vec}[m]$ that can be expressed concisely as a function in terms of conventional mathematical notation.
This will not always be the case and we should regard the assignment map $f:\mbf{vec}[m] \to \mbf{vec}[m]$ to be constructed in the more general context of an algorithm that can best be represented by a program.

When dealing with application specific assignment programs associated with some assignment map we can expect that the input list will often include the primary input vector variable along with some additional constant parameters that are employed in the program's internal algorithm that lead to the evaluation of the primary output vector variable.
If the assignment program is constructed as a program list we also need to include in the output list the variables associated with the intermediate calculations employed for the final calculation of the primary output variable.

Here we will assume that the assignment program
\be
f~[v]~[w], \quad v,w:\mbf{vec}[m],~m:\mbf{int1}
\ee
associated with the assignment map $f:\mbf{vec}[m] \to \mbf{vec}[m]$ is atomic.
The constant arrays and the variables employed in the intermediate computations are assigned within the program and discarded at the completion of the execution of the program.
If $f~[v]~[w]$ could otherwise be expressed as a program list involving atomic programs then we should regard it as a pseudo-atomic program (see Sections \ref{saps}). \index{pseudo-atomic program}
In either case it will be necessary to supply application specific axioms that make the main computational operations of the internal algorithm of $f~[v]~[w]$ recognizable to the machine.
If one prefers to express the assignment program associated with the map $f:\mbf{vec}[m] \to \mbf{vec}[m]$ as a program list then any expression of the form $f~[v]~[w]$ in the following sections may be replaced by that program list.

\newpage 

\section{Atomic iteration program.}

We associate the assignment map $f:\mbf{vec}[m] \to \mbf{vec}[m]$ with a program $f~[v]~[w]$ and abandon the notation of (\ref{int010}) by expressing the dynamical system by the program list
\ben \label{int020}
[f~[v^{(t-1)}]~[v^{(t)}]]_{t=1}^n, \quad v^{(t)}:\mbf{vec}[m],~n:\mbf{int0}
\een
with the understanding that the values of the initial condition, $v^{(0)}:\mbf{vec}[m]$, and the iteration number, $n$, have been assigned prior to entry into the program list (\ref{int020}).

In applications the input parameter $n$ is typically very large so that it is not feasible to store the entire program list (\ref{int020}).
One can construct an iteration assignment program\index{iteration program}
\ben \label{int040}
iterf~[v~n]~[w]
\een
where $v$ is the initial condition (equivalent to $v^{(0)}$ in (\ref{int020})) and $w$ is the value assigned output obtained after $n$ iterations of $f$ (equivalent to $v^{(n)}$ in (\ref{int020})).

We shall regard $iterf~[v~n]~[w]$ as an atomic program that is constructed by an imperative language using an iteration loop as outlined in the following pseudo-code.
\ben \label{int050}
\begin{array}{l}
\hline
\text{algorithm~} iterf~[v~n]~[w] \\
\hline
v:\mbf{vec}[m],~n:\mbf{int} \\
t:\mbf{int},~z:\mbf{vec}[m],~w:\mbf{vec}[m] \\
call~le~[0~n]~[~] \\
w:=v \\
do~t=1,\ldots,n \\
~~~z:=w \\
~~~call~f~[z]~[w] \\
end~do \\
\hline
\end{array}
\een
As with atomic vector programs we are assuming a dimension free formulation so that the parameter, $m:\mbf{int1}$, that represents the dimensions of the value assigned input state vector, $v$, is identified upon entry even though it does not appear in the input list of $iterf$.
Note that (\ref{int050}) represents a program constructed from an imperative language that allows each preceding solution of the iteration to be discarded through the reassignment $z:=w$.
Here the variables $t:\mbf{int}$ and $z:\mbf{vec}[m]$ are assigned internally and are released from memory storage once the program has been executed.
The $do$-loop is not activated when $n=0$, in which case the value assignment $w:=v$ is returned as output.
The atomic program $iterf~[v~n]~[w]$ will halt with an execution error if there is a type violation in $f~[z]~[w]$ during the iteration or $n$ has been assigned a negative value.

Sometimes one may be interested in storing intermediate steps.
In such a case we introduce the desired intermediate state vectors $v^{(l)}:\mbf{vec}[m]$, $l=0,1,\ldots,k$, for some $k:\mbf{int1}$, associated with the prescribed iteration numbers $n^{(l)}:\mbf{int1}$, $l=1,\ldots,k$, and construct the program
\be
[iterf~[v^{(l-1)}~n^{(l)}]~[v^{(l)}] ]_{l=1}^k
\ee
where each $v^{(l)}$ is evaluated after $n^{(l)}$ iterations from the starting value $v^{(l-1)}$.
Here $v^{(0)}$ is the prescribed initial state.
To avoid introducing too many variables we shall work with the iteration program (\ref{int050})

\underline{Fixed points of $\mbf{vec}[m]$.} As already mentioned, in contemporary analysis of real and discrete dynamical systems, the behavior of the solutions in the vicinity of fixed points are of particular interest.
The behavior of solutions about fixed points of fully discrete systems are less well known.
While our primary concern here is to establish computability of applications based on some assignment program $f~[v]~[w]$, we also desire tools that will allow us to investigate solution behaviors, often involving fixed points.

A fixed point of an assignment program $f~[v]~[w]$ can be defined as $v^*:\mbf{vec}[m]$ such that the program
\ben \label{int030}
[f~[v^*]~[w]~eqi~[w~v^*]~[~]]
\een 
is computable.

Since we are working with state vectors whose elements are type $\mbf{int}$, the regions containing fixed points satisfying (\ref{int030}) may not be the only ones of interest.
Regions containing neighboring points across which the assignment map $f:\mbf{vec}[m] \to \mbf{vec}[m]$ changes sign should be monitored during simulations to see if they take on properties that we associate with attractors.
More elusive are regions surrounding neighboring points where the assignment map $f:\mbf{vec}[m] \to \mbf{vec}[m]$ does not change sign but the assigned values of the state vectors are close to the zero vector. 
The behavior of the map $f$ in theses regions should also be monitored during simulations.

\section{Computability.}\label{saoc}

Much of the contemporary theories of continuous and discrete dynamical systems do not translate well when working with fully discrete dynamical systems in a machine environment, $\mathfrak{M}(mach)$.
Currently, a comprehensive review of fully discrete systems is not possible because the methods for such systems are as yet underdeveloped.
Here an attempt will be made to present some ideas that serve as a starting point from which a more detailed theory for the computability of fully discrete dynamical systems can be constructed.       

\underline{Application specific atomic programs.} 
We introduce the following application specific programs that define the fully discrete dynamical system.
As before, the dimensions $m:\mbf{int1}$ of the input vectors and boxes are assigned prior to entry into the atomic programs and are recognized within the atomic program.

\emph{Program type} $\mbf{asgn}$.

\begin{tabular}{|l|l|l|l|}
	\hline
	\textbf{Syntax} & \textbf{Type} & \textbf{Assignment} & \textbf{Type} \\
	& \textbf{checks} & \textbf{map} & \textbf{assignment} \\
	\hline \hline
	$f~[v]~[w]$ & $v:\mbf{vec}[m]$ & $f:\mbf{vec}[m] \to \mbf{vec}[m]$ & $w::\mbf{vec}[m]$ \\
	 &  & \text{application specific} &  \\
	\hline
	$iterf~[v~n]~[w]$ & $v:\mbf{vec}[m]$, & \text{see algorithm} & $w::\mbf{vec}[m]$ \\
	 & $n:\mbf{int0}$ &  (\ref{int050})      &             \\
	\hline
	$boundf~[p]~[q]$ & $p:\mbf{box}[m]$          & $q:=B(f,p)$ & $q::\mbf{box}[m]$ \\
	&  & \text{application specific} &  \\
	\hline
\end{tabular}

If it is feasible to obtain bounds on the primary output variable of the program $f~[v]~[w]$ over a discrete domain then computability may be established on that domain.
The program $boundf~[p]~[q]$ is a user supplied program that attempts to find a box $q=B(f,p)$ that is a sufficiently tight enclosure of the range $R(f,p)$.
We have discussed one possible method based on interval arithmetic, but this can only be applied if $f:\mbf{vec}[m] \to \mbf{vec}[m]$ can be readily expressed as a function of arithmetic operations.
Even in such a case, the tendency of interval arithmetic to overestimate the bounds of $R(f,p)$ might make it unsuitable as the method of choice for the construction of the box $q$.
Therefore, it is crucial that other methods be sought for this purpose.

We will leave the details of the internal algorithm of $boundf~[p]~[q]$ to be unspecified because it is likely that there are other methods for computing $B(f,p)$ that are more suited to the specific properties of the program $f~[v]~[w]$.
An exploration of general methods for constructing the bounds of an integer-valued map will not be pursued in detail here and is left open for future research.

\underline{Axioms of computability.}
Suppose that
\be\label{int300}
f~[v]~[w], \quad v,w:\mbf{vec}[m]
\ee
is the assignment program associated with some assignment map $f:\mbf{vec}[m] \to \mbf{vec}[m]$.
By construction the iteration assignment program $iterf~[v~n]~[w]$, defined by (\ref{int050}), satisfies the axioms
\be
\begin{array}{l}
	typev~[v]~[~] \\
	\hline
	iterf~[v~0]~[w] \\
\end{array}
\quad \textsf{axc1a} \hspace{10mm}
\begin{array}{l}
	iterf~[v~0]~[w] \\
	\hline
	eqv~[w~v]~[~] \\
\end{array}
\quad \textsf{axc1b}
\ee
These axioms reflect the property that the $do$-loop in (\ref{int050}) is not activated when $n=0$.

The iteration assignment program $iterf$ also obeys
\be
\begin{array}{l}
	f~[v]~[z] \\
	\hline
	iterf~[v~1]~[w] \\
\end{array}
\quad \textsf{axc2a} \hspace{10mm}
\begin{array}{l}
	iterf~[v~1]~[w] \\
	\hline
    f~[v]~[z] \\
\end{array}
\quad \textsf{axc2b}
\ee
\be
\begin{array}{l}
	f~[v]~[z] \\
	iterf~[v~1]~[w] \\
	\hline
	eqv~[z~w]~]~[~] \\
\end{array}
\quad \textsf{axc2c}
\ee
and
\be
\begin{array}{l}
	iterf~[v~n]~[s] \\
	iterf~[s~m]~[w] \\
	add~[n~m]~[l] \\
	iterf~[v~l]~[z] \\
	\hline
	eqv~[z~w]~[~] \\
\end{array}
\quad \textsf{axc3}
\ee

We observe that the iteration assignment program, $iterf~[v~n]~[w]$, can be associated with the map $\Phi(n,v)$ of (\ref{fddsmap}).
Our main objective here is to establish that given a discrete box, $p:\mbf{box}[m]$, if $v:\mbf{vec}[m]$ is contained in the box $p$ then the program $iterf~[v~n]~[w]$ is computable for all $n:\mbf{int0}$.
We can expect that computability will be guaranteed if the discrete range $R(f,p)$ is contained in the box $p$.

It should be noted that by definition of the assignment map, $f:\mbf{vec}[m] \to \mbf{vec}[m]$, the elements of the state vector, $v:\mbf{vec}[m]$, can only take on a finite number of values.
More precisely, the absolute value of each element of the state vector cannot exceed the machine integer, $nint$.     

We start by constructing the box $p:\mbf{box}[m]$ over which we wish to investigate the computability of our dynamical system.
Given $p:\mbf{box}[m]$, the user supplied program
\ben\label{int320}
boundf~[p]~[q]
\een
attempts to construct the box, $q:\mbf{box}[m]$, such that $q$ is a sufficiently tight box enclosure of $R(f,p)$.
If the program $boundf~[p]~[q]$ is defined as an atomic program then its internal algorithm is not directly visible to the machine so we must supply additional application specific axioms that reflect the methods by which the bounds of the box $q$ are obtained.
These bounds will be dependent on the bounds of the box $p$ and the specific internal algorithm of the program $f~[v]~[w]$.  
Whatever method we choose to construct $q$ from $p$ it must be soundly based such that the following axiom holds.
\be
\begin{array}{l}
	boundf~[p]~[q] \\
	eltbx~[v~p]~[~] \\
	f~[v]~[w] \\
	\hline
	eltbx~[w~q]~[~]\\
\end{array}
\quad \textsf{axc4}
\ee
This axiom states that given $p:\mbf{box}[m]$, if $v:\mbf{vec}[m]$ is an element contained in $p$ and $w:\mbf{vec}[m]$ is obtained from the evaluation $f~[v]~[w]$ then $w$ is an element contained in $q=B(f,p)$.
This is equivalent to the statement that $q$ will be a box that encloses $R(f,p)$.
(In some applications it may not be obvious that the internal algorithm of $boundf~[p]~[q]$ does indeed satisfy the axiom \textsf{axc4}.
In such cases it is better to do a preliminary analysis that proves \textsf{axc4} as a theorem based on the application specific axioms that are supplied to define the internal algorithms of the programs $f~[v]~[w]$ and $boundf~[p]~[q]$.)
  
The first task is to find suitable boxes $p$ and $q$ such that $q$ will be a sufficiently tight box enclosure of $R(f,p)$.
If in addition we can construct $q$ such that $p$ is a box enclosure of $q$ then we can apply the following axiom of computability.
\be
\begin{array}{l}
	boundf~[p]~[q] \\
	subbx~[q~p]~[~] \\
	eltbx~[v~p]~[~] \\
	le~[0~n]~[~] \\
	\hline
	iterf~[v~n]~[w] \\
\end{array}
\quad \textsf{axc5}
\ee
Axiom \textsf{axc5} states that given $p:\mbf{box}[m]$, if $q:\mbf{box}[m]$ is an enclosure of $R(f,p)$ such that $q \subseteq p$, then for any element $v:\mbf{vec}[m]$ contained in $p$ the iteration program $iterf~[v~n]~[w]$ will be computable for any $n:\mbf{int0}$.

\underline{Notes.}

\begin{itemize}
	\item In conventional mathematics we often desire a stronger result that proves that a predicate $P(n)$ is true for all $n$ in the set of the natural numbers, $\mb{N}$.
	Here we can only make the statement that $iterf~[v~n]~[w]$ will be computable for any $n:\mbf{int0}$ because we are working in an environment $\mathfrak{M}(mach)$.
	However, our machine, $\mathfrak{M}(mach)$, is arbitrary and the above axioms are valid for any machine large enough to construct a box $p$ such that $R(f,p) \subseteq p$.
	
\end{itemize}

\section{Future directions.}   

Of particular interest are applications of fully discrete dynamical systems on multidimensional lattices where the primary laws are governed by the conservation of information.
The simplest class of models are dynamical systems based on a binary cellular automata where the values of each dependent variable in each cell of a lattice can only acquire the value zero or one.
For a multidimensional lattice of $l$ cells and a model based on state variables of type $\mbf{vec}[k]$ in each cell the system state vector $v:\mbf{vec}[m]$ has dimensions $m=l*k$.
The domain $p=[a~b]:\mbf{box}[m]$ of a typical cellular automata map $f:\mbf{vec}[m] \to \mbf{vec}[m]$ has a simple structure where all of the elements of the lower box bound, $a:\mbf{vec}[m]$, are zero and all of the elements of the upper box bound, $b:\mbf{vec}[m]$, are one.
By construction we can expect that for any state vector $v:\mbf{vec}[m]$ contained in the box $p$ the program $f~[v]~[w]$ will always output a state vector $w:\mbf{vec}[m]$ whose elements are either zero or one. 
It follows that it will be sufficient for the program $boundf~[p]~[q]$ to make the simple assignment $q:=p$.
To make this visible to the machine we only need supply one application specific axiom
\be
\begin{array}{l}
	boundf~[p]~[q] \\
	\hline
	eqbx~[q~p]~[~] \\
\end{array}
\ee
This in combination with the box axioms, \textsf{bx1}-\textsf{bx5}, and theorem \textsf{thm50} will allow us to establish that \textsf{axc4} is satisfied and that the program $subbx~[q~p]~[~]$ appearing in the premise of \textsf{axc5} is computable.
Thus for any $v:\mbf{vec}[m]$ contained in $p$ and any $n:\mbf{int0}$ a computer model based on a typical binary cellular automata will always satisfy the axiom of computability \textsf{axc5}.

In this chapter we have examined computability of maps over domains represented by a single cuboid box with vector bounds.
There are many interesting systems where computability may be sought over domains that have a more complicated structure.
For such systems the theory of boxes as outlined in Section \ref{sintap} needs to be extended to include boxes with bounds that are best represented by arrays of higher rank.
Extensions could also include constructions that allow one to examine the discrete range, $R(f,p)$, for a map, $f:\mbf{vec}[m] \to \mbf{vec}[m]$, on domains represented by a box, $p$, that is the union of a number of mutually disjoint boxes.
Even for domains with these complicated structures, the axioms of computability, \textsf{axc1}-\textsf{axc5}, will still hold in their current form.

The axioms, \textsf{axc1}-\textsf{axc5}, are quite straight forward but are of little use unless one can construct a program $boundf~[p]~[q]$ that is based on an algorithm that obeys axiom \textsf{axc4}.
But satisfaction of axiom \textsf{axc4} on its own is insufficient because we must also construct $q$ such that it is a sufficiently tight enclosure of $R(f,p)$ so that if $p$ encloses $R(f,p)$ then $p$ also encloses $q$.

If the assignment program $f~[v]~[w]$ can be expressed in terms of basic operations of arithmetic we can construct the program $boundf~[p]~[q]$ such that it applies the rules of interval arithmetic.
For reasons discussed earlier, interval arithmetic may not be the best method to construct the tightest box $q$ that encloses $R(f,p)$.
Efficient algorithms for the program $boundf~[p]~[q]$ that are capable of achieving all of the goals stated above for general classes of integer maps are largely underdeveloped.
This indicates a need for a new area of research where methods are sought that focus on the determination of sufficiently tight bounds of integer vector functions over discrete boxes.

At this stage we cannot rule out the possibility that there will be applications where the bounds of an integer function will always be elusive.
In such cases the axioms, \textsf{axc4}-\textsf{axc5}, will only apply in principle and completely different approaches that establish computability of fully discrete dynamical systems need to be explored.

\chapter{Program Constructions as Proofs.}\label{cpcap}

\section{Human verses machine proofs.}

Proofs in contemporary mathematics are constructed from a language comprised of symbols and natural language and their merits often judged by their elegance.
This style of proof construction is natural to humans and has been accepted as the standard for strong reasons.
Purely symbolic proofs can be difficult to read and lack the expressiveness demanded by humans to satisfy their interpretation of understanding and meaning.
In this way there is an interplay between semantics and syntax where semantics takes the dominant role. 

Proofs are presented as an outline of a sequence of steps that are often bound together by trivial and tedious calculations.
The author of a proof attempts to provide the reader with an outline of the important steps leading to a conclusion by omitting the details of what may be regarded as obvious and trivial calculations.
Thus the reader is spared from the tedious details that can otherwise be a distraction from the main thrust of the proof.

For longer proofs elegance is difficult to maintain and can even be a challenge to read by experts in the particular subject area.
It is not uncommon for referees of mathematical proofs to call upon proof checking software to establish the correctness of a proof.
This raises the question as to the extent by which contemporary proofs are rigorous constructions and not merely outlines.
There is no defining line here and no formal boundaries exist that distinguish an outline of a proof from one that can be designated as rigorous. 

Machine proofs are uncompromising in rigor and demand the inclusion of even the most trivial calculations.
As a result machine proofs can be much longer than those written down by humans.
They are symbolic in structure, devoid of natural language and demand a completely different kind of interpretation.

Developers of proof assistance software often make some effort to provide an interface that allows the user to interact with the machine in the more familiar language of contemporary mathematics.
Here we have made very little effort in this regard.
This is a choice that is deliberate and is made to encourage the reader to acquire familiarity with machine proofs that are presented as a list of programs.
While such proofs might be unsightly at first, the reader needs to be assured that with some effort and experience they will find that this style of derivation will become no less natural than the more traditional style that they have been accustomed to.

Given current trends it does appear that efforts in acquiring familiarity with machine proofs, whether they be based on programs or any other machine language, are not wasted.
It is not unreasonable to anticipate that machine proofs will eventually become more widely used.
This will be especially beneficial for the construction and rigorous validation of computer models based on rule based algorithms and finite state arithmetic.
In a more general context, acquiring an understanding of machine proofs will be essential in efforts directed towards the ultimate goal of complete automation.

\underline{Soundness, consistency and completeness.}
The strength of classical formal systems are measured by their satisfaction of consistency, soundness, and completeness.
The standard systems of propositional logic and sequent calculus can be shown to satisfy these properties by employing a meta-theory of logic.
Crucial to the establishment of soundness and completeness are the notions of interpretations or models.

In the next chapter we will explore some necessary properties of applications of the formal system PECR that lead to completeness.
The main objective of PECR is to establish program computability for dynamical systems that are based on finite state arithmetic.
In a real world application sense, program computability is ultimately an empirical concept.
A program can be empirically tested for its computability with respect to a value assigned input by simply executing the program and observing whether it halts with an execution error or returns an output in a feasible time.

Empirical data play an important role in the scientific method.
This is an iterative process of self correction where theories are strengthened or replaced by continual revision based on new empirical evidence.
In the final chapter of this book we will explore these ideas in the context of our formal system, PECR.

\section{Higher order extensions.}\label{shop}

Up till now we have employed our formal system, PECR, to construct proofs for applications where the formal statements of the proof were zeroth-order programs.
For these applications the higher order atomic programs presented in Section \ref{hop} that appear in the constructions rules were understood to be first-order programs.

Here we will give a brief outline that extends our formal system, PECR, for general higher order constructs.\index{higher order constructs}
We write
\be
p^{(k)}:\mbf{prgm}^{(k)}, \quad k \geq 0
\ee
to mean that, $p^{(k)}$, has been assigned a value of a type $k$-order program.
For $k \geq 1$, $k$-order programs admit input variables that have been assigned values of type $(k-1)$-order programs.

We denote the program names of the type checking programs for $(k-1)$-order program extensions by $ext^{(k)}$ and the program names of the type checking programs for $(k-1)$-order false programs by $false^{(k)}$.
We have already been using the program names $ext$ for $ext^{(1)}$ and $false$ for $false^{(1)}$. 

The type checking program for program extensions can be generalized to $(k-1)$-order programs by
\ben\label{shop01}
ext^{(k)}~[p^{(k-1)}~c^{(k-1)}]~[~] , \quad k \geq 1
\een
Here $ext^{(k)}~[p^{(k-1)}~c^{(k-1)}]~[~]:\mbf{prgm}^{(k)}$ checks that the program $c^{(k-1)}:\mbf{prgm}^{(k-1)}$ has also been type assigned as an extension of the program $p^{(k-1)}:\mbf{prgm}^{(k-1)}$, i.e.
\be
c^{(k-1)}:\mbf{ext}^{(k)}[p^{(k-1)}]
\ee

The construction rules were stated as higher order irreducible extended programs.
In particular, when applied to applications based on zeroth-order programs the construction rules are themselves expressed in the form of a first-order extended program $[p^{(1)}~c^{(1)}]$, where $c^{(1)}:\mbf{ext}^{(2)}[p^{(1)}]$ and $p^{(1)},~c^{(1)}:\mbf{prgm}^{(1)}$ are programs based on the atomic programs of Section \ref{hop}.
(The construction rules are distinguished from the application specific axioms of the zeroth-order programs that are expressed as irreducible extended programs, $[p^{(0)}~c^{(0)}]$, where $c^{(0)}:\mbf{ext}^{(1)}[p^{(0)}]$ and $p^{(0)},~c^{(0)}:\mbf{prgm}^{(0)}$ are programs based on the atomic programs specific to the application.)

In general, the construction rules \textsf{per}, \textsf{cr1}-\textsf{cr9}, \textsf{flse1}-\textsf{flse2} and \textsf{dsj1}-\textsf{dsj6} along with the I/O type axioms and the substitution rule can be applied to applications based on programs of any order.
It follows that if the construction rules are employed to construct derivations for an application that comes with its own atomic programs of type $\mbf{prgm}^{(k-1)}$ then the atomic programs presented in Section \ref{hop} must be regarded as programs of type $\mbf{prgm}^{(k)}$.

\underline{Axioms of falsity.}
The format of derivations leading to proofs of theorems of falsity were outlined in Section \ref{satof}.
While the format differed slightly from the usual way in which derivations were presented, axioms/theorems of falsity can be incorporated into a general construct of type checking for program extensions.

The program
\be
false^{(k)}~[q^{(k-1)}]~[~], \quad k \geq 1
\ee
is computable if the program $q^{(k-1)}:\mbf{prgm}^{(k-1)}$ is a false $(k-1)$-order program.

Let us also denote the empty $k$-order program, $ep^{(k)}:\mbf{prgm}^{(k)}$ for $k \geq 0$.
We have already been using the name $ep$ for $ep^{(0)}$.
Axioms of falsity for zeroth-order programs have been expressed as
\be
\begin{array}{l}
	~ \\
	\hline
	false~[q]~[~] \\
\end{array}
\ee
where $q:\mbf{prgm}^{(0)}$ is regarded as a constant of the theory.
This is a higher order construct of a program extension where the premise is $ep^{(1)}$, the first-order empty program, and the conclusion is $false~[q]~[~]$.
Thus we have the higher order type checking program for axioms of falsity
\be
ext^{(2)}~[p^{(1)}~c^{(1)}]~[~]
\ee
where we have made the value assignments
\be
p^{(1)}:=ep^{(1)},~c^{(1)}:=false~[q]~[~]
\ee

We can generalize type checking of program extensions for axioms of falsity by making the value assignments
\be
p^{(k-1)}:=ep^{(k-1)},~c^{(k-1)}:=false^{(k-1)}~[q^{(k-2)}]~[~], \quad k \geq 2
\ee
in (\ref{shop01}). 
In this way we can regard (\ref{shop01}) as the general construct of type checking for program extensions that include statements of falsity.          

\section{Some properties of the construction rules.}\label{sspotcr}

The construction rules of PECR are presented as higher order irreducible extended programs.
In this section we will employ VPC as a self referencing tool to examine some properties of the construction rules themselves.
To this end the axioms that are supplied to the file \emph{axiom.dat} include the rules \textsf{per}, \textsf{cr1}-\textsf{cr9}, \textsf{flse1}-\textsf{flse2} and \textsf{dsj1}-\textsf{dsj6}.
To these construction rules we include I/O type axioms and the substitution rule.
These are automated within VPC.

Here, the construction rules can be regarded as the axioms of an application that we will refer to as a \emph{theory for the construction of programs as proofs}.
This is understood to be set in the context of the formal system PECR on which VPC is based.


In what follows is included some analogies that exist between PECR and the formal systems of contemporary proof theory.
In natural deduction, judgments take the form 
\ben\label{seq0}
\Pi \vdash C
\een
where $C$ is a single formula and
\be
\Pi = P_1,\ldots,P_n, \quad n \geq 0
\ee
is a sequent of formulas, $P_i,~i=1,\ldots,n$.
The turnstile symbol, $\vdash$, is used to represent entailment and we say that $\Pi$ entails $C$.
The semantics of the expression (\ref{seq0}) asserts that whenever all of the formulas $P_1,\ldots,P_n$ are true then $C$ is true.
In our analogies we use a program $p=[p_i]_{i=1}^n$ instead of $\Pi$ and replace the expression (\ref{seq0}) with the statement $c:\mbf{ext}[p]$.

Some of the analogies will be based on a comparison of the construction rules of PECR with the rules of sequent calculus (see for example \cite{buss}).
It should be noted that the sequent calculus comes in the two main versions of classical and intuitionistic logic.
For intuitionistic logic, only a single formula can appear on the right hand side of the turnstile whereas for classical logic the right hand side may include sequents.
In order that we retain a closer relevance to the inner workings of VPC we will avoid statements that are exclusive to classical logic and where appropriate present only the intuitionistic version.    

An important property of the expression (\ref{seq0}) is that the order of the formulas in the sequent $\Pi$ can be interchanged.
In the sequent calculus this is stated in the form of the left exchange rule
\be
\begin{array}{l}
	\Pi,A,B,\Lambda \vdash C \\
	\hline
	\Pi,B,A,\Lambda \vdash C \\
\end{array}
\ee
where $A,~B$ and $C$ are formulas and $\Pi$ and $\Lambda$ are sequents.
(There is a right exchange rule but this only applies to classical logic).
Our analogies might sometimes become a little stretched because there is no general rule in PECR that allows the reordering of subprograms in a program list.

The sequent calculus has extensions that include rules that involve quantifiers of formulas dependent on variables that are defined as terms.
These rules will not be presented here because they represent the most significant departure of the sequent calculus from the construction rules of PECR. 

\underline{Weakening.}
The weakening rule in the sequent calculus can be expressed as 
\ben\label{seq_weak}
\begin{array}{l}
	\Pi \vdash C \\
	\hline
	\Pi,A \vdash C \\
\end{array}
\een
where $A$ and $C$ are formulas and $\Pi$ is a sequent.
Here, $A$ can be any formula.
(Keep in mind that we will always present the intuitionistic version of the sequent calculus).

We can derive the following rule where the program $p$ of an extended program $[p~c],~c:\mbf{ext}[p]$, is weakened by the concatenation $r:=[p~a]$ for some program $a$.
We want to show that $c:\mbf{ext}[r]$.

In PECR there is a restriction because we cannot use any program $a$.
The variable names of the I/O lists of the introduced program, $a$, must be compatible with the variable names of the I/O lists of the programs $p$ and $c$ such that $r=[p~a]:\mbf{prgm}$ and $s=[r~c]:\mbf{prgm}$. 
As a consequence we must include in the premise of the following theorem the extra conditional statements $conc~[p~a]~[r]$ and $conc~[r~c]~[s]$.  
\begin{lstlisting}
Theorem thm1

ext [p c] [ ]
conc [p a] [r]
conc [r c] [s]
--------------
ext [r c] [ ]

Proof.
  1 ext [p c] [ ]
  2 conc [p a] [r]
  3 conc [r c] [s]
  4 sub [p r] [ ]            cr6a [2]
  5 aext [r c] [ ]           per [4 1 3]
  6 ext [r c] [ ]            cr1 [5]
\end{lstlisting}

As a consequence of the left exchange rule in the sequent calculus, the order of $\Pi,A$ on the left hand side of the turnstile can be interchanged.
Due to the restrictions imposed by the I/O dependency condition, in PECR we do not have a general rule that is analogous to the exchange rule so we need to check that weakening will also work for the concatenation $r=[a~p]$.
\begin{lstlisting}
Theorem thm2

ext [p c] [ ]
conc [a p] [r]
conc [r c] [s]
--------------
ext [r c] [ ]

Proof.
  1 ext [p c] [ ]
  2 conc [a p] [r]
  3 conc [r c] [s]
  4 sub [p r] [ ]            cr6b [2]
  5 aext [r c] [ ]           per [4 1 3]
  6 ext [r c] [ ]            cr1 [5]
\end{lstlisting}

\underline{Conjunction introduction.}
The left conjunction rules can be expressed as
\be
\begin{array}{ll}
	A,\Pi \vdash C \\
	\hline
	A \land B,\Pi \vdash C \\
\end{array}
\hspace{10mm}
\begin{array}{ll}
	B,\Pi \vdash C \\
	\hline
	A \land B,\Pi \vdash C \\
\end{array}
\ee
where $A,~B$ and $C$ are formulas and $\Pi$ is a sequent.

Theorem \textsf{thm3} is a derivation of an analogy of the first left conjunction rule.
We start with an extension $c:\mbf{ext}[r]$, where $r=[a~p]$, and express the analogy of a conjunction, $A \land B$, with the program concatenation $s=[a~b]$, for some program $b$.    
In the premise of the following theorem we must include two extra conditional statements $conc~[s~p]~[t]$ and $conc~[t~c]~[u]$ that reflect the requirement that the I/O lists of the introduced program $b$ must be compatible with the I/O lists of the programs $a$, $p$ and $c$ such that $t=[s~p]:\mbf{prgm}$ and $u=[t~c]:\mbf{prgm}$.
\begin{lstlisting}
Theorem thm3

conc [a p] [r]
ext [r c] [ ]
conc [a b] [s]
conc [s p] [t]
conc [t c] [u]
--------------
ext [t c] [ ]

Proof.
  1 conc [a p] [r]
  2 ext [r c] [ ]
  3 conc [a b] [s]
  4 conc [s p] [t]
  5 conc [t c] [u]
  6 sub [a s] [ ]            cr6a [3]
  7 sub [s t] [ ]            cr6a [4]
  8 sub [a t] [ ]            cr5c [6 7]
  9 sub [p t] [ ]            cr6b [4]
 10 sub [r t] [ ]            cr6c [1 8 9]
 11 aext [t c] [ ]           per [10 2 5]
 12 ext [t c] [ ]            cr1 [11]
\end{lstlisting}
Because we do not have an analogous rule for the exchange rule we need to check that conjunction introduction will also work for the concatenation $r=[p~a]$.
\begin{lstlisting}
Theorem thm4

conc [p a] [r]
ext [r c] [ ]
conc [a b] [s]
conc [s p] [t]
conc [t c] [u]
--------------
ext [t c] [ ]

Proof.
  1 conc [p a] [r]
  2 ext [r c] [ ]
  3 conc [a b] [s]
  4 conc [s p] [t]
  5 conc [t c] [u]
  6 sub [a s] [ ]            cr6a [3]
  7 sub [s t] [ ]            cr6a [4]
  8 sub [a t] [ ]            cr5c [6 7]
  9 sub [p t] [ ]            cr6b [4]
 10 sub [r t] [ ]            cr6c [1 9 8]
 11 aext [t c] [ ]           per [10 2 5]
 12 ext [t c] [ ]            cr1 [11]
\end{lstlisting}
Theorems \textsf{thm5} and \textsf{thm6}, respectively, are derivations that are analogous to the second left conjunction rule starting with the programs $[b~p]$ and $[p~b]$, respectively.
\begin{lstlisting}
Theorem thm5

conc [b p] [r]
ext [r c] [ ]
conc [a b] [s]
conc [s p] [t]
conc [t c] [u]
--------------
ext [t c] [ ]

Proof.
  1 conc [b p] [r]
  2 ext [r c] [ ]
  3 conc [a b] [s]
  4 conc [s p] [t]
  5 conc [t c] [u]
  6 sub [b s] [ ]            cr6b [3]
  7 sub [s t] [ ]            cr6a [4]
  8 sub [b t] [ ]            cr5c [6 7]
  9 sub [p t] [ ]            cr6b [4]
 10 sub [r t] [ ]            cr6c [1 8 9]
 11 aext [t c] [ ]           per [10 2 5]
 12 ext [t c] [ ]            cr1 [11]

Theorem thm6

conc [p b] [r]
ext [r c] [ ]
conc [a b] [s]
conc [s p] [t]
conc [t c] [u]
--------------
ext [t c] [ ]

Proof.
  1 conc [p b] [r]
  2 ext [r c] [ ]
  3 conc [a b] [s]
  4 conc [s p] [t]
  5 conc [t c] [u]
  6 sub [b s] [ ]            cr6b [3]
  7 sub [s t] [ ]            cr6a [4]
  8 sub [b t] [ ]            cr5c [6 7]
  9 sub [p t] [ ]            cr6b [4]
 10 sub [r t] [ ]            cr6c [1 9 8]
 11 aext [t c] [ ]           per [10 2 5]
 12 ext [t c] [ ]            cr1 [11]
\end{lstlisting}

One should note that, while subject to extra conditional constraints, the rules analogous to weakening and conjunction introduction are derivable in PECR, i.e. they are not axioms. 
It should also be observed that if we set $a$ to the empty program, i.e. $a=ep$, in theorems \textsf{thm3}-\textsf{thm4} we obtain theorems \textsf{thm1}-\textsf{thm2} that are similar to the left weakening rule.
Similarly, if we set $b$ to the empty program in theorems \textsf{thm5}-\textsf{thm6} we also obtain theorems \textsf{thm1}-\textsf{thm2}.
Thus, as well as being derivable, the rules of weakening and conjunction introduction in PECR are not independent.
This should not be surprising since, by our analogies, program concatenation cannot distinguish between weakening and conjunction introduction.

In the sequent calculus there is the right conjunction rule
\be
\begin{array}{l}
	\Pi \vdash A \hspace{5mm} \Pi \vdash B \\
	\hline
	\hspace{5mm} \Pi \vdash A \land B \\
\end{array} \\
\ee
where $A$ and $B$ are formulas and $\Pi$ is a sequent.
In PECR this is expressed in the form of the construction rule \textsf{cr9b}.

\underline{Extensions of equivalent premises.}
In VPC we always generate proofs such that the conclusion program of an extended program derivation is an atomic program.
This need not be the case in general and a conclusion program may have a list representation.
One can generate conclusions that are non-atomic by the recursive application of the construction rule \textsf{cr9b}.

The construction rule \textsf{cr9a} states that an equivalent program of an extension that need not be atomic can also be assigned the type of an extension.
A similar rule exists for program equivalent premises as the following derivation demonstrates.
\begin{lstlisting}
Theorem thm7

ext [q c] [ ]
eqp [q p] [ ]
--------------
aext [p c] [ ]

Proof.
  1 ext [q c] [ ]
  2 eqp [q p] [ ]
  3 conc [q c] [a]           cr2 [1]
  4 conc [p c] [b]           sr1 [3 2]
  5 sub [q p] [ ]            cr5a [2]
  6 aext [p c] [ ]           per [5 1 4]
\end{lstlisting}

While the substitution rule should not be applied as an axiom to the program $ext~[p~c]~[~]$, it is easy to see that it does satisfy the substitution rule as a derivation by simply applying the construction rule \textsf{cr2} separately to \textsf{cr9a} and theorem \textsf{thm7}.

\underline{The program extension rule.}
At each step of a proof construction a new statement, $c$, of the proof is generated from an extended program derivation
\ben\label{epd20}
[sub~[q~p]~[~]~ext~[q~c]~[~]~conc~[p~c]~[s]]
\een
where $p$ is the current derivation program of the proof and $q \subseteqq p$.
The process is one of finding an axiom/theorem whose premises can be matched with some sublist $q$ of $p$.
Crucial to identifying the program $[q~c]$ as an extended program, is to establish that $[q~c]$ is program and I/O equivalent to some known axiom/theorem.
Having achieved this, VPC then constructs the appropriate extended program derivation (\ref{epd20}).

In Section \ref{sof} the procedure for constructing $[q~c]$ as an extended program in an extended program derivation was outlined.
The procedure begins by finding a program $q1$ such that  $q1 \equiv q$ and $q1 \equiv_{io} q2$, where $[q2~c2],~c2:\mbf{iext}[q2]$, is an axiom/theorem that is stored in the file \emph{axiom.dat}.
We then construct $c1$ such that $[q1~c1] \equiv_{io} [q2~c2]$ and set $c:=c1$.
In the more general case of non-atomic extensions we can use any $c$ such that $c \equiv c1$.
The following theorem demonstrates that the program $[q~c]$ constructed in this way is indeed a program extension such that $c:\mbf{ext}[q]$.  

Note that for the program extension rule to hold we only require that $[q~c],~c:\mbf{ext}[p]$.
As outlined in Section \ref{sof}, there is no loss of generality in proof constructions by the restriction of matching $[q~c]$ to an irreducible extended program in the form of an axiom/theorem.
In the premise of theorem thm8 we demand only that $[q2~c2],~c2:\mbf{ext}[q2]$.

\begin{lstlisting}
Theorem thm8

eqp [q1 q] [ ]
conc [q1 c1] [u]
conc [q2 c2] [v]
ext [q2 c2] [ ]
eqio [q1 q2] [ ]
eqio [u v] [ ]
eqp [c1 c] [ ]
----------------
ext [q c] [ ]

Proof.
  1 eqp [q1 q] [ ]
  2 conc [q1 c1] [u]
  3 conc [q2 c2] [v]
  4 ext [q2 c2] [ ]
  5 eqio [q1 q2] [ ]
  6 eqio [u v] [ ]
  7 eqp [c1 c] [ ]
  8 aext [q1 c1] [ ]         cr3c [4 3 2 5 6]
  9 ext [q1 c1] [ ]          cr1 [8]
 10 aext [q c1] [ ]          thm7 [9 1]
 11 ext [q c1] [ ]           cr1 [10]
 12 aext [q c] [ ]           cr9a [11 7]
 13 ext [q c] [ ]            cr1 [12]
\end{lstlisting}
It should also be noted that if we insist that program extensions must always be atomic then the fourth statement in axiom \textsf{cr3c} and the fifth statement of theorem \textsf{thm8} can be removed.  

\underline{Disjunction distribution rules.}  
In classical logic the statements $P \land (A \lor B)$ and $(P \land A) \lor (P \land B)$ are logically equivalent.
In PECR there are similar rules that are reflected in the disjunction distribution rules.
When constructing disjunctions in VPC it is more convenient to employ the derivable rules presented below.    

Theorems \textsf{thm9}-\textsf{thm11} are derivations for these generalized distribution rules for disjunctions.
They make extensive use of the right and left disjunction distribution rules, \textsf{dsj3}-\textsf{dsj4}.

Theorem \textsf{thm9} shows that if $d =a~|~b$ and $u=[p~a~q]$ and $v=[p~b~q]$ are type $\mbf{prgm}$ then $e =u~|~v$ is type $\mbf{prgm}$.
This combines \textsf{dsj3a} and \textsf{dsj4a} into a single rule.
\begin{lstlisting}
Theorem thm9

disj [a b] [d]
conc [p a] [f]
conc [f q] [u]
conc [p b] [g]
conc [g q] [v]
--------------
disj [u v] [e]

Proof.
  1 disj [a b] [d]
  2 conc [p a] [f]
  3 conc [f q] [u]
  4 conc [p b] [g]
  5 conc [g q] [v]
  6 disj [f g] [c]           dsj3a [2 4 1]
  7 disj [u v] [e]           dsj4a [3 5 6]
\end{lstlisting}
Theorem \textsf{thm10} shows that if $d =a~|~b$ and $u=[p~a~q]$, $v=[p~b~q]$ and $r=[p~d]$ are type $\mbf{prgm}$ then $h=[r~q]=[p~d~q]$ is type $\mbf{prgm}$.
This combines \textsf{dsj3b} and \textsf{dsj4b} into a single rule.
\begin{lstlisting}
Theorem thm10

disj [a b] [d]
conc [p a] [f]
conc [f q] [u]
conc [p b] [g]
conc [g q] [v]
conc [p d] [r]
--------------
conc [r q] [h]

Proof.
  1 disj [a b] [d]
  2 conc [p a] [f]
  3 conc [f q] [u]
  4 conc [p b] [g]
  5 conc [g q] [v]
  6 conc [p d] [r]
  7 disj [f g] [c]           dsj3a [2 4 1]
  8 conc [c q] [e]           dsj4b [3 5 7]
  9 eqp [r c] [ ]            dsj3c [2 4 1 6 7]
 10 eqp [c r] [ ]            cr4b [9]
 11 conc [r q] [h]           sr1 [8 10]
\end{lstlisting}
Theorem \textsf{thm11} shows that if $d =a~|~b$ and $s=[p~d~q]$, $j =[p~a~q]~|~[p~b~q]$ are type $\mbf{prgm}$ then $j \equiv s$.
This combines \textsf{dsj3c} and \textsf{dsj4c} into a single rule.

\begin{lstlisting}
Theorem thm11

disj [a b] [d]
conc [p d] [r]
conc [r q] [s]
conc [p a] [f]
conc [f q] [u]
conc [p b] [g]
conc [g q] [v]
disj [u v] [j]
--------------
eqp [s j] [ ]

Proof.
  1 disj [a b] [d]
  2 conc [p d] [r]
  3 conc [r q] [s]
  4 conc [p a] [f]
  5 conc [f q] [u]
  6 conc [p b] [g]
  7 conc [g q] [v]
  8 disj [u v] [j]
  9 disj [f g] [c]           dsj3a [4 6 1]
 10 conc [c q] [e]           dsj4b [5 7 9]
 11 eqp [r c] [ ]            dsj3c [4 6 1 2 9]
 12 eqp [e s] [ ]            sr2 [3 11 10]
 13 eqp [s e] [ ]            cr4b [12]
 14 eqp [e j] [ ]            dsj4c [5 7 9 10 8]
 15 eqp [s j] [ ]            sr1 [13 14]
\end{lstlisting}

\underline{Disjunction contraction rule.}    
The contraction rule, \textsf{dsj1}, can be written in the more general form given by theorem \textsf{thm14}, below.
We first need to prove theorems \textsf{thm12} and \textsf{thm13}. 

Theorem \textsf{thm12} proves that if $d =a~|~b$, $s=[p~d~q]$, $j =[p~a~q]~|~[p~b~q]$ and $k=[j~c]$ are type $\mbf{prgm}$ then $e=[s~c]$ is type $\mbf{prgm}$.
\begin{lstlisting}
Theorem thm12

disj [a b] [d]
conc [p d] [r]
conc [r q] [s]
conc [p a] [f]
conc [f q] [u]
conc [p b] [g]
conc [g q] [v]
disj [u v] [j]
conc [j c] [k]
--------------
conc [s c] [e]

Proof.
  1 disj [a b] [d]
  2 conc [p d] [r]
  3 conc [r q] [s]
  4 conc [p a] [f]
  5 conc [f q] [u]
  6 conc [p b] [g]
  7 conc [g q] [v]
  8 disj [u v] [j]
  9 conc [j c] [k]
 10 eqp [s j] [ ]            thm11 [1 2 3 4 5 6 7 8]
 11 eqp [j s] [ ]            cr4b [10]
 12 conc [s c] [e]           sr1 [9 11]
\end{lstlisting}
Theorem \textsf{thm13} proves that if $d =a~|~b$, $s=[p~d~q]$, $j =[p~a~q]~|~[p~b~q]$, $k=[j~c]$ and $l=[s~c]$ are type $\mbf{prgm}$ then $l \equiv k$.
\begin{lstlisting}
Theorem thm13

disj [a b] [d]
conc [p d] [r]
conc [r q] [s]
conc [p a] [f]
conc [f q] [u]
conc [p b] [g]
conc [g q] [v]
disj [u v] [j]
conc [j c] [k]
conc [s c] [l]
--------------
eqp [l k] [ ]

Proof.
  1 disj [a b] [d]
  2 conc [p d] [r]
  3 conc [r q] [s]
  4 conc [p a] [f]
  5 conc [f q] [u]
  6 conc [p b] [g]
  7 conc [g q] [v]
  8 disj [u v] [j]
  9 conc [j c] [k]
 10 conc [s c] [l]
 11 disj [f g] [e]           dsj3a [4 6 1]
 12 conc [e q] [h]           dsj4b [5 7 11]
 13 eqp [r e] [ ]            dsj3c [4 6 1 2 11]
 14 eqp [h s] [ ]            sr2 [3 13 12]
 15 eqp [h j] [ ]            dsj4c [5 7 11 12 8]
 16 eqp [j h] [ ]            cr4b [15]
 17 conc [h c] [i]           sr1 [9 16]
 18 eqp [i k] [ ]            sr2 [9 16 17]
 19 eqp [l i] [ ]            sr2 [17 14 10]
 20 eqp [l k] [ ]            sr1 [19 18]
\end{lstlisting}
We can now prove theorem \textsf{thm14} that is the general version of the contraction rule.
It states that if $d =a~|~b$, $s=[p~d~q]$, $u=[p~a~q]$, $v=[p~b~q]$ and $j =u~|~v$ are type $\mbf{prgm}$ and $c:\mbf{ext}[u]$, $c:\mbf{ext}[v]$, then the type assignment $c::\mbf{ext}[s]$ is valid.
\begin{lstlisting}
Theorem thm14

disj [a b] [d]
conc [p d] [r]
conc [r q] [s]
conc [p a] [f]
conc [f q] [u]
conc [p b] [g]
conc [g q] [v]
disj [u v] [j]
ext [u c] [ ]
ext [v c] [ ]
--------------
aext [s c] [ ]

Proof.
  1 disj [a b] [d]
  2 conc [p d] [r]
  3 conc [r q] [s]
  4 conc [p a] [f]
  5 conc [f q] [u]
  6 conc [p b] [g]
  7 conc [g q] [v]
  8 disj [u v] [j]
  9 ext [u c] [ ]
 10 ext [v c] [ ]
 11 aext [j c] [ ]           dsj1 [9 10 8]
 12 ext [j c] [ ]            cr1 [11]
 13 eqp [s j] [ ]            thm11 [1 2 3 4 5 6 7 8]
 14 eqp [j s] [ ]            cr4b [13]
 15 aext [s c] [ ]           thm7 [12 14]
\end{lstlisting}

Analogies between disjunctions in PECR and those of formal systems in proof theory can sometimes become a little stretched.
With this in mind there are some observations that may be useful.

In Section \ref{sdisj} the procedures for disjunction splitting followed by a contraction were outlined.
This was presented in a form that employs the generalized contraction rule of theorem \textsf{thm14}.

We start with a program $s=[p~d~q]$, where $d= a~|~b$.
This is split into the two operand programs $u=[p~a~q]$ and $v=[p~b~q]$.
Independent derivations are then sought for each operand program that lead to the same conclusion, $c$.
Finally, we apply the generalized contraction rule of theorem \textsf{thm14} to obtain the extension $c:\mbf{ext}[s]$.

In the sequent calculus there is the left disjunction rule
\be
\begin{array}{l}
	\Pi,A \vdash C \hspace{5mm} \Pi,B \vdash C \\
	\hline
	\hspace{5mm} \Pi,A \lor B \vdash C \\
\end{array} \\
\ee
where $A,~B$ and $C$ are formulas and $\Pi$ is a sequent.
This can also be generalized by introducing an additional sequent that is appended to the sequents $\Pi,A$ and $\Pi,B$ in the first line.
In this generalized form it is easier to see that the left disjunction rule of the sequent calculus has some similarity with the generalized contraction rule of theorem \textsf{thm14}.

\underline{Extension disjunction introduction.}
In the sequent calculus there are the right disjunction introduction rules
\be
\begin{array}{ll}
	\Pi \vdash A \\
	\hline
	\Pi \vdash A \lor B \\
\end{array}
\hspace{10mm}
\begin{array}{ll}
	\Pi \vdash B \\
	\hline
	\Pi \vdash A \lor B \\
\end{array}
\ee
where $A$ and $B$ are formulas and $\Pi$ is a sequent.

In PECR we have the rule \textsf{disj6} that is analogous to the first of the right disjunction introduction rules.
Only one rule is needed because we can derive an analogy of the second of the right disjunction introduction rules as follows.
\begin{lstlisting}
Theorem thm15

ext [p b] [ ]
disj [a b] [d]
conc [p d] [s]
--------------
aext [p d] [ ]

Proof.
  1 ext [p b] [ ]
  2 disj [a b] [d]
  3 conc [p d] [s]
  4 disj [b a] [c]           dsj2a [2]
  5 eqp [d c] [ ]            dsj2b [4 2]
  6 eqp [c d] [ ]            cr4b [5]
  7 conc [p c] [e]           sr1 [3 5]
  8 aext [p c] [ ]           dsj6 [1 4 7]
  9 ext [p c] [ ]            cr1 [8]
 10 aext [p d] [ ]           cr9a [9 6]
\end{lstlisting}
The third statement, $conc~[p~d]~[s]$, in the premises of axiom \textsf{disj6} and theorem \textsf{thm15} is an extra conditional statement that reflects the requirement that the variable names of the I/O lists of any program that is introduced through a disjunction are compatible with those of the preceding programs such that $[p~d]:\mbf{prgm}$.    

\underline{Disjunction contraction rule 2.}
Theorem \textsf{thm16} shows that the contraction rule 2, introduced in Section \ref{sdisj}, is a theorem.
It states that if $d =a~|~b$ is type $\mbf{prgm}$, $c:\mbf{ext}[a]$ and $b:\mbf{false}$ then the type assignment $c::\mbf{ext}[d]$ is valid.
\begin{lstlisting}
Theorem thm16

disj [a b] [d]
ext [a c] [ ]
false [b] [ ]
--------------
aext [d c] [ ]

Proof.
  1 disj [a b] [d]
  2 ext [a c] [ ]
  3 false [b] [ ]
  4 eqp [d a] [ ]            dsj5 [1 3]
  5 eqp [a d] [ ]            cr4b [4]
  6 aext [d c] [ ]           thm7 [2 5]
\end{lstlisting}
Theorem \textsf{thm17} generalizes the contraction rule 2.
It states that if $d=a~|~b$, $s=[p~d~q]$, $u=[p~a~q]$ and $v=[p~d~q]$ are type $\mbf{prgm}$ and $c:\mbf{ext}[u]$, $v:\mbf{false}$, then the type assignment $c::\mbf{ext}[s]$ is valid.
\begin{lstlisting}
Theorem thm17

disj [a b] [d]
conc [p d] [r]
conc [r q] [s]
conc [p a] [f]
conc [f q] [u]
conc [p b] [g]
conc [g q] [v]
ext [u c] [ ]
false [v] [ ]
--------------
aext [s c] [ ]

Proof.
  1 disj [a b] [d]
  2 conc [p d] [r]
  3 conc [r q] [s]
  4 conc [p a] [f]
  5 conc [f q] [u]
  6 conc [p b] [g]
  7 conc [g q] [v]
  8 ext [u c] [ ]
  9 false [v] [ ]
 10 disj [u v] [e]           thm9 [1 4 5 6 7]
 11 eqp [s e] [ ]            thm11 [1 2 3 4 5 6 7 10]
 12 eqp [e u] [ ]            dsj5 [10 9]
 13 eqp [s u] [ ]            sr1 [11 12]
 14 eqp [u s] [ ]            cr4b [13]
 15 aext [s c] [ ]           thm7 [8 14]
\end{lstlisting}

\underline{Disjunction contraction rule 3.}
The following shows that the contraction rule 3, introduced in Section \ref{sdisj}, is a theorem. 
Theorem \textsf{thm18} states that if $d =a~|~b$ is type $\mbf{prgm}$ and $a,b:\mbf{false}$ then the type assignment $d::\mbf{false}$ is valid.
\begin{lstlisting}
Theorem thm18

disj [a b] [d]
false [a] [ ]
false [b] [ ]
--------------
afalse [d] [ ]

Proof.
  1 disj [a b] [d]
  2 false [a] [ ]
  3 false [b] [ ]
  4 eqp [d a] [ ]            dsj5 [1 3]
  5 eqp [a d] [ ]            cr4b [4]
  6 sub [a d] [ ]            cr5a [5]
  7 afalse [d] [ ]           flse1 [6 2]
\end{lstlisting}
Theorem \textsf{thm19} generalizes the contraction rule 3.
It states that if $d =a~|~b$, $s=[p~d~q]$, $u=[p~a~q]$ and $v=[p~d~q]$ are type $\mbf{prgm}$ and $u,v:\mbf{false}$ then the type assignment $s::\mbf{false}$ is valid.
\begin{lstlisting}
Theorem thm19

disj [a b] [d]
conc [p d] [r]
conc [r q] [s]
conc [p a] [f]
conc [f q] [u]
conc [p b] [g]
conc [g q] [v]
false [u] [ ]
false [v] [ ]
--------------
afalse [s] [ ]

Proof.
  1 disj [a b] [d]
  2 conc [p d] [r]
  3 conc [r q] [s]
  4 conc [p a] [f]
  5 conc [f q] [u]
  6 conc [p b] [g]
  7 conc [g q] [v]
  8 false [u] [ ]
  9 false [v] [ ]
 10 disj [u v] [c]           thm9 [1 4 5 6 7]
 11 eqp [s c] [ ]            thm11 [1 2 3 4 5 6 7 10]
 12 afalse [c] [ ]           thm18 [10 8 9]
 13 false [c] [ ]            flse2 [12]
 14 eqp [c s] [ ]            cr4b [11]
 15 sub [c s] [ ]            cr5a [14]
 16 afalse [s] [ ]           flse1 [15 13]
\end{lstlisting}

\textbf{Notes.}

\begin{itemize}  

	\item In applications of the sequent calculus the cut rule is used extensively in the construction of proofs.
	The cut rule can be expressed as
    \be
    \begin{array}{l}
    	\Gamma \vdash A \hspace{5mm} A,\Lambda \vdash B \\
    	\hline
    	\hspace{5mm} \Gamma,\Lambda \vdash B \\
    \end{array}
    \ee
    where $A$ and $B$ are formulas and $\Gamma$ and $\Lambda$ are sequents.
    In PECR there is no need for a similar rule when constructing proofs.
    However, when a proof is completed, theorem extraction relies on the elimination of intermediate statements of the proof.
    This is achieved by the connection list reduction algorithm of Section \ref{connection}.
    In PECR the algorithm of connection list reduction for the extraction of theorems from proofs can be thought of as analogous to an iterative application of the cut rule.
    
    \item There are two kinds of rules of the sequent calculus that have not been mentioned so far.
    The first of these are the negation rules.
    Negating formal statements presented as programs is often not meaningful so the negation rules of the sequent calculus have little relevance here.
    
    The second of these are the implication rules that include formulas involving the implication connective, $\to$.
    Analogies of program extensions and implications are too stretched to be of any use, mainly because of the range of interpretations that can be attached to an implication.
    It is worthwhile noting that the formula $A \to B$ is logically equivalent to the formula $\neg A \lor B$.
    The presence of a negation further demonstrates the difficulty in associating, by analogy, implications with any formal statements presented as programs.
    This is why it is better to focus on the stronger analogy between a program extension and a statement of entailment, (\ref{seq0}).

    \item Because there is no concise general rule for negating formal statements presented as programs it follows that there is no law of the excluded middle in the construction rules of PECR.      
    However, any formal system of proof theory, including the sequent calculus, can be implemented in applications of PECR that employ higher levels of abstractions.
    This will be discussed in more detail in the following chapter.          

\end{itemize}

\chapter{Abstractions.}\label{cabstract}

\section{Primary states of a theory.}\label{spsoat}\index{primary state}

So far a theory has been defined by the notation $S(atom,ax,cst,mach)$, where $atom$ is the list of atomic programs of the theory, $ax$ is the list of axioms, $cst$ is the list of constants and $mach=[nchar~nstr~nlst~nint]$ is a list of parameters that reflect the constraints of the theory in the machine environment, $\mathfrak{M}(mach)$.

Here we will discard the dependence of a theory, $S$, on the list of axioms, $ax$, and define substates of the theory that correspond to each specific choice of the list of axioms.
We shall also introduce an additional constraint on our theory that imposes on all irreducible extended programs a premise program list length that is bound by $nprem$, i.e. for any program $[p~c]$, $c:\mbf{iext}[p]$, we must have $length[p] \leq nprem$.

We now define a theory, $S$, by
\be
S(atom,cst,mach)
\ee
where the machine parameters list is given by
\be
mach=[nchar~nstr~nlst~nint~nprem]
\ee
This differs from our original definition of a theory by the exclusion of the dependence on the list of axioms, $ax$, and the inclusion of the bound, $nprem$, for any premise program list length.
We can think of the bound on the premise list length along with the other parameter constraints of the list $mach$ as representing the depth in which the theory, $S$, is to be explored.

The lists
\be
atom=[atom(i)]_{i=1}^{nat}, \quad nat:\mbf{int1},~atom(i):\mbf{atm}<:\mbf{prgm}
\ee
and
\be
\bal
& cst=[cst(i)]_{i=1}^{ncst}, \quad ncst:\mbf{int0},~cst(i):\mbf{char} \\
\eal
\ee
along with the additional constraints $mach$ are fixed.

An irreducible extended program takes the form $[p~c]$, where $c:\mbf{iext}[p]$ with the constraint, $length[p] \leq nprem$.
Let $s$ be the list of all irreducible extended programs of the theory, $S(atom,cst,mach)$.
A sublist $a \subseteqq s$ is said to be \emph{list of independent generators} of a list of theorems if no element of $a$ can be derived from other elements contained in $a$.
Any list of independent generators, $a$, can serve as the list of axioms of a theory provided that the list $a$ contains no redundant elements with respect to list of theorems that it generates, i.e. each element of $a$ is involved in the derivation of at least one theorem.   
Hence, the theory, $S$, can exist in any one of the finite primary states
\be
S^{(k)}, \quad k=1, \ldots ,ms 
\ee
where for each $S^{(k)}$ there corresponds a unique list of axioms
\be
ax^{(k)}=[ax^{(k)}(i)]_{i=1}^{nax^{(k)}} , \quad nax^{(k)}:\mbf{int0},~ax^{(k)}(i):\mbf{prgm}
\ee
Here each list $ax^{(k)}$, $k=1,\dots,ms$, is unique but we may have $ax^{(k)} \cap ax^{(l)} \ne [~]$, for some $l \ne k$.

Note that we are dealing with a finite system so that we can expect that there exist a possibly huge but finite number, $ms$, of primary states.
The list, $s$, of all irreducible extended programs of the theory, $S$, can be partitioned by each primary state, $S^{(k)}$, as
\ben\label{pttn01}
s \equiv [ax^{(k)}~th^{(k)}~ud^{(k)}]
\een
where
\be
th^{(k)} = [th^{(k)}(i)]_{i=1}^{nth^{(k)}} , \quad nth^{(k)}:\mbf{int1},~th^{(k)}(i):\mbf{prgm}
\ee
is the list of all irreducible extended programs that are generated from the axioms of $ax^{(k)}$ and
\be
ud^{(k)} = [ud^{(k)}(i)]_{i=1}^{nud^{(k)}} , \quad nud^{(k)}:\mbf{int0},~ud^{(k)}(i):\mbf{prgm}
\ee
is the list of all irreducible extended programs that cannot be derived from the axioms of $ax^{(k)}$.
The sum
\be
ns=nax^{(k)} + nth^{(k)} + nud^{(k)}
\ee
is a constant with respect to all primary states, $S^{(k)}$, $k=1,\ldots,ms$, and is equal to the fixed length of the list $s$.

When we say that elements of a list of theorems, $th$, are derivable or generated from the axioms of the list, $ax$, we will always mean that the theorem connection list of each element of the list of theorems, $th$, can be traced back to the labels of the individual axioms of $ax$ by the process of theorem connection list reduction as outlined in Section \ref{thmconn}.

It should be noted that while each element of the lists $atom$, $s$ and its partitions, $ax^{(k)},~th^{(k)}$ and $ud^{(k)}$, are type $\mbf{prgm}$ objects the lists themselves are not meant to represent type $\mbf{prgm}$ objects.
Throughout we have maintained a preference to represent collections of objects as lists rather than sets.
However, the lists $s$, $th^{(k)}$ and $ud^{(k)}$ typically have prohibitively large lengths and hence it will not be feasible to store them on a real world computer.
For this reason these lists should be regarded in an abstract sense.

It is desirable to construct primary states where the list lengths of $ax^{(k)}$ and $ud^{(k)}$ are minimal.
Of particular interest are primary states where $ud^{(k)}$ is the empty list.
Such primary states are said to be \emph{complete}.

In any primary state, $S^{(k)}$, every element of the list $[ax^{(k)}~ud^{(k)}]$ of the partition (\ref{pttn01}) is an irreducible extended program that cannot be derived in that primary state.
The elements of $ax^{(k)}$ are distinguished from elements of $ud^{(k)}$ in that $ax^{(k)}$ is a list of independent generators of the theorems list $th^{(k)}$.

\section{Completeness.}\index{completeness}

We will now explore some interesting properties of theories that lead to completeness.
The constructions of the previous section involved lists with lengths that are too large to be stored on a real world computer so we must regard them as abstract lists. 
Being forced into the realm of abstraction it will be more convenient to deal with sets rather than lists to take advantage of their nice property of automatically removing repetitions of elements under unions.

As described in the previous section, a theory, $S(atom,cst,mach)$, is defined by a fixed and finite collection of atomic programs and parameter constraints.
Let $uset$ be the universal set of all irreducible extended programs of the theory, $S$.
Each element of $uset$ is a program of the form $[p~c]$ such that $c:\mbf{iext}[p]$.
A primary state, $S^{(k)}$, $1 \leq k \leq ms$, of the theory, $S$, is defined by an independent set of axioms, $ax^{(k)}$, and a partition of the set $uset$ given by
\be
uset = ax^{(k)} \cup th^{(k)} \cup ud^{(k)}
\ee
where $th^{(k)}$ is the set of all theorems or irreducible extended programs that have proofs under the axioms $ax^{(k)}$ and $ud^{(k)}$ is the set of all irreducible extended programs that cannot be derived from the axioms of $ax^{(k)}$. 

\underline{Subset notation.}
In this and the remaining sections of this chapter the subset symbol $\subset$ will be used under the definition
\be
a \subset b \leftrightarrow \{ x \in a \to x \in b \}
\ee
Equality of sets, $a=b$, will hold if $a \subset b$ and $b \subset a$.

\underline{The $gen$ function.}
It will be necessary to be more precise as to what we mean by a generating set. We start by removing the restriction that the elements of a generating set are independent of one another. 

\begin{definition}\label{gen}(The $gen$ function.)\index{gen}
	A finite and nonempty set $t$ defined by
	\ben \label{gen10}
	t = gen[a]
	\een
	is the set of all irreducible extended programs that have proofs that can be derived from a set, $a$, such that $a \subset uset$ contains no redundant elements with respect to the derivations of elements of $t$, i.e. every element of $a$ is involved in the derivation of at least one element of $t$.
	We say that $t$ is a derivable set generated by the set $a$ and refer to $a$ as a set of generators of $t$.
\end{definition}

Let $t=\{ t(i) \}_{i=1}^n = gen[a]$, for some $n:\mbf{int1}$.
Each element $t(i) \in t$ has a proof that can be derived by a sequence of extended program derivations employing the elements contained in $a$ and elements of $t \setminus \{ t(i) \}$.
Hence each $t(i) \in t$ has a theorem connection list that contains the labels of elements of $a$ and $t \setminus \{ t(i) \}$.
While elements of $a$ may not be independent, the theorem connection list of $t(i)$ can be traced back to the elements of $a$ by employing the same process of theorem connection list reduction as outlined in Section \ref{thmconn} (in the algorithm of (\ref{algtclr}) simply replace the condition $b \setminus axl \neq [~]$ of the $if$ statement following the $do$ command of the outer loop with $b \setminus al \neq [~]$, where $al$ is the list of labels of the elements of the set of generators $a$).

A set, $a$, is an independent set of generators if no element of $a$ can be derived from the other elements of $a$.
There are five important properties of the $gen$ function that we will later express as axioms.

\underline{Properties of the $gen$ function.}

\begin{itemize}
	
	\item \textbf{Property 1.} For any set of generators, $a$, we say that $gen[a]$ is well defined if $gen[a]$ is nonempty and $a$ contains no redundant elements in the generation of $gen[a]$, i.e. each element of $a$ is involved in the derivation of at least one element of $gen[a]$.
	
	\item \textbf{Property 2.} For any set, $b$, such that $gen[b]$ is well defined there exists an independent set of generators, $a \subset b$, such that $gen[a]$ is well defined and $gen[a]=gen[b] \cup (b \setminus a)$.
	
	\item \textbf{Property 3.} If $gen[a]$ and $gen[b]$ are well defined then $gen[a \cup b]$ is also well defined.
	
	\item \textbf{Property 4.} $gen[a] \cup gen[b] \subset gen[a \cup b]$.
	
	\item \textbf{Property 5.} Proofs based on an identity are excluded so that for any set of generators, $a$,
	\be
	a \cap gen[a] = \emptyset
	\ee
	
\end{itemize}

Any independent set of generators can serve as a set of axioms that defines a primary state of the theory, $S$.
It is possible that there are primary states with an empty list of axioms.

It is important to keep in mind throughout that an independent set $a \subset uset$ is not necessarily a set of generators.
For an independent set $a$ to be a set of generators it must not contain any redundant elements with respect to the set $gen[a]$.
The redundancy condition is necessary to distinguish axioms of each primary state from its list of underivable irreducible extended programs.

\underline{Globally underivable axioms.} A theory may possess irreducible extended programs that will be underivable in all primary states but have the distinguishing property that they can serve as axioms.

\begin{definition}(Globally underivable axiom.)\index{globally underivable axiom}
	A \textbf{globally underivable axiom} of a theory, $S$, is an irreducible extended program that cannot be derived in any primary state but appears in the set of axioms of at least one primary state of the theory. 
\end{definition}

The set of all globally underivable axioms of a theory is denoted by
\be
gua= \{gua(i)\}_{i=1}^{ng}, \quad gua(i) \in uset,~ng:\mbf{int0}
\ee
We will not rule out the possibility that a theory has no globally underivable axioms, in which case we set $ng=0$ and $gua=\emptyset$.

By definition, for each globally underivable axiom, $gua(i)$, there exists a primary state, $S^{(k(i))}$, for some $k(i)$, $1 \leq k(i) \leq ms$, such that $gua(i) \in ax^{(k(i))}$, i.e. $gua(i)$ is involved in the derivation of at least one element of $th^{(k(i))}=gen[ax^{(k(i))}]$.

Thus
\be
gua \subset \cup_{i=1}^{ng} ax^{(k(i))}
\ee
Since each $gen[ax^{(k(i))}]$, $i=1,\ldots ng$, is well defined, we can repeatedly apply Property 3 and 4 of the $gen$ function to obtain
\be
\cup_{i=1}^{ng} gen(ax^{(k(i))}) \subset gen[\cup_{i=1}^{ng} ax^{(k(i))}]
\ee
Since $gen[\cup_{i=1}^{ng} ax^{(k(i))}]$ is also well defined it follows from Property 2 of the $gen$ function that there exists an independent set
\be
sgua \subset \cup_{i=1}^{ng} ax^{(k(i))}
\ee
such that $gen[sgua]$ is well defined and
\be
gen[sgua] =gen[\cup_{i=1}^{ng} ax^{(k(i))}] \cup (\cup_{i=1}^{ng} ax^{(k(i))} \setminus sgua) 
\ee 
Since $gua \subset \cup_{i=1}^{ng} ax^{(k(i))}$ and each element of $gua$ cannot belong to the derivable sets $gen[\cup_{i=1}^{ng} ax^{(k(i))}]$ and, in particular $\cup_{i=1}^{ng} ax^{(k(i))} \setminus sgua$, we must have
\be
gua \subset sgua
\ee
The set of independent generators, $sgua$, may not be unique.
We will always assume that we have chosen an independent set of generators, $sgua$, with the smallest cardinality that contains $gua$. 

\underline{$\beta$-completeness.} Irreducible extended programs of a theory may be underivable from the axioms in one primary state but derivable from the axioms of another primary state.
We will examine theories with the following property.

\begin{definition}($\beta$-complete.)
	A theory, $S$, is said to be \textbf{$\beta$-complete} if each irreducible extended program of the theory, $S$, that is not a globally underivable axiom has a proof in at least one primary state, i.e. if $c:\mbf{iext}[p]$ such that $[p~c] \notin gua$ then there exists a primary state, $S^{(k^\prime)}$, $1 \leq k^\prime \leq ms$, such that $[p~c] \in th^{(k^\prime)}$. 
\end{definition}

Let $u=\{u(i)\}_{i=1}^n \subset uset$ such that $u \cap gua=\emptyset$.
If a theory, $S$, is $\beta$-complete we must have for each $u(i)$ there exists a primary state, $S^{(l(i))}$, for some $l(i)$, $1 \leq l(i) \leq ms$, such that $u(i) \in th^{(l(i))}$.
Thus
\be
u \subset \cup_{i=1}^n th^{(l(i))} = \cup_{i=1}^n gen[ax^{(l(i))}] \subset gen[\cup_{i=1}^n ax^{(l(i))}]
\ee
where the last expression follows from repeated use of Property 3 and 4 of the $gen$ function.

Since $gen[\cup_{i=1}^n ax^{(l(i))}]$ is well defined it follows from Property 2 of the $gen$ function that there exists an independent set $b \subset \cup_{i=1}^n ax^{(l(i))}$ such that $gen[b]$ is well defined and
\be
gen[b] =gen[\cup_{i=1}^n ax^{(l(i))}] \cup (\cup_{i=1}^n ax^{(l(i))} \setminus b)
\ee

Thus it follows from the definition that if a theory, $S$, is $\beta$-complete, any set $u \subset uset$ such that $u \cap gua=\emptyset$ is a derivable set for which there exists an independent set of generators, $b$, such that
\be
u \subset gen[b]
\ee 

\underline{Complete primary states.} It trivially follows that a theory, $S$, that admits a primary state that is complete is $\beta$-complete.
To see this more clearly, let $S^{(k)}$ be a complete primary state of $S$ such that $uset = ax^{(k)} \cup th^{(k)}$.
By definition $th^{(k)}=gen[ax^{(k)}]$ so each element of $th^{(k)}$ is derivable in at least one primary state, namely $S^{(k)}$.

By definition, each element of $gua$ is not derivable in any primary state of the theory so is not derivable in $S^{(k)}$.
Since $ud^{(k)} = \emptyset$ we must have $gua \subset ax^{(k)}$.

An element of $ax^{(k)} \setminus gua$ is, by definition, not a globally underivable axiom and must therefore be derivable in at least one other primary state of the theory.

For the converse we make the following claim.

Claim: \emph{A theory that is $\beta$-complete admits a primary state that is complete.} 

While this claim might intuitively appear to be true, the construction of a complete primary state into a partition of an independent set of generators and its generated set is not trivial.
We will now construct an abstract application of PECR to generate a proof of this claim.

\section{Atomic programs.}

Our formal system PECR has been primarily designed to examine the computability of real world applications, especially applications posed as fully discrete dynamical systems.
However, it is possible to employ PECR to explore abstract theories by introducing objects that are subject to abstract type assignments.
It is important to note that we are building an abstract theory on top of a primitive language where inference is based on computability.
This forces us to find a way of associating the validity of our proofs based on abstractions with computability. 
It is in this way that, even under theories employing abstractions, we differ from contemporary proofs where reliance is placed on the semantics of truth value assignments.

As usual the I/O dummy variable names of programs are type $\mbf{char}$.
Here value assignments of I/O variables will be type $\mbf{term}$, where objects of type $\mbf{term}$ are alphanumeric character strings that also include the special characters $[~]$ (see Section \ref{sc}).

Recall that a term has the general representation
\be
\text{(term name)}[\text{list of variables}]
\ee
A term will appear in the form that we have been using for maps and functions.
For abstract applications, assignment programs simply construct terms that will represent functions or maps of sets to sets and predicates of sets.

Terms will be constructed from atomic programs whose output list contains a single element. 
It will be convenient to use the same name for a term as the name of the program that constructs that term as a character string, i.e. if $p~x~y$ is an atomic program that constructs a term we write $y:=p[x(1)~\ldots~x(nx)]$ to represent the value assignment of the output, $y$, as a character string of type $\mbf{term}$.
Note that we use spaces instead of commas to separate the list of arguments of the term.

From a machine perspective, terms are just character string assignments.
While a machine can recognize a term by its string structure the abstract properties that we wish to attach to a term can only take on some kind of meaning through an abstract type assignment combined with a collection of rules or axioms that reflect the properties of that term.

We shall work with sets.
Sets are collections of unique set elements that automatically remove repetitions under unions, e.g. $\{ a,b \} \cup \{ a,c \}= \{ a,b,c \}$.
This avoids the use of the $unique$ function that removes repetitions in lists.

We work with terms that can be assigned three main subtypes
\be
\bal
& \mbf{set} <: \mbf{term} \\
& \mbf{pred} <: \mbf{term} \\
& \mbf{flse} <: \mbf{term} \\
\eal
\ee
Here we will endow objects of type $\mbf{set}$ with the specific properties that we associate with elements that are irreducible extended programs.
Thus objects of type $\mbf{set}$ represent finite sets.

Objects of type $\mbf{pred}$ can be identified as predicates, similar to those of classical logic.
They will often be employed to express an abstract relation between two objects.
We can regard objects of type $\mbf{flse}$ as representing false sets and predicates.

The list of constants, $cst$, of the theory are
\be
\bal
& cst = [eset~uset~gua~sgua] \\
\eal
\ee
where the parameters $eset,uset,gua$ and $sgua$ represent fixed values of type $\mbf{set}$.
The constant $eset$ represents the empty set, the constant $uset$ represents the universal set of all irreducible extended programs of a theory and the constant $gua$ represents the set of globally underivable axioms.
The constant $sgua$ represents an object that is an independent set of generators with the smallest cardinality that contains $gua$ as outlined in the previous section.

In applications of PECR, atomic programs will halt prematurely and return an error message when a type violation is encountered.
Standard objects such as scalars and lists of the machine integers, fixed precision rational numbers, programs and terms have a well defined string structure as specified by their definitions and are recognized by the machine.
Recognizing an abstract object that is assigned a value of a subtype of a term can only be made through an abstract type assignment.
The following abstract type assignment and checking programs are given in the following tables.

\emph{Program type} $\mbf{tasgn}$.

\begin{tabular}{|l|l|l|}
	\hline
	\textbf{Syntax} & \textbf{Type checks} & \textbf{Type assignment} \\
	\hline \hline
	$aset~[a]~[~]$ & $a:\mbf{term}$ & $a::\mbf{set}$ \\
	\hline
	$apred~[a]~[~]$ & $a:\mbf{term}$ & $a::\mbf{pred}$ \\
	\hline
	$aflse~[a]~[~]$ & $a:\mbf{term}$ & $a::\mbf{flse}$ \\
	\hline
\end{tabular}

\emph{Program type} $\mbf{chck}$.

\begin{tabular}{|l|l|}
	\hline
	\textbf{Syntax} & \textbf{Type checks} \\
	\hline \hline
	$set~[a]~[~]$ & $a:\mbf{set}$ \\
	\hline
	$pred~[p]~[~]$ & $p:\mbf{pred}$ \\
	\hline
	$flse~[a]~[~]$ & $a:\mbf{flse}$ \\
	\hline
\end{tabular}

\newpage 

\underline{General value assignment programs.}

\emph{Program type} $\mbf{asgn}$.

\begin{tabular}{|l|l|l|l|l|}
	\hline
	\textbf{Syntax} & \textbf{Type} & \textbf{Value} & \textbf{Type} \\
	& \textbf{checks} & \textbf{assignment} & \textbf{assignment} \\
	\hline \hline
	$eqset~[a~b]~[p]$  & $a,b:\mbf{set}$ & $p:=eqset[a~b]$ & $p::\mbf{pred}$ \\
	\hline
	$eqpred~[a~b]~[p]$  & $a,b:\mbf{pred}$ & $p:=eqpred[a~b]$ & $p::\mbf{pred}$ \\
	\hline
	$neg~[p]~[q]$  & $p:\mbf{pred}$ & $q:=neg[p]$ & $q::\mbf{pred}$ \\
	\hline
	$subset~[a~b]~[p]$  & $a,b:\mbf{set}$ & $p:=subset[a~b]$ & $p::\mbf{pred}$ \\
	\hline
	$union~[a~b]~[c]$ & $a,b:\mbf{set}$ & $c:=union[a~b]$ & $c::\mbf{set}$ \\
	\hline
	$sint~[a~b]~[c]$ & $a,b:\mbf{set}$ & $c:=sint[a~b]$ & $c::\mbf{set}$ \\
	\hline
	$setm~[a~b]~[c]$ & $a,b:\mbf{set}$ & $c:=setm[a~b]$ & $c::\mbf{set}$ \\
	\hline
\end{tabular}

\underline{Specific set assignment programs.}

\emph{Program type} $\mbf{asgn}$.

\begin{tabular}{|l|l|l|l|l|}
	\hline
	\textbf{Syntax} & \textbf{Type} & \textbf{Value} & \textbf{Type} \\
	& \textbf{checks} & \textbf{assignment} & \textbf{assignment} \\
	\hline \hline
	$gen~[a]~[b]$ & $a:\mbf{set}$ & $b:=gen[a]$ & $b::\mbf{set}$ \\
	\hline
	$ipart~[a]~[b]$  & $a:\mbf{set}$ & $b:=ipart[a]$ & $b::\mbf{set}$ \\
	\hline
	$isgs~[a]~[b]$ & $a:\mbf{set}$ & $b:=isgs[a]$ & $b::\mbf{set}$ \\
	\hline
\end{tabular}

The programs $ipart~[a]~[b]$ and $isgs~[a]~[b]$ will be discussed in more detail later.

\underline{Equality/equivalence.} It is important to note that the equality/equivalence programs $eqset~[a~b]~[p]$ and $eqpred~[a~b]~[p]$ differ from the standard equality/equivalence programs that we have been using in that they construct predicates in the form of a term of type $\mbf{pred}$ that states an abstract equality/equivalence.
They are type $\mbf{asgn}$ objects as opposed to the standard equality/equivalence programs that are type $\mbf{chck}$ objects.

The application of the standard substitution rules, \textsf{sr1}-\textsf{sr2}, can only rely on equality/equivalence of value assignments that are recognized as identical or equivalent through their string structure.
Predicates of equality/equivalence can also be employed to define a substitution rule but only in an abstract sense.
This is discussed in more detail later.

\underline{Classical notation.}
The terms constructed by the above atomic value assignment programs can be recognized by the more familiar classical notations given in the tables below. 

\emph{Predicates.}

\begin{tabular}{|l|l|l|}
	\hline
	\textbf{Argument} & \textbf{Term} & \textbf{Classical} \\
	\textbf{types} &  & \textbf{notation} \\
	\hline \hline
	$a,b:\mbf{set}$ & $eqset[a~b]$ & $a=b$ \\
	\hline
	$a,b:\mbf{pred}$ & $eqpred[a~b]$ & $a \equiv b$ \\
	\hline
	$p:\mbf{pred}$ & $neg[p]$ & $\neg p$ \\
	\hline
	$a,b:\mbf{set}$ & $subset[a~b]$ & $a \subset b$ \\
	\hline
\end{tabular}

\emph{Sets.}

\begin{tabular}{|l|l|l|}
	\hline
	\textbf{Argument} & \textbf{Term} & \textbf{Classical} \\
	\textbf{types} &  & \textbf{notation} \\
	\hline \hline
	$a,b:\mbf{set}$ & $union[a~b]$ & $a \cup b$ \\
	\hline
	$a,b:\mbf{set}$ & $sint[a~b]$ & $a \cap b$ \\
	\hline
	$a,b:\mbf{set}$ & $setm[a~b]$ & $a \setminus b$ \\
	\hline
\end{tabular}

\section{Predicate equivalence.}\label{predequiv}

Predicate equivalence can be regarded as having some similarity with the classical \emph{if and only if} connective.
In the present context we are constructing predicates that have a simple structure.
The association of predicates with the \emph{if and only if} connective becomes more evident in abstract theories that include atomic programs that construct more complicated predicates.

In the next section we will present some axioms of subsets and set unions, intersections and subtractions.
These are by no means meant to represent the most fundamental axioms for basic set theory and should be regarded as a selection of rules that will be needed in our proofs.
These rules can be derived from a more fundamental theory of sets that includes a character string value assignment of type $\mbf{elt} <: \mbf{term}$, where the assigned type $x:\mbf{elt}$ associates the variable $x$ as an element of a set.
The construction of a predicate of the form $in[x~a]$, where $x:\mbf{elt}$ and $a:\mbf{set}$, can be associated with the classical statement $x \in a$.

Given two terms $p$ and $q$, one could include atomic programs that construct the predicates $and[p~q]$, $or[p~q]$, $imply[p~q]$ and $iff[p~q]$, respectively, that can be associated with the classical connectives of $\land,~\lor,~\to$ and $\leftrightarrow$, respectively.
Theories employing these connectives will be accompanied with the axioms of classical logic.

Further extensions could also include quantifiers of first-order logic where terms of the form $forall[x~p]$ and $exists[x~p]$ can be associated with the classical statements $\forall x~p$ and $\exists x~p$.
Here it is understood that the term $p$ has already been constructed with a dependence on $x$.
In PECR, statements of first-order logic can be expressed by way of programs that construct terms associated with quantifiers. 
For example, the following irreducible extended programs express the identities $\neg \forall x~p \leftrightarrow \exists x~\neg p$ and $\neg \exists x~p \leftrightarrow \forall x~\neg p$ of first order logic.
\be
\begin{array}{l}
	forall~[x~p]~[p1] \\
	neg~[p1]~[q1] \\
	neg~[p]~[p2] \\
	exists~[x~p2]~[q2] \\
	\hline
	eqpred~[q2~q1]~[r] \\
\end{array}
\quad \hspace{10mm}
\begin{array}{l}
	exists~[x~p]~[p1] \\
	neg~[p1]~[q1] \\
	neg~[p]~[p2] \\
	forall~[x~p2]~[q2] \\
	\hline
	eqpred~[q2~q1]~[r] \\
\end{array}
\ee
For theories involving quantifiers, additional axioms of abstract falsity should be included that identify the misuse of bound and free variables.

Throughout this book we have adopted a style of inference that is quantifier free.
This is maintained in the following sections of this chapter where we can still get by without quantifiers.
While this style has some similarities with logic programming the reader should note the special features that are necessarily introduced to adapt PECR to abstract applications where computability still replaces the semantics of truth value assignment.

\section{Axioms.} 

As already mentioned, the axioms presented below, particularly for subsets and set unions, intersections and subtractions, are not meant to represent the most fundamental axioms for basic set theory.
They mostly represent a selection of rules that will be adequate for the proofs that are constructed in the following section.

\underline{Abstract type assignments.} An object of type $\mbf{term}$ is immediately recognized by the machine from its string structure.
However, abstract objects that are subtypes of terms must be assigned an abstract subtype.
\be
\begin{array}{l}
	aset~[a]~[~] \\
	\hline
	set~[a]~[~] \\
\end{array}
\quad \textsf{aset} \hspace{5mm}
\begin{array}{l}
	apred~[a]~[~] \\
	\hline
	pred~[a]~[~] \\
\end{array}
\quad \textsf{apred} \hspace{5mm}
\begin{array}{l}
	aflse~[a]~[~] \\
	\hline
	flse~[a]~[~] \\
\end{array}
\quad \textsf{aflse}
\ee
We will have no need to apply these axioms directly in our proofs of the next section.
As always we can regard each derivation program to be at the core of a larger program that can be represented by the vertical list (\ref{core}).
The above axioms can be assumed to be activated by the program $read~[file1]~[piv[x]]$ where the primary input variables, $piv[x]$, of the derivation program are assigned their values along with their abstract subtype.
Assignment subprograms of the derivation program will assign the appropriate abstract type to the new variables that they introduce. 

\underline{Abstract axioms of falsity.}
There is an important distinction to be made between the use of the programs $false~[p]~[~]$ and $flse~[a]~[~]$.
The program $false~[p]~[~]$ is a higher order statement that indicates that a program, $p$, will halt with an error message for any value assigned input list.
Application of the program $false~[p]~[~]$ in our proofs requires a preliminary step involving the program $flse~[a]~[~]$.  

The program $aflse~[a]~[~]$ assigns the subtype $a::\mbf{flse} <: \mbf{term}$.
It overwrites an existing assigned subtype $a:\mbf{set}$ or $a:\mbf{pred}$.  
We can regard the statement $a:\mbf{flse}$ to mean that $a$ is a term that has been assigned a subtype that is a false set or a false predicate.

Suppose that $p~x~y$ is an atomic program such that $a \in [x~y]$ is a term that has been assigned the value subtype $a:\mbf{X}$, where $\mbf{X}$ represents the types $\mbf{set}$ or $\mbf{pred}$.
We have by the \textsf{aio} axiom
\be
\begin{array}{l}
	[p~x~y]_{a \in [x~y]} \\
	\hline
	X~[a]~[~] \\
\end{array}
\ee
where $X$ is a character string $set$ or $pred$.
Having stated the original assigned type of the variable $a$ we may apply the axiom of falsity
\be
\begin{array}{l}
	X~[a]~[~] \\
	flse~[a]~[~] \\
	\hline
	false \\
\end{array}
\quad \textsf{f0}
\ee
if we have also obtained the statement $flse~[a]~[~]$ from an abstract axiom of falsity in conjunction with axiom \textsf{aflse}.  

While there will be no need to employ axioms of falsity in the proofs presented in the next section we include the following abstract axioms of falsity as examples that are relevant to the theory being considered here.

\be
\begin{array}{l}
	eqset~[a~a]~[p] \\
	neg~[p]~[q] \\
	\hline
	aflse~[q]~[~] \\
\end{array}
\quad \textsf{f1}
\ee
The program $aflse~[q]~[~]$ indicates that the variable $q:\mbf{term}$ has been assigned the subtype $\mbf{flse}$.
It overwrites the original assigned subtype $q:\mbf{pred}$. 

For theories that are $\beta$-complete we have the following axiom that states that a set, $u$, is not derivable if $u \cap gua \ne \emptyset$.
\be
\begin{array}{l}
	sint~[u~gua]~[a] \\
	eqset~[a~eset]~[p] \\
	neg~[p]~[q] \\
	isgs~[u]~[b] \\
	\hline
	aflse~[b]~[~] \\
\end{array}
\quad \textsf{f2}
\ee
The program $isgs~[u]~[b]$ returns the value assignment of the term $b:=isgs[u]$ that represents an independent set of generators of $u$.
This will be defined in more detail below. 
The program $aflse~[b]~[~]$ indicates that the variable $b:\mbf{term}$ has been the assigned a value of subtype $\mbf{flse}$.

The empty set will be regarded as an underivable set so that
\be
\begin{array}{l}
	isgs~[eset]~[b] \\
	\hline
	aflse~[b]~[~] \\
\end{array}
\quad \textsf{f3}
\ee

A major shortcoming in any abstract application of PECR is that it is often difficult to provide an exhaustive collection of abstract axioms of falsity that detect all invalid statements of the theory.
In classical mathematics the situation is similar because theorems establishing the falsity of a premise each require an individual proof.
There is an important distinction to be made between classical proofs of falsity and those of PECR in that classical proofs of falsity rely on the law of the excluded middle.
The law of the excluded is not reflected in the construction rules of PECR although it can appear in abstract applications of PECR that employ the axioms of classical logic.
We will address this in a more general context in the next chapter.

\underline{Set equality.}

Symmetry and reflexivity. 
\be
\begin{array}{l}
	set~[a]~[~] \\
	\hline
	eqset~[a~a]~[p] \\
\end{array}
\quad \textsf{eqset1a} \hspace{10mm}
\begin{array}{l}
	eqset~[a~b]~[p] \\
	\hline
	eqset~[b~a]~[q] \\
\end{array}
\quad \textsf{eqset1b}
\ee

Transitivity of set equality follows from the abstract substitution rule (see below).

\underline{Equivalence of predicates.}

Symmetry and reflexivity.
\be
\begin{array}{l}
	pred~[a]~[~] \\
	\hline
	eqpred~[a~a]~[p] \\
\end{array}
\quad \textsf{eqpred1a} \hspace{5mm}
\begin{array}{l}
	eqpred~[a~b]~[p] \\
	\hline
	eqpred~[b~a]~[q] \\
\end{array}
\quad \textsf{eqpred1b} \hspace{10mm}
\ee
Transitivity of equivalence of predicates follows from the abstract substitution rule (see below).

As discussed in the previous section, the use of predicate equivalence becomes more important in theories that construct predicates with more complicated structures.
Most of the predicates of our theory are of a simple structure so there will be no need to employ predicate equivalence in the proofs of the next section.

\underline{The abstract substitution rule.}
In abstract applications, the substitution rules \textsf{sr1}-\textsf{sr2} are of little use since value assigned equality/equivalence can only be made if the machine can immediately identify value assignments by their string structures.
Equality/equivalence of two abstract objects that are attached to different character strings can only be made in an abstract sense.
Implementation of the substitution rule has to be incorporated into the abstraction. 

Let
\be
\bal
xi \in x,~xi:\mbf{X} \\
x^\prime=x(xi \to a),~a:\mbf{X} \\
y,y^\prime:\mbf{Y}
\eal
\ee
where $\mbf{X}$ and $\mbf{Y}$ represent the types $\mbf{set}$ or $\mbf{pred}$. 
The abstract substitution rules are
\be
\begin{array}{l}
	p~x~y  \\
	eqX~[xi~a]~[b] \\
	\hline
	p~x^\prime~y^\prime \\
\end{array}
\quad \textsf{sr1} \hspace{10mm}
\begin{array}{l}
	p~x~y \\
	eqX~[xi~a]~[b] \\
	p~x^\prime~y^\prime \\
	\hline
	eqY~[y^\prime~y]~[c] \\
\end{array}
\quad \textsf{sr2}
\ee
where $X$ and $Y$ represent the character strings $set$ or $pred$.
Note that abstract equality/equivalence programs differ from those of the standard equality/equivalence programs in that they have a nonempty output list.

\emph{We use the same labels \textsf{sr1}-\textsf{sr2} as for the standard substitution rules but it is important to keep in mind that any reference to these labels in the proofs of the next section are to be understood to refer to the abstract substitution rules stated above.}

\underline{Subsets.} The following axiom states that any object of type $\mbf{set}$ must be a subset of the universal set $uset$.
\be
\begin{array}{l}
	set~[a]~[~] \\
	\hline
	subset~[a~uset]~[p] \\
\end{array}
\quad \textsf{sub1}
\ee

We also have
\be
\begin{array}{l}
	eqset~[a~b]~[p] \\
	\hline
	subset~[b~a]~[q] \\
\end{array}
\quad \textsf{sub2a} \hspace{10mm}
\begin{array}{l}
	subset~[a~b]~[p] \\
	subset~[b~a]~[q] \\
	\hline
	eqset~[b~a]~[r] \\
\end{array}
\quad \textsf{sub2b}
\ee
\be
\begin{array}{l}
	subset~[a~b]~[p] \\
	subset~[b~c]~[q] \\
	\hline
	subset~[a~c]~[r] \\
\end{array}
\quad \textsf{sub2c}
\ee

Subsets also satisfy the property of symmetry
\be
\begin{array}{l}
	set~[a]~[~] \\
	\hline
	subset~[a~a]~[p] \\
\end{array}
\ee
This is not included as an axiom because it follows from \textsf{eqset1a} and \textsf{sub2a}.

\underline{Union of sets.}

\be
\begin{array}{l}
	set~[a]~[~] \\
	set~[b]~[~] \\
	\hline
	union~[a~b]~[c] \\
\end{array}
\quad \textsf{union1a} \hspace{10mm}
\begin{array}{l}
	union~[a~b]~[c] \\
	\hline
	union~[b~a]~[d] \\
\end{array}
\quad \textsf{union1b}
\ee
\be
\begin{array}{l}
	union~[a~b]~[c] \\
	union~[b~a]~[d] \\
	\hline
	eqset~[d~c]~[p] \\
\end{array}
\quad \textsf{union1c}
\ee

Axiom \textsf{union2b} states that the union of sets is associative.
\be
\begin{array}{l}
	union~[a~b]~[d] \\
	subset~[b~a]~[p] \\
	\hline
	eqset~[d~a]~[q] \\
\end{array}
\quad \textsf{union2a} \hspace{10mm}
\begin{array}{l}
	union~[a~b]~[d] \\
	union~[d~c]~[e] \\
	union~[b~c]~[f] \\
	union~[a~f]~[g] \\
	\hline
	eqset~[g~e]~[p] \\
\end{array}
\quad \textsf{union2b}
\ee

\be
\begin{array}{l}
	union~[a~b]~[c] \\
	\hline
	subset~[b~c]~[p] \\
\end{array}
\quad \textsf{union3a} \hspace{10mm}
\begin{array}{l}
	union~[a~b]~[d] \\
	union~[a~c]~[e] \\
	subset~[c~b]~[p] \\
	\hline
	subset~[e~d]~[q] \\
\end{array}
\quad \textsf{union3b}
\ee

\underline{Set intersection.}

\be
\begin{array}{l}
	set~[a]~[~] \\
	set~[b]~[~] \\
	\hline
	sint~[a~b]~[c] \\
\end{array}
\quad \textsf{sint1a} \hspace{10mm}
\begin{array}{l}
	sint~[a~b]~[c] \\
	eqset~[c~eset]~[p] \\
	subset~[d~b]~[q] \\
	sint~[a~d]~[e] \\
	\hline
	eqset~[e~eset]~[r] \\
\end{array}
\quad \textsf{sint1b}
\ee

\underline{Setminus.}

\be
\begin{array}{l}
	set~[a]~[~] \\
	set~[b]~[~] \\
	\hline
	setm~[a~b]~[c] \\
\end{array}
\quad \textsf{setm1a} \hspace{10mm}
\begin{array}{l}
	setm~[a~b]~[c] \\
	union~[b~c]~[d] \\
	\hline
	eqset~[d~a]~[p] \\
\end{array}
\quad \textsf{setm1b}
\ee
\be
\begin{array}{l}
	setm~[a~b]~[c] \\
	sint~[c~b]~[d] \\
	\hline
	eqset~[d~eset]~[p] \\
\end{array}
\quad \textsf{setm1c}
\ee

\underline{The $gen$ program.}
Axiom \textsf{gen1a} reflects Property 5 of the $gen$ function that states derivations exclude proofs based on an identity.
Axiom \textsf{gen1b} states that if $gen[a]$ and $gen[b]$ are well defined then $gen[a \cup b]$ is also well defined.
Axiom \textsf{gen1c} states that $gen[a] \cup gen[b] \subset gen[a \cup b]$.
\be
\begin{array}{l}
	gen~[a]~[c] \\
	sint~[a~c]~[d] \\
	\hline
	eqset~[d~eset]~[p] \\
\end{array}
\quad \textsf{gen1a} \hspace{10mm}
\begin{array}{l}
	gen~[a]~[c] \\
	gen~[b]~[d] \\
	union~[a~b]~[e] \\
	\hline
	gen~[e]~[f] \\
\end{array}
\quad \textsf{gen1b}
\ee
\be
\begin{array}{l}
	gen~[a]~[c] \\
	gen~[b]~[d] \\
	union~[a~b]~[e] \\
	gen~[e]~[f] \\
	union~[c~d]~[g] \\
	\hline
	subset~[g~f]~[p] \\
\end{array}
\quad \textsf{gen1c}
\ee

For any set of generators, $a$, there exists a set $b \subset a$ that is an independent set of generators with the properties outlined in Property 2 of the $gen$ function.  
The atomic program $ipart~[a]~[b]$ constructs the term $b:=ipart[a]$ that represents an independent subset of generators of the set of generators, $a$.
The properties of $ipart$ are reflected in the following axioms.
\be
\begin{array}{l}
	ipart~[a]~[b] \\
	\hline
	gen~[a]~[c] \\
\end{array}
\quad \textsf{ipart1a} \hspace{10mm}
\begin{array}{l}
	ipart~[a]~[b] \\
	\hline
	gen~[b]~[c] \\
\end{array}
\quad \textsf{ipart1b}
\ee
\be
\begin{array}{l}
	gen~[a]~[c] \\
	\hline
	ipart~[a]~[b] \\
\end{array}
\quad \textsf{ipart1c}
\ee

The following axiom states that if $a$ is an independent subset of the set of generators $b$ then $gen[a]=gen[b] \cup (b \setminus a)$.
\be
\begin{array}{l}
	gen~[b]~[d] \\
	ipart~[b]~[a] \\
	gen~[a]~[g] \\
	setm~[b~a]~[c] \\
	union~[d~c]~[e] \\
	\hline
	eqset~[e~g]~[p] \\
\end{array}
\quad \textsf{ipart2}
\ee

\underline{$\beta$-completeness.}
From the definition of $\beta$-completeness any set, $u$, such that $u \cap gua=\emptyset$, is a derivable set.
In the previous section it was shown that for any derivable set, $u$, there exists an independent set of generators, $b$, such that $u \subset gen[b]$. 
The program $isgs~[u]~[b]$ returns an object in the form of a term, $b:=isgs[u]$, that represents an independent set of generators such that $u \subset gen[b]$.
The properties of $b:=isgs[u]$ are reflected by the following axioms.
\be
\begin{array}{l}
	sint~[u~gua]~[c] \\
	eqset~[c~eset]~[p] \\	 
	\hline
	isgs~[u]~[b] \\
\end{array}
\quad \textsf{isgs1a} \hspace{10mm}
\begin{array}{l}
	isgs~[u]~[b] \\
	\hline
	gen~[b]~[e] \\
\end{array}
\quad \textsf{isgs1b}
\ee
\be
\begin{array}{l}
	isgs~[u]~[b] \\
	gen~[b]~[e] \\
	\hline
	subset~[u~e]~[p] \\
\end{array}
\quad \textsf{isgs1c}
\ee
Note that while $u \cap gua = \emptyset$, the set $b:=isgs[u]$ may contain elements of $gua$.
We do not rule out the possibility that a primary state has an empty set of generators in which case we may have $b= eset$ for any derivable set $u$. 

\underline{Globally underivable axioms.}
In the previous section we defined the set of independent generators, $sgua$, that contains the set of globally underivable axioms, $gua$, as a subset. 
The sets $gua$ and $sgua$ are regarded as constants of the theory.
The properties of the set $sgua$ are reflected in the following axioms.
They have an empty premise list.
\be
\begin{array}{l}
	~ \\
	\hline
	gen~[sgua]~[a] \\
\end{array}
\quad \textsf{sgua1a} \hspace{10mm}
\begin{array}{l}
	~ \\
	\hline
	subset~[gua~sgua]~[p] \\  
\end{array}
\quad \textsf{sgua1b}
\ee

The following axiom states that $sgua$ is an independent set of generators.
\be
\begin{array}{l}
	ipart~[sgua]~[a] \\
	\hline
	eqset~[a~sgua]~[p] \\
\end{array}
\quad \textsf{sgua2}   
\ee

\section{Complete primary states.}

Throughout we assume that we are working with a theory, $S$, that is $\beta$-complete so that any set, $u \subset uset$, such that $u \cap gua = \emptyset$ is a derivable set.
For any derivable set, $u$, there exists an independent set of generators, $b$, such that $u \subset gen[b]$.
The program $isgs~[u]~[b]$ constructs the value assignment $b:=isgs[u]$, where $isgs[u]:\mbf{set} <: \mbf{term}$.
The term $b:=isgs[u]$ represents an independent set of generators such that $u \subset gen[b]$.

We start by deriving a simple result that will be needed later.
The following lemma shows that given an independent set of generators, $b$, associated with some derivable set, $u$, the set $gen[sgua \cup b]$ is well defined.
\begin{lstlisting}
Lemma lem1.

isgs [u] [b]
union [sgua b] [c]
------------------
gen [c] [e]

Proof.
  1 isgs [u] [b]
  2 union [sgua b] [c]
  3 gen [sgua] [a]           sgua1a
  4 gen [b] [d]              isgs1b [1]
  5 gen [c] [e]              gen1b [3 4 2]
\end{lstlisting}

We need to check that our axioms will identify $uset \setminus sgua$ as a derivable set and hence is contained in the generated set of an independent set of generators.
\begin{lstlisting}
Lemma lem2.

setm [uset sgua] [u]
--------------------
isgs [u] [f]

Proof.
  1 setm [uset sgua] [u]
  2 set [u] [ ]              aio [1]
  3 set [sgua] [ ]           aio [1]
  4 sint [u sgua] [a]        sint1a [2 3]
  5 eqset [a eset] [b]       setm1c [1 4]
  6 subset [gua sgua] [c]    sgua1b
  7 set [gua] [ ]            aio [6]
  8 sint [u gua] [d]         sint1a [2 7]
  9 eqset [d eset] [e]       sint1b [4 5 8 6]
 10 isgs [u] [f]             isgs1a [8 9]
\end{lstlisting}

The following two lemmas will shorten later proofs.
Lemma \textsf{lem3} shows that if $a$ is an independent subset of the set of generators $b$ then $b \setminus a \subset gen[a]$.
This is followed by lemma \textsf{lem4} that shows $gen[b] \subset gen[a]$.
\begin{lstlisting}
Lemma lem3.

ipart [b] [a]
gen [a] [g]
setm [b a] [c]
----------------
subset [c g] [i]

Proof.
  1 ipart [b] [a]
  2 gen [a] [g]
  3 setm [b a] [c]
  4 gen [b] [d]              ipart1a [1]
  5 set [d] [ ]              aio [4]
  6 set [c] [ ]              aio [3]
  7 union [d c] [e]          union1a [5 6]
  8 subset [c e] [f]         union3a [7]
  9 eqset [e g] [h]          ipart2 [4 1 2 3 7]
 10 subset [c g] [i]         sr1 [8 9]

Lemma lem4.

gen [b] [d]
ipart [b] [a]
gen [a] [g]
----------------
subset [d g] [l]

Proof.
  1 gen [b] [d]
  2 ipart [b] [a]
  3 gen [a] [g]
  4 set [b] [ ]              aio [1]
  5 set [a] [ ]              aio [3]
  6 setm [b a] [c]           setm1a [4 5]
  7 set [d] [ ]              aio [1]
  8 set [c] [ ]              aio [6]
  9 union [d c] [e]          union1a [7 8]
 10 eqset [e g] [f]          ipart2 [1 2 3 6 9]
 11 union [c d] [h]          union1a [8 7]
 12 subset [d h] [i]         union3a [11]
 13 eqset [h e] [j]          union1c [9 11]
 14 eqset [h g] [k]          sr1 [13 10]
 15 subset [d g] [l]         sr1 [12 14]
\end{lstlisting}

Consider the partition $uset=sgua \cup u$, where $u =uset \setminus sgua$.
As demonstrated in the proof of lemma \textsf{lem2}, since $gua \subset sgua$ we have $u \cap gua = \emptyset$ so $u$ is a derivable set.
It is important to note that while $sgua$ is an independent set of generators, each element of the set $sgua \setminus gua$ is derivable in at least one primary state of the theory, $S$.

We shall start by examining some properties of the set $s=sgua \cup u \cup b$, where $b=isgs[u]$.
Since $b \subset uset$ we can later make use of the equality $s=uset$.

Theorem \textsf{thm1} shows that given $b$ is an independent set of generators associated with the derivable set $u$ then $sgua \cup u \cup b \subset sgua \cup b \cup gen[sgua \cup b]$.
As shown in the proof of lemma \textsf{lem1}, we are guaranteed that the set $gen[sgua \cup b]$ is well defined because both $gen[sgua]$ and $gen[b]$ are well defined.
\begin{lstlisting}
Theorem thm1.

union [sgua u] [z]
isgs [u] [b]
union [z b] [s]
union [sgua b] [c]
gen [c] [d]
union [c d] [e]
------------------
subset [s e] [c1]

Proof.
  1 union [sgua u] [z]
  2 isgs [u] [b]
  3 union [z b] [s]
  4 union [sgua b] [c]
  5 gen [c] [d]
  6 union [c d] [e]
  7 set [sgua] [ ]           aio [1]
  8 set [u] [ ]              aio [1]
  9 set [b] [ ]              aio [3]
 10 union [u b] [a]          union1a [8 9]
 11 union [b u] [f]          union1a [9 8]
 12 set [a] [ ]              aio [10]
 13 union [sgua a] [g]       union1a [7 12]
 14 eqset [a f] [h]          union1c [11 10]
 15 union [sgua f] [i]       sr1 [13 14]
 16 set [c] [ ]              aio [5]
 17 union [c u] [j]          union1a [16 8]
 18 eqset [i j] [k]          union2b [4 17 11 15]
 19 eqset [j i] [l]          eqset1b [18]
 20 eqset [i g] [m]          sr2 [13 14 15]
 21 eqset [j g] [n]          sr1 [19 20]
 22 eqset [g s] [o]          union2b [1 3 10 13]
 23 eqset [j s] [p]          sr1 [21 22]
 24 gen [sgua] [q]           sgua1a
 25 gen [b] [r]              isgs1b [2]
 26 set [q] [ ]              aio [24]
 27 set [r] [ ]              aio [25]
 28 union [q r] [t]          union1a [26 27]
 29 subset [r t] [v]         union3a [28]
 30 subset [u r] [w]         isgs1c [2 25]
 31 subset [u t] [x]         sub2c [30 29]
 32 subset [t d] [y]         gen1c [24 25 4 5 28]
 33 subset [u d] [a1]        sub2c [31 32]
 34 subset [j e] [b1]        union3b [6 17 33]
 35 subset [s e] [c1]        sr1 [34 23]
\end{lstlisting}

Theorem \textsf{thm2} shows that if $b$ is an independent set of generators associated with the derivable set $u$ and $a$ is an independent subset of the set of generators, $sgua \cup b$, then $sgua \cup u \cup b \subset sgua \cup b \cup gen[a]$.  
\begin{lstlisting}
Theorem thm2.

union [sgua u] [z]
isgs [u] [b]
union [z b] [s]
union [sgua b] [c]
gen [c] [d]
union [c d] [e]
ipart [c] [a]
gen [a] [g]
union [c g] [y]
------------------
subset [s y] [j]

Proof.
  1 union [sgua u] [z]
  2 isgs [u] [b]
  3 union [z b] [s]
  4 union [sgua b] [c]
  5 gen [c] [d]
  6 union [c d] [e]
  7 ipart [c] [a]
  8 gen [a] [g]
  9 union [c g] [y]
 10 subset [s e] [f]         thm1 [1 2 3 4 5 6]
 11 subset [d g] [h]         lem4 [5 7 8]
 12 subset [e y] [i]         union3b [9 6 11]
 13 subset [s y] [j]         sub2c [10 12]
\end{lstlisting}

Theorem \textsf{thm3} shows that if $a$ is an independent subset of a set of generators, $c$, then $c \cup gen[a] = a \cup gen[a]$.
\begin{lstlisting}
Theorem thm3.

ipart [c] [a]
gen [a] [g]
union [c g] [z]
union [a g] [w]
---------------
eqset [z w] [y]

Proof.
  1 ipart [c] [a]
  2 gen [a] [g]
  3 union [c g] [z]
  4 union [a g] [w]
  5 set [c] [ ]              aio [1]
  6 set [a] [ ]              aio [2]
  7 setm [c a] [b]           setm1a [5 6]
  8 set [b] [ ]              aio [7]
  9 union [a b] [d]          union1a [6 8]
 10 eqset [d c] [e]          setm1b [7 9]
 11 eqset [c d] [f]          eqset1b [10]
 12 union [d g] [h]          sr1 [3 11]
 13 eqset [h z] [i]          sr2 [3 11 12]
 14 set [g] [ ]              aio [3]
 15 union [b g] [j]          union1a [8 14]
 16 set [j] [ ]              aio [15]
 17 union [a j] [k]          union1a [6 16]
 18 eqset [k h] [l]          union2b [9 12 15 17]
 19 eqset [k z] [m]          sr1 [18 13]
 20 union [g b] [n]          union1a [14 8]
 21 eqset [n j] [o]          union1c [15 20]
 22 eqset [j n] [p]          eqset1b [21]
 23 union [a n] [q]          sr1 [17 22]
 24 eqset [q k] [r]          sr2 [17 22 23]
 25 eqset [q z] [s]          sr1 [24 19]
 26 subset [b g] [t]         lem3 [1 2 7]
 27 eqset [n g] [u]          union2a [20 26]
 28 eqset [w q] [v]          sr2 [23 27 4]
 29 eqset [w z] [x]          sr1 [28 25]
 30 eqset [z w] [y]          eqset1b [29]
\end{lstlisting}

Theorem \textsf{thm4} shows that $sgua \cup u \cup b \subset a \cup gen[a]$, where $b$ is an independent set of generators associated with a derivable set $u$.
\begin{lstlisting}
Theorem thm4.

union [sgua u] [z]
isgs [u] [b]
union [z b] [s]
union [sgua b] [c]
ipart [c] [a]
gen [a] [g]
union [a g] [w]
------------------
subset [s w] [j]

Proof.
  1 union [sgua u] [z]
  2 isgs [u] [b]
  3 union [z b] [s]
  4 union [sgua b] [c]
  5 ipart [c] [a]
  6 gen [a] [g]
  7 union [a g] [w]
  8 gen [c] [d]              ipart1a [5]
  9 set [c] [ ]              aio [5]
 10 set [d] [ ]              aio [8]
 11 union [c d] [e]          union1a [9 10]
 12 set [g] [ ]              aio [7]
 13 union [c g] [f]          union1a [9 12]
 14 subset [s f] [h]         thm2 [1 2 3 4 8 11 5 6 13]
 15 eqset [f w] [i]          thm3 [5 6 13 7]
 16 subset [s w] [j]         sr1 [14 15]
\end{lstlisting}

In lemma \textsf{lem2} it was shown that $uset \setminus sgua$ is a derivable set.
In the final step we construct the partition $uset=a \cup gen[a]$, where $a$ is an independent subset of the set of generators $sgua \cup b$, and $b$ is an independent set of generators associated with the derivable set $u=uset \setminus sgua$.
The construction $uset=a \cup gen[a]$ defines a primary state that is complete.
\begin{lstlisting}
Theorem thm5.

setm [uset sgua] [u]
isgs [u] [b]
union [sgua b] [c]
ipart [c] [a]
gen [a] [g]
union [a g] [w]
--------------------
eqset [uset w] [p]

Proof.
  1 setm [uset sgua] [u]
  2 isgs [u] [b]
  3 union [sgua b] [c]
  4 ipart [c] [a]
  5 gen [a] [g]
  6 union [a g] [w]
  7 set [w] [ ]              aio [6]
  8 subset [w uset] [d]      sub1 [7]
  9 set [sgua] [ ]           aio [1]
 10 set [u] [ ]              aio [2]
 11 union [sgua u] [e]       union1a [9 10]
 12 eqset [e uset] [f]       setm1b [1 11]
 13 set [e] [ ]              aio [12]
 14 set [b] [ ]              aio [3]
 15 union [e b] [h]          union1a [13 14]
 16 subset [b uset] [i]      sub1 [14]
 17 eqset [uset e] [j]       eqset1b [12]
 18 subset [b e] [k]         sr1 [16 17]
 19 eqset [h e] [l]          union2a [15 18]
 20 eqset [h uset] [m]       sr1 [19 12]
 21 subset [h w] [n]         thm4 [11 2 15 3 4 5 6]
 22 subset [uset w] [o]      sr1 [21 20]
 23 eqset [uset w] [p]       sub2b [8 22]
\end{lstlisting}

Lines 7-8 state the trivial result that $a \cup gen[a] \subset uset$ (the union of sets is a set).
Lines 9-22 establish that $uset \subset a \cup gen[a]$ from which it follows that $uset=a \cup gen[a]$. 


In the previous section we outlined how it trivially follows that a theory that has a primary state that is complete is $\beta$-complete.
A proof of the converse is given by theorem \textsf{thm5}.
Hence, in the classical sense, we have shown that a theory admits a primary state that is complete if and only if it is $\beta$-complete. 

While the above proofs might have greater appeal to those inclined towards contemporary mathematics they have limitations in real world computations.
A major drawback of our construction of a complete primary state is that it tells us nothing about the cardinality of its set of axioms.
An unmanageably large set of axioms is of little use in practice.

Rather than pursue this line of analysis any further we shall return our attention to real world computations to explore the primary states of a theory.

\chapter{Formal Systems in Science.}\label{cfsis}

\section{Introduction.}

In the early part of the 20th century, mathematicians set about to finally settle the issues surrounding the foundations of mathematics.
The aim was to remove the discourse out of the hands of the philosophers and by a process of self referencing carry out a formal study of the foundations of mathematics using the tools of mathematics itself.
While early efforts can be traced back to the work of Frege and others, the project was largely initiated by a series of lectures given by Hilbert, culminating in the work of Godel's incompleteness theorem.

Scientist, on the other hand, continue to carry out research into their special subject area with an acceptance that there is a well defined scientific method that they have an intuitive grasp of.
It is uncommon to find a formal course on the scientific method offered to undergraduates in any branch of the sciences.
Students are expected to acquire the rules of conduct when carrying out scientific research through general guidelines offered in the coursework of the various science disciplines that they have elected as part of their major.
It is ironic, then, that scientist have left the in-depth investigations of the scientific method to be carried out exclusively by philosophers.

In recent decades controversies over what actually constitutes scientific research have arisen with the ever increasing activity in peripheral areas such as the social sciences and related life sciences.  
It therefore seems timely that scientist make an effort similar to that made by mathematicians and examine the scientific method in a more formal sense.
To understand that such a project is possible one needs to recognize that the scientific method is a recursive self-correcting process that is essentially a dynamical system and hence can be posed as a problem in computation.

Here an attempt will be made to initiate this project by introducing some preliminary ideas based upon the tools that have been developed in the previous chapters of this book.
For this to make sense one must be receptive to the idea that scientific theories of the future will be expressed in a language of algorithms and programs.
Consequently, the status of a fully discrete computer model that can be derived from a theory is raised from one that is not just a useful research tool but in itself is the language that is used to define the laws of the application.

It should be noted that in the current paradigm the language of science is developed in a separate discipline, namely mathematics.
By adopting a language based on algorithms and programs the validation of a scientific theory automatically includes the validation of the language and formal system upon which it is based.     

\underline{Complexity.}
The scientific method relies heavily on data obtained from real world observations against which simulation results of application specific models are tested.
We will focus mainly on validating theories that include empirical checks of computability.
We do this because we are primarily interested here in employing some empirically based notion of soundness for applications of our formal system.    

In the sense of Chaitin-Kolmogorov complexity, a major objective is to construct the shortest program that represents a computer model.
This needs to be assessed with respect to the scope of applicability of the computer model.
Roughly speaking, the scope of applicability can be defined as the model's ability to generate solutions that simulate real world observations to within the experimental errors and confidence intervals of the widest range of observed data.
If we are to regard a computer model as defining a theory then we are moving towards some quantifiable way of assessing the elegance of a theory.
The objective, then, is to construct theories with minimal program complexity while possessing a maximal scope of applicability.

The notion of elegance of a theory based upon the language of program size has been explored by Chaitin, \cite{chait05}.
Chaitin constructed a classical proof demonstrating that there is no method that allows one to actually determine the size of the most elegant program that defines a given theory.
From this Chaitin concludes that a semi-empirical approach must be taken when constructing theories.
 
Here an attempt will be made to attack the scientific method in some formal sense while retaining an essential component of empiricism.
Real world measurements targeting specific models can be incorporated into the methods to be outlined.
In order that we remain focused on the main thrust of the method we omit details of how this could be done.
Because of this omission we are exploring a more general universe of valid computer models where the specific interpretation of a model and its scope of application is left unspecified.

\section{Empirical Computability Checking.}\index{empirical computability check}

From a strict formalist point of view the semantics of statements in a formal system are less of a concern than that of consistency.
In a computer environment a formal statement is expressed as a program whose internal computational operations are well defined.
In this context the interpretation of formal statements is unambiguous.

We have constructed our formal system, PECR, to be compatible with the constraints imposed by a machine environment, $\mathfrak{M}(mach)$.
In PECR the well formed formulas of classical logic are replaced by programs and the classical notion of attaching a truth value to a formal statement is now replaced by the property of computability.

While our primary objective is to establish the computability of programs by way of inference based upon a collection of construction rules, we can also check the computability of a program by empirical means.
This simply involves executing a program for a given value assigned input and observing whether it halts prematurely with an execution error or returns an output.
We shall often refer to this process as \emph{empirical computability checking}.

Of course, empirically checking for computability can only be useful if the program can be observed to either return an output or an error message in a reasonable time period.
Here we can be guided by a preliminary analysis of the internal algorithm of a program to establish whether it can be executed in polynomial time. 
Otherwise, what may be regarded as a reasonable time cannot be strictly defined and will be an arbitrary constraint imposed by a user.
  
For this reason establishing computability by inference is preferred because of its generality but there are situations where empirical computability checking will have an important role to play.

\underline{Consistency and soundness.}
In classical logic, consistency is defined in terms of formal statements and their negations.
Applications of the formal system PECR in its most primitive form do not make much use of negations so consistency in the conventional sense is not appropriate.
We can, however, approach the conventional notion of soundness as follows.

Suppose that the program $[p~c],~c:\mbf{iext}[p]$, has been supplied as an axiom or derived as a theorem.
Suppose further that by an empirical computability check the program $p$ is found to be computable for a given assigned input.
Soundness will be violated if by empirical means it is found that $[p~c]$ is not computable for the same value assigned input.

Similarly, suppose that the statement $p:\mbf{false}$ has been inferred or simply supplied as an axiom of falsity.
Another kind of violation of soundness may occur if by empirical computability checking it is found that there exists an assigned valued input such that the program $p$ is computable.

We can rewrite the two conditions for violation of soundness of a theory, $S$, as follows.
\begin{itemize}

\item $s=[p~c],~c:\mbf{iext}[p]$, is either an axiom or theorem of $S$ and we have by an empirical computability check that $p$ is computable for some value assignment of the primary input list, $piv[x_p]$, of $p$, and $s=[p~c]$ is not computable for the same value assigned input.

\item $p:\mbf{false}$ is an axiom or theorem of falsity of $S$ and we have by an empirical computability check that $p$ is computable for some value assignment of the primary input list, $piv[x_p]$, of $p$.

\end{itemize}
It should be noted that the second item is encompassed in the first item by the unification of higher order constructs of irreducible program extensions as outlined in Section \ref{shop}.   

By analogy with classical mathematical logic we are employing empirical computability checking to search for a counter example to a proposition asserting soundness.
Hence the above conditions are weak in the sense that they only address violations of soundness in an empirical sense and do not provide a formal procedure from which we can establish that a formal system is sound.
On the other hand, for programs that can be executed in a feasible time, computability is defined in an unambiguous way by empirical tests and can establish the computability of a program with respect to a given assigned input list with absolute certainty and hence requires no interpretation.
This reliance on empirical observations suggests a process closer to the scientific method rather than the higher goals demanded by conventional mathematics.
We are led to seriously consider the following.

\underline{Iterated axiomatic method.}\index{iterated axiomatic method}
The controversy surrounding the foundations of mathematics and formal systems in general are well known and remain a topic of serious debate.
Rather than attack this problem head on we may seek a path around it.
One approach is to accept a less ambitious form of inquiry that is closer to that found through the self correcting recursive process of the scientific method.
Consequently, the axiomatic method is weakened to incorporate some procedures that may be empirical.

First of all one concedes to the notion that, like postulates in science, laying down a collection of axioms to define a specific theory is a tentative process that is subject to modification.
It is through such a concession that an iterative mechanism is required for continual reevaluation and self correction.

One initiates an action of theorem mining by first laying down a collection of axioms for a theory, $S$.
By applying these axioms in conjunction with the construction rules, proofs are derived from which theorems are extracted as irreducible extended programs.
We concede that there may be irreducible extended programs of the theory, $S$, that may be missed by this process, i.e. irreducible extended programs that cannot be derived under the current collection of axioms.

Irreducible extended programs that cannot be derived are candidates for new axioms of the theory.
If by some means outside of the action of theorem mining a new irreducible extended program is found for which no derivation under the existing axioms is known then it can be appended to the collection of axioms.
In this way the theory under investigation can be built up with increasing scope of its theorem mining capabilities.

There is a point of caution here in that by simply appending a new irreducible extended program to the current list of axioms of a theory there is no guarantee that elements of the current list of axioms will remain independent.  
Taking this into account, in our recursive self improving procedure we include the following two actions that run concurrent to the action of theorem mining.

\begin{itemize}

\item Axioms are assumed to be irreducible extended programs until such time that they are found to violate soundness.
Violations of soundness can be detected through empirical computability checking.
When this occurs the offending programs that are stored as axioms are removed from storage along with all theorems whose derivations are dependent on them.
Such theorems can be identified by the procedure of theorem connection list reduction as outlined in Section \ref{thmconn}.  

\item If a derivation is found for an axiom then it is accessed from the file \emph{axiom.dat} and relabeled as a theorem.
This situation may occur when a program was incorrectly identified as an axiom from the start or a new axiom is introduced into the current collection of axioms.

\end{itemize}

\underline{Identifying new axioms.}
The actual task of identifying new axioms lies outside of the formal system in which they are employed.
At this stage such a task is largely a human enterprise but it is worthwhile to speculate that automation may be possible.

It is difficult to envisage a procedure of identifying axioms that can avoid some kind of empirical process.
This may involve a mechanism employing a targeted pattern recognition on permutations of lists of atomic programs subject to various combinations of variable bindings.
Immediate elimination of possible candidates can rely on the structural Conditions 1 and 2 of Definition \ref{ce}.  
Each of the remaining candidates of program lists will be subject to extensive testing with respect to a large range of prescribed value assigned inputs through empirical computability checking in combination with confidence valuation based on statistical analysis.
This empirically based procedure will largely test for violations of Condition 3 of Definition \ref{ce}.
  
Identifying new axioms in this way is another action that could be conducted concurrent to the main action of generating proofs and theorems.
Since our formal system is constrained under the machine environment $\mathfrak{M}(mach)$, we can expect that the empirical procedure just described may identify new axioms that are machine specific.
In a larger realm of investigation the machine specific parameters become variables that enter the self correcting recursive process.

\underline{Premature derivation halting.}
In any theorem mining activity there will always be a lack of certainty that all programs that are of type $\mbf{false}$ have been detected.
As a result we might extract theorems from derivations that have been halted prematurely with conclusions that do not state the falsity of their premise program.
However, such theorems that have been stored in \emph{axiom.dat} are benign in the sense that any proof construction starting from a premise program that is computable will never access such theorems.
For reasons outlined in Section \ref{ce} we do not regard the storage of these benign programs to be in violation of the formal definition of a program extension. 

Once a proof of a new theorem of falsity has been obtained it is stored in the file \emph{axiom.dat}.
A search can then be conducted on all axioms and theorems currently stored in \emph{axiom.dat} whose premise programs contain, as a sublist, the new false program associated with the new theorem of falsity.
When these are identified they are simply removed from storage along with all theorems whose theorem connection lists can be traced back to those programs that were stored as axioms/theorems through the process of theorem connection list reduction.

\newpage 

\vspace{5mm}

\underline{Notes.}

\begin{itemize}
	\item The self correcting procedure of the iterated axiomatic method has some similarity with belief revision theory.
	Belief revision theory began with the seminal paper \cite{agm} and remains the dominant paradigm of the subject to this present day.
	The theory is based on the so called AGM postulates that reflect the minimal change of a rational agent's belief state through the acquisition of new information.
	The three main actions of a change in a belief state are contraction, expansion and revision.
	
	The AGM paradigm draws heavily on conventional theories of logic and set theory and is not readily adapted to our formal system.
	Although the objectives of the AGM paradigm of belief revision appear to be related to the iterated axiomatic method there are properties of our formal system that require some significant departures.
	Keeping with our motivation for feasible computations on a real world computer we will take a more constructive approach to the self correcting process by way of the iterated axiomatic method.
	
\end{itemize}

\section{Primary and dynamic states of a theory.}\index{primary state}

In our formal system we need to distinguish between the construction rules, that can be regarded as the primitive inference rules, and the axioms associated with an application of a specific theory, $S$.
Under the action of theorem mining the construction rules are fixed, with the starting hypothesis that the formal system based upon the construction rules is sound. 
The application specific axioms are supplied by the user and serve as input to the proof assistance software (in our case VPC).
In the iterated axiomatic scheme the list of application specific axioms, that define the theory under investigation, can be modified under the self correcting process through the procedures that will now be outlined.

In the previous chapter we defined a theory as $S(atom,cst,mach)$, where
\be
atom=[atom(i)]_{i=1}^{nat} \quad nat:\mbf{int1},~atom(i):\mbf{atm}<:\mbf{prgm}
\ee
is the list of atomic programs of the theory,
\be
\bal
& cst=[cst(i)]_{i=1}^{ncst} \\
\eal
\ee
is the list of constants of the theory and
\be
mach=[nchar~nstr~nlst~nint~nprem]
\ee
is the list of machine parameters that define the depth in which the theory, S, is to be explored.
Any independent list of generators serves as a list of axioms.
Consequently a theory, $S$, can exist in any one of a finite number of primary states, where each primary state has a unique list of axioms.

An irreducible extended program takes the form $[p~c]$, where $c:\mbf{iext}[p]$ and $length[p] \leq nprem$.
Let $s$ be a list of all irreducible extended programs of the theory, $S$.
A sublist $a \subseteqq s$ is said to be \emph{list of independent generators} if no element of $a$ can be derived from other elements contained in $a$.
In addition, the list of independent generators, $a$, must not contain redundant terms with respect to the list of theorems that it generates, i.e. each element of $a$ is involved in the derivation of at least one theorem.   
Hence, the theory, $S$, can exist in any one of the primary states
\be
S^{(k)}, \quad k=1, \ldots ,ms 
\ee
where for each $S^{(k)}$ there corresponds a unique list of axioms
\be
ax^{(k)}=[ax^{(k)}(i)]_{i=1}^{nax^{(k)}}, \quad nax^{(k)}:\mbf{int0},~ax^{(k)}(i):\mbf{prgm} 
\ee
Here each list $ax^{(k)}$, $k=1,\dots,ms$, is unique but we may have $ax^{(k)} \cap ax^{(l)} \ne [~]$, for some $l \ne k$.

For each primary state, $S^{(k)}$, the list, $s$, of all irreducible extended programs of the theory, $S$, has the partition
\ben\label{pttn01}
s \equiv [ax^{(k)}~th^{(k)}~ud^{(k)}]
\een
where
\be
th^{(k)} = [th^{(k)}(i)]_{i=1}^{nth^{(k)}}, \quad nth^{(k)}:\mbf{int1},~th^{(k)}(i):\mbf{prgm}
\ee
is the list of all irreducible extended programs that can be derived from the axioms of $ax^{(k)}$ and
\be
ud^{(k)} = [ud^{(k)}(i)]_{i=1}^{nud^{(k)}}, \quad nud^{(k)}:\mbf{int0},~ud^{(k)}(i):\mbf{prgm}
\ee
is the list of all irreducible extended programs that cannot be derived from the axioms of $ax^{(k)}$.
The sum
\be
ns=nax^{(k)} + nth^{(k)} + nud^{(k)}
\ee
is a constant with respect to all primary states, $S^{(k)}$, $k=1,\ldots,ms$, and is equal to the fixed length of the list $s$.
The list $s$ and its partitions, $th^{(k)}$ and $ud^{(k)}$, typically have prohibitively large lengths so storing them on a real world computer is not feasible.
For this reason these lists should be regarded in an abstract sense.

It is desirable to construct primary states where the list lengths of $ax^{(k)}$ and $ud^{(k)}$ are minimal.
A primary state, $S^{(k)}$, is said to be \emph{complete} if $ud^{(k)}$ is the empty list.

In any primary state, $S^{(k)}$, every element of the list $[ax^{(k)}~ud^{(k)}]$ of the partition (\ref{pttn01}) is an irreducible extended program that cannot be derived in that primary state.
The elements of $ax^{(k)}$ are distinguished from elements of $ud^{(k)}$ in that $ax^{(k)}$ is a list of independent generators of the theorems list $th^{(k)}$.

Real world constructions of theories rely on empirical computations that check for violations of soundness.
Consequently, under the iterated axiomatic method we must regard the type assignment of an irreducible program extension that is associated with an axiom to be tentative and subject to empirical computability checking.

Whenever the system is in a primary state, $S^{(k)}$, for some $k$, and a proof is found under the axioms contained in the list of axioms, $ax^{(k)}$, the theorem is appended to the time dependent list of theorems
\be
th^{(k,t)} = [th^{(k)}(i)]_{i=1}^{t},~0 \leq t \leq nth^{(k)}
\ee
where the time parameter, $t$, is increased by one unit whenever a new theorem is added to the list.  
In this way each primary state, $S^{(k)}$, of the theory, $S$, can exist in the substate, $S^{(k,t)},~0 \leq t \leq nth^{(k)}$.
We will henceforth refer to each, $S^{(k,t)}$, as a \emph{dynamic state}\index{dynamic state} of the primary state, $S^{(k)}$.

\section{The iterated axiomatic method.}\index{iterated axiomatic method}

We now outline how to generate the sequence of primary states
\ben \label{pstate}
S^{(1)}, \ldots , S^{(ms)}
\een
Because of limitations in memory capacity on a real world computer it will not be feasible to store all of the data associated with every state $S^{(k)}$,~$k=1, \ldots, ms$.
We drop the superscripts ${(k)}$ and ${(k,t)}$ that refer to the primary and dynamic states of the theory and represent the iteration by the generation of a new state associated with the list $[ax~th]$, from a currently stored state, associated with the list $[ax0~th0]$.
Here $ax0$ and $th0$, respectively, are the lists of axioms and theorems, respectively, of the current state and $ax$ and $th$, respectively, are the lists of axioms and theorems, respectively, of the new state.

If at each update of a dynamic state we have $ax=ax0$ the current primary state remains the same.
Whenever there is a modification of the original list of axioms, i.e. $ax \neq ax0$, the system moves to a new primary state. If there is a modification of the current list of axioms based on a violation of soundness or there is an axiom that is relabeled as a theorem we must regard the current state to be a false primary state and exclude it as a member of the sequence (\ref{pstate}).

The iterated axiomatic method can be described by four main actions that are executed concurrently.
These actions are performed by the following programs.

\begin{tabular}{|l|l|l|}
	\hline
	\textbf{Program} & \textbf{Description} \\
	\hline \hline
	$ds~[ax0~th0]~[ax~th]$  & Update the dynamic state of the current \\
	 & primary state. \\
	\hline
	$fps~[ax0~th0]~[ax~th]$  & Falsity of the current primary state based \\
	 & on a violation of soundness. \\
	\hline
	$nps~[ax0~th0]~[ax~th]$  & Update to a new primary state when a new \\
	 & axiom is found. \\
	\hline
	$mps~[ax0~th0]~[ax~th]$  & Modify the current primary state when an \\
	 & axiom is relabeled as a theorem. \\
	\hline
\end{tabular}

There are four task specific programs that are called by the main programs of the above table.

\begin{tabular}{|l|l|}
		\hline
		\textbf{Program} & \textbf{Description} \\
		\hline \hline
		$tm~[ax0~th0]~[th1]$ & Searches for proofs of theorems from the input\\
		                     & lists $ax0$ and $th0$. \\
		                     & Called by $ds~[ax0~th0]~[ax~th]$. \\ 
		                     		                     
		\hline
		$axt~[ax0]~[ax1]$    & Empirically checks all elements of the current \\
		                     & list of axioms, $ax0$, for soundness. \\
                             & Called by $fps~[ax0~th0]~[ax~th]$. \\
		\hline
		$axs~[ax0~th0]~[ax1]$ & Searches for potential axioms. \\
		                      & Called by $nps~[ax0~th0]~[ax~th]$. \\
		\hline
		$md~[ax0~th0]~[th1]$ & Scans the list of theorems, $th0$, for equivalence \\
		& with axioms contained in the list $ax0$. \\
		                     & Called by $mps~[ax0~th0]~[ax~th]$. \\                     
		\hline			
\end{tabular}

More details of these programs are given below.

To each axiom and theorem is allocated a unique label.
In addition each theorem has a theorem connection list that contains all of the axiom/theorem labels used in its proof (see Section \ref{thmconn}).
We will assume that the lists of all axiom and theorem labels along with the list of all theorem connection lists are continually updated and stored in separate data files that can be accessed when needed by any program during its execution.

\underline{Theorem mining.} (Update the dynamic state.)
Theorem mining is the action of finding proofs of theorems.
We can assume that in the environment $\mathfrak{M}(mach)$ there exists a finite number, $ms:\mbf{int1}$, of primary states and that each primary state, $S^{(k)}$, can exist in a finite number, $nth^{(k)}:\mbf{int1}$, of dynamic states.

The following pseudo-code of the program $ds~[ax0~th0]~[ax~th]$ describes the process of incrementally updating the current dynamic state by finding a new theorem, $th1$, through the program $tm~[ax0~th0]~[th1]$.

\be
\begin{array}{l}
\hline
\text{algorithm~} ds~[ax0~th0]~[ax~th] \\
\hline
call~tm~[ax0~th0]~[th1] \\
halt[fps~nps~mps] \\
th:=[th0~th1] \\
ax:=ax0 \\
\hline
\end{array}
\ee
 
Note that $th0$ and $th$ are lists of theorems and $th1$ is a single theorem.

The theorem mining program, $tm~[ax0~th0]~[th1]$, searches for proofs of theorems from the input lists $ax0$ and $th0$.
Once it finds a proof of a new theorem, $th1$, it halts and outputs the theorem.
The execution of the concurrent programs, $fps,~nps,$ and $mps$ are halted by the command $halt[fps~nps~mps]$, represented here as a term.
The new axiom and theorem lists are updated and the execution of all concurrent programs, $ds,~fps,~nps$ and $mps$, are then restarted.

\underline{Testing axioms for soundness.} (Falsity of primary state)
We have no formal method for establishing with certainty that any asserted axiom satisfies the properties of an irreducible extended program.
One reliable property that is computationally feasible in the elimination process comes in the form of Conditions 1 and 2 of Definition \ref{ce}.
The challenge remains in satisfying Condition 3 of Definition \ref{ce}.

An axiom takes the form $[p~c]$, where $c:\mbf{iext}[p]$, with the constraint $length[p] \leq nprem$.
We can continually apply empirical computability checks on random samples of all possible primary input value assignments of $[p~c]$.
Soundness will be violated if we encounter a value assigned input of $p$ such that $c$ is not computable when $p$ is computable.
When this occurs the test is halted and the false axiom is discarded from the list of axioms.
All theorems whose theorem connection list reduction identifies the theorem's dependence on that false axiom are also discarded.
Theorem mining is restarted for a new primary state based on the reduced list of axioms and theorems.
 
In the absence of a violation of soundness through empirical computability checking there will always exist uncertainty, but the confidence that $[p~c]$ is an axiom increases as the sample size of the test increases.

The following pseudo-code of the program $fps$ describes the process of testing for violations of soundness through the axiom testing program $axt~[ax0]~[ax1]$.

\be
\begin{array}{l}
\hline
\text{algorithm~} fps~[ax0~th0]~[ax~th] \\
\hline
call~axt~[ax0]~[ax1] \\
halt[ds~nps~mps] \\
call~dep~[ax1~th0]~[th1] \\
th:=th0 \setminus th1 \\
ax:=ax0 \setminus [ax1] \\
\hline
\end{array}
\ee

The axiom testing program, $axt~[ax0]~[ax1]$, empirically checks all elements of the current list of axioms $ax0$ for soundness. If it encounters an axiom, $ax1 \in ax0$, that violates soundness it halts and outputs the false axiom $ax1$.

The program $dep~[ax1~th0]~[th1]$ finds the list of theorems $th1 \subseteqq th0$ whose proofs are dependent on the axiom $ax1$ by the process of theorem connection list reduction.
Here we have assumed that the theorem connection lists are stored in a data file that is updated every time the current theorem list is modified. 

\underline{Search for new axioms.} (New primary state)
Traditionally, the process of identifying axioms is a human activity that is often motivated by a preconceived idea of what properties one wishes to explore in a given theory.
The process of identifying axioms is poorly understood and proposing efficient algorithms that fully automate the procedure is a goal that reflects efforts in the wider area of artificial intelligence.

A primitive approach would involve searching for potential candidate premise programs, $p$, through all possible programs and then identifying potential conclusion programs, $c:\mbf{iext}[p]$.
Even though we are dealing with a finite system constrained by our machine environment, $\mathfrak{M}(mach)$, a blind search for potential candidate premise programs through all programs, $p:\mbf{prgm}[n]$, $length[p] \leq nprem$, might not be feasible.

It would make sense to start the premise program search by exhausting combinations of atomic programs with the smallest list lengths.
This would include trying out combinations of input variable names with respect to various bindings and constant assignments.
Conditions 1 and 2 of Definition \ref{ce} can easily be applied in this process to quickly eliminate candidates for axioms.  
Once a candidate axiom is identified the methods employed for testing the soundness of axioms by the program $axt$ can then be employed.

If we have sufficient confidence in the candidate axiom we can append it to the current list of axioms.
Theorem mining is restarted for a new primary state based on the new list of axioms.

The following pseudo-code of the program $nps$ describes the process of finding a new axiom through the axiom search program $axs$.
When a new axiom is found it is appended to the current list of axioms and theorem mining is continued for the new primary state.

\be
\begin{array}{l}
\hline
\text{algorithm~} nps~[ax0~th0]~[ax~th] \\
\hline
call~axs~[ax0~th0]~[ax1] \\
halt[ds~fps~mps] \\
ax:=[ax0~ax1] \\
th:=th0 \\
\hline
\end{array}
\ee

The axiom searching program, $axs~[ax0~th0]~[ax1]$, searches for potential axioms as irreducible extended programs from the list of atomic programs that are accessed from a fixed data file.
Once a candidate axiom is found the process of empirically testing that axiom for soundness is conducted by making use of the same algorithm employed by the program $axt$.
If the empirical test of the axiom, $ax1$, satisfies some confidence criteria the program $axs$ halts and outputs the axiom $ax1$.

The new axiom is appended to the current list of axioms $ax0$.
Upon restart the new axiom undergoes continued empirical testing for soundness through the program $axt$ under the concurrent action of the program $fps~[ax0~th0]~[ax~th]$.

\underline{Relabeling axioms as theorems.} (Modify primary state)
If under the action of theorem mining a proof of an axiom is found then it is removed from the list of axioms and relabeled as a theorem.    
Theorem mining is restarted for a new primary state based on the reduced list of axioms.

The following pseudo-code of the program $mps$ describes the process of modifying the current primary state when an axiom is found to have a proof by the program $md$.

\be
\begin{array}{l}
\hline
\text{algorithm~} mps~[ax0~th0]~[ax~th] \\
\hline
call~md~[ax0~th0]~[th1] \\
halt[ds~fps~nps] \\
ax:=ax0 \setminus th1 \\
th:=[th0~th1] \\
\hline
\end{array}
\ee

The program, $md~[ax0~th0]~[th1]$, simply scans the current list of theorems, $th0$, to see if any theorem is equivalent to an axiom contained in the list $ax0$.
When such a theorem, $th1$, is found it is removed from the list of axioms, $ax0$.
The relabeling of an axiom to a theorem means that the current primary state is false and the theorem mining is continued for the modified primary state.

\underline{Summary.}
The iteration of the concurrent executions of $ds$, $fps$, $nps$ and $mps$ is described by the following algorithm.
\ben\label{iteration}
\bal
\begin{array}{l}
	\hline
	\text{algorithm~for~the~iteration~of~concurrent~executions} \\
	\hline
	do \\
	\hspace{5mm} ax0:=ax \\
	\hspace{5mm} th0:=th \\
	\hspace{5mm} call~\left \{
	\begin{array}{l}
		ds~[ax0~th0]~[ax~th] \\
		fps~[ax0~th0]~[ax~th] \\
		nps~[ax0~th0]~[ax~th] \\
		mps~[ax0~th0]~[ax~th] \\
	\end{array}
	\right . \\
	end~do \\
	\hline
\end{array}
\eal
\een
The left brace, $\{$, after the $call$ statement indicates that the enclosed programs are executed concurrently.
When any one of the programs $ds$, $fps$, $nps$ and $mps$ updates the current primary or dynamic state the iteration is repeated with the new axiom and theorem lists.

The iteration is initiated by a read statement
\be
read~[file1]~[ax~th]
\ee
that accesses the initial list of axioms, $ax$, stored in the file, $file1$.
The list of theorems, $th$, is set to the empty list.
We may wish to start with an initial state $S^{(1,0)}$ that has no axioms, i.e. where $ax$ is the empty list.

In an ideal case, the four programs of $ds$, $fps$, $nps$ and $mps$ are fully automated.
We can expect that this is currently achievable for the actions of theorem mining and axiom testing.
The major challenge for automation remains with the action of searching for new axioms.
This task is performed by the program $axs~[ax0~th0]~[ax1]$ that is embedded within the program $nps~[ax0~th0]~[ax~th]$.
In the current state of development it is better to regard the execution of the program $axs~[ax0~th0]~[ax1]$ to be largely managed through the interaction of an external human agent.

Here, the issue of automation is one of finding best search algorithms for each task.
The theorem mining program, $tm~[ax0~th0]~[th1]$, is primarily based on the formal deductive methods of VPC but should also include best search algorithms of proofs that bypass the combinatorial explosion of proof trees.
More demanding is the automation of the program $axs~[ax0~th0]~[ax1]$ that essentially falls into the category of empirically based search algorithmic methods for irreducible extended programs that lay outside of the formal deductive method.

\underline{Notes.}

\begin{itemize}
	
	\item We can assume that upon initiation of the entire process, the fixed lists $atom$ and $cst$ are stored in data files and internally accessed by all programs $ds$, $fps$, $nps$ and $mps$, during their execution.
	Each new theorem comes with a theorem connection list that is appended to a dynamically updated data file that can be accessed by the program $dep$ when activated.
	
	\item All concurrent programs are repeatedly halted and restarted.
	To avoid repetitions in some of the computations, the programs $axt$ and $axs$ create additional files that continually update the history of the various ongoing axiom tests that are unaffected by the current changes.
	These files are accessed upon restart and the tests for the unaffected axioms are continued.
	
	\item While there is a finite number of primary and dynamic states, the complete length of the list of theorems, $nth^{(k)}$, for any primary state, $S^{(k)},~1 \leq k \leq ms$, is typically so large that it is unlikely that any individual primary state will be exhausted by the iteration in any feasible time. We can assume that the whole process can be run indefinitely until such time that the memory of the machine is used up.
	This is indicated in the pseudo-code (\ref{iteration}) where the $do$ loop is left open for indefinite iteration.
	This may be modified by inserting some condition such that if satisfied will exit the $do$ loop.   
	
	\item It is important to note that the programs $axs$ and $axt$, respectively, do not only search and test axioms, respectively.
	They more generally target irreducible extended programs that are not elements of the list of theorems.
	Therefore, it is possible that any current list of axioms, $ax^{(k)}$, of a primary state, $S^{(k)}$, might contain redundant elements, i.e. irreducible extended programs that should belong to the list $ud^{(k)}$.
	
\end{itemize}

\section{Concluding remarks.}

For each primary state, $S^{(k)}$, of a theory, $S$, we have partitioned the irreducible extended programs into the three distinct sublists, (\ref{pttn01}).
Elements of $ud^{(k)}$ are irreducible extended programs that cannot be derived from the axioms of the list $ax^{(k)}$.
We can expect that it is not uncommon that the length of the list $ud^{(k)}$ will be very large.
This stresses that in any scientific study the acquisition of knowledge for a given theory is not necessarily dominated by the formal deductive methods associated with the action of theorem mining but rather by the search for underivable irreducible extended programs.
This would essentially involve the same procedures associated with the search for new axioms as described in the previous sections.

It can then be argued that the action of searching for underivable irreducible extended programs should be regarded as an important component of any scientific research.
This is very much reflected by a recent movement of some mathematicians who are proponents of the idea that mathematics should place less emphasis on the axiomatic method and concentrate more on experimental mathematics (see for example \cite{zeil02}).
While acknowledging the merits of these arguments, the overall approach throughout this book has been to take the less extreme path by arguing that formal methods have an essential role to play alongside empirical methods. 

As has already been discussed, the process of identifying underivable irreducible extended programs is still far from being understood.
In our context we are generally interested in identifying irreducible extended programs outside of the formal deductive methods based on PECR. 
If we are to elevate the importance of this task in the sciences much more effort needs to be directed into understanding this process.
The hope is that the process can ultimately be described by empirically based algorithms, an effort that is very much associated with current research in artificial intelligence.

This book has focused on the mechanical aspects of deduction that are largely associated with the action of theorem mining.
In light of the above comments we could argue that this is perhaps the easiest part.
The development of methods that effectively move towards a goal of fully automating the action of identifying irreducible extended programs outside of formal deductive methods is an effort that would be an essential part of scientific research in the future.  

Ideally, the user has prescribed an initial collection of axioms for a given theory that is as concise as possible.
While soundness is the major objective, the hope that such a collection of axioms is exhaustive must often be abandoned.
Ultimately we are interested in primary states such that the lengths of the lists of axioms and underivable irreducible extended programs are minimized.

We may formulate the dynamical system as the combined actions of theorem mining, axiom soundness checking and new axiom searching.
Alternatively, we may regard the action of theorem mining as the sole defining process of the dynamical system and the concurrent actions of axiom soundness checking and new axiom searching as external sources of perturbations to theorem mining.
We can think of the dynamical system based on the action of theorem mining as a map that generates, for each primary state, $S^{(k)}$, a sequence of substates $S^{(k,t)}$, in the discrete time $t=0,\ldots,nth^{(k)}$.
The actions concurrent to theorem mining will provide a potential source of perturbations to our dynamical system that could knock a trajectory out of its current state to a new primary state.

The main thrust of this final chapter is to encourage the emergence of a new subject area that defines the scientific method as a dynamical system.
Such an area will be useful in uncovering such behavior as stability, sensitivity to initial conditions and other phenomena associated with dynamical systems.
Any knowledge obtained from this study will yield vital feedback on what limitations a user might expect from the initial data that is supplied to define a specific theory and possible procedures that could be applied in their selection that will ensure the most desired results.
We can anticipate that an exploration of the properties of this dynamical system is a field of study that will provide very useful insights that will eventually lay to rest many philosophical debates that currently surround the scientific method.

\printindex

\pagestyle{empty}

\backmatter

\end{document}